\documentclass[]{SciPost}

\usepackage{float}
\usepackage{blkarray}
\newcommand{\matindex}[1]{\mbox{\scriptsize#1}}
\usepackage{enumitem}
\usepackage{makecell}
\usepackage{bigints}
\usepackage{caption}
\numberwithin{equation}{section}
\usepackage[normalem]{ulem}
\usepackage{tensor}
\usepackage{multirow,booktabs} 

\usepackage[mathscr]{eucal}
\usepackage{tikz,tikz-3dplot,tikz-cd,circuitikz}
\usetikzlibrary{patterns,decorations.markings, decorations.pathmorphing, arrows,calc,knots,hobby,
	decorations.pathreplacing,
	shapes.geometric,
	calc, arrows, arrows.meta}
	
\newcommand\blfootnote[1]{%
  \begingroup
  \renewcommand\thefootnote{}\footnote{#1}%
  \addtocounter{footnote}{-1}%
  \endgroup
}
	
\tikzset{
	knot diagram/every strand/.append style={
		ultra thick,
		red
	},
	show curve controls/.style={
		postaction=decorate,
		decoration={show path construction,
			curveto code={
				\draw [blue, dashed]
				(\tikzinputsegmentfirst) -- (\tikzinputsegmentsupporta)
				node [at end, draw, solid, red, inner sep=2pt]{};
				\draw [blue, dashed]
				(\tikzinputsegmentsupportb) -- (\tikzinputsegmentlast)
				node [at start, draw, solid, red, inner sep=2pt]{}
				node [at end, fill, blue, ellipse, inner sep=2pt]{}
				;
			}
		}
	},
	show curve endpoints/.style={
		postaction=decorate,
		decoration={show path construction,
			curveto code={
				\node [fill, blue, ellipse, inner sep=2pt] at (\tikzinputsegmentlast) {}
				;
			}
		}
	}
}

\tikzset{
	cross/.pic = {
		\draw[rotate = 45] (-#1,0) -- (#1,0);
		\draw[rotate = 45] (0,-#1) -- (0, #1);
	}
}

\newcommand{\StringRed}{
\begin{tikzpicture}
[scale=0.5,baseline={([yshift=-.5ex]current bounding box.center)}]
\foreach \x in {5,9} {
    \foreach \y in {1,5} {
        \path[draw, white, fill=gray!20] (\x,\y) rectangle (\x+2,\y+2);
    }
}
\foreach \x in {3,7} {
    \foreach \y in {3,7} {
        \path[draw,white, fill=gray!20] (\x,\y) rectangle (\x+2,\y+2);
    }
}

\foreach \x in {3,5,7,9} {
    \foreach \y in {1,3,5,7} {
        \path[draw,thick] (\x,\y) rectangle (\x+2,\y+2);
    }
}

\foreach \x in {3,5,7,9,11} {
    \path[draw,dashed,thick,->] (\x,9) -- (\x,10);
    \path[draw,dashed,thick,<-] (\x,0) -- (\x,1);
    \path[draw,dashed,thick,->] (11,\x-2) -- (12,\x-2);
    \path[draw,dashed,thick,<-]  (2,\x-2) -- (3,\x-2);
}

\path[draw, thick,->,dashed, red] (10,6) -- (9.5,5.5);
\path[draw, thick,->,dashed, red] (5,7) -- (4.5,7.5);
\path[draw, thick,-,dashed, red,rounded corners=20pt] (10,6) -- (8,4) -- (6,2) -- (4,4) -- (6,6) -- (4,8);

\draw[draw, red!50, ultra thick, fill=red!50,circle] (10,6) circle (0.2cm);
\draw[draw, red!50, ultra thick, fill=red!50,circle] (4,8) circle (0.2cm);
\draw (10,6) node {\scalebox{0.8}{$+$}};
\draw (4,8) node {\scalebox{0.8}{$-$}};

\end{tikzpicture}
}

\newcommand{\StringBlue}{
\begin{tikzpicture}
[scale=0.5,baseline={([yshift=-.5ex]current bounding box.center)}]
\foreach \x in {5,9} {
    \foreach \y in {1,5} {
        \path[draw, white, fill=gray!20] (\x,\y) rectangle (\x+2,\y+2);
    }
}
\foreach \x in {3,7} {
    \foreach \y in {3,7} {
        \path[draw, white, fill=gray!20] (\x,\y) rectangle (\x+2,\y+2);
    }
}

\foreach \x in {3,5,7,9} {
    \foreach \y in {1,3,5,7} {
        \path[draw,thick] (\x,\y) rectangle (\x+2,\y+2);
    }
}

\foreach \x in {3,5,7,9,11} {
    \path[draw,dashed,thick,->] (\x,9) -- (\x,10);
    \path[draw,dashed,thick,<-] (\x,0) -- (\x,1);
    \path[draw,dashed,thick,->] (11,\x-2) -- (12,\x-2);
    \path[draw,dashed,thick,<-]  (2,\x-2) -- (3,\x-2);
}

\path[draw, thick,->,dashed, blue] (8,2) -- (8.5,2.5);
\path[draw, thick,->,dashed, blue] (5,7) -- (5.5,7.5);
\path[draw, thick,-,dashed, blue,rounded corners=20pt] (8,2) -- (10,4) -- (8,6) -- (6,4) -- (4,6) -- (6,8);

\draw[draw, blue!50, ultra thick, fill=blue!50,circle] (8,2) circle (0.2cm);
\draw[draw, blue!50, ultra thick, fill=blue!50,circle] (6,8) circle (0.2cm);
\draw (8,2) node {\scalebox{0.8}{$+$}};
\draw (6,8) node {\scalebox{0.8}{$-$}};

\end{tikzpicture}
}

\newcommand{\StringRedBlue}{
\begin{tikzpicture}
[scale=0.5,baseline={([yshift=-.5ex]current bounding box.center)}]
\foreach \x in {5,9} {
    \foreach \y in {1,5} {
        \path[draw, white, fill=gray!20] (\x,\y) rectangle (\x+2,\y+2);
    }
}
\foreach \x in {3,7} {
    \foreach \y in {3,7} {
        \path[draw, white, fill=gray!20] (\x,\y) rectangle (\x+2,\y+2);
    }
}

\foreach \x in {3,5,7,9} {
    \foreach \y in {1,3,5,7} {
        \path[draw,thick] (\x,\y) rectangle (\x+2,\y+2);
    }
}

\foreach \x in {3,5,7,9,11} {
    \path[draw,dashed,thick,->] (\x,9) -- (\x,10);
    \path[draw,dashed,thick,<-] (\x,0) -- (\x,1);
    \path[draw,dashed,thick,->] (11,\x-2) -- (12,\x-2);
    \path[draw,dashed,thick,<-]  (2,\x-2) -- (3,\x-2);
}

\draw[draw, green!80,  ultra thick] (3,7) -- (5,7);
\draw[draw, green!80,  ultra thick] (7,3) -- (7,1);

\path[draw, thick,->,dashed, red] (6,2) -- (6.5,2.5);
\path[draw, thick,->,dashed, red] (5,7) -- (4.5,7.5);
\path[draw, thick,-,dashed, red,rounded corners=20pt] (6,2) -- (8,4) -- (10,6) -- (8,8) -- (6,6) -- (4,8);

\path[draw, thick,->,dashed, blue] (8,2) -- (8.5,2.5);
\path[draw, thick,->,dashed, blue] (5,7) -- (4.5,6.5);
\path[draw, thick,-,dashed, blue,rounded corners=20pt] (8,2) -- (10,4) -- (6,8) -- (4,6);

\draw[draw, red!50, ultra thick, fill=red!50,circle] (6,2) circle (0.2cm);
\draw[draw, red!50, ultra thick, fill=red!50,circle] (4,8) circle (0.2cm);
\draw (6,2) node {\scalebox{0.8}{$+$}};
\draw (4,8) node {\scalebox{0.8}{$-$}};

\draw[draw, blue!50, ultra thick, fill=blue!50,circle] (8,2) circle (0.2cm);
\draw[draw, blue!50, ultra thick, fill=blue!50,circle] (4,6) circle (0.2cm);
\draw (8,2) node {\scalebox{0.8}{$+$}};
\draw (4,6) node {\scalebox{0.8}{$-$}};

\end{tikzpicture}
}

\newcommand{\Oop}[1]{
\begin{tikzpicture}
[scale=0.28,baseline={([yshift=-.5ex]current bounding box.center)}]
\path[draw,thick,gray!70] (-1,-1) rectangle (1,1);
\draw (0,0) node {$#1$};
\begin{scope}[rounded corners=12pt,thick,rotate=50,decoration={
    markings,
    mark=at position 0.5 with {\arrow{>}}}
    ] 
    \path[draw,postaction={decorate}] (-2,0) -- (0,2) -- (2,0) -- (0,-2) -- cycle;
    \path[draw,postaction={decorate}] (0,-2) -- (-2,0) -- (0,2) -- (2,0) -- cycle;
    \path[draw,postaction={decorate}] (2,0) -- (0,-2) -- (-2,0) -- (0,2) -- cycle;
    \path[draw,postaction={decorate}] (0,2) -- (2,0) -- (0,-2) -- (-2,0) -- cycle;        
\end{scope}

\end{tikzpicture}
}

\newcommand{\OopDagger}[1]{
\begin{tikzpicture}
[scale=0.28,baseline={([yshift=-.5ex]current bounding box.center)}]
\path[draw,thick,gray!70] (-1,-1) rectangle (1,1);
\draw (0,0) node {$#1$};
\begin{scope}[rounded corners=12pt,thick,rotate=50,decoration={
    markings,
    mark=at position 0.5 with {\arrow{>}}}
    ] 
    \path[draw,postaction={decorate}] (0,-2) -- (2,0) -- (0, 2) -- (-2,0) -- cycle;
     \path[draw,postaction={decorate}] (-2,0) -- (0,-2) -- (2,0) -- (0, 2)  -- cycle;
     \path[draw,postaction={decorate}] (0, 2) -- (-2,0) -- (0,-2) -- (2,0)  -- cycle;
      \path[draw,postaction={decorate}] (2,0) -- (0, 2) -- (-2,0) -- (0,-2)  -- cycle;        
\end{scope}

\end{tikzpicture}
}

\newcommand{\OopNumb}[1]{
\begin{tikzpicture}
[scale=0.28,baseline={([yshift=-.5ex]current bounding box.center)}]
\path[draw,thick,gray!70] (-1,-1) rectangle (1,1);
\draw (0,0) node {$#1$};
\draw (-1.3,-1.3) node {\scalebox{0.8}{$4$}};
\draw (-1.3,1.3) node {\scalebox{0.8}{$1$}};
\draw (1.3,-1.3) node {\scalebox{0.8}{$3$}};
\draw (1.3,1.3) node {\scalebox{0.8}{$2$}};

\begin{scope}[rounded corners=12pt,thick,rotate=50,decoration={
    markings,
    mark=at position 0.5 with {\arrow{>}}}
    ] 
    \path[draw,postaction={decorate}] (-2,0) -- (0,2) -- (2,0) -- (0,-2) -- cycle;
    \path[draw,postaction={decorate}] (0,-2) -- (-2,0) -- (0,2) -- (2,0) -- cycle;
    \path[draw,postaction={decorate}] (2,0) -- (0,-2) -- (-2,0) -- (0,2) -- cycle;
    \path[draw,postaction={decorate}] (0,2) -- (2,0) -- (0,-2) -- (-2,0) -- cycle;        
\end{scope}

\end{tikzpicture}
}

\newcommand{\Z}[1]{
\begin{tikzpicture}
[scale=0.28,baseline={([yshift=-.5ex]current bounding box.center)},rounded corners=12pt,thick,decoration={
    markings,
    mark=at position 0.5 with {\arrow{>}}}]
    \path[draw,postaction={decorate}] (0,0) -- (2,2);
    \draw (0,0) node[right]{\scalebox{0.8}{$#1$}};
\end{tikzpicture}
}

\newcommand{\Zd}[1]{
\begin{tikzpicture}
[scale=0.28,baseline={([yshift=-.5ex]current bounding box.center)},rounded corners=12pt,thick,decoration={
    markings,
    mark=at position 0.5 with {\arrow{>}}}]
    \path[draw,postaction={decorate}] (2,2) -- (0,0);
    \draw (0,0) node[right]{\scalebox{0.8}{$#1$}};
\end{tikzpicture}
}

\newcommand{\Xd}[1]{
\begin{tikzpicture}
[scale=0.28,baseline={([yshift=-.5ex]current bounding box.center)},rounded corners=12pt,thick,decoration={
    markings,
    mark=at position 0.5 with {\arrow{>}}}]
    \path[draw,postaction={decorate}] (2,0) -- (0,2);
    \draw (2,0) node[right]{\scalebox{0.8}{$#1$}};
\end{tikzpicture}
}

\newcommand{\X}[1]{
\begin{tikzpicture}
[scale=0.28,baseline={([yshift=-.5ex]current bounding box.center)},rounded corners=12pt,thick,decoration={
    markings,
    mark=at position 0.5 with {\arrow{>}}}]
    \path[draw,postaction={decorate}] (0,2) -- (2,0);
    \draw (2,0) node[right]{\scalebox{0.8}{$#1$}};
\end{tikzpicture}
}

\newcommand{\XZ}{
\begin{tikzpicture}
[scale=0.28,baseline={([yshift=-.5ex]current bounding box.center)},rounded corners=12pt,thick,decoration={
    markings,
    mark=at position 0.8 with {\arrow{>}}}]
    \path[draw,postaction={decorate}] (0,0) -- (2,2);
    
 \path[draw,white,double=black] (0,2) -- (2,0);
 \path[draw,postaction={decorate}] (0,2) -- (2,0);
\end{tikzpicture}
}

\newcommand{\ZX}{
\begin{tikzpicture}
[scale=0.28,baseline={([yshift=-.5ex]current bounding box.center)},rounded corners=12pt,thick,decoration={
    markings,
    mark=at position 0.8 with {\arrow{>}}}]
    \path[draw,postaction={decorate}] (0,2) -- (2,0);
    
 \path[draw,white,double=black] (0,0) -- (2,2);
 \path[draw,postaction={decorate}] (0,0) -- (2,2);
\end{tikzpicture}
}

\newcommand{\XdZ}{
\begin{tikzpicture}
[scale=0.28,baseline={([yshift=-.5ex]current bounding box.center)},rounded corners=12pt,thick,decoration={
    markings,
    mark=at position 0.8 with {\arrow{>}}}]
    \path[draw,postaction={decorate}] (0,0) -- (2,2);
    
 \path[draw,white,double=black] (2,0) -- (0,2);
 \path[draw,postaction={decorate}] (2,0) -- (0,2);
\end{tikzpicture}
}

\newcommand{\ZXd}{
\begin{tikzpicture}
[scale=0.28,baseline={([yshift=-.5ex]current bounding box.center)},rounded corners=12pt,thick,decoration={
    markings,
    mark=at position 0.8 with {\arrow{>}}}]
    \path[draw,postaction={decorate}] (2,0) -- (0,2);
    
 \path[draw,white,double=black] (0,0) -- (2,2);
 \path[draw,postaction={decorate}] (0,0) -- (2,2);
\end{tikzpicture}
}

\newcommand{\XZd}{
\begin{tikzpicture}
[scale=0.28,baseline={([yshift=-.5ex]current bounding box.center)},rounded corners=12pt,thick,decoration={
    markings,
    mark=at position 0.8 with {\arrow{>}}}]
    \path[draw,postaction={decorate}] (2,2) -- (0,0);
    
 \path[draw,white,double=black] (0,2) -- (2,0);
 \path[draw,postaction={decorate}] (0,2) -- (2,0);
\end{tikzpicture}
}

\newcommand{\ZdX}{
\begin{tikzpicture}
[scale=0.28,baseline={([yshift=-.5ex]current bounding box.center)},rounded corners=12pt,thick,decoration={
    markings,
    mark=at position 0.8 with {\arrow{>}}}]
    \path[draw,postaction={decorate}] (0,2) -- (2,0);
    
 \path[draw,white,double=black] (2,2) -- (0,0);
 \path[draw,postaction={decorate}] (2,2) -- (0,0);
\end{tikzpicture}
}

\newcommand{\XdZd}{
\begin{tikzpicture}
[scale=0.28,baseline={([yshift=-.5ex]current bounding box.center)},rounded corners=12pt,thick,decoration={
    markings,
    mark=at position 0.8 with {\arrow{>}}}]
    \path[draw,postaction={decorate}] (2,2) -- (0,0);
    
 \path[draw,white,double=black] (2,0) -- (0,2);
 \path[draw,postaction={decorate}] (2,0) -- (0,2);
\end{tikzpicture}
}

\newcommand{\ZdXd}{
\begin{tikzpicture}
[scale=0.28,baseline={([yshift=-.5ex]current bounding box.center)},rounded corners=12pt,thick,decoration={
    markings,
    mark=at position 0.8 with {\arrow{>}}}]
    \path[draw,postaction={decorate}] (2,0) -- (0,2);
    
 \path[draw,white,double=black] (2,2) -- (0,0);
 \path[draw,postaction={decorate}] (2,2) -- (0,0);
\end{tikzpicture}
}

\newcommand{\OopRed}[1]{
\begin{tikzpicture}
[scale=0.28,baseline={([yshift=-.5ex]current bounding box.center)}]
\path[draw,thick,gray!70] (-1,-1) rectangle (1,1);
\draw (0,0) node {$#1$};
\begin{scope}[rounded corners=12pt,thick,red,rotate=50,decoration={
    markings,
    mark=at position 0.5 with {\arrow{>}}}
    ] 
    \path[draw,postaction={decorate}] (-2,0) -- (0,2) -- (2,0) -- (0,-2) -- cycle;
    \path[draw,postaction={decorate}] (0,-2) -- (-2,0) -- (0,2) -- (2,0) -- cycle;
    \path[draw,postaction={decorate}] (2,0) -- (0,-2) -- (-2,0) -- (0,2) -- cycle;
    \path[draw,postaction={decorate}] (0,2) -- (2,0) -- (0,-2) -- (-2,0) -- cycle;        
\end{scope}

\end{tikzpicture}
}

\newcommand{\OopBlue}[1]{
\begin{tikzpicture}
[scale=0.28,baseline={([yshift=-.5ex]current bounding box.center)}]
\path[draw,thick,gray!70,fill=gray!20] (-1,-1) rectangle (1,1);
\draw (0,0) node {$#1$};
\begin{scope}[rounded corners=12pt,thick,blue,rotate=50,decoration={
    markings,
    mark=at position 0.5 with {\arrow{>}}}
    ] 
    \path[draw,postaction={decorate}] (-2,0) -- (0,2) -- (2,0) -- (0,-2) -- cycle;
    \path[draw,postaction={decorate}] (0,-2) -- (-2,0) -- (0,2) -- (2,0) -- cycle;
    \path[draw,postaction={decorate}] (2,0) -- (0,-2) -- (-2,0) -- (0,2) -- cycle;
    \path[draw,postaction={decorate}] (0,2) -- (2,0) -- (0,-2) -- (-2,0) -- cycle;        
\end{scope}

\end{tikzpicture}
}

\newcommand{\EdgeRed}[1]{
\begin{tikzpicture}
[scale=0.28,baseline={([yshift=-.5ex]current bounding box.center)}]
\path[draw,thick,white] (-1,-1) rectangle (1,1);
\path[draw,thick,gray!70,fill=gray!20] (-3,-1) rectangle (-1,1);
\draw (0,0) node {$#1$};
\draw[very thick,black!80,dashed] (-1,-1.8) -- (-1,1.8);
\begin{scope}[rounded corners=12pt,thick,red,rotate=0,decoration={
    markings,
    mark=at position 0.52 with {\arrow{>}}}
    ] 
    \path[draw,postaction={decorate}] (0,-2) -- (-2,0) -- (0,2);       
\end{scope}

\end{tikzpicture}
}

\newcommand{\EdgeBlue}[1]{
\begin{tikzpicture}
[scale=0.28,baseline={([yshift=-.5ex]current bounding box.center)}]
\path[draw,thick,white,fill=gray!20] (-1,-1) rectangle (1,1);
\path[draw,thick,gray!70] (-3,-1) rectangle (-1,1);
\draw (0,0) node {$#1$};
\draw[very thick,black!80,dashed] (-1,-1.8) -- (-1,1.8);
\begin{scope}[rounded corners=12pt,thick,blue,rotate=0,decoration={
    markings,
    mark=at position 0.52 with {\arrow{>}}}
    ] 
    \path[draw,postaction={decorate}] (0,-2) -- (-2,0) -- (0,2);       
\end{scope}

\end{tikzpicture}
}

\newcommand{\GammaRed}{
\begin{tikzpicture}
[scale=0.22,baseline={([yshift=-.5ex]current bounding box.center)}]
\path[draw,thick,white,fill=gray!20] (-1.5,-3.5) rectangle (-0.5,-2.5);
\path[draw,thick,white,fill=gray!20] (-1.5,-1.5) rectangle (-0.5,-0.5);
\path[draw,thick,white,fill=gray!20] (-0.5,-2.5) rectangle (0.5,-1.5);
\path[draw,thick,white,fill=gray!20] (-0.5,-0.5) rectangle (0.5,0.5);
\begin{scope}[rounded corners=1.5pt,thick,red,rotate=0,decoration={
    markings,
    mark=at position 0.55 with {\arrow{>}}}
    ] 
    \path[draw,postaction={decorate}] (0,-4) -- (-1,-3) -- (0,-2) -- (-1,-1) -- (0,0) -- (-1,1);       
\end{scope}
\end{tikzpicture}
}

\newcommand{\GammaBlue}{
\begin{tikzpicture}
[scale=0.22,baseline={([yshift=-.5ex]current bounding box.center)}]
\path[draw,thick,white,fill=gray!20] (-1.5,-2.5) rectangle (-0.5,-1.5);
\path[draw,thick,white,fill=gray!20] (-1.5,-0.5) rectangle (-0.5,0.5);
\path[draw,thick,white,fill=gray!20] (-0.5,-3.5) rectangle (0.5,-2.5);
\path[draw,thick,white,fill=gray!20] (-0.5,-1.5) rectangle (0.5,-0.5);
\begin{scope}[rounded corners=1.5pt,thick,blue,rotate=0,decoration={
    markings,
    mark=at position 0.55 with {\arrow{>}}}
    ] 
    \path[draw,postaction={decorate}] (0,-4) -- (-1,-3) -- (0,-2) -- (-1,-1) -- (0,0) -- (-1,1);       
\end{scope}
\end{tikzpicture}
}

\newcommand{\LatticeEdge}{
\begin{tikzpicture}
[scale=0.5,baseline={([yshift=-.5ex]current bounding box.center)}]

\foreach \x in {1,5,9} {
    \foreach \y in {1,5,9} {
        \path[draw, white, fill=gray!20] (\x,\y) rectangle (\x+2,\y+2);
    }
}
\foreach \x in {-1,3,7,11} {
    \foreach \y in {-1,3,7} {
        \path[draw, white,fill=gray!20] (\x,\y) rectangle (\x+2,\y+2);
    }
}
\foreach \x in {-1,1,3,5,7,9} {
    \foreach \y in {1,3,5,7,9} {
        \path[draw,thick] (\x,\y) rectangle (\x+2,\y+2);
    }
}

\foreach \x in {-1,1,3,5,7,9,11} {
    \path[draw,dashed,thick] (\x,-1) -- (\x,1);
}
\foreach \x in {1,3,5,7,9,11} {
    \path[draw,dashed,thick,->]  (-1,\x) -- (-2,\x);
}
\draw[ultra thick] (11,1) -- (11,11);

\foreach \x in {0,4,8} {
\draw[dashed, green,thin,fill=green!10,fill opacity=0.5] (11,\x) ellipse (0.8cm and 1cm); 
}

\begin{scope}[rounded corners=20pt,thick,rotate=0,decoration={
    markings,
    mark=at position 0.52 with {\arrow{>}}}
    ] 
    \path[blue,draw,postaction={decorate}] (12,6) -- (10,8) -- (12,10);
    \path[red, draw,postaction={decorate}] (12,0) -- (10,2) -- (12,4);       
\end{scope}
\begin{scope}[rounded corners=20pt,thick,red,rotate=0,decoration={
    markings,
    mark=at position 0.14 with {\arrow{>}}}
    ] 
    \path[draw,postaction={decorate}] (4,4) -- (6,6) -- (8,4) -- (6,2) -- cycle;
    \path[draw,postaction={decorate}] (6,2) -- (4,4) -- (6,6) -- (8,4) -- cycle;
    \path[draw,postaction={decorate}] (8,4) --(6,2) -- (4,4) -- (6,6) -- cycle;
    \path[draw,postaction={decorate}] (6,6) -- (8,4) --(6,2) -- (4,4) --  cycle;       
\end{scope}
\begin{scope}[rounded corners=20pt,thick,blue,rotate=0,decoration={
    markings,
    mark=at position 0.14 with {\arrow{>}}}
    ] 
    \path[draw,postaction={decorate}] (0,6) -- (2,8) -- (4,6) -- (2,4) -- cycle;
    \path[draw,postaction={decorate}] (2,4) -- (0,6) -- (2,8) -- (4,6) -- cycle;
    \path[draw,postaction={decorate}] (4,6) --(2,4) -- (0,6) -- (2,8) -- cycle;
    \path[draw,postaction={decorate}] (2,8) -- (4,6) --(2,4) -- (0,6) --  cycle;   
\end{scope}

\draw (14,10) node {$1$};
\draw (14,8) node {$\tilde 1$};
\draw (14,5) node {\rotatebox{90}{$\dots$}};
\draw (14,2) node {$\frac{L_y}2$};
\draw (14,0) node {$\frac{\tilde L_y}2$};

\end{tikzpicture}
}

\newcommand{\WRed}{
\begin{tikzpicture}
[scale=0.22,baseline={([yshift=-.5ex]current bounding box.center)},rotate=90]
\path[draw,thick,white,fill=gray!20] (-1.5,-3.5) rectangle (-0.5,-2.5);
\path[draw,thick,white,fill=gray!20] (-1.5,-1.5) rectangle (-0.5,-0.5);

\path[draw,thick,white,fill=gray!20] (-0.5,-4.5) rectangle (0.5,-3.5);
\path[draw,thick,white,fill=gray!20] (-0.5,-2.5) rectangle (0.5,-1.5);
\path[draw,thick,white,fill=gray!20] (-0.5,-0.5) rectangle (0.5,0.5);

\draw[thick, dashed] (-2.2,-3.5) -- (1.2,-3.5);

\begin{scope}[rounded corners=1.5pt,thick,red,rotate=0,decoration={
    markings,
    mark=at position 0.55 with {\arrow{>}}}
    ] 
    \path[draw,postaction={decorate}] (0,-4) -- (-1,-3) -- (0,-2) -- (-1,-1) -- (0,0) -- (-1,1);       
\end{scope}
\end{tikzpicture}
}

\newcommand{\WBlue}{
\begin{tikzpicture}
[scale=0.22,baseline={([yshift=-.5ex]current bounding box.center)},rotate=90]
\path[draw,thick,white,fill=gray!20] (-1.5,-4.5) rectangle (-0.5,-3.5);
\path[draw,thick,white,fill=gray!20] (-1.5,-2.5) rectangle (-0.5,-1.5);
\path[draw,thick,white,fill=gray!20] (-1.5,-0.5) rectangle (-0.5,0.5);

\path[draw,thick,white,fill=gray!20] (-0.5,-3.5) rectangle (0.5,-2.5);
\path[draw,thick,white,fill=gray!20] (-0.5,-1.5) rectangle (0.5,-0.5);

\draw[thick, dashed] (-2.2,-3.5) -- (1.2,-3.5);

\begin{scope}[rounded corners=1.5pt,thick,blue,rotate=0,decoration={
    markings,
    mark=at position 0.55 with {\arrow{>}}}
    ] 
    \path[draw,postaction={decorate}] (0,-4) -- (-1,-3) -- (0,-2) -- (-1,-1) -- (0,0) -- (-1,1);       
\end{scope}
\end{tikzpicture}
}

\newcommand{\TwoBlocks}[1]{
\begin{tikzpicture}
[scale=0.5,baseline={([yshift=-.5ex]current bounding box.center)}]
\ifnum#1=1  
        \path[draw,white,fill=gray!20] (0,0) rectangle (0+1,0+1);
        \path[draw,white] (0,1) rectangle (0+1,1+1);
        \draw (.5,.5) node {$\tilde p$};
        \draw (.5,1.5) node {$p$};
        \draw[thick, dotted] (0,-.5) -- (0,2.25);
    \else
        \path[draw,white] (0,0) rectangle (0+1,0+1);
        \path[draw,white,fill=gray!20] (0,1) rectangle (0+1,1+1);
        \draw (.5,.5) node {$p$};
        \draw (.5,1.5) node {$\tilde p$};
        \draw[thick, dotted] (0,-.25) -- (0,2.5);
\fi

\draw[ultra thick] (0,0) -- (0,2);

\end{tikzpicture}
}

\newcommand{\ComEdge}[3]{
\begin{tikzpicture}
[scale=0.22,baseline={([yshift=-.5ex]current bounding box.center)}]
\begin{scope}[rounded corners=12pt,thick,rotate=0,decoration={
    markings,
    mark=at position 0.52 with {\arrow{>}}}
    ]
\ifnum#1=1 
    \path[draw,#2,postaction={decorate}] (0,-2) -- (-2,0) -- (0,2);
    \path[draw,white,double=black] (0,-4) -- (-2,-2) -- (0,0);
    \path[draw,#3,postaction={decorate}] (0,-4) -- (-2,-2) -- (0,0);
\else
    \path[draw,#3,postaction={decorate}] (0,-4) -- (-2,-2) -- (0,0);
    \path[draw,white,double=black] (0,-2) -- (-2,0) -- (0,2);
    \path[draw,#2,postaction={decorate}] (0,-2) -- (-2,0) -- (0,2);
\fi
\end{scope}
\end{tikzpicture}
}

\newcommand{\StateChoice}{
\begin{tikzpicture}
[scale=0.4,baseline={([yshift=-.5ex]current bounding box.center)}]
\path[draw,thick,gray!70] (-1,-1) rectangle (1,1);
\path[draw,thick,white,fill=gray!20] (-3+0.85,-1) rectangle (-1,1);
\path[draw,thick,white,fill=gray!20] (1,-1) rectangle (3-0.85,1);
\path[draw,thick,white,fill=gray!20] (-1,1) rectangle (1,3-0.85);
\path[draw,thick,white,fill=gray!20] (-1,-3+0.85) rectangle (1,-1);
\draw (-1+0.2,1) node {\scalebox{0.8}{$\ket 0_z$}};
\draw (1+0.2,1) node {\scalebox{0.8}{$\ket 0_x$}};
\draw (1+0.2,-1) node {\scalebox{0.8}{$\ket 0_z$}};
\draw (-1+0.2,-1) node {\scalebox{0.8}{$\ket 0_x$}};

\end{tikzpicture}
}

\definecolor{lightgray}{RGB}{203,203,203}
\definecolor{newblue}{RGB}{42, 47, 118}
\definecolor{newred}{RGB}{214, 38, 49}
\definecolor{newgreen}{RGB}{0, 151, 76}

\newcommand{\ThreeDBoundaryDisk}{
\begin{tikzpicture}
[scale=0.48,baseline={([yshift=-.5ex]current bounding box.center)},rounded corners=12pt,thick,decoration={
    markings,
    mark=at position 0.8 with {\arrow{}}}]
    \clip (0,0) circle (10cm);
 \draw[fill=lightgray, lightgray] (0,0) circle (10cm);
 \foreach \x in {-10, -5, 0, 5, 10}{
   \foreach \y in {-10, -5, 0, 5, 10}{
  	\path [draw=none,fill=newblue, fill opacity = 0.5,even odd rule]  (\x,\y) circle (1) (\x,\y) circle (1/2);
    	\path [draw=none,fill=newred, fill opacity = 0.5,even odd rule]  (\x,\y) circle (3/2) (\x,\y) circle (1);
    	\path [draw=none,fill=newgreen, fill opacity = 0.5,even odd rule] (\x,\y) circle (2) (\x,\y) circle (3/2);
	
	\draw[newblue, line width=0.4, dashed] (\x,\y) circle (1-1/4);
	\draw[newred, line width=0.4, dashed] (\x,\y) circle (3/2-1/4);
	\draw[newgreen, line width=0.4, dashed] (\x,\y) circle (2-1/4);
	\draw[black, line width=0.4, dashed] (\x,\y) circle (5/2-1/4);
	}
    }
    
     \foreach \t in {0, 90}{%
     	\foreach \x in {-10, -5, 0, 5, 10}{
	  \foreach \y in {-10, -5, 0, 5, 10}{
		\ifnum \t=0
        			\draw[black, thin, decorate, decoration={snake, segment length=0.65mm, amplitude=0.3mm}, rounded corners=0pt, line width=0.3, shift={(\x,\y)}] (0,0) -- (2,2);
		\fi
    	 	\draw[newblue, thin, decorate, decoration={snake, segment length=0.9mm, amplitude=0.3mm}, rounded corners=0pt, line width=0.3, rotate=\t, shift={(\x,\y)}] (0,-0.75) arc (45:-45:2.50);
    	 	\draw[newred, thin, decorate, decoration={snake, segment length=0.85mm, amplitude=0.3mm}, rounded corners=0pt, line width=0.3, rotate=\t, shift={(\x,\y)}] (0,-1.25) arc (30:-30:2.5);
    	  	\draw[newgreen, thin, decorate, decoration={snake, segment length=0.8mm, amplitude=0.3mm}, rounded corners=0pt, line width=0.3, rotate=\t, shift={(\x,\y)}] (0,-1.75) arc (-25:25:-1.80);
		
		 	\draw[fill=newblue, newblue, shift={(\x,\y)}] (0,-0.75) circle (0.04);
    		 	\draw[fill=newblue, newblue, shift={(\x,\y)}] (0,-4.25) circle (0.04);
			\draw[fill=newblue, newblue, shift={(\x,\y)}] (0.75,0) circle (0.04);
			\draw[fill=newblue, newblue, shift={(\x,\y)}] (4.25,0) circle (0.04);
			
			\draw[fill=newred, newred, shift={(\x,\y)}] (0,-1.25) circle (0.04);
    		 	\draw[fill=newred, newred, shift={(\x,\y)}] (0,-3.75) circle (0.04);
			\draw[fill=newred, newred, shift={(\x,\y)}] (1.25,0) circle (0.04);
			\draw[fill=newred, newred, shift={(\x,\y)}] (3.75,0) circle (0.04);
			
			\draw[fill=newgreen, newgreen, shift={(\x,\y)}] (0,-1.75) circle (0.04);
    		 	\draw[fill=newgreen, newgreen, shift={(\x,\y)}] (0,-3.25) circle (0.04);
			\draw[fill=newgreen, newgreen, shift={(\x,\y)}] (1.75,0) circle (0.04);
			\draw[fill=newgreen, newgreen, shift={(\x,\y)}] (3.25,0) circle (0.04);
			
			\draw[fill=black, black, shift={(\x,\y)}] (0,0) circle (0.04);
			\draw[fill=black, black, shift={(\x,\y)}] (2,2) circle (0.04);
		}
    	} 
    }

\end{tikzpicture}
}

\newcommand{\BasicStringAlgebra}[6]{ %
\begin{tikzpicture}[scale=0.25,baseline={([yshift=-2.0ex] current bounding box.center)}] %

\begin{scope}[rounded corners=0pt,thick,rotate=0,decoration={
    markings,
    mark=at position 0.55 with {\arrow{>}}}
    ] 
    \foreach \x/\y/\length/\label in {#1/0/#2/#5, #3/2.5/#4/#6}{
      \draw[yshift=.2cm] (\x+\length*0.5,\y+1) node {$\label$};
      \draw[black,fill=black] (\x,\y) circle (.3);
      \draw[black,fill=black] (\x+\length,\y) circle (.3);
      \path[draw,postaction={decorate}] (\x,\y) -- (\x+\length,\y);}  
\end{scope}

\end{tikzpicture}
}

\newcommand{\SingleStringLine}[2]{ %
\begin{tikzpicture}[scale=0.25,baseline={([yshift=-1.75ex] current bounding box.center)}] %

\begin{scope}[rounded corners=0pt,thick,rotate=0,decoration={
    markings,
    mark=at position 0.55 with {\arrow{>}}}
    ] 
      \draw[yshift=.2cm] (#1*0.5,1) node {$#2$};
      \draw[black,fill=black] (0,0) circle (.3);
      \draw[black,fill=black] (#1,0) circle (.3);
      \path[draw,postaction={decorate}] (0,0) -- (#1,0);
\end{scope}

\end{tikzpicture}
}

\newcommand{\GeneralOperatorConstuction}{ %
\vspace{-.5cm}
\begin{tikzpicture}[scale=0.28,baseline={([yshift=-2.0ex] current bounding box.center)}] %
\def\length{11}
\def\yspace{3}

\begin{scope}[rounded corners=0pt,thick,rotate=0,decoration={
    markings,
    mark=at position 0.55 with {\arrow{}}}
    ] 
    \foreach \x in {1, 2, 3, 4}{
      \draw[black,fill=black] (\x*\length,0) circle (.3);
      \path[draw,postaction={decorate}] (\x*\length,0) -- (\x*\length+\length,0);
      \draw (\x*\length,1.5) node {$\mathcal O_{\alpha_{\x}}\mathcal O_{\ms g_{\x}}$};
      }
      \draw (5*\length,1.5) node {$\mathcal O_{\alpha_{5}}\mathcal O_{\ms g_{5}}$};
      \draw[black,fill=black] (5*\length,0) circle (.3);
      \path[draw,postaction={decorate}] (5*\length,0) -- (5*\length+\length/3,0);
      \draw (5*\length+\length/2,0) node {$\cdots$};
\end{scope}

\draw[line width=1pt,->] (33,-1) -- (33,-5) node[right,yshift=.6cm] {decomposition};

\begin{scope}[rounded corners=0pt,thick,rotate=0,decoration={
    markings,
    mark=at position 0.55 with {\arrow{>}}}, yshift=-1cm
    ] 
    \foreach \x/\y in {1/-1.5, 2/-3, 3/-4.5}{
      \draw[black,fill=black] (\x*\length,\y*\yspace) circle (.3);
      \draw[black,fill=black] (\x*\length+\length,\y*\yspace) circle (.3);
      \path[draw,postaction={decorate}] (\x*\length,\y*\yspace) -- (\x*\length+\length,\y*\yspace);
      }  
      \draw[yshift=.2cm,xshift=-.2cm] (1*\length,-1.5*\yspace+1) node {$\mathcal O_{\alpha_1}\mathcal O_{\ms g_1}$};
      \draw[yshift=.2cm, xshift=.2cm] (2*\length,-1.5*\yspace+1) node {$\mathcal O^\dagger_{\alpha_1}$};

      \draw[yshift=.2cm] (2*\length,-3*\yspace+1) node {$\mathcal O_{\alpha_1}\mathcal{O}_{\alpha_2}\mathcal O_{\ms g_2}$};
      
      \draw[yshift=.2cm, xshift=-.2cm] (3*\length,-3*\yspace+1) node {$\mathcal O^\dagger_{\alpha_2}\mathcal{O}^\dagger_{\alpha_1}$};

      \draw[yshift=.2cm] (3*\length,-4.5*\yspace+1) node {$\mathcal O_{\alpha_1}\mathcal{O}_{\alpha_2}\mathcal{O}_{\alpha_3}\mathcal O_{\ms g_3}$};
      \draw[yshift=.2cm] (4*\length,-4.5*\yspace+1) node {$\mathcal O^\dagger_{\alpha_3}\mathcal{O}^\dagger_{\alpha_2}\mathcal{O}^\dagger_{\alpha_3}$};

      \draw[yshift=.2cm] (4.75*\length,-5.5*\yspace+1) node {$\cdots\cdots$};
    
\end{scope}

\end{tikzpicture}
}

\newcommand{\LocalWWCrossingAlgebraOne}{
\begin{tikzpicture}
[scale=0.28,baseline={([yshift=-.5ex]current bounding box.center)},rounded corners=12pt,very thick,decoration={
    markings,
    mark=at position 0.8 with {\arrow{>}}}]
    
    \def\radius{2.5}
    \def\sqrtTwo{1.41421}
    \def\radiusSqTwo{\radius / \sqrtTwo}
    
    \filldraw[dashed, color=black!60, fill=gray!8, very thick](0,0) circle (\radius);
    
    \path[color=black, draw,postaction={decorate}] (-\radiusSqTwo,-\radiusSqTwo) -- (\radiusSqTwo,\radiusSqTwo);
    
    \fill[fill=gray!8, very thick](0,0) circle (\radius/5);
    
    \path[color=black, draw,postaction={decorate}] (-\radiusSqTwo,\radiusSqTwo) -- (\radiusSqTwo,-\radiusSqTwo);
     
     \draw[dashed, color=black!60, very thick](0,0) circle (\radius);
     
     \node[] at (-\radiusSqTwo-1,\radiusSqTwo+1) {\scriptsize$(\ms g,\alpha)$};
     
     \node[] at (-\radiusSqTwo-1,-\radiusSqTwo-1) {\scriptsize$(\ms g',\alpha')$};
\end{tikzpicture}
}

\newcommand{\LocalWWCrossingAlgebraTwo}{
\begin{tikzpicture}
[scale=0.28,baseline={([yshift=-.5ex]current bounding box.center)},rounded corners=12pt,very thick,decoration={
    markings,
    mark=at position 0.8 with {\arrow{>}}}]
    
    \def\radius{2.5}
    \def\sqrtTwo{1.41421}
    \def\radiusSqTwo{\radius / \sqrtTwo}
    
    \filldraw[dashed, color=black!60, fill=gray!8, very thick](0,0) circle (\radius);
    
    \path[color=black, draw,postaction={decorate}] (-\radiusSqTwo,\radiusSqTwo) -- (\radiusSqTwo,-\radiusSqTwo);
    
    \fill[fill=gray!8, very thick](0,0) circle (\radius/5);
    
    \path[color=black, draw,postaction={decorate}] (-\radiusSqTwo,-\radiusSqTwo) -- (\radiusSqTwo,\radiusSqTwo);
     
     \draw[dashed, color=black!60, very thick](0,0) circle (\radius);
     
     \node[] at (\radiusSqTwo+1,\radiusSqTwo+1) {\scriptsize$(\ms g,\alpha)$};
     
     \node[] at (\radiusSqTwo+1,-\radiusSqTwo-1) {\scriptsize$(\ms g',\alpha')$};
\end{tikzpicture}
}

\newcommand{\LocalRibbonLoop}{
\begin{tikzpicture}
[scale=0.28,baseline={([yshift=-.5ex]current bounding box.center)},rounded corners=4pt, thick,decoration={
    markings,
    mark=at position 0.9 with {\arrow{>}}}]
    
    \def\radius{2.5}
    \def\sqrtTwo{1.41421}
    \def\radiusSqTwo{\radius / \sqrtTwo}
    
    \filldraw[dashed, color=black!60, fill=gray!8, very thick](0,0) circle (\radius);
    
    \begin{knot}[clip width=10, clip radius=3pt, consider self intersections, end tolerance=3pt,background color=gray!8]
	    \strand[color=black, draw, preaction={decorate},line width=1.2pt] (0,-\radius) -- (0,-0.8)  .. controls +(0.8*\radius,1.8*\radius) and +(0.8*\radius,-1.8*\radius) .. (0, 0.8) -- (0,\radius);
	\end{knot}
     
     \draw[dashed, color=black!60, very thick](0,0) circle (\radius);
\end{tikzpicture}
}

\newcommand{\LocalRibbonUnloop}{
\begin{tikzpicture}
[scale=0.28,baseline={([yshift=-.5ex]current bounding box.center)},rounded corners=4pt,very thick,decoration={
    markings,
    mark=at position 0.6 with {\arrow{>}}}]
    
    \def\radius{2.5}
    \def\sqrtTwo{1.41421}
    \def\radiusSqTwo{\radius / \sqrtTwo}

    \filldraw[dashed, color=black!60, fill=gray!8, very thick](0,0) circle (\radius);
    
    \path[color=black, draw,postaction={decorate}] (0,-\radius) -- (0,\radius);

     \draw[dashed, color=black!60, very thick](0,0) circle (\radius);

\end{tikzpicture}
}

\newcommand{\LocalLinesUnfused}{
\begin{tikzpicture}[scale=0.28,baseline={([yshift=-.5ex]current bounding box.center)},rounded corners=4pt,very thick,decoration={
    markings,
    mark=at position 0.6 with {\arrow{>}}}]
    
    \def\radius{2.5}
    \def\sqrtTwo{1.41421}
    \def\radiusSqTwo{\radius / \sqrtTwo}
    
    \filldraw[dashed, color=black!60, fill=gray!8, very thick](0,0) circle (\radius);
    
    \begin{scope}
        \clip (0,0) circle (\radius);

        \path[color=black, draw,postaction={decorate}] (-0.5*\radius,-\radius) -- (-0.5*\radius,\radius);
        \path[color=black, draw,postaction={decorate}] (0.5*\radius,-\radius) -- (0.5*\radius,\radius);
    \end{scope}

     \draw[dashed, color=black!60, very thick](0,0) circle (\radius);
     
     \node[] at (0,0) {$\times$};
     \node[] at (-0.8*\radius,\radiusSqTwo+1.4) {\scriptsize$(\ms g_1,\alpha_1)$};
     \node[] at (0.8*\radius,\radiusSqTwo+1.4) {\scriptsize$(\ms g_2,\alpha_2)$};
    
\end{tikzpicture}
}

\newcommand{\LocalLinesfused}{
\begin{tikzpicture}
[scale=0.28,baseline={([yshift=-.5ex]current bounding box.center)},rounded corners=4pt,very thick,decoration={
    markings,
    mark=at position 0.6 with {\arrow{>}}}]
    
    \def\radius{2.5}
    \def\sqrtTwo{1.41421}
    \def\radiusSqTwo{\radius / \sqrtTwo}
    
    \filldraw[dashed, color=black!60, fill=gray!8, very thick](0,0) circle (\radius);
    
    \begin{scope}
        \clip (0,0) circle (\radius);

        \path[color=black, draw,postaction={decorate}] (0,-\radius) -- (0,\radius);
    \end{scope}

     \draw[dashed, color=black!60, very thick](0,0) circle (\radius);
     
     \node[] at (0,\radiusSqTwo+1.4) {\scriptsize$(\ms g_1+\ms g_2,\alpha_1+\alpha_2)$};
    
\end{tikzpicture}
}

\newcommand{\GammaEdgeCommutatorOne}{
\begin{tikzpicture}
[scale=0.26,baseline={([yshift=-.5ex]current bounding box.center)}, rounded corners=1.5pt,thick,black,rotate=0,decoration={
    markings,
    mark=at position 0.55 with {\arrow{>}}}]

\draw[very thick,black!80,dashed] (-0.5,-2.8) -- (-0.5,2.8);
\begin{knot}[clip width=10, clip radius=3pt, end tolerance=3pt]
    \strand[draw,preaction={decorate},black,line width=1.2pt] (-2,-3.5) -- (-2,3.5);
    \strand[draw,preaction={decorate},black,line width=1.2pt] (0,-2) .. controls +(-6,-1.5) and +(-6,1.5) .. (0,2);
\end{knot}
\end{tikzpicture}
}

\newcommand{\GammaEdgeCommutatorTwo}{
\begin{tikzpicture}
[scale=0.26,baseline={([yshift=-.5ex]current bounding box.center)}, rounded corners=1.5pt,thick,black,rotate=0,decoration={
    markings,
    mark=at position 0.55 with {\arrow{>}}}]

\draw[very thick,black!80,dashed] (-0.5,-2.8) -- (-0.5,2.8);
\begin{knot}[clip width=10, clip radius=3pt, end tolerance=3pt]
    \strand[draw,preaction={decorate},black,line width=1.2pt] (0,-2) .. controls +(-6,-1.5) and +(-6,1.5) .. (0,2);
    \strand[draw,preaction={decorate},black,line width=1.2pt] (-2,-3.5) -- (-2,3.5);
\end{knot}
\end{tikzpicture}
}

\newcommand{\SymmetryIsTopological}{
	\begin{tikzpicture}[scale=1.1,baseline=(current bounding box.base)]
		\draw[line width=1.5pt,blue!70,, opacity = 1.0,rounded corners] (-2,0) .. controls (-1.5,1.5) and (-1,.75) .. (0,.75) .. controls (1.5,1.6) .. (2,0);
		\draw[line width=1.5pt,blue!70,, opacity = 1.0,rounded corners] (-2,0) -- (2,0);
		
		\fill[pattern=north west lines, pattern color=blue,opacity=0.5] (-2,0) .. controls (-1.5,1.5) and (-1,.75) .. (0,.75) .. controls (1.5,1.6) .. (2,0) -- (-2,0);

		\path (-1.2,1.5) pic[red,line width=2pt] {cross=3pt}node[left]{$\mcal{O}_1$};
		\path (.2,1.5) pic[red,line width=2pt] {cross=3pt}node[left]{$\mcal{O}_2$};
		\path (1.3,-.5) pic[red,line width=2pt] {cross=3pt}node[left]{$\mcal{O}_3$};
		
		\draw[line width=2pt] (-2,-1) -- (-2,2) -- (2,2) -- (2,-1) -- (-2,-1);
		
		\node[] at (0.4,0.4) {$Y_{d+1}$};
		\node[] at (-1.5,-0.3) {$X_d$};
		\node[] at (1.5, 1.5) {$X'_d$};
		
		\draw[->,line width=1pt] (-2.3,-1.5) -- (-2.3,-0.3)node[xshift=-.55cm,yshift=-.7cm]{time};
		\draw[->,line width=1pt] (-2.6,-1.3) -- (-1.1,-1.3)node[xshift=-.8cm,yshift=-.3cm]{space};
	\end{tikzpicture}
}

\newcommand{\ActionOfZeroFormSymmetry}{
	\begin{tikzpicture}[scale=.8]
	    \path (0,0) pic[red,line width=2pt] {cross=3pt}node[xshift=0cm,right]{$\mcal{O}_{\alpha}$};
		
		\node at (0,2.5) {$\mcal{U}_{\ms g}[X_d]$};
		\shade[ball color = ProcessBlue!70, opacity = 0.4] (0,0) circle (2cm);
		\draw (0,0) circle (2cm);
		\draw (-2,0) arc (180:360:2 and 0.6);
		\draw[dashed] (2,0) arc (0:180:2 and 0.6);
		
		\node at (3,0) {$=\ms R_\alpha(\ms g)$};
		\path (4.1,0) pic[red,line width=2pt] {cross=3pt}node[xshift=0cm,right]{$\mcal{O}_{\alpha}$};
	\end{tikzpicture}
}

\newcommand{\SymmetryOperatorTimeSlice}{
\begin{tikzpicture}[scale=.8]
		\path[shade,fill=ProcessBlue!70, fill opacity=0.4,rounded corners=4pt] (-2,-2) -- (0,0) -- (4,0) -- (2,-2) -- (-2,-2) -- (0,0);
		\path (1,-1) pic[red,line width=2pt] {cross=3pt}node[xshift=.1cm,right]{$\mcal{O}_{\alpha}$};
	\end{tikzpicture}
}

\newcommand{\ActionOfZeroFormSymmetryOnOneFormSymmetry}{
\begin{tikzpicture}[scale=.8]
\begin{scope}[rounded corners=1.5pt,very thick,black,rotate=0,decoration={
    markings,
    mark=at position 0.55 with {\arrow{>}}}
    ]
        \draw[color=red!80, very thick,postaction={decorate}] (-3,2.8) -- (0.7,2.5);
		\path[thick, black,shade,fill=ProcessBlue!70, fill opacity=0.6,rounded corners=4pt] (0,0) -- (0,4) -- (1.7,5.5) -- (1.7,1.5) -- (0,0) -- (0,4);
		\draw[color=blue!80, very thick,postaction={decorate}] (0.7,2.5) -- (4.1,2.2);
		
		\node at (-1.05,3.3) {$\ms g$};
     \node at (1.0,3.8) {$\dual\sigma$};
	\node at (2.55,3) {$\sigma\cdot \ms g$};
\end{scope}
	
	\end{tikzpicture}
}

\newcommand{\LineGoingThroughDefectTimeSlice}{
\begin{tikzpicture}[scale=0.26,baseline={([yshift=-.5ex]current bounding box.center)},rounded corners=5pt,very thick,black,rotate=0,decoration={
    markings,
    mark=at position 0.55 with {\arrow{>}}}]

    \draw (0, 3.0) node {$\dual\sigma$};
    \draw (-4,2.2) node {\scriptsize$\mscr{W}_{(\ms g,\alpha)}$};
    \draw (4,2.2) node {\scriptsize$\mscr{W}_{\sigma\cdot(\ms g,\alpha)}$};
    
    \path[draw,color=red,postaction={decorate}] (-8,0) -- (0,0);
    \path[draw,color=green!40!blue,postaction={decorate}] (0,0) -- (8,0);   
    \path[draw,color=Aquamarine,line width = 3, dash pattern=on 1 off 0.5] (0,2) -- (0,-2);   

\end{tikzpicture}
}

\newcommand{\TwistDefectImage}{
\begin{tikzpicture}[scale=0.26,baseline={([yshift=-.5ex]current bounding box.center)},rounded corners=5pt,very thick,black,rotate=0,decoration={
    markings,
    mark=at position 0.55 with {\arrow{>}}}]
    \path[draw,color=Aquamarine,line width = 3, dash pattern=on 1 off 0.5] (-6,0) -- (6,0);
    \draw (0, 1.5) node {$\dual\sigma$};
    \fill[fill=Orange] (-6,-0) circle (10pt);
    \fill[fill=Orange] (6,-0) circle (10pt);

\end{tikzpicture}
}

\newcommand{\ActionOfBulkDefectOnBoundaryOperators}{
\begin{tikzpicture}
[scale=0.26,baseline={([yshift=-.5ex]current bounding box.center)},rounded corners=5pt,very thick,black,rotate=0]

\begin{scope}
    \draw[very thick,black!80,dashed] (-0.5,-2.8) -- (-0.5,2.8);
    \path[draw,postaction={decorate}, color=green!40!blue] (0,-2) .. controls +(-5,-1.5) and +(-5,1.5) .. (0,2);
\path[draw,color=Aquamarine,line width = 3, dash pattern=on 1 off 0.5] (-5,4) -- (-5,-4);   
\end{scope}

\begin{scope}[decoration={markings,mark=at position 1 with {\arrow{>}}}]
    \path[draw,postaction={decorate}] (2,0) -- (4,0);
    \path[draw,postaction={decorate}] (12,0) -- (14,0);
\end{scope}

\begin{scope}[shift={(10,0)}] 
    \draw[very thick,black!80,dashed] (-0.5,-2.8) -- (-0.5,2.8);

    \begin{scope}
        \clip (-2,4) -- (-2,-4) -- (-6,-4) -- (-6, 4) -- (-2, 4);
        \path[draw,postaction={decorate},color=red] (0,-2) .. controls +(-5,-1.5) and +(-5,1.5) .. (0,2);
    \end{scope}
    \begin{scope}
        \clip (-2,4) -- (-2,-4) -- (2,-4) -- (2, 4) -- (-2, 4);
        \path[draw,postaction={decorate},color=green!40!blue] (0,-2) .. controls +(-5,-1.5) and +(-5,1.5) .. (0,2);
    \end{scope}
    
\path[draw,color=Aquamarine,line width = 3, dash pattern=on 1 off 0.5] (-2,4) -- (-2,-4);   
\end{scope}

\begin{scope}[shift={(20,0)}] 
    \path[draw,color=Aquamarine,line width = 3, dash pattern=on 1 off 0.5] (-0.5,4) -- (-0.5,-4);   
    \draw[very thick,black!80,dashed] (-0.5,-2.8) -- (-0.5,2.8);

     \path[draw,postaction={decorate},color=red] (0,-2) .. controls +(-5,-1.5) and +(-5,1.5) .. (0,2);
\end{scope}

\end{tikzpicture}
}

\newcommand{\AnyonsInTheBulk}{
\begin{tikzpicture}[scale=.53]

    \draw[dashed,line width=1pt,rounded corners=0.5pt] (-11,-2) -- (-8,1) -- (-8,-3) -- (-11,-6) -- (-11,-2);
    \draw[dashed,line width=1pt,rounded corners=0.5pt] (-8,-3) -- (1,-3);
	    
    \begin{scope}[shift={(0,-0.8)}]
        \path[line width=2pt, rounded corners=2pt,fill=Dandelion!20] (1,-1) -- (-2,-4) -- (-11,-4) -- (-8,-1) -- (1,-1);
    	 
        \begin{scope}[rounded corners=0.1pt,thick,rotate=0,decoration={ markings, mark=at position 0.5 with {\arrow{>}}},line width=1.5pt] 
            \draw[red,xshift=1cm] (-9,-3.5) .. controls (-10,-3) and (-6,-2) .. (-8,-1.5);
        	\fill[fill=red,xshift=1cm] (-9,-3.5) circle (3pt);
        	\fill[fill=red,xshift=1cm] (-8,-1.5) circle (3pt);
        	\draw[red,xshift=1cm,dotted, line width=0.9] (-9,-3.5) -- (-9,-2);
        	\draw[red,xshift=1cm,dotted, line width=0.9] (-8,-1.5) -- (-8,0);
    		\draw[Blue,xshift=1cm] (-6,-3) .. controls (-5,-4.5) and (-4.5,-0.5) .. (-3.5,-2.);
        	\fill[fill=Blue,xshift=1cm] (-6,-3) circle (3pt);
    		\fill[fill=Blue,xshift=1cm] (-3.5,-2.) circle (3pt);
    		\draw[Blue,xshift=1cm,dotted, line width=0.9] (-6,-3) -- (-6,-1.5);
        	\draw[Blue,xshift=1cm,dotted, line width=0.9] (-3.5,-2) -- (-3.5,-0.5);
		\end{scope}

    	\node[red] at (-8.0,-1.9) {\scriptsize $(\ms g,\alpha)$};
    	\node[Blue] at (-2.9,-2.8) {\scriptsize $(\ms g',\alpha')$};
	\end{scope}
		
	\begin{scope}[all/.style={fill=blue!10,fill opacity=0.2,line width=1pt,rounded corners=0.5pt},all,every node/.append style={all}]
        \filldraw (-2,-2) -- (1,1) -- (1,-3) -- (-2,-6) -- (-2,-2);
        \filldraw (-2,-2)-- (-11,-2) -- (-11,-6) -- (-2,-6) -- (-2,-2);
    	\filldraw (-11,-2) -- (-8,1) -- (1,1) -- (-2,-2);
	\end{scope}
		
	\begin{scope}[line width=0.8pt,yshift=0.2cm]
    	\draw[->] (1.4,1.5) to [out=-30,in=0] (1.1,-1.8);
        \node[align=center] at (0.0,1.5) {\scriptsize{Time-slice}};
        		
        \draw[->] (-6.0,-0.8) to[out=-60,in=10] (-6.3,-2.7);
        \node[align=center] at (-6.5, 0.) {\scriptsize{Bulk line}};
        \node[align=center] at (-6.5,-0.5) {\scriptsize{operators}};
        
        \draw[->] (-2.5,-0.8) to[out=1000,in=100] (-5.05,-3.7);
        \node[align=center] at (-2.5, 0.) {\scriptsize{Anyonic}};
        \node[align=center] at (-2.5,-0.5) {\scriptsize{excitations}};
	\end{scope}	
	\end{tikzpicture}
}

\newcommand{\ActionOfSymmetryOperatorsOnBulkLinesModified}{
\begin{tikzpicture}[scale=.53]

    \draw[dashed,line width=1pt,rounded corners=0.5pt] (-11,-2) -- (-8,1) -- (-8,-3) -- (-11,-6) -- (-11,-2);
    \draw[dashed,line width=1pt,rounded corners=0.5pt] (-8,-3) -- (1,-3);
	    
    \begin{scope}[shift={(0,-0.8)}]
        \path[line width=2pt, rounded corners=2pt,fill=Dandelion!20] (1,-1) -- (-2,-4) -- (-11,-4) -- (-8,-1) -- (1,-1);
    	 
    	 \begin{scope}[rounded corners=0.1pt,thick,rotate=0,decoration={ markings, mark=at position 0.5 with {\arrow{>}}},line width=1.5pt,xshift=3cm] 
            \draw[red] (-9,-3.5) .. controls (-10,-3) and (-6,-2) .. (-8,-1.5);
        	\fill[fill=red] (-9,-3.5) circle (3pt);
        	\fill[fill=red] (-8,-1.5) circle (3pt);
        	\draw[red, line width=1.5] (-9,-3.5) -- (-9,-2.5);
        	\draw[red, line width=1.5] (-8,-1.5) -- (-8,-0.5);
		\end{scope}
    	 
    	\path[line width=2pt, rounded corners=4pt,pattern=north west lines, pattern color=Aquamarine,opacity=0.8,scale=1.0, shift={(0,1.2)}] (1,-1) -- (-2,-4) -- (-11,-4) -- (-8,-1) -- (1,-1);
    	 
        \begin{scope}[rounded corners=0.1pt,thick,rotate=0,decoration={ markings, mark=at position 0.5 with {\arrow{>}}},line width=1.5pt,xshift=3cm] 
        	\draw[Blue, line width=1.5] (-9,-2.5) -- (-9,-1.5);
        	\draw[Blue, line width=1.5] (-8,-0.5) -- (-8,0.5);
        	\draw[Blue, dashed, line width=1.2, yshift=2cm] (-9,-3.5) .. controls (-10,-3) and (-6,-2) .. (-8,-1.5);
		\end{scope}

    	\node[red] at (-7,-3.1) {\tiny $(\ms g,\alpha)$};
    	\node[Blue] at (-7.0,-2.0) {\tiny $\sigma\cdot (\ms g,\alpha)$};
    	
    	\node at (-9.2,-2.3) {\scriptsize $\dual\sigma$};
	\end{scope}
		
	\begin{scope}[all/.style={fill=blue!10,fill opacity=0.2,line width=1pt,rounded corners=0.5pt},all,every node/.append style={all}]
        \filldraw (-2,-2) -- (1,1) -- (1,-3) -- (-2,-6) -- (-2,-2);
        \filldraw (-2,-2)-- (-11,-2) -- (-11,-6) -- (-2,-6) -- (-2,-2);
    	\filldraw (-11,-2) -- (-8,1) -- (1,1) -- (-2,-2);
	\end{scope}
		
	\begin{scope}[line width=0.8pt,yshift=0.2cm]
        \draw[->] (-6.7,-0.8) .. controls (-7.0,-1) and (-7.0,-2) .. (-7.0,-2.7);
        \node[align=center] at (-6.5, 0.) {\scriptsize{Permutation of}};
        \node[align=center] at (-6.5,-0.5) {\scriptsize{charges}};
        
        \draw[->] (-2.5,-0.8) to[out=1000,in=100] (-2.65,-2.8);
        \node[align=center] at (-2.5, 0.) {\scriptsize{Anyonic}};
        \node[align=center] at (-2.5,-0.5) {\scriptsize{symmetries}};
	\end{scope}
	
    \draw [decorate, decoration = {brace},  thick] (1.2,-0.5) --  (1.2,-2);
    \node at (1.8,-1.25) {\footnotesize $\epsilon$};
        
	\end{tikzpicture}
}

\newcommand{\IntersectionOfSymmetryOperatorInTheBulk}{
\begin{tikzpicture}[scale=.53]

    \draw[dashed,line width=1pt,rounded corners=0.5pt] (-11,-2) -- (-8,1) -- (-8,-3) -- (-11,-6) -- (-11,-2);
    \draw[dashed,line width=1pt,rounded corners=0.5pt] (-8,-3) -- (1,-3);
	    
    \begin{scope}[shift={(0,-0.8)}]
        \path[line width=2pt, rounded corners=2pt,fill=Dandelion!20] (1,-1) -- (-2,-4) -- (-11,-4) -- (-8,-1) -- (1,-1);
    	
    	\draw[red,rounded corners=0.1pt,thick,rotate=0,decoration={ markings, mark=at position 0.5 with {\arrow{>}}},line width=1.5pt,shift={(-1,0)}] (-7.0,-2.5) --  (-4.5,-2.5);
    	 
    	 \path[line width=2pt, rounded corners=4pt,pattern=north west lines, pattern color=Aquamarine,opacity=0.8,shift={(-1,0.8)}] (-6,-2) -- (-3,1) -- (-3,-3) -- (-6,-6) -- (-6,-2);
    	\begin{scope}[shift={(-1,0)}]
    	    \clip (1,-1) -- (-2,-4)  -- (-6,-4) -- (-3,-1) -- (1,-1);
	        \path[line width=2pt, rounded corners=2pt,fill=Dandelion!20] (1,-1) -- (-2,-4) -- (-11,-4) -- (-8,-1) -- (1,-1);
    	\end{scope}
    	\draw[dashed,color=Aquamarine,line width=1.6pt] (-4,-1) -- (-7,-4);
    	 
        \begin{scope}[rounded corners=0.1pt,thick,rotate=0,decoration={ markings, mark=at position 0.5 with {\arrow{>}}},line width=1.5pt,shift={(-1,0)}] 
        	\fill[fill=red] (-7.0,-2.5) circle (3pt);
    		\draw[Blue]  (-4.5,-2.5) -- (-2.5,-2.5);
        	\fill[fill=Blue] (-2.5,-2.5) circle (3pt);
        	\fill[fill=Aquamarine]  (-4.5,-2.5)  circle (3pt);
        	
        	\node[red] at (-7,-2.1) {\tiny $(\ms g,\alpha)$};
    	    \node[Blue] at (-2.6,-2.1) {\tiny $\sigma\cdot (\ms g,\alpha)$};
		\end{scope}
    	
    	\node at (-4.6,0.3) {\scriptsize $\dual\sigma$};
	\end{scope}
		
	\begin{scope}[all/.style={fill=blue!10,fill opacity=0.2,line width=1pt,rounded corners=0.5pt},all,every node/.append style={all}]
        \filldraw (-2,-2) -- (1,1) -- (1,-3) -- (-2,-6) -- (-2,-2);
        \filldraw (-2,-2)-- (-11,-2) -- (-11,-6) -- (-2,-6) -- (-2,-2);
    	\filldraw (-11,-2) -- (-8,1) -- (1,1) -- (-2,-2);
	\end{scope}
	
	\path[draw, red,rounded corners=7.0pt,rotate=0,decoration={ markings, mark=at position 0.5 with {\arrow{>}}},line width=1.0pt,shift={(-1,-3)}] (0,0) -- (-0.4,1.2) -- (1,2) -- (1.5, 1) -- (0.6,-0.2) -- (0,0);

	\begin{scope}[line width=0.8pt]
        		
        \draw[->] (-6.5,1.2) to[out=1000,in=100] (-5.05,-2.6);
        \node[align=center] at (-6.5, 2.) {\scriptsize{Transparent}};
        \node[align=center] at (-6.5, 1.5) {\scriptsize{domain wall}};
        
        \draw[->] (-1.,1.2) to[out=1000,in=100] (-3.6,-2.6);
        \node[align=center] at (-1., 2.) {\scriptsize{Permutation of}};
        \node[align=center] at (-1., 1.5) {\scriptsize{anyonic charges}};
	\end{scope}	
	\end{tikzpicture}
}

\newcommand{\TwistDefectNoninvertibleDefectLine}{
\begin{tikzpicture}[scale=.53]

    \draw[dashed,line width=1pt,rounded corners=0.5pt] (-11,-2) -- (-8,1) -- (-8,-3) -- (-11,-6) -- (-11,-2);
    \draw[dashed,line width=1pt,rounded corners=0.5pt] (-8,-3) -- (1,-3);
	    
    \begin{scope}[shift={(0,-0.8)}]
        \path[line width=2pt, rounded corners=2pt,fill=Dandelion!20,opacity=0.85] (1,-1) -- (-2,-4) -- (-11,-4) -- (-8,-1) -- (1,-1);

    	\begin{scope}[shift={(-7.,0)}]
    	    \draw[color=blue,line width=1.6pt,shift={(0.,-1.3)}] (-0.515075,-2.51508) -- (-0.515075,0.3);
    	 
        	\foreach \y in {14,...,19}{
        	    \draw[red,line width=0.8pt] (-0.515075-0.1,0.3+0.1 - 4/20*\y ) -- (-0.515075+0.1,0.3-0.1 - 4/20*\y );
        	}
    	\end{scope}
    	 
    	 \path[line width=2pt, rounded corners=4pt,pattern=north west lines, pattern color=Aquamarine,opacity=0.8,shift={(0.707107*2.1,0.707107*2.1+0.8)}] (-2,-2)-- (-9,-2) -- (-9,-6) -- (-2,-6) -- (-2,-2);
    	\begin{scope}
    	    \clip (-0.515075,-2.51508) -- (-9.51508,-2.51508) -- (-11,-4) -- (-2,-4) -- (-0.515075,-2.51508);
	        \path[line width=2pt, rounded corners=2pt,fill=Dandelion!20,opacity=0.85] (1,-1) -- (-2,-4) -- (-11,-4) -- (-8,-1) -- (1,-1);
    	\end{scope}
    	\draw[dashed,color=Aquamarine,line width=1.6pt,shift={(0.707107*2.1,0.707107*2.1)}] (-2,-4) -- (-9,-4);
    	
    	\draw[color=blue,line width=1.6pt] (-0.515075,-3.7) -- (-0.515075,0.3);
    	 
    	\foreach \y in {1,...,19}{
    	    \draw[red,line width=0.8pt] (-0.515075-0.1,0.3+0.1 - 4/20*\y ) -- (-0.515075+0.1,0.3-0.1 - 4/20*\y );
    	}
    	\fill[fill=Orange] (-0.515075,-2.51508) circle (5pt);
    	
    	\begin{scope}[shift={(-7.,0)}]
    	    \draw[color=blue,line width=1.6pt] (-0.515075,-2.51508) -- (-0.515075,0.3);
    	 
        	\foreach \y in {1,...,14}{
        	    \draw[red,line width=0.8pt] (-0.515075-0.1,0.3+0.1 - 4/20*\y ) -- (-0.515075+0.1,0.3-0.1 - 4/20*\y );
        	}
        	\fill[fill=Orange] (-0.515075,-2.51508) circle (5pt);
    	\end{scope}
    	 
    	\node at (-4.2,-0.5) {\scriptsize $\dual\sigma$};
	\end{scope}
		
	\begin{scope}[all/.style={fill=blue!10,fill opacity=0.2,line width=1pt,rounded corners=0.5pt},all,every node/.append style={all}]
        \filldraw (-2,-2) -- (1,1) -- (1,-3) -- (-2,-6) -- (-2,-2);
        \filldraw (-2,-2)-- (-11,-2) -- (-11,-6) -- (-2,-6) -- (-2,-2);
    	\filldraw (-11,-2) -- (-8,1) -- (1,1) -- (-2,-2);
	\end{scope}
		
	\begin{scope}[line width=0.8pt]
        		
        \draw[->] (-8.4,1.4) to[out=-140,in=-180] (-7.815075,-3.3);
        \node[align=center] at (-6.5, 2.) {\scriptsize{Extrinsic}};
        \node[align=center] at (-6.5, 1.5) {\scriptsize{twist defects}};
        
        \draw[->] (1.,1.85) to[out=-10,in=0] (-0.25,-2.4);
        \node[align=center] at (-1.5, 2.) {\scriptsize{Non-invertible}};
        \node[align=center] at (-1.5, 1.5) {\scriptsize{topological line defect}};
	\end{scope}	
	\end{tikzpicture}
}

\newcommand{\BulkSymmetryOperatorBroughtToTheBoundary}{
\begin{tikzpicture}[scale=.53]

    \draw[dashed,line width=1pt,rounded corners=0.5pt] (-11,-2) -- (-8,1) -- (-8,-3) -- (-11,-6) -- (-11,-2);
    \draw[dashed,line width=1pt,rounded corners=0.5pt] (-8,-3) -- (1,-3);
	    
    \begin{scope}[shift={(0,-0.8)}]
        \path[line width=2pt, rounded corners=2pt,fill=Dandelion!20] (1,-1) -- (-2,-4) -- (-11,-4) -- (-8,-1) -- (1,-1);
    	 
    	\begin{scope}[shift={(-1,0)}]
        	 \path[line width=2pt, rounded corners=4pt,pattern=north west lines, pattern color=Aquamarine,opacity=0.8,shift={(-1,0.8)}] (-6,-2) -- (-3,1) -- (-3,-3) -- (-6,-6) -- (-6,-2);
        	\begin{scope}[shift={(-1,0)}]
        	    \clip (1,-1) -- (-2,-4)  -- (-6,-4) -- (-3,-1) -- (1,-1);
    	        \path[line width=2pt, rounded corners=2pt,fill=Dandelion!20] (1,-1) -- (-2,-4) -- (-11,-4) -- (-8,-1) -- (1,-1);
        	\end{scope}
        	\draw[dashed,color=Aquamarine,line width=1.6pt] (-4,-1) -- (-7,-4);
        	\node at (-5.5,-1.8) {\scriptsize $\dual\sigma$};
    	\end{scope}
    	
    	\begin{scope}[shift={(5,0)}]
        	 \path[line width=2pt, rounded corners=4pt,pattern=north west lines, pattern color=Aquamarine,opacity=0.8,shift={(-1,0.8)}] (-6,-2) -- (-3,1) -- (-3,-3) -- (-6,-6) -- (-6,-2);
        	\draw[dashed,color=Aquamarine,line width=1.6pt] (-4,-1) -- (-7,-4);
        	\node at (-5.5,-1.8) {\scriptsize $\dual\sigma$};
    	\end{scope}
	\end{scope}
	
	\node[align=center] at (-3.25, -0.4) {\tiny{Push towards}};
	\node[align=center] at (-3.25, -0.85) {\tiny{boundary}};
	\draw[line width = 1pt, ->] (-4.9,-1.2) -- (-1.49,-1.2);
		
	\begin{scope}[all/.style={fill=blue!10,fill opacity=0.2,line width=1pt,rounded corners=0.5pt},all,every node/.append style={all}]
        \filldraw (-2,-2) -- (1,1) -- (1,-3) -- (-2,-6) -- (-2,-2);
        \filldraw (-2,-2)-- (-11,-2) -- (-11,-6) -- (-2,-6) -- (-2,-2);
    	\filldraw (-11,-2) -- (-8,1) -- (1,1) -- (-2,-2);
	\end{scope}
		
	\begin{scope}[line width=0.8pt]
        		
        \node[align=center] at (-3.5, 2.) {\scriptsize{Bulk 0-form symmetry}};
        \node[align=center] at (-3.5, 1.5) {\scriptsize{becomes generalized boundary KW-duality}};
        
        \node[align=center,xslant=-.5, yslant=0.65] at (0.5-0.707107*2, -0.707107*2) {\tiny{Boundary}};
        \node[align=center,xslant=-.5, yslant=0.65] at (0.5-0.707107*0.9, -0.707107*0.9-0.5) {\tiny{Space-time}};
        
        \node[align=center] at (-4., -3.) {\tiny{Acts on full}};
        \node[align=center] at (-4., -3.5) {\tiny{boundary space-time}};
	\end{scope}	
	\end{tikzpicture}
}

\newcommand{\FusionOfSurfacesOne}{
\begin{tikzpicture}[scale=.4, baseline={([yshift=-.5ex]current bounding box.center)}]
	    
    \fill[line width=0pt, rounded corners=4pt, color=Black!10,opacity=1] (1,-1) -- (-2,-4) -- (-11,-4) -- (-8,-1) -- cycle;
    \draw[line width=1pt, rounded corners=4pt, color=Black] (1,-1) -- (-2,-4) -- (-11,-4) -- (-8,-1) -- cycle;
    
    \node at (-5.6,-2.0) {\scriptsize $\dual{\sigma_1}$};
    
    \fill[line width=0pt, rounded corners=4pt, color=Black!10,opacity=0.6,shift={(0,1.5)}] (1,-1) -- (-2,-4) -- (-11,-4) -- (-8,-1) -- cycle;
    \draw[line width=1pt, rounded corners=4pt, color=Black,shift={(0,1.5)}] (1,-1) -- (-2,-4) -- (-11,-4) -- (-8,-1) -- cycle;

    \node at (-4.6,-0.5) {\scriptsize $\dual{\sigma_2}$};

	\end{tikzpicture}
}

\newcommand{\FusionOfSurfacesTwo}{
\begin{tikzpicture}[scale=.4, baseline={([yshift=-.5ex]current bounding box.center)}]
	    
    \fill[line width=0pt, rounded corners=4pt, color=Black!10,opacity=1] (1,-1) -- (-2,-4) -- (-11,-4) -- (-8,-1) -- cycle;
    \draw[line width=1pt, rounded corners=4pt, color=Black] (1,-1) -- (-2,-4) -- (-11,-4) -- (-8,-1) -- cycle;
    
    \node at (-5.3,-2.5) {\scriptsize $\dual{\sigma_1\sigma_2}$};
	\end{tikzpicture}
}

\newcommand{\GammaRedCylinder}{
\begin{tikzpicture}
[scale=0.26,baseline={([yshift=-.5ex]current bounding box.center)}]

    \begin{scope}[rounded corners=1.5pt,very thick,red,dashed, decoration={
        markings,
        mark=at position 0.55 with {\arrow{>}}}, shift={(1.5,0)}]
        \path[clip] (-1.5,2) -- (-1.5,-2) -- (1.5,-2) -- (1.5,2) -- cycle;
        \draw[postaction={decorate}](-1.5,0) ellipse (1cm and 2cm);
    \end{scope}
    \begin{scope}[rounded corners=1.5pt,very thick,red, decoration={
        markings,
        mark=at position 0.5 with {\arrow{<}}}, shift={(1.5,0)}]
        \path[clip] (-1.5,2) -- (-1.5,-2) -- (-3.5,-2) -- (-3.5,2) -- cycle;
        \draw[postaction={decorate}](-1.5,0) ellipse (1cm and 2cm);
    \end{scope}
    
    \draw[dotted] (3,0) ellipse (1cm and 2cm);
    \draw[dotted] (-2.75, 2) -- (2.75, 2);
    \draw[dotted] (-2.75, -2) -- (2.75, -2);
    
\end{tikzpicture}
}

\newcommand{\GammaBlueCylinder}{
\begin{tikzpicture}
[scale=0.26,baseline={([yshift=-.5ex]current bounding box.center)}]

    \begin{scope}[rounded corners=1.5pt,very thick,blue,dashed, decoration={
        markings,
        mark=at position 0.55 with {\arrow{>}}}, shift={(1.5,0)}]
        \path[clip] (-1.5,2) -- (-1.5,-2) -- (1.5,-2) -- (1.5,2) -- cycle;
        \draw[postaction={decorate}](-1.5,0) ellipse (1cm and 2cm);
    \end{scope}
    \begin{scope}[rounded corners=1.5pt,very thick,blue, decoration={
        markings,
        mark=at position 0.5 with {\arrow{<}}}, shift={(1.5,0)}]
        \path[clip] (-1.5,2) -- (-1.5,-2) -- (-3.5,-2) -- (-3.5,2) -- cycle;
        \draw[postaction={decorate}](-1.5,0) ellipse (1cm and 2cm);
    \end{scope}
    
    \draw[dotted] (3,0) ellipse (1cm and 2cm);
    \draw[dotted] (-2.75, 2) -- (2.75, 2);
    \draw[dotted] (-2.75, -2) -- (2.75, -2);
    
\end{tikzpicture}
}

\newcommand{\WRedCylinder}{
\begin{tikzpicture}
[scale=0.26,baseline={([yshift=-.5ex]current bounding box.center)}]

    \draw[dotted] (3,0) ellipse (1cm and 2cm);
    \draw[dotted] (-2.75, 2) -- (2.75, 2);
    \draw[dotted] (-2.75, -2) -- (2.75, -2);
    
    \begin{scope}[rounded corners=1.5pt,very thick,red, decoration={
        markings,
        mark=at position 0.6 with {\arrow{>}}}]
        \path[draw,postaction={decorate}] (2., 0) -- (-2.75, 0);  
    \end{scope}
    
\end{tikzpicture}
}

\newcommand{\WBlueCylinder}{
\begin{tikzpicture}
[scale=0.26,baseline={([yshift=-.5ex]current bounding box.center)}]

    \draw[dotted] (3,0) ellipse (1cm and 2cm);
    \draw[dotted] (-2.75, 2) -- (2.75, 2);
    \draw[dotted] (-2.75, -2) -- (2.75, -2);
    
    \begin{scope}[rounded corners=1.5pt,very thick,blue, decoration={
        markings,
        mark=at position 0.6 with {\arrow{>}}}]
        \path[draw,postaction={decorate}] (2., 0) -- (-2.75, 0);  
    \end{scope}
    
\end{tikzpicture}
}

\newcommand{\GammaWCommutatorCylinderOne}{
\begin{tikzpicture}
[scale=0.26,baseline={([yshift=-.5ex]current bounding box.center)}]

    \begin{scope}[rounded corners=1.5pt,very thick,red,dashed, decoration={
        markings,
        mark=at position 0.55 with {\arrow{>}}}, shift={(1.5,0)}]
        \path[clip] (-1.5,2) -- (-1.5,-2) -- (1.5,-2) -- (1.5,2) -- cycle;
        \draw[postaction={decorate}](-1.5,0) ellipse (1cm and 2cm);
    \end{scope}
    
    \begin{scope}[rounded corners=1.5pt,very thick,blue, decoration={
        markings,
        mark=at position 0.4 with {\arrow{>}}}]
        \path[draw,postaction={decorate}] (2., 0) -- (-2.75, 0);  
    \end{scope}
    
    \draw[fill, white] (-1,0) circle (0.3cm);
    
    \begin{scope}[rounded corners=1.5pt,very thick,red, decoration={
        markings,
        mark=at position 0.4 with {\arrow{<}}}, shift={(1.5,0)}]
        \path[clip] (-1.5,2) -- (-1.5,-2) -- (-3.5,-2) -- (-3.5,2) -- cycle;
        \draw[postaction={decorate}](-1.5,0) ellipse (1cm and 2cm);
    \end{scope}

    \draw[dotted] (3,0) ellipse (1cm and 2cm);
    \draw[dotted] (-2.75, 2) -- (2.75, 2);
    \draw[dotted] (-2.75, -2) -- (2.75, -2);
    
\end{tikzpicture}
}

\newcommand{\GammaWCommutatorCylinderTwo}{
\begin{tikzpicture}
[scale=0.26,baseline={([yshift=-.5ex]current bounding box.center)}]

    \begin{scope}[rounded corners=1.5pt,very thick,red,dashed, decoration={
        markings,
        mark=at position 0.55 with {\arrow{>}}}, shift={(1.5,0)}]
        \path[clip] (-1.5,2) -- (-1.5,-2) -- (1.5,-2) -- (1.5,2) -- cycle;
        \draw[postaction={decorate}](-1.5,0) ellipse (1cm and 2cm);
    \end{scope}

    \begin{scope}[rounded corners=1.5pt,very thick,red, decoration={
        markings,
        mark=at position 0.4 with {\arrow{<}}}, shift={(1.5,0)}]
        \path[clip] (-1.5,2) -- (-1.5,-2) -- (-3.5,-2) -- (-3.5,2) -- cycle;
        \draw[postaction={decorate}](-1.5,0) ellipse (1cm and 2cm);
    \end{scope}
    
    \draw[fill, white] (-1,0) circle (0.3cm);
    
    \begin{scope}[rounded corners=1.5pt,very thick,blue, decoration={
        markings,
        mark=at position 0.4 with {\arrow{>}}}]
        \path[draw,postaction={decorate}] (2., 0) -- (-2.75, 0);  
    \end{scope}

    \draw[dotted] (3,0) ellipse (1cm and 2cm);
    \draw[dotted] (-2.75, 2) -- (2.75, 2);
    \draw[dotted] (-2.75, -2) -- (2.75, -2);
    
\end{tikzpicture}
}

\newcommand{\GammaWCommutatorCylinderThree}{
\begin{tikzpicture}
[scale=0.26,baseline={([yshift=-.5ex]current bounding box.center)}]

    \begin{scope}[rounded corners=1.5pt,very thick,blue,dashed, decoration={
        markings,
        mark=at position 0.55 with {\arrow{>}}}, shift={(1.5,0)}]
        \path[clip] (-1.5,2) -- (-1.5,-2) -- (1.5,-2) -- (1.5,2) -- cycle;
        \draw[postaction={decorate}](-1.5,0) ellipse (1cm and 2cm);
    \end{scope}
    
    \begin{scope}[rounded corners=1.5pt,very thick,red, decoration={
        markings,
        mark=at position 0.4 with {\arrow{>}}}]
        \path[draw,postaction={decorate}] (2., 0) -- (-2.75, 0);  
    \end{scope}
    
    \draw[fill, white] (-1,0) circle (0.3cm);
    
    \begin{scope}[rounded corners=1.5pt,very thick,blue, decoration={
        markings,
        mark=at position 0.4 with {\arrow{<}}}, shift={(1.5,0)}]
        \path[clip] (-1.5,2) -- (-1.5,-2) -- (-3.5,-2) -- (-3.5,2) -- cycle;
        \draw[postaction={decorate}](-1.5,0) ellipse (1cm and 2cm);
    \end{scope}

    \draw[dotted] (3,0) ellipse (1cm and 2cm);
    \draw[dotted] (-2.75, 2) -- (2.75, 2);
    \draw[dotted] (-2.75, -2) -- (2.75, -2);
    
\end{tikzpicture}
}

\newcommand{\GammaWCommutatorCylinderFour}{
\begin{tikzpicture}
[scale=0.26,baseline={([yshift=-.5ex]current bounding box.center)}]

    \begin{scope}[rounded corners=1.5pt,very thick, blue,dashed, decoration={
        markings,
        mark=at position 0.55 with {\arrow{>}}}, shift={(1.5,0)}]
        \path[clip] (-1.5,2) -- (-1.5,-2) -- (1.5,-2) -- (1.5,2) -- cycle;
        \draw[postaction={decorate}](-1.5,0) ellipse (1cm and 2cm);
    \end{scope}

    \begin{scope}[rounded corners=1.5pt,very thick,blue, decoration={
        markings,
        mark=at position 0.4 with {\arrow{<}}}, shift={(1.5,0)}]
        \path[clip] (-1.5,2) -- (-1.5,-2) -- (-3.5,-2) -- (-3.5,2) -- cycle;
        \draw[postaction={decorate}](-1.5,0) ellipse (1cm and 2cm);
    \end{scope}
    
    \draw[fill, white] (-1,0) circle (0.3cm);
    
    \begin{scope}[rounded corners=1.5pt,very thick,red, decoration={
        markings,
        mark=at position 0.4 with {\arrow{>}}}]
        \path[draw,postaction={decorate}] (2., 0) -- (-2.75, 0);  
    \end{scope}

    \draw[dotted] (3,0) ellipse (1cm and 2cm);
    \draw[dotted] (-2.75, 2) -- (2.75, 2);
    \draw[dotted] (-2.75, -2) -- (2.75, -2);
    
\end{tikzpicture}
}

\newcommand{\DualityDefectDraggedOverLocalOperatorBulkPicture}{
\begin{tikzpicture}[scale=.53]

    \draw[dashed,line width=1pt,rounded corners=0.5pt] (-11,-2) -- (-8,1) -- (-8,-3) -- (-11,-6) -- (-11,-2);
    \draw[dashed,line width=1pt,rounded corners=0.5pt] (-8,-3) -- (1,-3);
	    
    \begin{scope}[shift={(0,-0.8)}]
        \path[line width=2pt, rounded corners=2pt,fill=Dandelion!20] (1,-1) -- (-2,-4) -- (-11,-4) -- (-8,-1) -- (1,-1);

    	 \path[line width=2pt, rounded corners=4pt,pattern=north west lines, pattern color=Aquamarine,opacity=0.8,shift={(0,0.8)}] (-9,-2)-- (-11,-2) -- (-11,-6) -- (-9,-6) -- (-9,-2);
    	 \draw[color=blue,line width=1.6pt,shift={(0,0.8)}] (-9,-2) -- (-9,-6);
    	\foreach \y in {1,...,19}{
    	    \draw[red,line width=0.8pt,shift={(0,0.8)}] (-9-0.1,-6+0.1 + 4/20*\y ) -- (-9+0.1,-6-0.1 + 4/20*\y );
    	}
    	 
    	\begin{scope}[shift={(-3,0)}]
    	    \clip (1,-1) -- (-2,-4)  -- (-6,-4) -- (-3,-1) -- (1,-1);
	        \path[line width=2pt, rounded corners=2pt,fill=Dandelion!20] (1,-1) -- (-2,-4) -- (-11,-4) -- (-8,-1) -- (1,-1);
    	\end{scope}

    	\path[draw,blue,dashed, rounded corners=0.1pt,thin,,decoration={ markings, mark=at position 0.5 with {\arrow{>}}},line width=1.5pt] (-4,-4) .. controls +(1,1.0) and +(-0.5,0.5) .. (-7,-4);
        \fill[fill=blue] (-7,-4) circle (4pt);
        \fill[fill=blue] (-4,-4) circle (4pt);
    	
    	\node at (-10,-2) {\scriptsize $\dual\sigma$};
	\end{scope}
		
	\begin{scope}[all/.style={fill=blue!10,fill opacity=0.2,line width=1pt,rounded corners=0.5pt},all,every node/.append style={all}]
        \filldraw (-2,-2) -- (1,1) -- (1,-3) -- (-2,-6) -- (-2,-2);
        \filldraw (-2,-2)-- (-11,-2) -- (-11,-6) -- (-2,-6) -- (-2,-2);
    	\filldraw (-11,-2) -- (-8,1) -- (1,1) -- (-2,-2);
	\end{scope}

	\begin{scope}[shift={(14,0)}]
    \draw[dashed,line width=1pt,rounded corners=0.5pt] (-11,-2) -- (-8,1) -- (-8,-3) -- (-11,-6) -- (-11,-2);
    \draw[dashed,line width=1pt,rounded corners=0.5pt] (-8,-3) -- (1,-3);
	    
    \begin{scope}[shift={(0,-0.8)}]
        \path[line width=2pt, rounded corners=2pt,fill=Dandelion!20] (1,-1) -- (-2,-4) -- (-11,-4) -- (-8,-1) -- (1,-1);
    	 
    	 \path[line width=2pt, rounded corners=4pt,pattern=north west lines, pattern color=Aquamarine,opacity=0.8,shift={(0,0.8)}] (-5.5,-2)-- (-11,-2) -- (-11,-6) -- (-5.5,-6) -- (-5.5,-2);
    	 \draw[dashed,color=Aquamarine,line width=1.6pt,shift={(0,0.8)}] (-5.5,-2) -- (-5.5,-6);
    	 \draw[color=blue,line width=1.6pt,shift={(0,0.8)}] (-5.5,-2) -- (-5.5,-6);
    	\foreach \y in {1,...,19}{
    	    \draw[red,line width=0.8pt,shift={(0,0.8)}] (-5.5-0.1,-6+0.1 + 4/20*\y ) -- (-5.5+0.1,-6-0.1 + 4/20*\y );
    	}
    	 
    	\begin{scope}[shift={(0,0)}]
    	    \clip (1,-1) -- (-2,-4)  -- (-4,-4) -- (-1,-1) -- (1,-1);
	        \path[line width=2pt, rounded corners=2pt,fill=Dandelion!20] (1,-1) -- (-2,-4) -- (-11,-4) -- (-8,-1) -- (1,-1);
    	\end{scope}

        \path[draw,red, rounded corners=0.1pt,thin,,decoration={ markings, mark=at position 0.5 with {\arrow{>}}},line width=1.5pt] (-5.5,-4) -- (-7,-4);
    	\path[draw,blue,dashed, rounded corners=0.1pt,thin,,decoration={ markings, mark=at position 0.5 with {\arrow{>}}},line width=1.5pt] (-4,-4) -- (-5.5,-4);
        \fill[fill=blue] (-7,-4) circle (4pt);
        \fill[fill=blue] (-4,-4) circle (4pt);
    	
    	\node at (-10,-2) {\scriptsize $\dual\sigma$};
	\end{scope}

	\begin{scope}[all/.style={fill=blue!10,fill opacity=0.2,line width=1pt,rounded corners=0.5pt},all,every node/.append style={all}]
        \filldraw (-2,-2) -- (1,1) -- (1,-3) -- (-2,-6) -- (-2,-2);
        \filldraw (-2,-2)-- (-11,-2) -- (-11,-6) -- (-2,-6) -- (-2,-2);
    	\filldraw (-11,-2) -- (-8,1) -- (1,1) -- (-2,-2);
	\end{scope}
	\end{scope}
	
		\node at (1.5,-4) {\large$=$};
		
	\end{tikzpicture}
}

\newcommand{\CriticalPointMappingFigure}
    {
    \begin{tikzpicture}[scale=0.26,baseline={([yshift=-.5ex]current bounding box.center)},rounded corners=5pt,very thick,decoration={
		markings,
		mark=at position 0.75 with {\arrow{>}}}]

	\path[draw,line width=2pt,color=black] (-10,0) -- (10,0);

	\draw[line width=2pt,Green,postaction={decorate}] (-7,5) to[out=-20,in=120] (-5,0) to[out=310,in=10] (-7,-5);
	
	\draw[line width=2pt,Green,dashed] (-7,5) -- (-10,5.5);
	\draw[line width=2pt,Green,dashed] (-7,-5) -- (-10,-5.5);
	
	\begin{scope}[rotate=180]
		\draw[line width=2pt,Green,postaction={decorate}] (-7,5) to[out=-20,in=120] (-5,0) to[out=310,in=10] (-7,-5);
		
		\draw[line width=2pt,Green,dashed] (-7,5) -- (-10,5.5);
		
		\draw[line width=2pt,Green,dashed] (-7,-5) -- (-10,-5.5);
	\end{scope}

	\filldraw[color=red] (-5,0) circle (0.5);
	\filldraw[color=red] (5,0) circle (0.5);
	
	\node at (0,+2.5) {$\mcal{L}_1$};
	\node at (0,-2.5) {$\mcal{L}_2$};

\end{tikzpicture}
    }

\newcommand{\DualityDefectDraggedOverLocalOperatorOneNEW}{
\begin{tikzpicture}
[scale=0.26,baseline={([yshift=-.5ex]current bounding box.center)},rounded corners=5pt,very thick,black,decoration={
        markings,
        mark=at position 0.5 with {\arrow{>}}}]

\draw[red, line width=0.2pt,dashed,postaction={decorate}] (-10,0) -- (1.5,0);
\draw[black,fill=black] (1.5,0) circle (.35);
\draw[black,fill=black] (-10,0) circle (.35);
\draw[yshift=.2cm] (1.5,1.5) node {$\mathcal O_{\alpha,i}$};
\draw[yshift=.2cm] (-10,1.5) node {$\mathcal O_{-\alpha,j}$};
\draw[yshift=.2cm] (-3.5,-1.5) node {\footnotesize$S_{(0,\alpha)}$};

\draw[yshift=.2cm] (1,5) node {$\sigma$};

\path[draw,dashed, color=Aquamarine] (-0,-6) -- (0,-3) .. controls +(5,-3) and +(5,3) .. (0,3) -- (0,6);
\end{tikzpicture}
}

\newcommand{\DualityDefectDraggedOverLocalOperatorTwoNEW}{
\begin{tikzpicture}
[scale=0.26,baseline={([yshift=-.5ex]current bounding box.center)},rounded corners=5pt,very thick,black,decoration={
        markings,
        mark=at position 0.25 with {\arrow{>}}}]

\begin{scope}
        \clip (0,-6) -- (0,-3) .. controls +(-5,-3) and +(-5,3) .. (0,3) -- (0,6) -- (-10,6) -- (-10, -6) -- (0,-6);
        \draw[red, line width=0.2pt,dashed,postaction={decorate}] (-10,0) -- (1.5,0);
\end{scope}

\begin{scope}[decoration={
        markings,
        mark=at position 0.75 with {\arrow{>}}}]
        \clip (0,-6) -- (0,-3) .. controls +(-5,-3) and +(-5,3) .. (0,3) -- (0,6) -- (4,6) -- (4, -6) -- (0,-6);
        \draw[blue, line width=2.pt,postaction={decorate}] (-10,0) -- (1.5,0);
\end{scope}

\draw[black,fill=black] (1.5,0) circle (.35);
\draw[black,fill=black] (-10,0) circle (.35);
\draw[yshift=.2cm] (0.5,1.5) node {\footnotesize $S_{\sigma\cdot(0,\alpha)}$};
\draw[yshift=.2cm] (-10,1.5) node {$\mathcal O_{-\alpha,j}$};
\draw[yshift=.2cm] (-6,-1.5) node {\footnotesize$S_{(0,\alpha)}$};

\draw[yshift=.2cm] (1,5) node {$\sigma$};

\path[draw,dashed, color=Aquamarine] (-0,-6) -- (0,-3) .. controls +(-5,-3) and +(-5,3) .. (0,3) -- (0,6);

\end{tikzpicture}
}

\newcommand{\GammaWPenetration}{
\begin{tikzpicture}
[scale=0.26,baseline={([yshift=-.5ex]current bounding box.center)}]

    \begin{scope}[rounded corners=1.5pt,very thick,blue,dashed, decoration={
        markings,
        mark=at position 0.55 with {\arrow{>}}}, shift={(4.5,0)}]
        \path[clip] (-1.5,2) -- (-1.5,-2) -- (1.5,-2) -- (1.5,2) -- cycle;
        \draw[postaction={decorate}](-1.5,0) ellipse (1cm and 2cm);
    \end{scope}
    
    \begin{scope}[rounded corners=1.5pt,very thick,red, decoration={
        markings,
        mark=at position 0.4 with {\arrow{>}}}]
        \path[draw,postaction={decorate}] (3., 0) -- (-2.75, 0);  
    \end{scope}
    
    \draw[fill, red] (3,0) circle (0.3cm);
    \draw[fill, white] (2,0) circle (0.3cm);
    
    \begin{scope}[rounded corners=1.5pt,very thick,blue, decoration={
        markings,
        mark=at position 0.4 with {\arrow{<}}}, shift={(4.5,0)}]
        \path[clip] (-1.5,2) -- (-1.5,-2) -- (-3.5,-2) -- (-3.5,2) -- cycle;
        \draw[postaction={decorate}](-1.5,0) ellipse (1cm and 2cm);
    \end{scope}
    
    \draw(2.5,3) node {\footnotesize$\mathcal U_{\ms g}$};
    \draw(-0.2,1) node {\footnotesize$\alpha$};
    \draw(2.5,-3) node {\footnotesize$\;$};

\end{tikzpicture}
}

\newcommand{\GammaWNoPenetration}{
\begin{tikzpicture}
[scale=0.26,baseline={([yshift=-.5ex]current bounding box.center)}]

    \begin{scope}[rounded corners=1.5pt,very thick,red, decoration={
        markings,
        mark=at position 0.4 with {\arrow{>}}}]
        \path[draw,postaction={decorate}] (3., 0) -- (-2.75, 0);  
    \end{scope}
    
    \draw[fill, red] (3,0) circle (0.3cm);
    
    \draw(2.5,3) node {\footnotesize$\;$};
    \draw(-0.2,1) node {\footnotesize$\alpha$};
     \draw(2.5,-3) node {\footnotesize$\;$};

\end{tikzpicture}
}

\newcommand{\GammaRedCylinderDecomposedIntoBoundaryOperators}{
\begin{tikzpicture}
[scale=0.26,baseline={([yshift=-.5ex]current bounding box.center)}]

    \begin{scope}[rounded corners=0.5pt,red, line width=0.8]
        \path[clip] (-2.75, 2) -- (2.75, 2)  .. controls +(-1,-0.8) and +(-1,0.8) ..  (2.75, -2) -- (-2.75, -2);
        \foreach \x in {0,1, 2}{
            \draw[postaction={decorate}](2+\x*0.2,\x) ellipse (0.5cm and 0.5cm);
        }
        \foreach \x in {-2,-1}{
            \draw[postaction={decorate}](2-\x*0.2,\x) ellipse (0.5cm and 0.5cm);
        }
    \end{scope}

    \draw[dotted] (3,0) ellipse (1cm and 2cm);
    \draw[dotted] (-2.75, 2) -- (2.75, 2);
    \draw[dotted] (-2.75, -2) -- (2.75, -2);
    
\end{tikzpicture}
}

\newcommand{\GammaBlueCylinderDecomposedIntoBoundaryOperators}{
\begin{tikzpicture}
[scale=0.26,baseline={([yshift=-.5ex]current bounding box.center)}]
    \begin{scope}[rounded corners=0.5pt,blue, line width=0.8]
        \path[clip] (-2.75, 2) -- (2.75, 2)  .. controls +(-1,-0.8) and +(-1,0.8) ..  (2.75, -2) -- (-2.75, -2);
        \foreach \x in {0,1, 2}{
            \draw[postaction={decorate}](2+\x*0.2,\x) ellipse (0.5cm and 0.5cm);
        }
        \foreach \x in {-2,-1}{
            \draw[postaction={decorate}](2-\x*0.2,\x) ellipse (0.5cm and 0.5cm);
        }
    \end{scope}

    \draw[dotted] (3,0) ellipse (1cm and 2cm);
    \draw[dotted] (-2.75, 2) -- (2.75, 2);
    \draw[dotted] (-2.75, -2) -- (2.75, -2);
    
\end{tikzpicture}
}

\newcommand{\BoundaryOperatorsOnCylinder}{
\begin{tikzpicture}
[scale=0.3,baseline={([yshift=-.5ex]current bounding box.center)}]
    \begin{scope}[rounded corners=0.5pt,black, line width=1.3,decoration={
    markings,
    mark=at position 0.55 with {\arrow{>}}}]
        \path[clip] (-2.75, 2) -- (2.75, 2)  .. controls +(-1,-0.8) and +(-1,0.8) ..  (2.75, -2) -- (-2.75, -2);
        \path[draw,postaction={decorate}] (3,-1.3) .. controls +(-2.5,-0.3) and +(-2.5,0.3)  .. (3,1.3);
    \end{scope}
    
    \draw (2.9,-1.3) node {\scriptsize $a$};
    \draw (2.9,1.3) node {\scriptsize $b$};
    
    \draw[dotted] (3,0) ellipse (1cm and 2cm);
    \draw[dotted] (-2.75, 2) -- (2.75, 2);
    \draw[dotted] (-2.75, -2) -- (2.75, -2);
    
\end{tikzpicture}
}

\newcommand{\TriangulationGaugeTransformation}{
\begin{tikzpicture}
[scale=0.28,baseline={([yshift=-.5ex]current bounding box.center)},rounded corners=12pt,very thick,decoration={
    markings,
    mark=at position 0.5 with {\arrow{>}}}]
    
    \def\radius{7}
    \def\delta{0.5}

    \begin{scope}
        \begin{scope}
            \clip (2,2) circle (\radius);
            \foreach \x in {-2, -1,0,1, 2}{
                \foreach \y in {-2,-1,0,1, 2}{
                    \path[color=black, draw] (-2+4*\x,-2+4*\y) -- (-2+4*\x,2+4*\y);
                    \path[color=black, draw] (-2+4*\x,-2+4*\y) -- (2+4*\x,-2+4*\y);
                    \path[color=black, draw] (-2+4*\x,-2+4*\y) -- (2+4*\x,2+4*\y);
                }
            }
            \path[color=green, draw] (-2,-2-4) -- (-2,2-4);
            \foreach \x in {0,1, 2}{
                    \path[color=green, draw] (-2,-2+4*\x) -- (2,-2+4*\x);
                    \path[color=green, draw] (-2,-2+4*\x) -- (2, 2+4*\x);
            }
            
            \path[color=red, draw] (-3, -5) -- (1,-1) -- (-1, 1) -- (1, 3) -- (-1, 5) -- (1,7) -- (-1,9);
        \end{scope}
        
        \draw[fill, color=blue!60, very thick](2,2) circle (0.4);
         
         \draw[dashed, color=black!60, very thick](2,2) circle (\radius);
         
    \end{scope}
    
    \node at (12,2) {$\longrightarrow$};
    
    \begin{scope}[shift={(20,0)}]
        \begin{scope}
            \clip (2,2) circle (\radius);
            \foreach \x in {-2, -1,0,1, 2}{
                \foreach \y in {-2,-1,0,1, 2}{
                    \path[color=black, draw] (-2+4*\x,-2+4*\y) -- (-2+4*\x,2+4*\y);
                    \path[color=black, draw] (-2+4*\x,-2+4*\y) -- (2+4*\x,-2+4*\y);
                    \path[color=black, draw] (-2+4*\x,-2+4*\y) -- (2+4*\x,2+4*\y);
                }
            }
            \path[color=green, draw] (-2+4,-2) -- (-2+4, 2);
             \path[color=green, draw] (-2+4,-2+4) -- (-2+4, 2+4);
            
            \path[color=green, draw] (-2,-2-4) -- (-2,2-4);
            
            \path[color=green, draw] (-2,-2) -- (2,-2);
            \path[color=green, draw] (-2,-2+8) -- (2,-2+8);
            \path[color=green, draw] (-2+4,-2+4) -- (2+4,-2+4);
    
            \path[color=green, draw] (-2,-2+4) -- (2, 2+4);
            \path[color=green, draw] (-2+4,-2+4) -- (2+4, 2+4);
            \path[color=green, draw] (-2,-2+8) -- (2, 2+8);
            
            \path[color=red, draw] (-3, -5) -- (3,1) -- (5, 3) -- (3, 5) -- (1, 3) -- (-1,5) -- (1,7) -- (-1,9);
        \end{scope}

         \draw[dashed, color=black!60, very thick](2,2) circle (\radius);
         
    \end{scope}
\end{tikzpicture}
}

\newcommand{\CocyleDiscreteGaugeField}{
\begin{tikzpicture}
		
	\draw[rounded corners=4pt,fill=Gray!50!,line width=2pt] (-2.9,0) -- (3.1,0) -- (0,{3*sqrt(3)}) -- (-3,0) -- (3,0);
	
	\path[line width=2pt, postaction={decoration={markings,mark=at position .3 with {\arrow{>}}},decorate}] (-2.9,0) -- (3.1,0);
	
	\path[line width=2pt, postaction={decoration={markings,mark=at position .3 with {\arrow{>}}},decorate}] (3.1,0) -- (0,{3*sqrt(3)});
	
	\path[line width=2pt, postaction={decoration={markings,mark=at position .3 with {\arrow{<}}},decorate}] (0,{3*sqrt(3)}) -- (-3,0);
	
	\draw[line width=3pt,Green,postaction={decoration={markings,mark=at position .6 with {\arrow{>}}},decorate}] (3/2,{3/2*sqrt(3)}) node [right,black] {$A_{jk}$} -- (0,{3/sqrt(3)});
	\draw[line width=3pt,Green,postaction={decoration={markings,mark=at position .55 with {\arrow{<}}},decorate}] (-3/2,{3/2*sqrt(3)})node [left,black] {$A_{ik}$} -- (0,{3/sqrt(3)});
	\draw[line width=3pt,Green,postaction={decoration={markings,mark=at position .6 with {\arrow{>}}},decorate}] (0,0) node [below,black] {$A_{ij}$} -- (0,{3/sqrt(3)});
	
	\draw[red,fill] (0,{3/sqrt(3)}) circle (5pt);
	
	\node at (0,-1) {$A_{ij}+A_{jk}=A_{ik}$};
\end{tikzpicture}
}

\newcommand{\VariousTypesOfBulkLineOperators}{

\begin{tikzpicture}[scale=.8]
	
	\draw[fill=Gray!50!,line width=2pt] (3,0) ellipse (.5cm and 2cm);
	
	\draw[line width=2pt] (3,2) -- (-6,2);
	\draw[line width=2pt] (3,-2) -- (-6,-2);
	
	\draw[Maroon,line width=1pt,dashed] (-1.5,0) ellipse (.5cm and 2cm);
	
	\begin{scope}
		\clip (-1.5,2) -- (-2.5,2) -- (-2.5,-2) -- (-1.5,-2);
		\draw[Maroon,line width=2pt,postaction={decoration={markings,mark=at position .6 with {\arrow{>}}},decorate}] (-1.5,0) ellipse (.5cm and 2cm);
	\end{scope}

	\draw[decorate,
	decoration={snake,amplitude=1mm,segment length=10mm},line width=2pt,Green,postaction={decoration={markings,mark=at position .6 with {\arrow{>}}},decorate}] (2.5,0) -- (-1.9,0);
	
	\draw[decorate,
	decoration={snake,amplitude=1mm,segment length=10mm},line width=2pt,Green,postaction={decoration={markings,mark=at position .6 with {\arrow{>}}},decorate}] (-2.1,0) -- (-5.5,0);
	
	\draw[looseness=1.25,line width=2pt,Magenta,postaction={decoration={markings,mark=at position .6 with {\arrow{>}}},decorate}] (2.65,1.2) to[out=180,in=90] (1,.05);
	\draw[looseness=1.25,line width=2pt,Magenta,postaction={decoration={markings,mark=at position .5 with {\arrow{<}}},decorate}] (2.65,-1.2) to[out=180,in=270] (1,-.25);
	
	\draw[Magenta,fill] (2.6,1.2)node[right,black]{$a$} circle (3pt);
	\draw[Magenta,fill] (2.6,-1.2)node[right,black]{$b$} circle (3pt);
	
	\node[Maroon] at (-1.5,-2.5) {$\Gamma_d$};
	\node[Green] at (-4,.5) {$\mscr{Y}_d$};
	\node[Magenta] at (.5,-1.3) {$S_{d}(a,b)$};

\end{tikzpicture}

}

\newcommand{\StringOperatorsBetweenLatticeSites}{
    \begin{tikzpicture}
	
	\draw[line width=1pt,dashed] (-6,0) -- (-4.5,0);
	\draw[line width=2pt] (-4.5,0) -- (4.5,0);
	\draw[line width=1pt,dashed] (4.5,0) -- (6,0);
	  
	\foreach \x in {-4,-2,0,2}{
	\draw[fill,MidnightBlue] (\x,0) circle (3pt);
	\draw[fill,MidnightBlue] (-\x,0) circle (3pt);
	\draw[looseness=2,line width=1.5pt,MidnightBlue] (\x,0) to[out=90,in=90] (\x+2,0);
	}	

	\foreach \x in {-3,-1,1,3}
	\draw[fill,Maroon] (\x,0) circle (3pt);
	
	\begin{scope}
		\clip (3.5,-.5) -- (3.5,1.5) -- (4.5,1.5) -- (4.5,-.5) -- (3.5,-.5);
		\draw[looseness=2,line width=1.5pt,MidnightBlue] (4,0) to[out=90,in=90] (6,0);
	\end{scope}	
	
	\begin{scope}[xscale=-1]
		\clip (3.5,-.5) -- (3.5,1.5) -- (4.5,1.5) -- (4.5,-.5) -- (3.5,-.5);
		\draw[looseness=2,line width=1.5pt,MidnightBlue] (4,0) to[out=90,in=90] (6,0);
	\end{scope}

	\foreach \x in {-2,0,2}{
	\begin{scope}
	\clip (-\x-.53,1.7) -- (-\x-.53,-.5) -- (-\x+.53,-.5) -- (-\x+.53,1.7) -- (-\x+.53,1.7);
	\draw[fill,Maroon] (\x-1,0) circle (3pt);
	\draw[looseness=2.25,line width=1.5pt,Maroon] (-\x-1,0) to[out=90,in=90] (-\x+1,0);
	\end{scope}}	

	\foreach \x in {-3,-1,1}{
		\begin{scope}
			\clip (-.3+\x,-.5) -- (-.3+\x,1.5) -- (\x+.35,1.5) -- (\x+.35,-.5) -- (-.3+\x,-.5);
		
			\draw[fill,Maroon] (\x,0) circle (3pt);
	
			\draw[looseness=2.25,line width=1.5pt,Maroon] (\x,0) to[out=90,in=90] (\x+2,0);
				
	\end{scope}}

	\begin{scope}[xscale=-1]
		\foreach \x in {-3,-1,1}{
			\begin{scope}
				\clip (-.3+\x,-.5) -- (-.3+\x,1.5) -- (\x+.35,1.5) -- (\x+.35,-.5) -- (-.3+\x,-.5);
				
				\draw[fill,Maroon] (\x,0) circle (3pt);
				
				\draw[looseness=2.25,line width=1.5pt,Maroon] (\x,0) to[out=90,in=90] (\x+2,0);
				
		\end{scope}}	
	\end{scope}
	
	\begin{scope}
		\clip (3,-.5) -- (3,1.5) -- (3.35,1.5) -- (3.35,-.5) -- (3,-.5);
		\draw[looseness=2.25,line width=1.5pt,Maroon] (3,0) to[out=90,in=90] (5,0);
	\end{scope}
	
	\begin{scope}
		\clip (3.45,-.5) -- (3.45,1.5) -- (4.6,1.5) -- (4.6,-.5) -- (3.45,-.5);
		\draw[looseness=2.25,line width=1.5pt,Maroon] (3,0) to[out=90,in=90] (5,0);
	\end{scope}

	\begin{scope}[xscale=-1]
		\begin{scope}
			\clip (3,-.5) -- (3,1.5) -- (3.35,1.5) -- (3.35,-.5) -- (3,-.5);
			\draw[looseness=2.25,line width=1.5pt,Maroon] (3,0) to[out=90,in=90] (5,0);
		\end{scope}
		
		\begin{scope}
			\clip (3.45,-.5) -- (3.45,1.5) -- (4.6,1.5) -- (4.6,-.5) -- (3.45,-.5);
			\draw[looseness=2.25,line width=1.5pt,Maroon] (3,0) to[out=90,in=90] (5,0);
		\end{scope}
	\end{scope}
	
	\foreach \x in {-4,...,-1}
	\node[below] at (\x,-.1) {\tiny $2i \x$};
	
	\foreach \x in {+1,+2,+3,+4}
	\node[below] at (\x,-.1) {\tiny $2i \x$};
	
	\node[below] at (0,-.1) {\tiny $2i$};
	
	\foreach \x in {-3,-1,1,3}
	\node at (\x,1.4) {\fontsize{8}{8}\selectfont $S_{(0,\alpha)}$};
	
	\foreach \x in {-4,-2,0,2,4}
	\node at (\x,1.55) {\fontsize{8}{8}\selectfont $S_{(\msf g,0)}$};

	\foreach \x in {-3,-1,1,3,5}
	\draw [
	line width=1.5pt,Gray,
	decoration={
		brace,
		raise=0cm
	},
	decorate
	] (\x-.1,-.5) -- (\x-1.9,-.5);
	
	\foreach \x in {-2,-1,+1,+2}
	\node[below,Gray] at (2*\x,-.5)  {\small $i\x$};
	\node[below,Gray] at (0,-.6)  {\small $i$};
\end{tikzpicture}
    }

\newcommand{\PseudoSpinWithAandBOperators}{
    \begin{tikzpicture}
	\draw[line width=1pt,dashed] (-6,0) -- (-4.5,0);
	\draw[line width=2pt] (-4.5,0) -- (4.5,0);
	\draw[line width=1pt,dashed] (4.5,0) -- (6,0);

	\foreach \x in {-4,-2,0,2,4}{
	\draw[line width=8.2pt,Gray,opacity=.5] (\x-.55,0) -- (\x+.55,0);
	\draw[fill,MidnightBlue] (\x,0) circle (3pt);
	\draw[fill,Maroon] (\x-.4,0) circle (3pt);
	\draw[fill,Maroon] (\x+.4,0) circle (3pt);
	
	}
	
	\foreach \x in {-4,-2,0,2,4}{
	\begin{scope}
		\clip (\x-.6,-.5) -- (\x-.2,-.5) -- (\x-.2,1.5) -- (\x-.6,1.5) -- (\x-.6,-.5);
		\draw[line width=1.5pt,Maroon,looseness=3] (\x-.4,0) to[out=90,in=90] (\x+.4,0);
	\end{scope}
	}

	\foreach \x in {-4,-2,0,2,4}{
		\begin{scope}[xscale=-1]
			\clip (\x-.6,-.5) -- (\x-.2,-.5) -- (\x-.2,1.5) -- (\x-.6,1.5) -- (\x-.6,-.5);
			\draw[line width=1.5pt,Maroon,looseness=3] (\x-.4,0) to[out=90,in=90] (\x+.4,0);
		\end{scope}
	}

	\foreach \x in {-4,-2,0,2,4}{
		\begin{scope}
			\clip (\x-.1,-.5) -- (\x+.1,-.5) -- (\x+.1,1.5) -- (\x-.1,1.5) -- (\x-.1,-.5);
			\draw[line width=1.5pt,Maroon,looseness=3] (\x-.4,0) to[out=90,in=90] (\x+.4,0);
		\end{scope}
	}

	\foreach \x in {-4,-2,0,2}
	\draw[looseness=2,line width=1.5pt,MidnightBlue] (\x,0) to[out=90,in=90] (\x+2,0);
	
	\begin{scope}
		\clip (4,-.5) -- (4,1.5) -- (4.5,1.5) -- (4.5,-.5) -- (4,-.5);
		\draw[looseness=2,line width=1.5pt,MidnightBlue] (4,0) to[out=90,in=90] (4+2,0);
	\end{scope}

	\begin{scope}[xscale=-1]
		\clip (4,-.5) -- (4,1.5) -- (4.5,1.5) -- (4.5,-.5) -- (4,-.5);
		\draw[looseness=2,line width=1.5pt,MidnightBlue] (4,0) to[out=90,in=90] (4+2,0);
	\end{scope}
	
	\foreach \x [count=\i, evaluate=\i as \f using {2*\i-6.05}] in {$A^{\msf g}_{j-2}$, $A^{\msf g}_{j-1}$, $A^{\msf g}_{j}$, $A^{\msf g}_{j+1}$,$A^{\msf g}_{j+2}$}
	\node[Maroon] at (\f,1) {\tiny \x};
	
	\foreach \x [count=\i, evaluate=\i as \f using {2*\i-5}] in {$B^\alpha_{j-2,j-1}$, $B^\alpha_{j-1,j}$, $B^\alpha_{j,j+1}$, $B^\alpha_{j+1,j+2}$}
	\node[MidnightBlue] at (\f,1.35) {\tiny \x};
	
	\foreach \x [count=\i, evaluate=\i as \f using {2*\i-6}] in {$i-2$, $i-1$, $i$, $i+1$, $i+2$}
	\node[] at (\f,-.3) {\small\x};
\end{tikzpicture}
    
    }

\newcommand{\PseudoSpinChainForDyonicsOperators}{
	\begin{tikzpicture}
	\draw[line width=1pt,dashed] (-6,0) -- (-4.5,0);
	\draw[line width=2pt] (-4.5,0) -- (4.5,0);
	\draw[line width=1pt,dashed] (4.5,0) -- (6,0);

	\foreach \x in {-4,-2,0,2,4}{
	    \draw[line width=8.2pt,Gray,opacity=.5] (\x-.55,0) -- (\x+.55,0);
		\draw[fill,MidnightBlue] (\x,0) circle (3pt);
		\draw[fill,Maroon] (\x-.4,0) circle (3pt);
		\draw[fill,Maroon] (\x+.4,0) circle (3pt);
		
	}
	
	\foreach \x in {-2,2}{
		\begin{scope}
			\clip (\x-.6,-.5) -- (\x-.2,-.5) -- (\x-.2,1.5) -- (\x-.6,1.5) -- (\x-.6,-.5);
			\draw[line width=1.5pt,Maroon,looseness=3] (\x-.4,0) to[out=90,in=90] (\x+.4,0);
		\end{scope}
	}
	
	\foreach \x in {-2,2}{
		\begin{scope}[xscale=-1]
			\clip (\x-.6,-.5) -- (\x-.2,-.5) -- (\x-.2,1.5) -- (\x-.6,1.5) -- (\x-.6,-.5);
			\draw[line width=1.5pt,Maroon,looseness=3] (\x-.4,0) to[out=90,in=90] (\x+.4,0);
		\end{scope}
	}
	
	\foreach \x in {-2,2}{
		\begin{scope}
			\clip (\x-.1,-.5) -- (\x+.1,-.5) -- (\x+.1,1.5) -- (\x-.1,1.5) -- (\x-.1,-.5);
			\draw[line width=1.5pt,Maroon,looseness=3] (\x-.4,0) to[out=90,in=90] (\x+.4,0);
		\end{scope}
	}

	\foreach \x in {-2,0,2}
	\draw[looseness=2,line width=1.5pt,MidnightBlue] (\x,0) to[out=90,in=90] (\x+2,0);
	\draw[looseness=3,line width=1.5pt,MidnightBlue] (0,0) to[out=90,in=97] (2,0);
	
	\begin{scope}
		\clip (-3.35,1.03) -- (-4.2,1.03) -- (-4.2,-.5) -- (-3.5,-.5) -- (-3.5,.8) -- (-3.35,1.03);
		\draw[looseness=2,line width=1.5pt,MidnightBlue] (-4,0) to[out=90,in=90] (-2,0);
	\end{scope}
	
	\begin{scope}
		\clip (-3.35,1.5) -- (-3.35,-.5) -- (-1,-.5) -- (-1,1.5) -- (-3.35,1.5);
		\draw[looseness=2,line width=1.5pt,MidnightBlue] (-4,0) to[out=90,in=90] (-2,0);
	\end{scope}

	\begin{scope}
		\clip (-2.8,-.5) -- (-2.8,1.05) -- (-2.6,1.05) -- (-2.6,.5) -- (-2.4,.5) -- (-2.4,-.5) -- (-2.8,-.5);
		\draw[looseness=7,line width=1.5pt,Maroon] (-2.4,0) to[out=130,in=50] (-1.6,0);
	\end{scope}
	\begin{scope}[]
		\clip (-1.2,-.5) -- (-1.2,1.05) -- (-1.4,1.05) -- (-1.4,.5) -- (-1.6,.5) -- (-1.6,-.5) -- (-1.2,-.5);
		\draw[looseness=7,line width=1.5pt,Maroon] (-2.4,0) to[out=130,in=50] (-1.6,0);
	\end{scope}
	\begin{scope}[]
		\clip (-2.52,1.5) -- (-1.49,1.5) -- (-1.49,.8) -- (-2.52,.8) -- (-2.52,1.5);
		\draw[looseness=7,line width=1.5pt,Maroon] (-2.4,0) to[out=130,in=50] (-1.6,0);
	\end{scope}
	
	\begin{scope}
		\clip (-1,1.05) -- (-.7,1.05) -- (-.7,1.2) -- (-.9,1.2) -- (-.9,1.6) -- (.2,1.6) -- (.2,1) -- (.25,1) -- (.25,1.2) -- (.36,1.2) -- (.36,.95) -- (.3,.95) -- (.3,.85) -- (.6,.85) -- (.6,-.5) -- (-1,-.5) -- (-1,1.05) ;
		\draw[looseness=7,line width=1.5pt,Maroon] (-.4,0) to[out=130,in=80] (.4,0);
	\end{scope}

	\draw[looseness=7,line width=1.5pt,Maroon] (-4.4,0) to[out=120,in=60] (-3.6,0);
	
    \node[Maroon] at (-4,1.6) {\tiny $A^{\msf g_1}$};
    \node[Maroon] at (-2,.95) {\tiny $A^{\msf g_2}$};
    \node[Maroon] at (-2,1.45) {\tiny $A^{\msf g_3}$};
    \node[Maroon] at (-.2,1.6) {\tiny $A^{\msf g_4}$};
    \node[Maroon] at (2.1,.95) {\tiny $A^{\msf g_4}$};
    
    \node[MidnightBlue] at (-3,1.32) {\tiny $B^{\alpha_1}$};
    \node[MidnightBlue] at (-1,1.32) {\tiny $B^{\alpha_2}$};
    \node[MidnightBlue] at (1,1.32) {\tiny $B^{\alpha_3}$};
    \node[MidnightBlue] at (.8,1.9) {\tiny $B^{\alpha_4}$};
    \node[MidnightBlue] at (3,1.32) {\tiny $B^{\alpha_4}$};
\end{tikzpicture}
}

\newcommand{\SectorChangingYOperatorActingOnAPseudoSpinChain}{

\begin{tikzpicture}

	\draw[line width=1pt,dashed] (-3.5,0) -- (-3,0);
	\draw[line width=2pt] (-3,0) -- (3,0);
	\draw[line width=1pt,dashed] (3,0) -- (3.5,0);

	\foreach \x in {-2,0,2}{
	    \draw[line width=8.2pt,Gray,opacity=.5] (\x-.55,0) -- (\x+.55,0);
		\draw[fill,MidnightBlue] (\x,0) circle (3pt);
		\draw[fill,Maroon] (\x-.4,0) circle (3pt);
		\draw[fill,Maroon] (\x+.4,0) circle (3pt);
	}
	
	\node[below,yshift=-.1cm] at (-2,0) {$j=L-1$};
	\node[below,yshift=-.1cm] at (0,0) {$j=L$};
	\node[below,yshift=-.1cm] at (2,0) {$j=1$};
	
	\begin{scope}
	 \clip (0,0) -- (.19,.9) -- (.55,.7) -- (.53,2) -- (1.3,1.9)  -- (1.3,2.1) -- (1.1,2.05) -- (1.1,4.1) -- (2.5,3) -- (2.4,-.5) -- (0,0);
	 \draw[line width=1pt, MidnightBlue,decorate,
	decoration={snake,amplitude=1mm,segment length=10mm}] (1.7,3) -- (0,0);
	\end{scope}

	\draw[line width=2pt, looseness=7, Maroon] (-2.4,0) to[out=120,in=60] (-1.6,0);
	 
	\begin{scope}
	   \clip (-2.2,-.5) -- (-2.2,1.22) -- (-1.5,1.22) -- (-1.5,1.7) -- (.5,1.7) -- (.5,-.5) -- (-2.2,-.5);
	   \draw[line width=2pt, looseness=2.5, MidnightBlue] (-2,0) to[out=100,in=80] (0,0); 
	\end{scope}

	\begin{scope}
	   \clip (-1,-.5) -- (-1,1.19) -- (-.4,1.19) -- (-.3,1.19) -- (-.3,1.35) -- (-.6,1.35) -- (-.6,2) -- (.8,2) -- (.3,1.2) -- (.35,1) --  (.8,1.8) -- (1.1,1.3) --  (.8,.9) -- (.8,-.5) -- (-1,-.5);
	   \draw[line width=2pt, looseness=8, Maroon] (-.4,0) to[out=100,in=80] (+.4,0); 
	\end{scope}
	
	\begin{scope}
	\clip (2,.025) -- (-.5,.025) -- (.415,.85) -- (.415,.7) -- (.53,.7) -- (.53,1.93) -- (1.25,1.93) -- (1.25,2.05) -- (1.1,2.05) -- (1.1,2.975) -- (2.5,2.975) -- (2.5,2.07) -- (1.35,2.07) -- (1.35,1.955) -- (2,1.955) -- (2,.025);
	\draw[
	decoration={snake,amplitude=1mm,segment length=10mm},line width=1pt] decorate{(1.7,3)-- (0,0)} -- (.4,0) decorate{(.4,0) -- (2.1,3)} -- (1.7,3);
	\end{scope}
	
	\node[] at (1.95,3.3) {$\mscr{Y}_{d'}$};
	
	\begin{scope}
	\clip (2,.025) -- (-.5,.025) -- (.415,.85) -- (.415,.7) -- (.53,.7) -- (.53,1.93) -- (1.25,1.93) -- (1.25,2.05) -- (1.1,2.05) -- (1.1,2.975) -- (2.5,2.975) -- (2.5,2.07) -- (1.35,2.07) -- (1.35,1.955) -- (2,1.955) -- (2,.025);
	\draw[line width=1.1pt,
	decoration={snake,amplitude=1mm,segment length=10mm},decorate,MidnightBlue] (1.7,3) -- (0,0);
	\end{scope}
	
	\begin{scope}
	\clip (2,.025) -- (-.5,.025) -- (.415,.85) -- (.415,.7) -- (.53,.7) -- (.53,1.93) -- (1.25,1.93) -- (1.25,2.05) -- (1.1,2.05) -- (1.1,2.975) -- (2.5,2.975) -- (2.5,2.07) -- (1.35,2.07) -- (1.35,1.955) -- (2,1.955) -- (2,.025);
	\draw[line width=1.1pt,
	decoration={snake,amplitude=1mm,segment length=10mm},decorate,Maroon] (.4,0) -- (2.1,3);
	\end{scope}
	
	\path[pattern=north west lines, pattern color=OliveGreen,opacity=.4,
	decoration={snake,amplitude=1mm,segment length=10mm},line width=1pt] decorate{(1.7,3) to[color=MidnightBlue] (0,0)} -- (.4,0) decorate{(.4,0) to[color=Maroon](2.1,3)} -- (1.7,3);
	
    \begin{scope}
	    \clip (1.9,1.5) -- (2,1.1) -- (2.13,1.1) -- (2,1.5) -- (3,1) -- (3,-.5) -- (1,-.5)-- (1,1.5) -- (1.9,1.5);
	    \draw[line width=2pt, looseness=6, Maroon] (1.6,0) to[out=120,in=60] (+2.4,0);
	\end{scope}
	
	\draw[line width=2pt, looseness=3.5, MidnightBlue] (0,0) to[out=80,in=80] (2,0);

	\node[Maroon] at (-1.9,1.65) {\tiny $A^{\msf g}_{L-1}$};
	\node[Maroon] at (0,2.1) {\tiny $A^{\msf g}_{L}$};
	\node[Maroon,right] at (2,1.35) {\tiny $A^{\msf g}_{1}$};
	
	\node[MidnightBlue,right] at (-3.4,2.2) {\tiny $B^{\alpha}_{L-2,L-1}$};
	\node[MidnightBlue] at (-1,1.65) {\tiny $B^{\alpha}_{L-1,L}$};
	\node[MidnightBlue,right] at (1.7,1.9) {\tiny $B^{\alpha}_{L,1}$};
	\node[MidnightBlue,right] at (2.6,1.6) {\tiny $B^{\alpha}_{1,2}$};
	
	\begin{scope}
	    \clip (-1.8,0) -- (-2.15,1.35) -- (-2.4,1.25) -- (-2.4,1.37) -- (-2.1,1.47) -- (-3,3) -- (-3,-.5) -- (-1.8,0);
	    \draw[line width=2pt, looseness=3.5, MidnightBlue] (-2,0) to[out=95,in=70] (-4,0);
	\end{scope}
	
	\begin{scope}
	    \clip (1.8,0) -- (2.4,1) -- (2.55,.8) -- (2.65,.9) -- (2.5,1.07) -- (3,2) -- (3,-.5) -- (1.8,0);
	    \draw[line width=2pt, looseness=2.5, MidnightBlue] (2,0) to[out=70,in=90] (4,0);
	\end{scope}
	
\end{tikzpicture}
}

\newcommand{\ExampleOfTwoSimplex}{
    \begin{tikzpicture}
    \draw[line width=1.5pt,->] (0,0) -- (2,0);
    \draw[line width=1.5pt,->] (0,0) -- (0,2);
    \draw[line width=1.5pt,->] (0,0) -- (-1.5,-1.5);
    
    \path [pattern=north west lines, pattern color=DarkOrchid] (0,1.5) -- (1.5,0) -- (-1.1,-1.1);
    
    \draw[line width=2pt, Green,,postaction={decoration={markings,mark=at position .5 with {\arrow{>}}},decorate}] (0,1.5) node[left,black]{$v_0$} -- (1.5,0) node[above,black]{$v_1$};
    \draw[line width=2pt, Green,,postaction={decoration={markings,mark=at position .5 with {\arrow{>}}},decorate}] (0,1.5) -- (-1.1,-1.1) node[above,xshift=-.2cm,black]{$v_2$};
    \draw[line width=2pt, Green,,postaction={decoration={markings,mark=at position .5 with {\arrow{>}}},decorate}] (1.5,0) -- (-1.1,-1.1);
    
    \draw[fill,red] (0,1.45) circle (3pt);
    \draw[fill,red] (1.45,0) circle (3pt);
    \draw[fill,red] (-1.05,-1.05) circle (3pt);

    \end{tikzpicture}
}

\newcommand{\ProperSimplicialComplex}{

\begin{tikzpicture}

    \path[pattern=north west lines, pattern color=Maroon] (0,0) -- (-1,1) -- (-3,-1) -- cycle;
    
    \path[line width=.5pt, pattern=north west lines, pattern color=Blue] (0,0) -- (-1,1) -- (2,0) -- cycle;
    
    \draw[line width=2pt,Green,postaction={decoration={markings,mark=at position .5 with {\arrow{>}}},decorate}] (0,0)  -- (-3,-1);
    \draw[line width=2pt,Green,postaction={decoration={markings,mark=at position .5 with {\arrow{>}}},decorate}] (-1,1) -- (-3,-1);

    \draw[line width=2pt,Green,,postaction={decoration={markings,mark=at position .5 with {\arrow{>}}},decorate}] (0,0) -- (2,0);
    \draw[line width=2pt,Green,postaction={decoration={markings,mark=at position .5 with {\arrow{>}}},decorate}] (-1,1) -- (2,0);

    \draw[line width=2pt,Green,postaction={decoration={markings,mark=at position .55 with {\arrow{>}}},decorate}] (0,0) -- (-1,1);
    
    \draw[red,fill] (-.05,0)node[below,black] {$v_0$} circle (3pt);
    
    \draw[red,fill] (-1,1) node[above,black] {$v_1$} circle (3pt);
    
    \draw[red,fill] (-3,-1) node[left,black] {$v_2$} circle (3pt);
    
    \draw[red,fill] (2,0) node[right,black] {$v_3$} circle (3pt);
    
\end{tikzpicture}
}

\newcommand{\NonProperSimplicialComplex}{

\begin{tikzpicture}

    \path[pattern=north west lines, pattern color=Maroon] (0,0) -- (-1,1) -- (-3,-1) -- cycle;
    
    \path[line width=.5pt, pattern=north west lines, pattern color=Blue] (.5,-.5) -- (-.5,.5) -- (3,0) -- cycle;

    \draw[line width=2pt,Green,postaction={decoration={markings,mark=at position .5 with {\arrow{>}}},decorate}] (0,0)  -- (-3,-1);
    \draw[line width=2pt,Green,postaction={decoration={markings,mark=at position .5 with {\arrow{>}}},decorate}] (-1,1) -- (-3,-1);

    \draw[line width=2pt,Green,,postaction={decoration={markings,mark=at position .5 with {\arrow{>}}},decorate}] (.5,-.5) -- (3,0);
    \draw[line width=2pt,Green,postaction={decoration={markings,mark=at position .5 with {\arrow{>}}},decorate}] (-.5,.5) -- (3,0);
    \draw[line width=2pt,Green,postaction={decoration={markings,mark=at position .8 with {\arrow{<}}},decorate}] (-.5,.5) -- (.5,-.5);

    \draw[line width=2pt,Green,postaction={decoration={markings,mark=at position .8 with {\arrow{>}}},decorate}] (0,0) -- (-1,1);
    
    \draw[red,fill] (-.05,0)node[below,black] {$v_0$} circle (3pt);
    
    \draw[red,fill] (-1,1) node[above,black] {$v_1$} circle (3pt);
    
    \draw[red,fill] (-3,-1) node[left,black] {$v_2$} circle (3pt);

    \draw[red,fill] (.5,-.5) node[below,black] {$v_3$} circle (3pt);
    
    \draw[red,fill] (-.5,.5) node[above,black,xshift=.3cm] {$v_4$} circle (3pt);
    
    \draw[red,fill] (3,0) node[right,black] {$v_5$} circle (3pt);

\end{tikzpicture}

}

\newcommand{\SimplicialComplexForACylinder}{

\begin{tikzpicture}
    
    \draw[line width=2pt, Maroon,postaction={decoration={markings,mark=at position .43 with {\arrow{<}}},decorate},dashed] (2,-2) -- (-2,-2);
    
    \draw[line width=2pt, Green,postaction={decoration={markings,mark=at position .55 with {\arrow{<}}},decorate},dashed] (2,2) -- (-2,-2);

    \path[pattern=north east lines, pattern color=Maroon] (-2,-2) -- (2,-2) -- (0,-5);
    
    \path[pattern=north west lines, pattern color=Green] (2,2) -- (-2,-2) -- (2,-2) -- cycle;
    
    \path[pattern=north west lines, pattern color=Green] (-2,2) -- (-2,-2) -- (2,2) -- cycle;
    
    \path[pattern=north west lines, pattern color=Green] (2,2) -- (2,-2) -- (0,-5) -- cycle;
    \path[pattern=north west lines, pattern color=Green] (-2,2) -- (-2,-2) -- (0,-5) -- cycle;
    
    \path[pattern=north west lines, pattern color=Green] (2,2) -- (0,-1) -- (0,-5) -- cycle;
    
    \path[pattern=north west lines, pattern color=Green] (-2,2) -- (0,-1) -- (0,-5) -- cycle;
    
    \path[pattern=north east lines, pattern color=Maroon] (-2,2) -- (2,2) -- (0,-1) -- cycle;

    \draw[line width=2pt,Green,postaction={decoration={markings,mark=at position .55 with {\arrow{>}}},decorate}] (-2,-2) -- (-2,2);
    
    \draw[line width=2pt, Green,postaction={decoration={markings,mark=at position .55 with {\arrow{>}}},decorate}] (2,2) -- (2,-2);
    
    \draw[line width=2pt, Green,postaction={decoration={markings,mark=at position .55 with {\arrow{<}}},decorate}] (0,-5) -- (0,-1);
    
    \draw[line width=2pt, Green,,postaction={decoration={markings,mark=at position .55 with {\arrow{>}}},decorate}] (2,2) -- (0,-5);
    \draw[line width=2pt, Green,,postaction={decoration={markings,mark=at position .55 with {\arrow{>}}},decorate}] (-2,2) -- (0,-5);
    
    \draw[line width=2pt, Maroon,,postaction={decoration={markings,mark=at position .55 with {\arrow{>}}},decorate}] (-2,2) -- (2,2);
    
    \draw[line width=2pt, Maroon,postaction={decoration={markings,mark=at position .55 with {\arrow{>}}},decorate}] (2,2) -- (0,-1);
    \draw[line width=2pt, Maroon,postaction={decoration={markings,mark=at position .55 with {\arrow{>}}},decorate}] (-2,2) -- (0,-1);
    
    \draw[line width=2pt, Maroon,postaction={decoration={markings,mark=at position .55 with {\arrow{>}}},decorate}] (2,-2) -- (0,-5);
    \draw[line width=2pt, Maroon,postaction={decoration={markings,mark=at position .55 with {\arrow{>}}},decorate}] (-2,-2) -- (0,-5);

    \draw[red,fill] (-2,-2) node[left,black]{$v_0$} circle (3pt);
    \draw[red,fill] (-2,2) node[left,black]{$v_1$} circle (3pt);
    \draw[red,fill] (2,2) node[right,black]{$v_2$} circle (3pt);
    \draw[red,fill] (2,-2) node[right,black]{$v_3$} circle (3pt);
    \draw[red,fill] (0,-5) node[below,black]{$v_4$} circle (3pt);
    \draw[red,fill] (0,-1) node[right,black]{$v_5$} circle (3pt);
    
\end{tikzpicture}

}

\newcommand{\ZTwoSqPhaseDiagram}{
\begin{tikzpicture}
[scale=0.8, every node/.style={scale=0.8},baseline={([yshift=-.5ex]current bounding box.center)},rounded corners=1pt,very thick,black,decoration={
        markings,
        mark=at position 0.5 with {\arrow{>}}}]
        
        \begin{scope}
            \coordinate (A) at (-3,0);
            \coordinate (B) at (3,0);
            \coordinate (C) at (0,3*1.73205); %
            \coordinate (AB) at ($(A)!0.5!(B)$);
            \coordinate (AC) at ($(A)!0.5!(C)$);
            \coordinate (BC) at ($(B)!0.5!(C)$);
            \coordinate (ABC) at ($1/3*(A)+1/3*(B)+1/3*(C)$);
            
            \path[fill=blue!20,draw] (AB) -- (ABC) -- (AC) -- (A) -- (AB);
            \path[fill=green!20,draw] (AB) -- (ABC) -- (BC) -- (B) -- (AB);
            \path[fill=yellow!20,draw] (BC) -- (ABC) -- (AC) -- (C) -- (BC);
            
            \draw[line width=2pt] (A) -- (B) -- (C) -- (A);
            
            \node[circle,black,fill=blue!20] (CircleNode) at (A) {\scriptsize{6}};
            \node[circle,black,fill=green!40] (CircleNode) at (B) {\scriptsize{5}};
            \node[circle,black,fill=yellow!40] (CircleNode) at (C) {\scriptsize{1}};
            
            \draw[red,fill=red] (AB) circle (.15);
            \draw[red,fill=red] (AC) circle (.15);
            \draw[red,fill=red] (BC) circle (.15);
            \draw[red,fill=red] (ABC) circle (.08);
            
            \draw[red!80,line width=2.4pt] (AB) -- (ABC);
            \draw[red!80,line width=2.4pt] (AC) -- (ABC);
            \draw[red!80,line width=2.4pt] (BC) -- (ABC);
            
            \node[] at ($(A)-(0.5,0.5)$) {SPT${}_0$};
            \node[] at ($(B)+(0.5,-0.5)$) {SPT${}_1$};
            \node[] at ($(C)+(0,0.5)$) {SSB};
            
            \node[] at ($(AB)-(0,0.4)$) {$\mscr{C}_{56}$};
            \node[] at ($(AC)-(0.4,-0.3)$) {$\mscr{C}_{16}$};
            \node[] at ($(BC)+(0.4,0.3)$) {$\mscr{C}_{15}$};
            \node[] at ($(ABC)+(0,0.5)$) {$\mscr{C}_{156}$};

            \draw[line width=1pt,<->, Blue!80,looseness=.8] (2.8,-.3) to[out=210,in=-30] (-2.8,-.3);
            
            \node[Blue!80] at ($(AB)+(0,-1.3)$) {$k_1$};
    
            \draw[line width=1pt,<->,Blue!80,looseness=1.1] ($(AC)+(0.5,0.25)$) to[out=30,in=150] ($(BC)+(-0.5,0.25)$)node[above,xshift=-1cm,yshift=0.3cm]{$k_1$};
            
            \draw[line width=1pt, <->, Green,looseness=.8] (.4,{3*sqrt(3)}) to [out=-20, in=80] (3,0.4);
            \node[Green] at (3,3) {$k_2$};
            
            \draw[line width=1pt, <->, Green,looseness=1.2] (-.45,.4) to[out=160,in=265] (-1.4,2);
            \node[Green] at (-1.6,.9) {$k_2$};
            
            \begin{scope}[xscale=-1]
                \draw[line width=1pt, <->, Brown,looseness=.8] (.4,{3*sqrt(3)}) to [out=-20, in=80] (3,0.4);
            \node[Brown] at (3,3) {$k_3$};
            
            \draw[line width=1pt, <->, Brown,looseness=1.2] (-.45,.4) to[out=160,in=265] (-1.4,2);
            \node[Brown] at (-1.6,.9) {$k_3$};
            \end{scope}
            
        \end{scope}
        
        \draw[line width=1pt, <->, Red] (3,4) -- (6,4) ;
        \node[] at (4.5,4.35) {$\sigma_L$};
        
        \begin{scope}[xshift=9cm]
            \coordinate (A) at (-3,0);
            \coordinate (B) at (3,0);
            \coordinate (C) at (0,3*1.73205); %
            \coordinate (AB) at ($(A)!0.5!(B)$);
            \coordinate (AC) at ($(A)!0.5!(C)$);
            \coordinate (BC) at ($(B)!0.5!(C)$);
            \coordinate (ABC) at ($1/3*(A)+1/3*(B)+1/3*(C)$);
            
            \path[fill=cyan!20,draw] (AB) -- (ABC) -- (AC) -- (A) -- (AB);
            \path[fill=orange!20,draw] (AB) -- (ABC) -- (BC) -- (B) -- (AB);
            \path[fill=magenta!20,draw] (BC) -- (ABC) -- (AC) -- (C) -- (BC);
            
            \draw[line width=2pt] (A) -- (B) -- (C) -- (A);
            
            \node[circle,black,fill=cyan!40] (CircleNode) at (A) {\scriptsize{3}};
            \node[circle,black,fill=orange!40] (CircleNode) at (B) {\scriptsize{4}};
            \node[circle,black,fill=magenta!40] (CircleNode) at (C) {\scriptsize{2}};
            
            \draw[red,fill=red] (AB) circle (.15);
            \draw[red,fill=red] (AC) circle (.15);
            \draw[red,fill=red] (BC) circle (.15);
            \draw[red,fill=red] (ABC) circle (.08);
            
            \draw[red!80,line width=2.4pt] (AB) -- (ABC);
            \draw[red!80,line width=2.4pt] (AC) -- (ABC);
            \draw[red!80,line width=2.4pt] (BC) -- (ABC);
            
            \node[] at ($(A)-(0.5,0.5)$) {PSB${}_R$};
            \node[] at ($(B)+(0.5,-0.5)$) {PSB${}_D$};
            \node[] at ($(C)+(0,0.5)$) {PSB${}_L$};
            
            \node[] at ($(AB)-(0,0.4)$) {$\mscr{C}_{34}$};
            \node[] at ($(AC)-(0.4,-0.3)$) {$\mscr{C}_{23}$};
            \node[] at ($(BC)+(0.4,0.3)$) {$\mscr{C}_{24}$};
            \node[] at ($(ABC)+(0,0.5)$) {$\mscr{C}_{234}$};

            \draw[line width=1pt,<->, Blue!80,looseness=.8] (2.8,-.3) to[out=210,in=-30] (-2.8,-.3);
            
            \node[Blue!80] at ($(AB)+(0,-1.3)$) {$h_1$};

            \draw[line width=1pt,<->,Blue!80,looseness=1.1] ($(AC)+(0.5,0.25)$) to[out=30,in=150] ($(BC)+(-0.5,0.25)$)node[above,xshift=-1.1cm,yshift=0.3cm]{$h_1$};    
            
            \draw[line width=1pt, <->, Green,looseness=.8] (.4,{3*sqrt(3)}) to [out=-20, in=80] (3,0.4);
            \node[Green] at (3,3) {$h_2$};
            
            \draw[line width=1pt, <->, Green,looseness=1.2] (-.45,.4) to[out=160,in=265] (-1.4,2);
            \node[Green] at (-1.6,.9) {$h_2$};
            
            \begin{scope}[xscale=-1]
                \draw[line width=1pt, <->, Brown,looseness=.8] (.4,{3*sqrt(3)}) to [out=-20, in=80] (3,0.4);
            \node[Brown] at (3,3) {$h_3$};
            
            \draw[line width=1pt, <->, Brown,looseness=1.2] (-.45,.4) to[out=160,in=265] (-1.4,2);
            \node[Brown] at (-1.6,.9) {$h_3$};
            \end{scope}
        \end{scope}
    
    \end{tikzpicture}
}

\newcommand{\ZTwoSqPhaseDiagramSelfDualLines}{
\begin{tikzpicture}
[scale=0.8, every node/.style={scale=0.8},baseline={([yshift=-.5ex]current bounding box.center)},rounded corners=1pt,very thick,black,decoration={
        markings,
        mark=at position 0.5 with {\arrow{>}}}]
        
        \coordinate (A) at (-3,0);
        \coordinate (B) at (3,0);
        \coordinate (C) at (0,3*1.73205); %
        \coordinate (AB) at ($(A)!0.5!(B)$);
        \coordinate (AC) at ($(A)!0.5!(C)$);
        \coordinate (BC) at ($(B)!0.5!(C)$);
        \coordinate (ABC) at ($1/3*(A)+1/3*(B)+1/3*(C)$);
        
        \draw[line width=2pt] (A) -- (B) -- (C) -- (A);
        
        \draw[yellow!80,line width=2.4pt] (AB) -- (ABC);
        \draw[yellow!80,line width=2.pt,dashed] (C) -- (ABC);
        
         \draw[green!80,line width=2.4pt] (AC) -- (ABC);
         \draw[green!80,line width=2.pt,dashed] (B) -- (ABC);
        
         \draw[blue!80,line width=2.4pt] (BC) -- (ABC);
         \draw[blue!80,line width=2.4pt,dashed] (A) -- (ABC);
        
        \node[circle,black,fill=blue!40] (CircleNode) at (A) {\scriptsize{3}};
        \node[circle,black,fill=green!40] (CircleNode) at (B) {\scriptsize{4}};
        \node[circle,black,fill=yellow!40] (CircleNode) at (C) {\scriptsize{2}};
        
        \draw[yellow!40,fill=yellow!40] (AB) circle (.15);
        \draw[green!40,fill=green!40] (AC) circle (.15);
        \draw[blue!40,fill=blue!40] (BC) circle (.15);
        \draw[red!90,fill=red!90] (ABC) circle (.15);

        \node[] at ($(A)-(0.5,0.5)$) {PSB$_R$};
        \node[] at ($(B)+(0.5,-0.5)$) {PSB$_D$};
        \node[] at ($(C)+(0,0.5)$) {PSB$_L$};
        
        \node[] at ($(AB)-(0,0.4)$) {$\mscr{C}_{34}$};
        \node[] at ($(AC)-(0.4,-0.3)$) {$\mscr{C}_{23}$};
        \node[] at ($(BC)+(0.4,0.3)$) {$\mscr{C}_{24}$};
 
            \draw[line width=1pt,<->, Blue!80,looseness=.8] (2.8,-.3) to[out=210,in=-30] (-2.8,-.3);
            
            \node[Blue!80] at ($(AB)+(0,-1.3)$) {$h_1$};

            \draw[line width=1pt, <->, Green,looseness=.8] (.4,{3*sqrt(3)}) to [out=-20, in=80] (3,0.4);
            \node[Green] at (3,3) {$h_2$};

            \begin{scope}[xscale=-1]
                \draw[line width=1pt, <->, Brown,looseness=.8] (.4,{3*sqrt(3)}) to [out=-20, in=80] (3,0.4);
            \node[Brown] at (3,3) {$h_3$};
            
            \end{scope}
    \end{tikzpicture}
}

\newcommand{\ZTwoSqAbstractDualityActionOnLs}{
\begin{tikzpicture}
[scale=0.8, every node/.style={scale=0.8},baseline={([yshift=-.5ex]current bounding box.center)},rounded corners=1pt,very thick,black,decoration={
        markings,
        mark=at position 0.5 with {\arrow{>}}}]
        
        \begin{scope}
            \coordinate (A) at (-3,0);
            \coordinate (B) at (3,0);
            \coordinate (C) at (0,3*1.73205); %
            \coordinate (AB) at ($(A)!0.5!(B)$);
            \coordinate (AC) at ($(A)!0.5!(C)$);
            \coordinate (BC) at ($(B)!0.5!(C)$);
            \coordinate (ABC) at ($1/3*(A)+1/3*(B)+1/3*(C)$);
      
            \draw[line width=2pt] (A) -- (B) -- (C) -- (A);
            
            \node[circle,black,fill=blue!20] (CircleNode) at (A) {\scriptsize{6}};
            \node[circle,black,fill=green!40] (CircleNode) at (B) {\scriptsize{5}};
            \node[circle,black,fill=yellow!40] (CircleNode) at (C) {\scriptsize{1}};

            \node[] at ($(A)-(0.5,0.5)$) {$\mathcal L_6$};
            \node[] at ($(B)+(0.5,-0.5)$) {$\mathcal L_5$};
            \node[] at ($(C)+(0,0.65)$) {$\mathcal L_1$};

            \draw[line width=1pt,<->, Blue!80,looseness=.8] (2.8,-.3) to[out=210,in=-30] (-2.8,-.3);
            \node[Blue!80] at ($(AB)+(0,-1.3)$) {$k_1$};

            \draw[line width=1pt, <->, Green,looseness=.8] (.4,{3*sqrt(3)}) to [out=-20, in=80] (3,0.4);
            \node[Green] at (3,3) {$k_2$};

            \begin{scope}[xscale=-1]
                \draw[line width=1pt, <->, Brown,looseness=.8] (.4,{3*sqrt(3)}) to [out=-20, in=80] (3,0.4);
            \node[Brown] at (3,3) {$k_3$};
            
            \end{scope}
            
        \end{scope}
        
        \draw[line width=1pt, <->, Red] (3,4) -- (6,4) ;
        \node[] at (4.5,4.35) {$\sigma_L$};
        
        \begin{scope}[xshift=9cm]
            \coordinate (A) at (-3,0);
            \coordinate (B) at (3,0);
            \coordinate (C) at (0,3*1.73205); %
            \coordinate (AB) at ($(A)!0.5!(B)$);
            \coordinate (AC) at ($(A)!0.5!(C)$);
            \coordinate (BC) at ($(B)!0.5!(C)$);
            \coordinate (ABC) at ($1/3*(A)+1/3*(B)+1/3*(C)$);

            \draw[line width=2pt] (A) -- (B) -- (C) -- (A);
            
            \node[circle,black,fill=cyan!40] (CircleNode) at (A) {\scriptsize{3}};
            \node[circle,black,fill=orange!40] (CircleNode) at (B) {\scriptsize{4}};
            \node[circle,black,fill=magenta!40] (CircleNode) at (C) {\scriptsize{2}};
     
            \node[] at ($(A)-(0.5,0.5)$) {$\mathcal L_3$};
            \node[] at ($(B)+(0.5,-0.5)$) {$\mathcal L_4$};
            \node[] at ($(C)+(0,0.65)$) {$\mathcal L_2$};

            \draw[line width=1pt,<->, Blue!80,looseness=.8] (2.8,-.3) to[out=210,in=-30] (-2.8,-.3);
            
            \node[Blue!80] at ($(AB)+(0,-1.3)$) {$h_1$};

            \draw[line width=1pt, <->, Green,looseness=.8] (.4,{3*sqrt(3)}) to [out=-20, in=80] (3,0.4);
            \node[Green] at (3,3) {$h_2$};

            \begin{scope}[xscale=-1]
                \draw[line width=1pt, <->, Brown,looseness=.8] (.4,{3*sqrt(3)}) to [out=-20, in=80] (3,0.4);
            \node[Brown] at (3,3) {$h_3$};
            
            \end{scope}
        \end{scope}
    
    \end{tikzpicture}
}

\newcommand{\LineBetweenTriangles}[2]{
\tdplotsetmaincoords{#1}{#2}
\begin{tikzpicture}
[scale=0.8,every node/.style={scale=0.8},baseline={([yshift=-.5ex]current bounding box.center)},rounded corners=1pt,very thick,black,decoration={
        markings,
        mark=at position 0.5 with {\arrow{>}}}]
        
        \begin{scope}[tdplot_main_coords,shift={(0,4,0)}]
            \coordinate (A) at (-3,0,0);
            \coordinate (B) at (3,0,0);
            \coordinate (C) at (0,0,3*1.73205); %
            \coordinate (AB) at ($(A)!0.5!(B)$);
            \coordinate (AC) at ($(A)!0.5!(C)$);
            \coordinate (BC) at ($(B)!0.5!(C)$);
            \coordinate (ABC) at ($1/3*(A)+1/3*(B)+1/3*(C)$);

            \draw[red!80,line width=1.6pt] (AB) -- (ABC);
            \draw[red!80,line width=1.6pt] (AC) -- (ABC);
            \draw[red!80,line width=1.6pt] (BC) -- (ABC);

            \draw[line width=2pt] (A) -- (B) -- (C) -- (A);

            \draw[red!80,line width=2.0pt] (ABC) -- ($(ABC)+(0,-8,0)$);
            \draw[black!80,line width=1.0pt,dotted] (A) -- ($(A)+(0,-8,0)$);
            \draw[black!80,line width=1.0pt,dotted] (B) -- ($(B)+(0,-8,0)$);

             \node[circle,black,fill=blue!20] (CircleNode) at (A) {\scriptsize{6}};
             \node[circle,black,fill=green!40] (CircleNode) at (B) {\scriptsize{5}};
             \node[circle,black,fill=yellow!40] (CircleNode) at (C) {\scriptsize{1}};

             \node[] at ($(ABC) + (0.3,0,0.66)$) {\footnotesize{$\mscr{C}_{156}$}};

        \end{scope}

         \begin{scope}[tdplot_main_coords]
            \coordinate (A) at (-3,0,0);
            \coordinate (B) at (3,0,0);
            \coordinate (C) at (0,0,3*1.73205); %
            \coordinate (AB) at ($(A)!0.5!(B)$);
            \coordinate (AC) at ($(A)!0.5!(C)$);
            \coordinate (BC) at ($(B)!0.5!(C)$);
            \coordinate (ABC) at ($1/3*(A)+1/3*(B)+1/3*(C)$);

            \draw[line width=2pt,dashed] (A) -- (B) -- (C) -- (A);

        \end{scope}

        \node[] at ($(ABC)+(0.2,0.7,0.6)$) {\footnotesize{$\mscr{C}_{123456}$}};
        \draw[red,fill=red] (ABC) circle (0.2);
        \draw[orange,fill=orange] (A) circle (0.15);
        \draw[orange,fill=orange] (B) circle (0.15);
        \draw[purple!80,fill=purple!80] (AB) circle (0.15);
        \node[] at ($(A)-(0.2,0.7,0.6)$) {\footnotesize{$\mscr{C}_{36}$}};
        \node[] at ($(B)-(0.2,0.7,0.6)$) {\footnotesize{$\mscr{C}_{45}$}};
        \node[] at ($(AB)-(0,0,1)$) {\footnotesize{$\mscr{C}_{3456}$}};

        \begin{scope}[tdplot_main_coords,shift={(0,-4,0)}]
            \coordinate (A) at (-3,0,0);
            \coordinate (B) at (3,0,0);
            \coordinate (C) at (0,0,3*1.73205); %
            \coordinate (AB) at ($(A)!0.5!(B)$);
            \coordinate (AC) at ($(A)!0.5!(C)$);
            \coordinate (BC) at ($(B)!0.5!(C)$);
            \coordinate (ABC) at ($1/3*(A)+1/3*(B)+1/3*(C)$);

            \draw[red!80,line width=1.6pt] (AB) -- (ABC);
            \draw[red!80,line width=1.6pt] (AC) -- (ABC);
            \draw[red!80,line width=1.6pt] (BC) -- (ABC);

            \node[] at ($(ABC) + (0.3,0,0.66)$) {\footnotesize{$\mscr{C}_{234}$}};

            \draw[line width=2pt] (A) -- (B) -- (C) -- (A);
            
            \node[circle,black,fill=cyan!40] (CircleNode) at (A) {\scriptsize{3}};
            \node[circle,black,fill=orange!40] (CircleNode) at (B) {\scriptsize{4}};
            \node[circle,black,fill=magenta!40] (CircleNode) at (C) {\scriptsize{2}};

        \end{scope}

    \end{tikzpicture}
}

\newcommand{\SquareDuality}[2]{
\tdplotsetmaincoords{#1}{#2}
\begin{tikzpicture}
[scale=0.5,every node/.style={scale=0.8},baseline={([yshift=-.5ex]current bounding box.center)},rounded corners=1pt,very thick,black,decoration={
        markings,
        mark=at position 0.5 with {\arrow{>}}}]
        
        \begin{scope}[tdplot_main_coords,shift={(0,2,0)}]
            \coordinate (A) at (-3,0,0);
            \coordinate (B) at (3,0,0);
            \coordinate (C) at (0,0,3*1.73205); %
            \coordinate (AB) at ($(A)!0.5!(B)$);
            \coordinate (AC) at ($(A)!0.5!(C)$);
            \coordinate (BC) at ($(B)!0.5!(C)$);
            \coordinate (ABC) at ($1/3*(A)+1/3*(B)+1/3*(C)$);

            \draw[red!80,line width=1.6pt] (AB) -- (ABC);
            \draw[red!80,line width=1.6pt] (AC) -- (ABC);
            \draw[red!80,line width=1.6pt] (BC) -- (ABC);

            \draw[line width=2pt] (A) -- (B) -- (C) -- (A);

            \draw[black!80,line width=1.0pt,dotted] (B) -- ($(B)+(0,-4,0)$);
            \draw[black!80,line width=1.0pt,dotted] (A) -- ($(A)+(0,-4,0)$);
            \draw[black!80,line width=1.0pt,dotted] (C) -- ($(C)+(0,-4,0)$);

             \node[circle,black,fill=blue!20] (CircleNode) at (A) {\scriptsize{6}};
             \node[circle,black,fill=green!40] (CircleNode) at (B) {\scriptsize{5}};
             \node[circle,black,fill=yellow!40] (CircleNode) at (C) {\scriptsize{1}};
           
        \end{scope}

        \tdplotdrawarc[tdplot_main_coords,->,Green!80]{(0,0,0)}{1.5}{50}{350}{anchor=north}{};
        \path[tdplot_main_coords,draw,<->,Green!80] ($(C)+(0,0,0.5)$) .. controls ($(C)+(0,-1,1.8)$) and ($(C)+(0,-3,1.8)$) .. ($(C)+(0,-4,0.5)$);

        \begin{scope}[tdplot_main_coords,shift={(0,-2,0)}]
            \coordinate (A) at (-3,0,0);
            \coordinate (B) at (3,0,0);
            \coordinate (C) at (0,0,3*1.73205); %
            \coordinate (AB) at ($(A)!0.5!(B)$);
            \coordinate (AC) at ($(A)!0.5!(C)$);
            \coordinate (BC) at ($(B)!0.5!(C)$);
            \coordinate (ABC) at ($1/3*(A)+1/3*(B)+1/3*(C)$);

            \draw[red!80,line width=1.6pt] (AB) -- (ABC);
            \draw[red!80,line width=1.6pt] (AC) -- (ABC);
            \draw[red!80,line width=1.6pt] (BC) -- (ABC);

            \draw[line width=2pt] (A) -- (B) -- (C) -- (A);
            
            \node[circle,black,fill=cyan!40] (CircleNode) at (A) {\scriptsize{3}};
            \node[circle,black,fill=orange!40] (CircleNode) at (B) {\scriptsize{4}};
            \node[circle,black,fill=magenta!40] (CircleNode) at (C) {\scriptsize{2}};

        \end{scope}

    \end{tikzpicture}
}

\newcommand{\ZthreeZthreeTetrahedron}[2]{
\tdplotsetmaincoords{#1}{#2}
\begin{tikzpicture}
[scale=1.4,every node/.style={scale=0.8},baseline={([yshift=-.5ex]current bounding box.center)},rounded corners=1pt,very thick,black,decoration={
        markings,
        mark=at position 0.5 with {\arrow{>}}}]
        
        \begin{scope}[tdplot_main_coords]
            \coordinate (A) at (1,1,1);
            \coordinate (B) at (-1,-1,1);
            \coordinate (C) at (-1,1,-1);
            \coordinate (D) at (1,-1,-1);
        
            \draw[black!80,fill=Emerald!50,opacity=0.4,line width=1.2pt] (A) -- (B) -- (C) -- (A);
            \draw[black!80,fill=Emerald!50,opacity=0.4,line width=1.2pt] (A) -- (D) -- (C) -- (A);
            \draw[black!80,fill=Emerald!50,opacity=0.4,line width=1.2pt] (B) -- (C) -- (D) -- (B);
            
            \node[circle,black,fill=blue!20] (CircleNode) at (A) {\scriptsize{7}};
            \node[circle,black,fill=green!40] (CircleNode) at (B) {\scriptsize{8}};
            \node[circle,black,fill=yellow!40] (CircleNode) at (C) {\scriptsize{1}};
            \node[circle,black,fill=orange!20] (CircleNode) at (D) {\scriptsize{6}};

        \end{scope}

        \begin{scope}[shift={(4,0)},tdplot_main_coords]
            \coordinate (A) at (1,1,1);
            \coordinate (B) at (-1,-1,1);
            \coordinate (C) at (-1,1,-1);
            \coordinate (D) at (1,-1,-1);
        
            \draw[black!80,fill=Fuchsia!40,opacity=0.4,line width=1.2pt] (A) -- (B) -- (C) -- (A);
            \draw[black!80,fill=Fuchsia!40,opacity=0.4,line width=1.2pt] (A) -- (D) -- (C) -- (A);
            \draw[black!80,fill=Fuchsia!40,opacity=0.4,line width=1.2pt] (B) -- (C) -- (D) -- (B);
            
            \node[circle,black,fill=red!30] (CircleNode) at (A) {\scriptsize{2}};
            \node[circle,black,fill=purple!60] (CircleNode) at (B) {\scriptsize{3}};
            \node[circle,black,fill=cyan!40] (CircleNode) at (C) {\scriptsize{4}};
            \node[circle,black,fill=magenta!20] (CircleNode) at (D) {\scriptsize{5}};

        \end{scope}

    \end{tikzpicture}
}

\newcommand{\TMatrixFromThreeStrings}{
    \begin{tikzpicture}[scale=1.2]
    
    \begin{scope}[rotate=-90]
        \draw[line width=2pt,looseness=1.75,color=Maroon,postaction={decoration={markings,mark=at position .5 with {\arrow{<}}},decorate}] (-3,2.5) to[out=210,in=150] (-3,-2);
        \draw[line width=2pt,looseness=1.5,color=NavyBlue,postaction={decoration={markings,mark=at position .45 with {\arrow{>}}},decorate}] (-3,1.75) to[out=210,in=145] (-3,-2);
        \draw[line width=2pt,looseness=1.5,color=OliveGreen,postaction={decoration={markings,mark=at position .3 with {\arrow{<}}},decorate}] (-3,.5) to[out=210,in=145] (-3,-2);
        
        \draw[line width=5pt,Gray!40] (-3,3) -- (-3,-3);
    \end{scope}
    
    \node[color=Maroon] at (1.01,4.36) {\scriptsize $S_d(\ell_3)$};
    \node[color=NavyBlue] at (0.46,3.9)  {\scriptsize $S_{d}(\ell_2)$};
    \node[color=OliveGreen] at (-0.4,3.35) {\scriptsize $S_d(\ell_1)$};
    \end{tikzpicture}
}

\newcommand{\InsertionOfSymmetryOperators}{
\begin{tikzpicture}[scale=.85]
    
    \draw[line width=2pt,fill=gray!20] (-3,3) -- (3,3) -- (3,-3) -- (-3,-3) -- (-3,3);
    
    \draw[line width=2pt,color=Maroon,postaction={decoration={markings,mark=at position .25 with {\arrow{>}}},decorate}] (-3,0) -- (3,0);
    
    \draw[line width=2pt,color=NavyBlue,postaction={decoration={markings,mark=at position .25 with {\arrow{>}}},decorate}] (0,-3) -- (0,3);
    
    \node[color=Maroon] at (-1.5,.5) {$\mcal{U}_{\msf h}$};
    
    \node[color=NavyBlue] at (.5,-1.5) {$\mcal{U}_{\msf g}$};
    
    \begin{scope}[shift={(-3.4,-3.4)}]
        \draw[line width=1pt,->] (0,0) -- (1.5,0) node [below,xshift=-.75cm]{space};
        \draw[line width=1pt,->] (0,0) -- (0,1.5)node [left,xshift=-.3cm,yshift=-.2cm,rotate=90]{time};
    \end{scope}
\end{tikzpicture}
}

\newcommand{\InsertionOfSymmetryOperatorUg}{
\begin{tikzpicture}[scale=.70]
    
    \draw[line width=2pt,fill=gray!20] (-3,3) -- (3,3) -- (3,-3) -- (-3,-3) -- (-3,3);
    
    \draw[line width=2pt,color=NavyBlue,postaction={decoration={markings,mark=at position .5 with {\arrow{>}}},decorate}] (0,-3) -- (0,3);

    \begin{scope}[shift={(-3.4,-3.4)}]
        \draw[line width=1pt,->] (0,0) -- (1.5,0) node [below,xshift=-.75cm]{\footnotesize space};
        \draw[line width=1pt,->] (0,0) -- (0,1.5)node [left,xshift=-.3cm,yshift=-.2cm,rotate=90]{\small time};
    \end{scope}
    
    \draw[fill] (-1.5,1.5) circle (3pt);
    \draw[fill] (1.5,1.5) circle (3pt);
    
    \draw[line width=.5pt,dashed,postaction={decoration={markings,mark=at position .25 with {\arrow{>}}},decorate},postaction={decoration={markings,mark=at position .75 with {\arrow{>}}},decorate}] (-1.5,1.5) to (1.5,1.5);
    
    \node[color=NavyBlue] at (.75,0.0) {\small $\mcal{U}_{\msf g}$};
    
    \node[above] at (-1.5,1.5) {\footnotesize $\mcal{O}_{\alpha}$};
    \node[above] at (1.5,1.5) {\footnotesize $\msf{R}_\alpha(\msf g)\cdot \mcal{O}_{\alpha}$};
\end{tikzpicture}
}

\newcommand{\SymmetryBreakingWormholes}{
    \begin{tikzpicture}[scale=.65]
    
        \path[fill=Gray!50!] (-2,0) to[out=30,in=180] (+1,1) to[out=0,in=210] (+4,0) to[out=30,in=90] (+9,-1) to[out=-90,in=0] (4,-4) to[out=180,in=-30] (0,-3) to[out=150,in=210] (-2,0);
        
        \draw[dashed] (2.5,-1) to[out=245,in=300,looseness=.5] (-.5,-1)     to[out=65,in=115,looseness=.5] (2.5,-1);
        \path[fill=MidnightBlue!70!] (2.5,-1) to[out=245,in=300,looseness=.5] (-.5,-1)     to[out=65,in=115,looseness=.5] (2.5,-1);
            
        \draw[dashed] (7.5,-1.5) to[out=245,in=300,looseness=.5] (4.5,-1.5)     to[out=65,in=115,looseness=.5] (7.5,-1.5);
        \path[fill=MidnightBlue!70!] (7.5,-1.5) to[out=245,in=300,looseness=.5] (4.5,-1.5)     to[out=65,in=115,looseness=.5] (7.5,-1.5);

        \draw[fill=MidnightBlue!50!] (+3,-2.8) ellipse (1.5cm and .5cm);

        \begin{scope}
         \clip[] (-.5,-1) -- (-.5,-2) -- (2.5,-2) -- (2.5,-1) -- (-.5,-1);
        \draw[] (2.5,-1) to[out=245,in=300,looseness=.5] (-.5,-1)     to[out=65,in=115,looseness=.5] (2.5,-1);
        \end{scope}
        
        \begin{scope}
         \clip[] (4.5,-1.5) -- (4.5,-2.5) -- (7.5,-2.5) -- (7.5,-1.5) -- (4.5,-1.5);
        \draw[] (7.5,-1.5) to[out=245,in=300,looseness=.5] (4.5,-1.5)     to[out=65,in=115,looseness=.5] (7.5,-1.5);
        \end{scope}

        \path[fill=LimeGreen!60!,opacity=.5] (4.5,-1.5) to [out=90,in=90,looseness=6] (2.5,-1) to[out=245,in=300,looseness=.5] (-.5,-1)    to[out=90,in=90,looseness=3] (7.5,-1.5) to[out=245,in=300,looseness=.5] (4.5,-1.5);
        
        \draw[opacity=.6] (4.5,-1.5) to[out=90,in=90,looseness=6] (2.5,-1);
        \draw[opacity=.6] (7.5,-1.5) to[out=90,in=90,looseness=3] (-.5,-1);
        
        \draw[->,line width=1pt] (-4,1) to[out=210,in=180,looseness=1.5] (-2.5,-.5);
        \node[align=center] at (-4,1.7) {\small $\msf G$ symmetric theories};
        
        \draw[->,line width=1pt] (10.5,2) to[out=-30,in=0,looseness=1.5] (7.4,1);
        \node[align=center] at (10.5,2.6) {\small$\msf G'$ symmetric theories};
        
        \node[] at (+.9,-1) {\scriptsize$\mcal{L}$};
        \node[] at (+5.95,-1.5) {\scriptsize$\mcal{L}'$};
        \node[] at (+3,-2.8) {\scriptsize$\mcal{L}''$};
        \node[align=center] at (+3.4,3.6) {\footnotesize Symmetry-breaking \\ \footnotesize region};
    \end{tikzpicture}
    }

\newcommand{\ket}[1]{| #1\rangle}

\newcommand{\bs}{\boldsymbol}
\newcommand{\ms}{\mathsf}
\newcommand{\wh}{\widehat}
\newcommand{\mc}{\mathcal}

\newcommand{\wt}{\widetilde}

\newcommand{\pd}{\phantom{\dagger}}

\newcommand{\tenofo}[1]{\text{\normalfont #1}}

\newcommand{\mfk}[1]{\mathfrak{#1}}
\newcommand{\mbb}[1]{\mathbb{#1}}
\newcommand{\mcal}[1]{\mathcal{#1}}
\newcommand{\mscr}[1]{\mathscr{#1}}
\newcommand{\msf}[1]{\mathsf{#1}}
\newcommand{\mbs}[1]{\boldsymbol{#1}}
\newcommand{\sd}{\mathsf{d}}
\newcommand{\msg}{\ms{g}}
\newcommand{\msh}{\ms{h}}

\newcommand{\dual}[1]{\mathscr D_{#1}}

\newcommand{\dualhd}[1]{\widehat{\mathscr D}_{#1}\cdot}

\newcommand{\boldit}[1]{\textit{\textbf{#1}}}

\newcommand{\newl}{\medskip\noindent}

\newcommand{\U}{\tenofo{U}}

\hypersetup{
    colorlinks = true, linktoc=page,
    linkcolor={ao!80!black},
    citecolor={red!50!black},
    urlcolor={dcyan!80!black}
}
\definecolor{bleu}{rgb}{0.0, 0.5, 0.69}
\definecolor{bred}{rgb}{0.8, 0.25, 0.33}
\definecolor{cordovan}{rgb}{0.54, 0.25, 0.27}
\definecolor{ceru}{rgb}{0.03, 0.27, 0.49}
\definecolor{dcyan}{rgb}{0.0, 0.55, 0.55}
\definecolor{chestnut}{rgb}{0.73, 0.31, 0.28}
\definecolor{indigo}{rgb}{0.0, 0.25, 0.42}
\definecolor{dmb}{rgb}{0.38, 0.51, 0.71}
\definecolor{brred}{rgb}{0.8, 0.25, 0.33}
\definecolor{agreen}{rgb}{0.55, 0.71, 0.0}
\definecolor{ao}{rgb}{0.0, 0.5, 0.0}

\usepackage{colortbl}
\DeclareSymbolFont{usualmathcal}{OMS}{cmsy}{m}{n}
\DeclareSymbolFontAlphabet{\mathcal}{usualmathcal}

\begin{document}

\begin{center}
    \Large\bf {Topological Holography}: Towards a Unification of \\  Landau and Beyond-Landau Physics
\end{center}

\begin{center}
{\bf Heidar Moradi},\textsuperscript{\it a,b,c}
{\bf Seyed Faroogh Moosavian},\textsuperscript{\it d} and
{\bf Apoorv Tiwari}\textsuperscript{\it e}\blfootnote{\hspace{-.65cm}\href{mailto:h.moradi@kent.ac.uk}{h.moradi@kent.ac.uk} \\ \href{mailto:sfmoosavian@gmail.com}{sfmoosavian@gmail.com} \\ \href{mailto:t.apoorv@gmail.com}{t.apoorv@gmail.com}}
\end{center}

\begin{enumerate}[label=\alph*,itemsep=.5pt]
    {\small\it 
    \item Physics and Astronomy, Division of Natural Sciences, University of Kent, Canterbury CT2 7NZ, United Kingdom
    
    \item Department of Applied Mathematics and Theoretical Physics, Centre for Mathematical Sciences, Wilberforce Road, Cambridge CB3 0WA, United Kingdom

    \item The Cavendish Laboratory, Department of Physics,
   19 J J Thomson Avenue, Cambridge CB3 0HE, United Kingdom
   
   \item Department of Physics, McGill University, Ernest Rutherford Physics Building, 3600 Rue University, Montr\'eal, QC H3A 2T8, Canada
   
   \item Department of Physics, KTH Royal Institute of Technology, Stockholm, 106 91 Sweden}
\end{enumerate}

\abstract{
We outline a holographic$^\star$ framework that attempts to unify Landau and beyond-Landau paradigms of quantum phases and phase transitions.
Leveraging a modern understanding of symmetries as topological defects/operators, the framework uses a topological order to organize the space of quantum systems with a global symmetry in one lower dimension.
The global symmetry naturally serves as an input for the topological order. 
In particular, we holographically construct a {\it String Operator Algebra} (SOA) which is the building block of symmetric quantum systems with a given symmetry $\ms G$ in one lower dimension. 
This exposes a vast web of dualities which act on the space of $\ms G$-symmetric quantum systems. 
The SOA facilitates the classification of gapped phases as well as their corresponding order parameters and fundamental excitations, while dualities help to navigate and predict various corners of phase diagrams and analytically compute universality classes of phase transitions.
A novelty of the approach is that it treats conventional Landau and unconventional topological phase transitions on an equal footing, thereby providing a holographic unification of these seemingly-disparate domains of understanding.
We uncover a new feature of gapped phases and their multi-critical points, which we dub \textit{fusion structure}, that encodes information about which phases and transitions can be dual to each other.
Furthermore, we discover that self-dual systems typically posses emergent non-invertible, i.e., beyond group-like symmetries.
We apply these ideas to $1+1d$ quantum spin chains with finite Abelian group symmetry, using topologically-ordered systems in $2+1d$.
We predict the phase diagrams of various concrete spin models, and analytically compute the full conformal spectra of non-trivial quantum phase transitions, which we then verify numerically.

\newl {\scriptsize $\star$ Not to be confused with the holographic principle in contexts such as the AdS/CFT Correspondence.}
}
\pagenumbering{gobble}

\newpage

\pagenumbering{arabic}

\tableofcontents

\section{Introduction}
\label{sec:intro}

Condensed-matter physics is the study of phases of matter and phase transitions between them. 
A fundamental endeavor of condensed-matter theory is to develop a framework within which possible phases of matter can be classified and characterized. 
Historically, the most successful framework in this regard has been the so-called {\it Landau theory}\cite{Landau1, Ginzburg2009, landau1958annd} within which symmetry and its spontaneous-breaking patterns play a key role. 
The Landau-Ginzburg-Wilson mean field theory provides a formalism capable of describing many aspects of conventional phases based on the existence of local order parameters. 
Corresponding quantum phase transitions occur between ordered and disordered phases where the symmetry of the ground state of the ordered phase is necessarily a subgroup of that of the disordered phase.
Typical examples of such transitions are the $\mathbb Z_{2}$ and $\ms{U}(1)$ symmetry breaking Ising and $\ms{XY}$ transitions, respectively \cite{sachdev_2011}.

\smallskip The last four decades have seen a rapidly growing understanding of quantum phases and phase transitions that lie beyond Landau's framework.
Among these, an important class are so-called topological phases of matter, which are equivalence classes of locally indistinguishable gapped quantum systems (more precisely quantum liquids \cite{ZengWen201406}) that are distinguished by their global or topological properties.
Topological phases fall into two broad categories known as topologically ordered and symmetry protected topological (SPT) phases of matter respectively.
The fractional quantum Hall phase \cite{FQH_1982, Laughlin_1983, Kalmeyer_Laughlin} is a well-known example of the former category.
Such phases have highly entangled liquid-like ground states which evade conventional ordering at zero temperature.
They are instead distinguished by their patterns of long range entanglement \cite{Kitaev_preskill_TEE_2006, Levin_Wen_TEE_2006, Dong_2008, Ye_2018}  which manifest in fractionalized (anyonic) excitations with topological correlations \cite{Wen_Zee_1992, Levin_Wen_2005, Tiwari_2017, Putrov:2016qdo}.
A defining property of topologically ordered systems is the existence of ground state degeneracy when put on a topologically non-trivial manifold \cite{Wen:1989iv,Wen:1989zg,WenZee1998}.
The multiplet of ground states form projective representations of the mapping class (modular) group of the spatial manifold, which can serve as a diagnostic of the phase \cite{Keski-VakkuriWen1993,GroverTurnerEtal2012,MoradiWen2015,Moradi_2014}. 
A paradigmatic example of a topologically ordered system is the deconfined phase of the $\mbb Z_2$ gauge theory \cite{Wegner:1971app, Fradkin_Shenker_1979, Kogut_RMP_1979, Kitaev199707}, which is expected to be realized in frustrated magnets \cite{Read_Sachdev_1991, Wen_1991, Sachdev_1992, Savary_2016} and play a crucial role in the physics of high temperature superconductors \cite{Anderson_1987,Kivelson_1987, Baskaran_1988, Rokhsar_Kivelson_1988, Read_Chakraborty_1989, Senthil_2000}.
This is the simplest example of a general class of topological orders known as topological finite group gauge theories \cite{DijkgraafWitten199004, deWildPropitius:1995cf, Yidun_2013,KongLanWenZhang202003}.

\smallskip On the other hand, SPTs---as the name suggests---realize global symmetries in topologically distinct ways which is manifest in their bulk topological response to background symmetry gauge fields \cite{SPT_chiu, ChenGuLiuWen201106} and in the appearance of anomalous quantum systems on their boundary\cite{ryu_anomaly, LevinGu_2012, Shinsei_2013, Hsieh_2016, BoHan_2017, Tiwari_2018, witten2016fermion}.
These phases are trivial if symmetry is explicitly broken.
The earliest and perhaps best known examples of SPTs are free fermion topological insulators and superconductors \cite{Kane_Mele_2005, Kane_MeleII_2005, Bernevig_Hughes_Zhang_2006, Fu_Kane}, which are classified using K-theory in what is known as the ten-fold way classification \cite{Kitaev:2009mg, Schnyder_2008, Ryu_2010}. 
Interactions can alter the non-interacting classification  \cite{FidkowskiKitaevSPT2011, Fidkowski_2010, Senthil_Wang_2014, Mudry_2015}.
The interacting classification of fermionic topological phases is formulated in terms of intricate mathematical structures such as super-group cohomology \cite{GuWen_2014, Bhardwaj_2017} and cobordism theory \cite{kapustin2014symmetry, kapustin2015fermionic}.
On the other hand, since bosons generically condense into the lowest energy level, interactions are necessary to stabilize many-body gapped bosonic systems. %
In $1+1d$, bosonic SPTs are classified by projective representations of the symmetry group \cite{ PollmannSPT2009A2012, Chen1604, ChenGuWen201004}, which can be generalized to higher dimensions using the cohomology of groups \cite{ChenGuLiuWen201106}.
The low-energy properties of topological phases of matter are expected to be described by topological quantum field theories.

\smallskip Within the Landau paradigm, second-order or continuous quantum phase transitions occur between two phases such that the symmetry preserved by the (partially) ordered phase is necessarily a subgroup of the symmetry preserved by the disordered phase.
It is now known that various kinds of non-Landau transitions may occur in quantum systems.
These could happen if the phases on either side of the transition are themselves beyond the Landau classification scheme \cite{bais2001quantum, bais2009condensate, DH_Lee, ChenWangLuLee201302} or if the symmetry group on one side of the transition is not a subgroup of the symmetry on the other side.
The former kind of transitions can be realized for instance between topological phases or between topological and non-topological phases of matter.
The latter are known as deconfined quantum critical (DQC) transitions \cite{DQCP1, DQCP2}. 
Such transitions take place when defects or the disorder parameters of the broken symmetry on one side of the transition host the order parameter of the symmetry which is broken on the other side of the transition \cite{levin2004deconfined}.
While the phases on either side of such a transition are captured by Landau's classification scheme, the transition itself is beyond Landau.
A prototypical example of the deconfined transition is realized between the valence bond solid and antiferromagnetic Neel state in a spin-rotation-symmetric spin model on a square lattice.
The valence bond solid state breaks spatial four-fold rotation symmetry while the Neel state breaks spin-rotation symmetry. 
From a modern perspective, such transitions are understood to be realized in quantum systems that have mixed anomalies between internal and spatial symmetries, i.e., Lieb-Schultz-Mattis constraints \cite{LSM_1961, Oshikawa_2000, Hastings_2004 ,Shinsei_LSM, watanabe_anomaly_LSM, Aksoy:2021uxb}.

\smallskip As mentioned above, symmetry plays a very important role in organizing our understanding about the kinds of gapped phases and phase transitions realized in quantum systems.
Despite its long and illustrious history, the understanding of symmetries in quantum matter is very much an evolving subject.
In fact, there have been numerous developments over the past decade, that have led to new perspectives on the question: \emph{what are global symmetries of quantum systems?} \cite{gaiotto2015generalized, Kapustin:2013uxa, thorngren2019fusion}. 
Conventionally, a global symmetry is defined via a collection of operators which (i) act on all of space, (ii) commute with the Hamiltonian and (iii) satisfy group-like {fusion}/composition rules.
Mathematically, such symmetries are described by groups.
Operators charged under such a symmetry are zero-dimensional (point-like) and transform in representations of the symmetry group.
The low-energy description of quantum systems are usually formulated in terms of Euclidean quantum field theories, which possess emergent spacetime rotation symmetry in addition to other global symmetries. 
Within the low-energy description, there is no preferred time direction and hence no natural notion of all of space.
The symmetry operators may therefore be defined on any codimension-1 (i.e., $\#$ of spacetime dimensions $-1$) hypersurface in spacetime. 
The condition of commutativity with the Hamiltonian generalizes to `commutativity' with the stress tensor---this amounts to the general property that \emph{symmetry operators are topological}. 
This perspective begets the generalization that any topological subsector (subalgebra) of operators in a quantum system embodies a generalized symmetry structure. 
The topological operators within a theory need not be limited to codimension-1 operators that satisfy fusion rules representative of a group. 
Instead, in general, the set of topological operators may have higher codimensions \cite{gaiotto2015generalized}  and satisfy fusion rules that vastly generalize the notion of groups \cite{Kapustin:2013uxa, Cordova2019_1, Cordova2019_2, Delcamp:2018wlb, Delcamp:2019fdp, PetkovaZuber2001, Fuchs:2002, Fuchs:2007tx, BhardwajTachikawa201704, Wen_higher, thorngren2019fusion, ChangYingHsuanShaoWangYin201802, Thorngren_anyon, Kaidi:2021xfk, Choi:2021kmx, Yang:2021xob, Heidenreich:2021xpr, Roumpedakis:2022aik, Bhardwaj:2022yxj, Choi:2022zal}.
Furthermore, global symmetries may have 't Hooft anomalies, which put strong non-perturbative constraints on the low energy phases/ground states realizable within the quantum system. 
Collectively, the mathematical framework that describes topological operators and therefore generalized symmetries in quantum field theory is that of higher fusion categories \cite{Bhardwaj:2022yxj}.

\smallskip This modern perspective of global symmetries provides a number of insights applicable to the program of classifying and characterizing quantum matter.
For instance, Landau's symmetry breaking paradigm can be expanded to incorporate breaking of generalized symmetries \cite{gaiotto2015generalized,Lake201802,McGreevy202204}.
In particular, a so-called $p$-form symmetry corresponding to a finite Abelian group $\ms G$ contains co-dimension-$(p+1)$ topological operators that act on $p$-dimensional charged objects which transform in non-trivial representations of $\ms G$.
A system with spontaneously-broken $0$-form $\ms G$ symmetry has a ground state degeneracy of $|\ms G|$ on a connected spatial manifold.
Similarly, a system with spontaneously broken $1$-form $\ms G$ symmetry has a ground state degeneracy of $b_1^{|\ms G|}$ where $b_1$ is the number of independent non-contractible loops (1st Betti number) of the spatial manifold. This is the well-known topological ground-state degeneracy in topologically-ordered systems related to anyonic excitations.
For this reason, it has been noted \cite{NussinovOrtiz200605,NussinovOrtiz200702,Lake201802,Hsin_Lam_Seiberg} that 1-form symmetry breaking is closely related to topological order. Recently, these ideas have been used to propose a partial incorporation of topological order into Landau's classification scheme \cite{McGreevy202204}. Technically, SPT phases do not fit into the Landau paradigm as there are no spontaneously symmetry-breaking (even for generalized symmetries).

\smallskip In this paper, we describe a different approach toward the unification of Landau and beyond-Landau paradigms. 
Our approach exploits the topological nature of global symmetries to decouple the global-symmetry-related features of a symmetric quantum system from its local physics/dynamics.
The symmetry operators and their action on charged operators can be  holographically encapsulated in a topologically-ordered system that lives in one higher dimension.
The action of symmetry operators on charged operators is encoded in the braiding of topological defects (e.g., anyons in $2+1$ dimensions) of the topologically-ordered system.
From this point of view, different phases such as Landau symmetry-breaking phases, SPTs as well as (Abelian) topologically-ordered phases all arise on the same footing, i.e, as gapped boundaries of a topological order in one higher dimension.
Furthermore, the bulk topological order can be thought of as a theoretical gadget that allows one to conveniently discover numerous non-perturbative statements about the space of symmetric quantum systems.
These include the classification of gapped phases, order parameters of each phase as well as non-trivial dualities that act on the space of theories.
The dualities descend from the internal symmetries of the bulk topological order, which act on boundary theories through topological domain walls, and reveal rich hidden structures of the space of theories.
We develop tools to exploit these dualities to constrain and construct the phase diagram and compute its various features efficiently through a study of the associated topological order.   
In order to provide concrete realizations that exemplify the utility of this approach in the simplest physically-relevant setting, we restrict ourselves to the study of $1+1d$ quantum systems with finite Abelian group symmetry.

\smallskip Ideas similar to the ones developed in this work have been around for quite some time. %
For example, in \cite{Witten200307} it was shown that there is an $\tenofo{SL}(2,\mathbb Z)$ action on the space of three-dimensional conformal field theories with global $\U(1)$ symmetry, reminiscent of the dualities discussed in this paper. It was shown that these dualities can be understood as originating from the $\tenofo{SL}(2,\mathbb Z)$ electric-magnetic dualities in a four-dimensional $U(1)$ gauge theory \cite{KapustinTikhonov200904, GaiottoWitten200807, FuchsSchweigertWaldorf200703,Fuchs:2007tx,SarkissianSchweigert200810}. More concretely, these dualities are implemented as topological domains walls that are brought to the boundary, giving rise to a mapping of boundary theories.

\smallskip More recent ideas related to using topological orders in one higher dimension to study symmetric quantum systems have appeared in a number of recent works \cite{Witten199812, Freed:2018cec, GaiottoKulp202008, Wen_higher, Wen_Fusion,Ji:2019ugf, Thorngren_anyon, Chatterjee:2022kxb, Chatterjee:2022tyg, Apruzzi:2021nmk, Bhardwaj:2020ymp, Apruzzi:2022dlm,Lootens:2021tet, Aasen_2016, Aasen:2020jwb,HuWan202007,HuHuangHungWan202109,KongZheng201705,KongZheng201905,KongZheng201912,KongWenZheng202108,XuZhang202205,KaidiOhmoriZheng202209}. 
Below, we compare and contrast our work with these related directions.

\smallskip %
In \cite{Wen_higher, Wen_Fusion, Ji:2019ugf, Chatterjee:2022jll, Chatterjee:2022kxb, Chatterjee:2022tyg}, the notion of categorical symmetry was introduced which combines a finite symmetry and its dual symmetry which typically arises via the process of gauging the original symmetry.
Relatedly, these approaches treat the order and disorder parameter on an equal footing.
The similarity with the present work stems from the fact that the categorical symmetry can be organized into a topological order in one dimension higher. 
An important difference between the approach employed in these works and ours is that we do not consider the notion of ``dual symmetry".
Instead, we simply fix a finite global symmetry and study the space/algebra of local symmetric operators with respect to the given symmetry.
Upon doing so, we naturally discover the algebra of topological operators in a topological order in one higher dimension.
Furthermore, we describe a complimentary approach wherein we start from a topological order in 2+1 dimensions and derive symmetric pseudo-spin chains on its boundary. 
We show that the space/algebra of local operators acting on the pseudo-spin chain is precisely the algebra of local symmetric operators with respect to a symmetry structure that is read-off from the 2+1 dimensional topological order.
This algebra is related to the algebra of patch operators discussed in \cite{Wen_Fusion, Chatterjee:2022kxb}.

\smallskip %
Another related direction is the so-called Topological Wick Rotation \cite{KongZheng201705,KongZheng201905,KongZheng201912,KongWenZheng202108, Kong:2020iek}. The physical picture of Topological Wick rotation is related to considering a topological order in $d+1$-dimensions and restricting the theory to a $d$-dimensional spatial slice at a fixed time. Moreover the spatial slice is considered to be open. Then the authors argue that one may ``Wick rotate" one of the spatial directions and dualize the topological order to a quantum liquid in $(d-1)+1$ dimensions. The category of  operators then has the same algebraic (more precisely categorical) structure as that of the operators in the topological order.
Although our work bears some similarities with this series of works, the framework we use, the conceptual picture and approach are very different and do not rely on such a construction.

\smallskip %
Recently, there has been a lot of interesting progress in a direction related to so-called symmetry topological field theories (SymTFTs) \cite{Freed:2012bs,Freed:2018cec, GaiottoKulp202008, Apruzzi:2021nmk, Apruzzi:2022dlm, Burbano:2021loy, Freed:2022qnc, Freed:2022iao, KaidiOhmoriZheng202209, Kaidi:2023maf}.
The main idea is that a $d$-dimensional theory $\mathfrak T$ on a manifold $M$ with a global symmetry $\mathcal C$ can be constructed as topological theory in $d+1$-dimensions defined on $M\times [0,1]$ with the original theory $\mathfrak{T}$ on one boundary say $M\times \left\{0\right\}$ and a topological boundary condition on the other boundary $M\times \left\{1\right\}$.
An attractive feature of this construction is that the symmetry related properties can be dealt with by simply manipulating the topological boundary condition.
Indeed various dualities can be implemented by simply altering boundary conditions.
In our work, we do not consider an interval compactification with a topological boundary condition on one boundary and a generic non-topological boundary condition on the other.
Instead, we consider the bulk to be a semi-infinite cylinder, which allows us to access all the symmetry twisted symmetry sectors directly using the semi-infinite line operator with one end on the boundary to toggle between these different sectors.
Furthermore, our method is more oriented towards studying concrete lattice spin models as compared with general applications of symmetry TFTs.
A condensed-matter-oriented study of symmetry TFTs was developed in the work of Lichtman et al \cite{Thorngren_anyon}. 
In this work, the authors use a finite cylinder spatial geometry with a Dirichlet boundary condition for the electric line operators on one boundary.
The other boundary is left free and the bulk magnetic line operator wrapping the non-contractible cycle projects on to the dynamical (i.e., non-topological boundary) as a symmetry operator.
In this picture when the non-topological boundary is an electric condensate, one obtains the spontaneous symmetry-breaking phase, while when the boundary is a magnetic condensate, one obtains the symmetric phase.

\smallskip The present work provides a complimentary approach to these past works
and contextualizes these approaches to concrete lattice models that are fairly standard in quantum magnetism. 
Moreover, we lay out a detailed and practical framework to use such abstract notions to study the generically complex phase diagrams of these models.

\subsection*{Summary of results}

In this work, we detail a holographic approach that uses a topological order to systematically study the global symmetry aspects of quantum systems in one lower dimension.
The framework is sufficiently general and can be applied to the study of symmetric quantum systems in general spacetime dimensions.
In order to illustrate the construction in a simple and explicit setting, in this work we restrict ourselves to anomaly-free $1+1d$ quantum systems with finite Abelian group symmetry (or equivalently, $2+1d$ topological orders corresponding to Dijkgraaf-Witten theories with finite Abelian groups $\ms G$). 
The paper is organized into three main parts.
Below we summarize the main results and contributions of each of these parts.

\paragraph{Framework of topological holography:} 
In Section \ref{sec:SymmetryAsTopologicalManifolds}, we review the concept of (higher-form) symmetries as topological operators/defects.
In Section \ref{Sec:framework}, we describe the framework of topological holography.
In particular, we review the symmetries of a $2+1d$ Abelian $\ms G$ topological order.
These comprise of two parts---(i) $1$-form symmetries $\mcal{A}\simeq\ms G\times\ms G$ corresponding to anyonic Wilson lines, and (ii) $0$-form symmetries $\mcal{G}[\ms G]$ corresponding to permutation of anyons that preserve the topological correlations of the anyons.
We show that the algebra of $\ms G$-symmetric operators of $1+1d$ theories, which we denote as the String Operator Algebra $\mbb{SOA}[\ms G]$, can be constructed from the bulk $1$-form symmetries of the higher dimensional topological order.
We consider representations of these algebras on generalized spin chains.
Using the SOA, we parameterize the space of $\ms G$-symmetric (spin-chain) Hamiltonians. 
Furthermore, we show that the bulk $0$-form symmetries act as dualities on the space of $1+1d$ $\ms G$-symmetric theories, mapping between different symmetric Hamiltonians.
The duality group $\mcal{G}[\ms G]$ is in general non-Abelian and very large even for small finite Abelian groups $\ms G$. For example, in the case of $\ms G=\mbb{Z}_2^4$, there are almost 350 million dualities. 

\smallskip We develop a detailed bulk-boundary dictionary for topological holography. Amongst other things, this holographic realization allows us to classify gapped phases of $\ms G$-symmetric theories and construct fixed-point spin-chain Hamiltonians for each phase. 
From the point of view of topological order, each gapped phase is related to the condensation of a subset of anyons $\mcal{L}$.
From the condensation of anyons, we construct spin-chain order parameters for each gapped phase. 
We also argue how non-Abelian and sometimes non-invertible symmetries can emerge at self-dual regions in the space of symmetric models.

\paragraph{Applying topological holography:} In Section \ref{topological holography: application}, we outline how topological holography might be applied to study the space of symmetric quantum systems. In particular, the existence of dualities, the classification of gapped phases and the transformations under dualities can be used to strongly constrain and sometimes construct phase diagrams of concrete models.
We derive several powerful and general results, useful in the study of phase diagrams of concrete Hamiltonian models.
A summary of some of these results are
\begin{enumerate}
    \item Properties such as a Hamiltonian $H$ being gapped, gapless, critical, or at a first-order transition is preserved under duality transformations.
    
    \item The action of the bulk 0-form symmetry group $\mcal{G}[\ms G]$ on the set of gapped phases gives rise to a vast \boldit{web of dualities} encoded in a graph whose vertices are gapped phases and edges are dualities.
    
    \item Each gapped phase $\mcal{L}$ is characterized by a pair $(\ms H_{\mcal{L}},\psi_{\mcal{L}})$, corresponding to a spontaneous symmetry-breaking down to a subgroup $\ms H_{\mcal{L}}\subset \ms G$ together with an SPT twist $\psi_{\mcal{L}}\in H^2(\ms H_{\mcal{L}},\ms U(1))$. We show how the data $(\ms H_{\mcal{L}},\psi_{\mcal{L}})$ can be extracted from the anyons in $\mc L$.

    \item {\it Obstruction to duality (gapped)}: Each gapped phase is characterized by a \boldit{fusion structure} corresponding to the algebra of condensed order parameters. There cannot exist dualities  between two gapped phases $\mcal{L}$ and $\mcal{L}'$ unless they have isomorphic fusion structures. The web of dualities splits into connected components, one for each type of fusion structure. 
    
    \item{\it Emergent symmetries}: If a Hamiltonian $H$ is self-dual under a subgroup $\mscr{S}$ of dualities $\mcal{G}[\ms G]$, then this implies that it has new emergent symmetries. These are sometimes generalized non-invertible symmetries, in particular when they descend from bulk electric-magnetic 0-form symmetries. %

    \item{\it Implication of self-duality}: Given a Hamiltonian $H$ self-dual under a subgroup $\mscr{S}$, we give criteria to determine when the self-duality is an obstruction to $H$ being in a gapped phase. In such a situation, the Hamiltonian $H$ must be either at a phase transition or in a gapless phase.   
    
    \item {\it Multi-criticality}: Given a critical point $\mscr{C}$ self-dual under subgroup $\mscr{S}$ of dualities $\mcal{G}[\ms G]$, we show how to determine the lower bound on the degree of multi-criticality of $\mscr{C}$.
    
    \item {\it Obstruction to duality (critical)}: critical points are characterized by a set of fusion structures corresponding to the gapped phases attached to them. Two critical points $\mscr{C}$ and $\mscr{C}'$ cannot be dual to each other unless their fusion structures are the same. This implies the decomposition of the space of $\ms G$-symmetric critical points into blocks labeled by fusion structures. Dualities can only act within each block. 
\end{enumerate}
Using these results, one can significantly constrain and potentially determine the phase diagrams of spin-chain models with very simple calculations. Topological holography is especially powerful for symmetry groups $\ms G$ that have large duality groups $\mcal{G}[\ms G]$ and few possible fusion structures. 

\smallskip One of our most powerful applications of topological holography is to determine the complete conformal spectra of non-trivial critical points. This we achieve by lifting dualities from Hamiltonians (spatial) to the level of partition functions (spacetime). Given a (generalized symmetry-twisted) thermal partition function $\mcal{Z}_{\ms g,\ms h}$, the partition function of a dual theory is given by
\begin{equation}
    \mcal{Z}^\vee_{\ms g^\vee,\ms h^\vee}=\frac{1}{|\ms G|} \sum_{\mathsf g,\mathsf h\in \mathsf G} \eta\left(\mathsf g,\mathsf h,\mathsf g^\vee,\mathsf h^\vee\right)\;\mathcal{Z}_{\mathsf g,\mathsf h}.
\end{equation}
We find an explicit formula for the phases $\eta\left(\mathsf g,\mathsf h,\mathsf g^\vee,\mathsf h^\vee\right)$ in terms of dualities. Physically, these factors are related to topological partition functions of certain SPT phases living on duality domain-walls in the bulk. Finally, we give an interpretation of these dualities from the perspective of gauging global symmetries.

\paragraph{Study of spin chains using topological holography:} In Section \ref{sec:examples}, we illustrate the theory developed in prior sections for specific examples of $\ms G$. 

\smallskip First, we discuss the simplest example $\ms G=\mbb{Z}_2$. Next, we briefly discuss the $\ms G=\mbb{Z}_N$ example, where topological holography is not very powerful due to multiple fusion structures and a small duality group. Next, we move on to examples where topological holography is very powerful such as $\ms G=\mbb{Z}_2\times\mbb{Z}_2$, and $\ms G=\mbb{Z}_3\times\mbb{Z}_3$. In particular, using topological holography we classify possible gapped phases, their associated set of condensed bulk anyons, the duality group, and the corresponding web of dualities. We then construct explicit spin-chain representations of the $\mbb{SOA}[\ms G]$ algebras. For each gapped phase, we construct fixed-point Hamiltonians. From the set of condensed anyons, we determine the corresponding spin-chain order parameters for each gapped phase, and from the set of confined anyons, we determine the fundamental spin-chain excitations. 

\smallskip To further exemplify our approach, we define so-called minimal Hamiltonians which are a linear combinations of fixed-point Hamiltonians and study their phase diagrams. 
These give us a rich set of models with rich phase diagrams containing all gapped phases and many non-trivial phase transitions. Using the general results mentioned above, we map out phase diagrams, identifying gapped regions, critical points/lines/surfaces, first-order transitions, etc. It turns out that these phase diagrams are surprisingly geometrical as we determine using dualities. Using Density Matrix Renormalization Group (DMRG) method and the order parameters provided by topological holography, we numerically verify the predicted phase diagrams. By computing entanglement entropy numerically, we confirm the predicted central charges along various points and critical lines. 

\smallskip For many non-trivial critical points, we explicitly compute the full conformal spectra using analytical methods from topological holography. The only extra input are the partition functions of simple order-disorder transitions such as the $c=\frac{1}{2}$ Ising CFT. From this, we compute the spectrum of several interesting ``{\it Landau-forbidden}" transitions such as {\it topological phase transitions} with no symmetry breaking between SPT phases, {\it topological spontaneous symmetry-breaking transitions} between a non-trivial SPT phase and a symmetry-broken phase, and {\it topological deconfined transitions}. From the action of global symmetry $\ms G$ on primary fields, we highlight the difference between  topological spontaneous symmetry-breaking transitions and conventional Landau-allowed spontaneous symmetry-breaking transitions. Our approach makes explicit the relation between symmetry-enriched CFTs  \cite{VerresenThorngrenJonesPollmann201905} and SPT invariants. Using Exact Diagonalization techniques, we numerically confirm these analytical predictions of conformal spectra. 

\smallskip More details about construction of generalized spins chain from holographic perspective, duality groups, symmetry-twisted boundary conditions, conformal spectra of various transitions, and some relevant background material are collected in the appendices. A simple and concrete realization of topological holography in an exactly-solvable lattice model is given in Appendix \ref{App: Wen-plaquette}.

\bigskip 
{\bf Note:} The idea that anyons and anyonic symmetries are related to global symmetries and dualities in one lower dimension has been discussed in \cite{WeiMoradi2014} and also more thoroughly in \cite{MoradiWei2014} (see chapter 7 and section 7.5.5 of \cite{MoradiPhD2018}). 
Furthermore,  a relation between TQFTs in $d+1$-dimensions and generalized Kramers-Wannier dualities in $d$-dimensions was stated as an on-going work (see section 7.6 of \cite{MoradiPhD2018}).
This old unfinished work was revisited by us recently which the current paper (and other on-going works) is a result of. 
During the preparation of this paper, we became aware of several works \cite{Freed:2018cec, GaiottoKulp202008, Wen_higher, Wen_Fusion,Ji:2019ugf, Thorngren_anyon, Chatterjee:2022kxb, Chatterjee:2022tyg, Apruzzi:2021nmk, Bhardwaj:2020ymp,Lootens:2021tet,Aasen_2016, Aasen:2020jwb} that appeared in the recent past which contain related insights.
While this paper some conceptual overlap with this literature, we believe that it is a good complement to the ongoing progress in the field and contains many new insights.

\section{Symmetry structures
as topological operators}
\label{sec:SymmetryAsTopologicalManifolds}

The concept of symmetry plays a fundamental role in this work. Therefore, we start with a brief overview of what we mean by symmetries in generic quantum systems. We then explain how symmetries can be used to decompose the space of local operators.

\subsection{Symmetries are topological}
 A quantum system is typically defined by a Hamiltonian acting on a Hilbert space which admits a tensor product decomposition into local (on-site) Hilbert spaces.
 Likewise, the Hamiltonian admits a decomposition into a sum of local terms each of which acts on the corresponding local Hilbert space.
 Global symmetry operators act on all of space and are required to commute with the Hamiltonian.
 They typically satisfy a group composition law, however in certain quantum systems they can satisfy a more general mathematical structure such as a fusion algebra which is not isomorphic to any group \cite{PetkovaZuber2001, Fuchs:2002, Fuchs:2007tx, BhardwajTachikawa201704, Wen_higher, thorngren2019fusion, ChangYingHsuanShaoWangYin201802, Thorngren_anyon, Kaidi:2021xfk, Choi:2021kmx, Yang:2021xob, Heidenreich:2021xpr, Roumpedakis:2022aik, Bhardwaj:2022yxj, Choi:2022zal}. 
 For this paper, it is instructive to review the concept of symmetry from a modern perspective. %
\begin{figure}[t!]
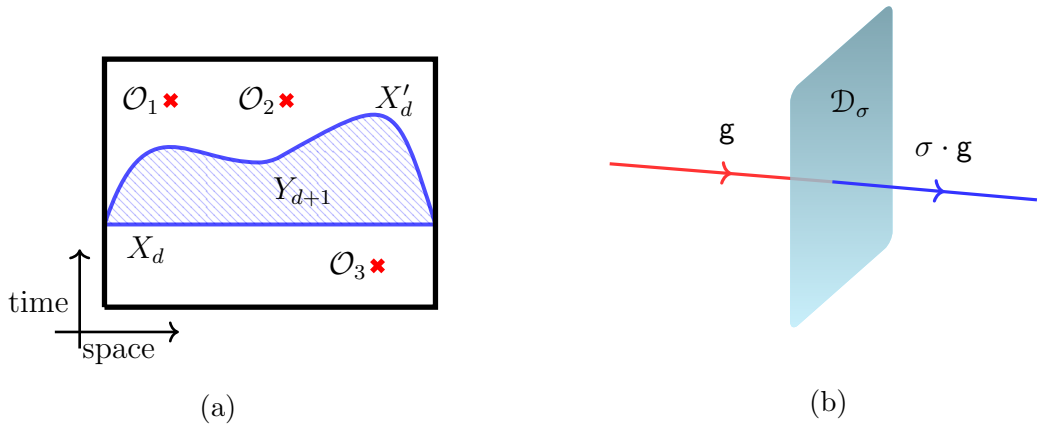
\centering
    \hspace{-1cm}\begin{subfigure}{.39\textwidth}\centering
    \SymmetryIsTopological
    \subcaption{}
    \label{fig:SymmetryIsTopological}
    \end{subfigure}
    ~
    \hspace{+1cm}\begin{subfigure}{.39\textwidth}\centering
    \ActionOfZeroFormSymmetryOnOneFormSymmetry\vspace*{.5cm}
    \subcaption{}
    \label{fig:ActionOfZeroFormSymmetryOnOneFormSymmetry}%
    \end{subfigure}

    \caption{a) Symmetry operators are topological $\mathcal U_{\ms{g}}[X_{d}]=\mathcal U_{\ms{g}}[X_{d}']$. This is due to the conserved current $d{\star}j=0$ implying $\int_{Y_{d+1}}d{\star}j = \int_{X_d
}{\star}j-\int_{X'_d}{\star}j=0$, where $Y_{d+1}$ is a manifold bounded by $X_d$ and $X_{d'}$. b) When both $0$-form (here a surface) and $1$-form (here a line) symmetries are present, the $0$-form symmetry can act on the $1$-form symmetry to form a higher mathematical structure called a $2$-group.}
\end{figure}

\medskip \noindent When a quantum system is described by a Euclidean quantum field theory, there is no preferred spatial hypersurface therefore the symmetry operators may be defined on any codimmension-1 surface embedded in spacetime. Consequently, the condition of commutativity with the Hamiltonian generalizes to commutativity with the energy-momentum tensor which implies topological invariance of the symmetry operators.
More precisely, any correlation function or observable of a quantum field theory is invariant under smooth  (topological) deformations of the support of the symmetry operators of the theory.
A well-known example is a $d+1$-dimensional quantum field theory invariant under a global $\ms{U}(1)$ symmetry. Noether's theorem mandates the existence of a conserved $1$-form current $j$, satisfying the conservation law $\sd{\star}j\sim \partial_\mu j^\mu=0$, associated to this symmetry. The global symmetry operators take the form 
\begin{align}
\mathcal U_{\ms{g}}[X_{d}]:=\exp\left\{i\ms{g}\int_{X_{d}}{\star}j\right\},
\end{align}
where $\ms{g}\in \mathbb R/2\pi \mathbb Z \simeq \ms{U}(1)$, $X_{d}$ is a given $d$-dimensional hypersurface embedded in the $d+1$-dimensional spacetime. The exponent is often called the Noether charge and is the generator of infinitesimal symmetry transformations. These operators satisfy the $\ms{U}(1)$ group composition rule 
\begin{align}
\mathcal U_{\ms{g}}[X_{d}]\times \mathcal U_{\ms{g}'}[X_{d}] =  \mathcal U_{\ms{g}+\ms{g}'}[X_{d}].
\end{align}
Additionally, as a consequence of current conservation ($\sd{\star}j=0$), these operators can be moved around freely within correlation functions as long as they don't pierce any local operator $\mc{O}_i$ that carries a non-trivial $\ms{U}(1)$ charge i.e.
\begin{align}
    \Big \langle \mathcal U_{\ms{g}}[X_{d}]\prod_{i}\mathcal O_i(x_i)\Big\rangle =     \Big\langle \mathcal U_{\ms{g}}[X'_{d}]\prod_{i}\mathcal O_i(x_i)\Big\rangle,
\end{align}
when there exists a closed $d+1$-dimensional  region which does not contain any of the points $x_i$ and whose boundary is the union of $X_{d}$ and $X_{d}'$ (see Figure \ref{fig:SymmetryIsTopological}). This means that symmetry operators in a quantum theory are topological. We emphasize that even for non-topological systems, the symmetry structure is always topological. 

\begin{figure}
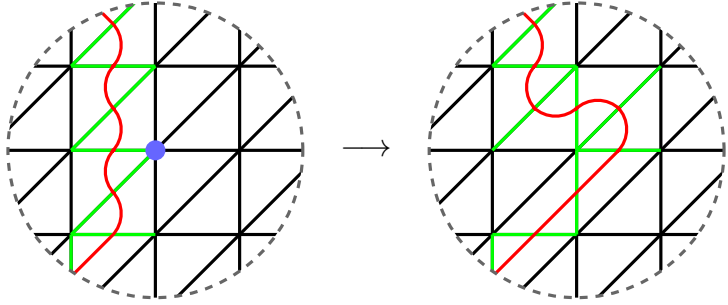

    \centering
    \TriangulationGaugeTransformation
    \caption{ A gauge field corresponding to a finite Abelian group can be defined on a triangulated spacetime and  is equivalent to collection of symmetry defects. 
    Gauge transformations implement topological re-configurations of the symmetry defects.
    The figure illustrates a $\mbb Z_2$ gauge field $A$.
    The edges (1-simplices) in green and black have $A_{ij}=1$ and $0$ respectively. 
    Equivalently, this gauge field configuration can be described by a single symmetry defect denoted in red.
    A gauge transformation $\phi_i=1$ on a single site (in blue) and 0 everywhere else transforms the gauge field and correspondingly topologically deforms the symmetry defect.  
    }
    \label{fig:TriangulationGaugeTransformation}
\end{figure}

\medskip \noindent An alternative way of seeing why symmetry operators are topological is to note that they, by definition, commute with the Hamiltonian and hence are conserved under time-evolution.
They can therefore be deformed freely inside the partition function without any obstruction from the Hamiltonian.
This point of view applies to finite symmetries as well, where there are no conserved Noether currents.
More generally, similar to the $\ms{U}(1)$ case, gauge invariance of the background gauge field for a finite group implies topological invariance of the discrete symmetry structure (see Fig.~\ref{fig:TriangulationGaugeTransformation}).
\begin{figure}[t!]
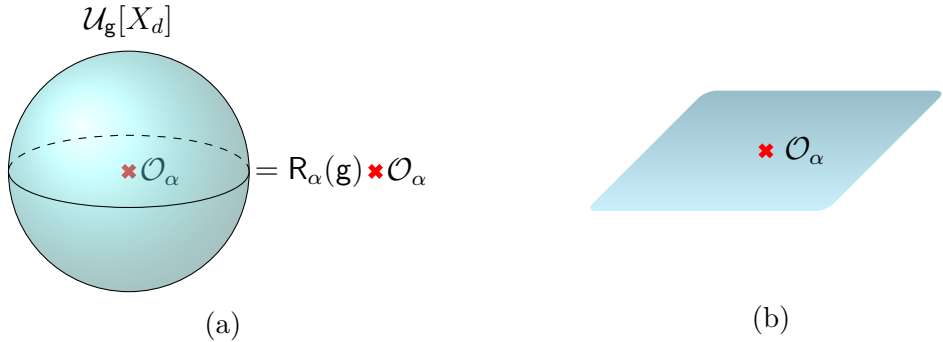
\centering
    \begin{subfigure}{.4\linewidth}\centering
    \ActionOfZeroFormSymmetry
    \subcaption{}
    \label{fig:ActionOfZeroFormSymmetry}
    \end{subfigure}
    ~
    \begin{subfigure}{.4\linewidth}\centering
    \SymmetryOperatorTimeSlice\vspace{1.0cm}
        \subcaption{}\vspace{-1.2cm}
        \label{fig:SymmetryOperatorTimeSlice}
    \end{subfigure}
    \caption{\textbf{(a)} A ($0$-form) global symmetry is associated with a $d$-dimensional surface $X_d$ in a $d+1$-dimensional spacetime. A $0$-dimensional (local) operator will transform under some representation $\alpha$ when crossing this hypersurface. If $X_d$ wraps around the point $x$, then by shrinking the surface $\mc U_{\ms g}[X_d]\mathcal O_{\alpha}(x) = \ms R_\alpha(\ms g)O_{\alpha}(x)$. \textbf{(b)} If $X_d$ is a time-slice associated with a Hilbert space then we have $\mc U_{\ms g}[X_d]\mathcal O_{\alpha}(x) = \ms R_\alpha(\ms g)O_{\alpha}(x)\mc U_{\ms g}[X_d]$, which is the standard transformation law in a Hilbert space.}\label{fig:0-Form_Operator_Vs_Local_Operator}
\end{figure}

\smallskip Ordinary global symmetries like $\mcal U_{\ms{g}}[X_{d}]$ naturally act on $0$-dimensional operators. In particular local operators that are {\it charged} under these symmetries transform under a representation $\alpha$ of the corresponding symmetry group $\ms{G}$ if passed through the hypersurface $X_d$. Let $X_d$ wrap a point $x$. Since the symmetry operators are topological, we can deform $X_d$, and in particular, we can shrink it. As a result of shrinking the surface, we have 
\begin{equation}\label{eq:U[X]_surface_around_O(x)}
    \mc U_{\ms g}[X_d]\mathcal O_{\alpha}(x) = \ms R_\alpha(\ms g)\mcal{O}_{\alpha}(x),
\end{equation}
where $\ms R_\alpha(\ms g)$ is the representation matrix of $\ms{G}$ (see Figure \ref{fig:ActionOfZeroFormSymmetry}).
If $X_d$ is a time-slice and $x\in X_d$, then
\begin{equation}\label{eq:U[X]_surface_timeslice_O(x)}
    \mc U_{\ms g}[X_d]\mathcal O_{\alpha}(x)\mc U_{\ms g}^{\dagger}[X_d] = \ms R_\alpha(\ms g)\mcal{O}_{\alpha}(x)\,,
\end{equation}
which is the standard way symmetries act on operators at the level of the Hilbert space of theory (see Figure \ref{fig:SymmetryOperatorTimeSlice}). Since these ordinary symmetries act on $0$-dimensional operators, they are sometimes called $0$-form symmetries and the corresponding symmetry operators are supported on a $d$-dimensional (or codimension-$1$\footnote{In this work, we often use the standard terminology codimension. A codimension-$p$ submanifold in a $d+1$-dimensional spacetime has $d+1-p$ spacetime dimensions.}) manifolds $X_d$ in the ambient $d+1$-dimensional spacetime. We will call manifolds associated with symmetries, operators or defects interchangeably. More precisely these are referred to as operators when the surface is spacelike and defects otherwise.

\medskip \noindent As mentioned, this is not restricted to continuous symmetries. Similarly, discrete symmetries can also be associated with manifolds $X_d$. 
A well-known example in this regard is a global $\mathbb Z_{2}$ operator in spin-1/2 systems defined as $\prod_{i}\sigma_i^{x}$ and the product runs over the spin sites. In Euclidean field theory, a suitably chosen collection of symmetry operators can be used to simulate a given $\mathbb Z_{2}$ background connection. Invariance of the background connection under $\mathbb Z_2$ gauge transformations translates to topological invariance of the symmetry operators. 

\medskip \noindent Recent years have seen various important generalizations of symmetry structures in quantum systems. One of these is the notion of generalized global symmetries which are symmetry structures under which extended objects are charged \cite{gaiotto2015generalized}. More specifically, in a $d+1$-dimensional spacetime, there can exist $d+1-q-1=d-q$ dimensional symmetry defects $\mc U_{\ms g}[X_{d-q}]$ which act on operators $\mathcal O[\Sigma_q]$ supported on $q$-dimensional submanifolds $\Sigma_q$. Again, the symmetry operators $\mc U_{\ms g}[X_{d-q}]$ are topological and can be freely deformed within correlation functions, although the charged operators $\mathcal O[\Sigma_q]$ need not be topological.
Equations similar to \eqref{eq:U[X]_surface_around_O(x)} and \eqref{eq:U[X]_surface_timeslice_O(x)} still hold in this case \cite{gaiotto2015generalized}. For example

\begin{equation}\label{eq:U[X]_linking_O(Sigma)}
    \mc U_{\ms g}[X_{d-q}]\mathcal O_{\alpha}[\Sigma_q] = 
        \ms R_\alpha(\ms g)^{\texttt{Link}[X_{d-q},\Sigma_q]}\;\mcal O_{\alpha}[\Sigma_q],
\end{equation}
where $\texttt{Link}[X_{d-q},\Sigma_q]$ is the linking number between the two manifolds. And if both $X_{d-q}$ and $\Sigma_q$ are in a time-slice, we find the usual Hilbert space action of symmetry
\begin{equation}\label{eq:U[X]_intersecting_O(Sigma)}
    \mc U_{\ms g}[X_{d-q}]\mathcal O_{\alpha}[\Sigma_q] \mc U_{\ms g}^{-1}[X_{d-q}] = \ms R_\alpha(\ms g)^{\texttt{Intersect}[X_{d-q},\Sigma_q]}\;\mcal O_{\alpha}[\Sigma_q],
\end{equation}
where $\texttt{Intersect}[X_{d-q},\Sigma_q]$ is the intersection number. Such symmetries are called $q$-form symmetries, where $q=0$ corresponds to what is conventionally known as global symmetry. Continuous $q$-form symmetries also lead to conservation laws similar to $\sd\star j=0$, but now conservation of $q$-dimensional charges.
Note that if both operators $\mcal U_{\ms g}[X_{d-q}]$ and $\mathcal O_{\alpha}[\Sigma_q]$ are topological, this implies that both are higher-form symmetry generators.
A non-trivial $\ms R_\alpha$ in \eqref{eq:U[X]_linking_O(Sigma)} and \eqref{eq:U[X]_intersecting_O(Sigma)} gives rise to higher-form 't Hooft anomalies, as it serves as an obstruction to gauging both of these higher form symmetries together \cite{gaiotto2015generalized}. Regarding the application of these types of symmetries, we notice that extended objects can commonly emerge in condensed matter systems, despite the microscopic theory only having point-like degrees of freedom. Examples include Wilson operators in gauge-theoretic descriptions of spin systems and vortices in superfluids and superconductors.

\smallskip When multiple higher-form symmetries are present simultaneously, one might wonder which mathematical structure they form together. Interestingly, it turns out that $0$-form symmetries $\mc U[X_d]$ can act on $1$-form (or higher-form) symmetries $\mc U[Y_{d-1}]$ when $Y_{d-1}$ pierces through $X_d$ as in Figure \ref{fig:ActionOfZeroFormSymmetryOnOneFormSymmetry}. This leads to a mathematical structure called a $2$-group (and more generally to $n$-groups)\cite{Kapustin:2013uxa, Cordova2019_1,Cordova2019_2, Delcamp:2018wlb, Delcamp:2019fdp }. Such a situation is common in topologically-ordered systems in $2+1d$ and they will play the key role in this paper. The above mentioned notion of non-invertible symmetries can also be thought of within this framework \cite{PetkovaZuber2001, Fuchs:2002, Fuchs:2007tx, BhardwajTachikawa201704, Wen_higher, thorngren2019fusion, ChangYingHsuanShaoWangYin201802, Thorngren_anyon, Kaidi:2021xfk, Choi:2021kmx, Yang:2021xob, Heidenreich:2021xpr, Roumpedakis:2022aik, Bhardwaj:2022yxj, Choi:2022zal}. We notice that systems with these types of generalized symmetries can be accommodated into the framework presented here for symmetries with group-like structure. Other new notions of symmetry exist which for now is outside of the conceptual framework of this paper. For example subsystem symmetries, which play an important role in fracton topological order and can give rise to unusual phenomena like IR-UV mixing \cite{GorantlaLamSeibergShao202108}.

\medskip \noindent Let us briefly say few words about  anomalies associated to global symmetries, i.e., 't Hooft anomalies.
Certain symmetric quantum systems might have quantum anomalies which, being an renormalization group invariant property \cite{tHooft1979}, strongly constrain their dynamics. 
If a system has a quantum anomaly in the microscopic/ultraviolet description, its low-energy/infra-red description can only realize phases which saturate the anomaly.
Concretely, for a given quantum system, anomalies can be diagnosed by lack of gauge invariance of the gauged partition function or equivalently by the lack of topological invariance of a certain network of symmetry operators.
Two condensed matter contexts where quantum anomalies arise are $(i)$ the surface of symmetry-protected topological (SPT) phases \cite{ryu_anomaly, shinsei_anomaly2, BoHan_2017} and $(ii)$ in low-energy descriptions of quantum systems with Lieb-Schultz-Mattis constraints \cite{Shinsei_LSM, Thorngren_LSM, watanabe_anomaly_LSM, Aksoy:2021uxb}.

\smallskip Since the concept of symmetry is inherently topological in any quantum system, it is natural to ask whether the symmetry aspects of a theory can be studied and understood using a purely topological theory? In this paper, we are particularly interested in studying $1+1d$ quantum systems with a finite Abelian global symmetry group $\ms G$. 
The fundamental theorem of finite Abelian groups determines the structure of such groups: any finite Abelian group is isomorphic to a direct product of cyclic groups whose orders are powers of primes\footnote{See section 2.1 of \cite{KurzweilStellmacher2004} for detailed discussion of this result.}
\begin{equation}\label{eq:the generic finite Abelian group}
    \ms G=\prod_a\mbb{Z}_{N_a}, \qquad n_a=p_a^{k_a}
\end{equation}
where $N_a$s and $k_a$s are positive integers and $p_a$s are (not necessarily distinct) prime numbers.
It turns out that systems with such symmetry groups can be analyzed as the boundary of an auxiliary $2+1d$ topologically ordered system, described by a topological gauge theory with gauge group $\ms G$. 
As we will illustrate in detail, such theories encode all the physically-important features related to the global symmetry of the lower dimensional theory.
This will give us powerful tools to study phase diagrams and phase transitions of $1+1d$ $\ms G$-symmetric theories. 
The restriction of dimension and group structure is merely done to illustrate the power of this construction in a simple setting.
We expect these concepts to naturally generalize to higher dimensions. We leave such explorations for future work. 

\subsection{Building blocks of symmetric local operators} \label{sec:SymmetryStructureAndBuildingBlocksofLocalOperators}
To establish some notation, we will briefly study the space of linear operators acting on a Hilbert space. Consider a Hilbert space which admits a local tensor product decomposition $\mathcal H = \otimes_i\mathcal H_i$ and the corresponding space of linear operators $\mathcal L(\mc H)=\otimes_i\mcal{L}(\mathcal H_i)$. A (usual $0$-form) symmetry acts on local operators $\mathcal O_i\in\mathcal L(\mathcal H_i)$ as $\mathcal O_i\rightarrow \mathcal U_{\ms g}\mathcal O_i\mathcal U_{\ms g}^\dagger$ and we assume that the operator $\mc{U}_{\ms{g}}$ has the following tensor product decomposition
\begin{equation}\label{eq:SymmetryOperatorGroupTensorDecomposition}
    \mathcal U_{\ms g} = \bigotimes_i \mathcal O_{\ms g,i},
\end{equation}
with the property $\mathcal O_{\ms g_1,i}\mathcal O_{\ms g_2,i} = \mathcal O_{\ms g_1\ms g_2,i}$. In other words, we assume that the system has an on-site symmetry. Since the space of local operators is a representation space for the symmetry, we can decompose it into irreducible representations as
\begin{equation}
\mathcal L(\mathcal H_i) = \bigoplus_{\alpha\in\text{Rep}(\ms G)} n_{\alpha,i} \mathcal L_{\alpha, i},
\end{equation}
where $\text{Rep}(\ms G)$ is the set of irreducible representations of the symmetry group $\ms G$ and $n_{\alpha,i}$ is the multiplicity of the representation $\alpha$ at the site $i$. For simplicity of notation, we will suppress the multiplicity of a representation. The irreducible local operators $\mathcal O_{\alpha,i}\in\mathcal L_{\alpha,i}$ have well-defined \textit{charges} under the action of the symmetry group $\ms G$ 
\begin{equation}\label{eq:symmetry_action}
    \mathcal U_{\ms g}\mathcal O_{\alpha,i}\mathcal U_{\ms g}^\dagger  = \ms R_{\alpha}(\ms g)\cdot\mathcal O_{\alpha,i},
\end{equation}
where $\ms R_{\alpha}(\ms g)$ is the matrix representation of the group element $\ms g$ in the representation $\alpha$. This relation is nothing but \eqref{eq:U[X]_surface_timeslice_O(x)}. This implies the following local operator identity 
\begin{equation}\label{eq:local symmetry action}
    \mathcal O_{\ms g,i}\mathcal O_{\alpha,i}\mathcal O_{\ms g,i}^\dagger  = \ms R_{\alpha}(\ms g)\cdot\mathcal O_{\alpha,i}.
\end{equation}
The set of operators $\left\{\mc{O}_{\ms{g},i},\mc{O}_{\alpha,i}\right\}$ for $\ms g\in\ms G$ and $\alpha\in\text{Rep}(\ms G)$,
are the building blocks of all other local operators and in particular the subalgebra of $\ms G$-symmetric operators.
For example, for a finite Abelian group $\ms G$, a symmetric operator with finite support\footnote{Here by finite support, we mean that the set $\{i_1,\cdots,i_n\}$ which labels the local sites is a finite set.} takes the form
\begin{align}\label{eq:SymmetricOperator_prototype}
    \mc K = \mc O_{\ms \alpha_1, i_1}\mc O_{\ms g_1, i_1} \dots \mc O_{\ms \alpha_n, i_n}\mc O_{\ms g_n, i_n},
\end{align}
with the constraint that the product of local representations $\alpha_{i_k}$ over $k$ is the trivial representation in $\text{Rep}(\ms G)$.

\smallskip As the simplest example, for $\ms G = \mathbb Z_2 = \{0,1\}$, we have two non-trivial representations $\tenofo{Rep}(\mathbb Z_2) = \{0,1\}$ given by $\ms R_{\alpha=0}(\ms g) = 1$ and $\ms R_{\alpha=1}(\ms g) = (-1)^{\ms g}$. The two corresponding local operators $\mathcal O_{\ms g,i}=(\sigma_i^x)^{\ms g}$ and $\mathcal O_{\alpha,i}=(\sigma_i^z)^{\alpha}$ satisfy
\begin{equation}
    \mathcal U_{\ms g}\sigma^x_i\mathcal U_{\ms g}^\dagger = \sigma^x_i, \qquad \mathcal U_{\ms g}\sigma^z_i\mathcal U_{\ms g}^\dagger = (-1)^{\ms g} \sigma^z_i,
\end{equation}
where the symmetry operator is given by $\mcal U_{\ms g} = \left(\prod_i\sigma^x_i\right)^{\ms g}$ and $\sigma^x$ and $\sigma^z$ denote the standard Pauli matrices.

\smallskip For $\ms G = \mathbb Z_N = \{0, 1, \dots, N-1\}$, we have the representations $\tenofo{Rep}(\mathbb Z_N) = \{0,1, \dots, N-1\}$ given by $\ms R_\alpha(\ms g) = e^{\frac{2\pi i}{N}\alpha\ms g}$ and the operators $\mathcal O_{\ms g} = X^{\ms g}$ and $\mathcal O_{\alpha} = Z^{\alpha}$, satisfying
\begin{equation}
    \mathcal U_{\ms g}X_i^{\ms g}\mathcal U_{\ms g}^\dagger = X_i^{\ms g}, \qquad \mathcal U_{\ms g}Z_i^\alpha\mathcal U_{\ms g}^\dagger = e^{\frac{2\pi i}{N}\alpha\ms g} Z_i^\alpha,
\end{equation}
where the symmetry operator is given by $\mcal{U}_{\ms g} = \left(\prod_iX_i\right)^{\ms g}$. Here $X$ and $Z$ are the $\mathbb Z_N$ generalization of Pauli matrices satisfying the algebra
\begin{equation}
    X^{\ms g}Z^\alpha = e^{\frac{2\pi i}{N}\alpha\ms g}\, Z^\alpha X^{\ms g}.
\end{equation}
In this work, we are mostly interested in finite Abelian groups of the form $\ms G=\mbb{Z}_{N_1}\times\cdots\times\mbb{Z}_{N_n}$. For this group, the general formula for representation $\ms R_{\alpha}(\cdot)$ is given by
\begin{equation}\label{eq:the formula for representation labeled by alpha for a general finite Abelian group}
    \ms{R}_\alpha(\ms g)=\exp\left(2\pi\mfk{i}\sum_{a=1}^n\frac{\alpha_a\ms g_a}{N_a}\right), \qquad (\ms g_a,\alpha_a)\in \mbb{Z}_{N_a}\times\tenofo{Rep}(\mbb{Z}_{N_a}).
\end{equation}

\section{Topological holography: Framework}
\label{Sec:framework}

As discussed in the previous section, symmetry operators are topological in nature.
Therefore, we can pose the following question: can symmetry-related aspects of a $1+1d$ theory be described by a topologically-ordered system in one higher dimension?
If so, which topological order would be a candidate?
One way to think about this is the following: 0-form symmetry in $1+1d$ are given by a collection of line-like operators, therefore which $2+1d$ topological orders have a (subset of) similar line-like topological operators?
Such topological line operators are in fact ubiquitous in $2+1d$ topological orders and correspond to anyonic excitation in the topological phase.
From the perspective of symmetry, each Abelian anyon generates a  $1$-form global symmetry.
In other words, in order to describe a $\ms G$-symmetric $1+1d$ theory we need to look at theories in $2+1d$ with  1-form $\ms G$-symmetries.
This is exactly the case for topological gauge theories with gauge group $\ms G$.\footnote{More precisely, $2+1d$ topological gauge theories with gauge group $\ms G$ always have a non-anomalous 1-form symmetry subgroup $\ms G$.}
This is somewhat reminiscent of the AdS/CFT correspondence, where the gauge symmetry of the bulk gravitational theory, is dual to the global symmetry of the theory at its boundary\footnote{Recently, this has be modified to a more refined statement: the global symmetry of the boundary theory is actually dual to what has been dubbed as a long-range gauge symmetry of the bulk theory \cite{HarlowOoguri201810a,HarlowOoguri201810b}.}\cite{Witten199802}.
Therefore the reason gauge symmetry in the bulk becomes global ($0$-form) symmetry of the boundary originates from the $1$-form symmetry of the bulk and the 0-form symmetry of the boundary being the same.
Note that this reasoning does not straightforwardly generalize to non-Abelian groups since 1-form symmetries cannot be non-Abelian.

\smallskip In the following sections we will describe the (higher) symmetries of topological gauge theories with finite Abelian gauge group $\ms G$.
We then describe the space of $1+1d$ theories that can be realized at its boundary.
Since this construction attempts to understand $1+1d$ theories from a topological theory in higher dimensions, we call it \textit{Topological Holography}.\footnote{This should not be confused with the notion of topological holography in the context of topological string theory \cite{CostelloLi201606,Costello201610,Costello201705,IshtiaqueMoosavianZhou201809,CostelloGaiotto201812}.}

\subsection{Symmetries of topologically-ordered phases}

In this section we describe (generalized) symmetries that topologically-ordered phases have. For simplicity we illustrate our points using untwisted Dijkgraaf-Witten theories corresponding to finite Abelian gauge groups. However, the concepts are general. More complete treatment of symmetries of $2+1d$ topologically-ordered phases can be found in \cite{Barkeshli_2019,Teo_2015,CarquevilleRunkelSchaumann201710}.

\begin{figure}[t!]
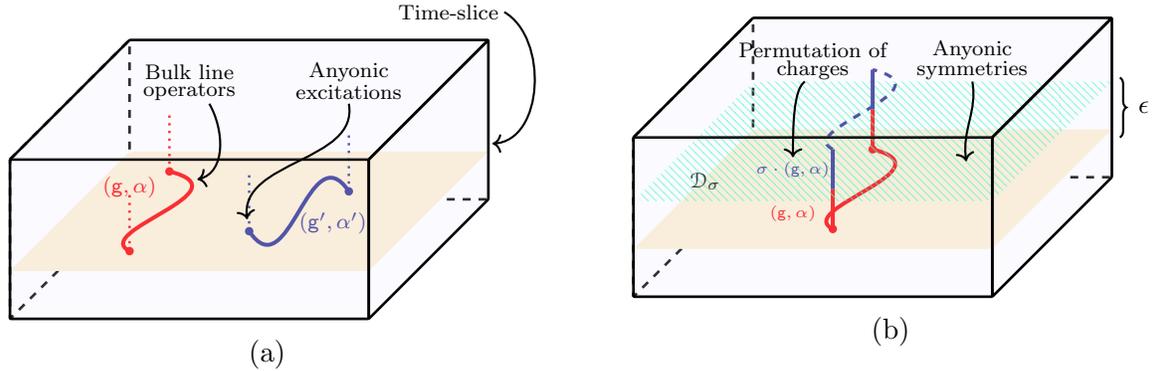
\centering
    \begin{subfigure}{.4\linewidth}\centering
    \AnyonsInTheBulk
    \subcaption{}
    \label{fig:AnyonsInTheBulk}
    \end{subfigure}\hspace{1cm}
    ~
    \begin{subfigure}{.4\linewidth}\centering
    \ActionOfSymmetryOperatorsOnBulkLinesModified
        \subcaption{}
        \label{fig:ActionOfSymmetryOperatorsOnBulkLines}
    \end{subfigure}
    \caption{Illustration of the 2+1 dimensional spacetime of a topological gauge theory. The orange surface is a time-slice associated with a Hilbert space.
    \textbf{(a)}
    The theory has topological line operators labeled by $d=(\ms g,\alpha)$. The set of these operators are $\mathcal A$, is the group of $1$-form symmetries. In a topological phase, open line operators create anyonic excitations at the end-points of the line. Under time evolution, these end-point evolve into world-lines (dotted lines).
    \textbf{(b)}
    Surface operators associated to global ($0$-form) symmetries of the theory $\sigma\in\mathcal G[\ms G]$. When the surface corresponding to the symmetry $\dual\sigma$ is put on the time-slice it acts on the Hilbert space as a conventional symmetry operator, transforming the charges $(\ms g,\alpha)$ to $\sigma\cdot(\ms g,\alpha)\in\mathcal A$ with isomorphic operator algebras.
    }
\end{figure}

\smallskip Topological gauge theories, also known as Dijkgraaf-Witten theories \cite{DijkgraafWitten199004} or Quantum Double models $\mathcal D(\ms G)$ \cite{Kitaev199707} have two types of topological operators or symmetries that play a role in our construction. 
First, they are equipped with a set of line operators $\mscr{W}_d[\gamma]$ supported on any curve $\gamma$ and are labeled by $d=(\ms g, \alpha)$, where $\ms g\in \ms G$ and $\alpha\in\text{Rep}(\ms G)$ correspond to the magnetic and electric charge of the line operator, respectively. 
These are often called Wilson lines.
From the point of view of topological order, the end-points of these line operators correspond to anyonic excitations.
The set of labels $(\ms g,\alpha)$ (or equivalently the set of line operators) form a group, denoted by $\mc A$, where the group multiplication is given by the fusion of quasiparticles (anyons). In the case of Abelian theories that we are considering in this paper, the fusion is simply given by $\mscr{W}_{d_1}[\gamma]
\times \mscr{W}_{d_2}[\gamma]=\mscr{W}_{d_1+d_2}[\gamma]$ or written graphically
\begin{equation}\label{eq:LocalFusionOfLineOperators}
    \LocalLinesUnfused = \LocalLinesfused.
\end{equation}
When acting on a Hilbert space conjugation is defined as $\mscr{W}^\dagger_d[\gamma] = \mscr{W}_{-d}[\gamma]$, making them unitary. The anyonic charge $d$ is defined with respect to the orientation of $\gamma$, changing the orientation flips the charge $\mscr{W}_d[\gamma] = \mscr{W}_{-d}[\bar\gamma]$, where $\bar\gamma$ is $\gamma$ with opposite orientation.
These line operators are topological, i.e., they can be freely deformed. 
Therefore in the language of higher-symmetries, we can think of the group of line operators/anyonic excitations $\mathcal A$, as the group of $1$-form symmetries of a topologically ordered system (see Figure \ref{fig:AnyonsInTheBulk}).

\smallskip The second type of topological operators or symmetries in a $2+1d$ topologically-ordered phase are $\dual \sigma[\Sigma]$ for any oriented surface $\Sigma$, labeled by a group element $\sigma\in\mathcal G[\ms G]$. In the literature on topologically-order phases, the symmetries in $\mathcal G[\ms G]$ are known as {\it anyonic symmetries} since they permute all anyons without changing their fusion rules or braiding phases \cite{Bombin2010, Barkeshli_2013, Barkeshli_2013b, Teo_2015, Barkeshli_2019}. We will define them more precisely in Section \ref{subsec:from bulk symmetry to boundary dualities}.  We note that $\mathcal G[\ms G]$ can be non-Abelian even if $\ms G$ is Abelian. In this paper we will see that $\mathcal G[\ms G]$ can contain hundreds of millions of elements, even for extremely simple groups like $\ms G= (\mathbb Z_2)^4$ (see Appendix \ref{appsec:computation of duality group}). Again the element $\sigma$ is defined relative to the orientation of $\Sigma$. Here an orientation of $\Sigma$ is given by its normal vector. Flipping the direction of the normal vector inverses the group element: $\dual\sigma[\Sigma] = \dual{\sigma^{-1}}[\bar\Sigma]$, where $\bar\Sigma$ is $\Sigma$ with opposite normal vector. On the other hand, conjugation is defined as usual, i.e. $\dual\sigma^\dagger[\Sigma] =\dual{\sigma^{-1}}[\Sigma]$. Since they are the same dimension as a spatial time-slice, they correspond to conventional ($0$-form) symmetries of the topological phase. These surfaces also have a fusion product on a fixed oriented surface $\Sigma$, $\dual{\sigma_1}[\Sigma]\times \dual{\sigma_2}[\Sigma] = \dual{\sigma_1\sigma_2}[\Sigma]$. This can be represented graphically as 
\begin{equation}\label{eq:fusion of surface operators figure}
    \FusionOfSurfacesOne = \FusionOfSurfacesTwo.
\end{equation}
Note that since $\mathcal G[\ms G]$ can be non-Abelian, this product must be defined with a convention relative to the choice of orientation/normal direction. Such a choice can always be made for $d$ dimensional manifolds in $d+1$-dimensional spacetimes, which is why $0$-form symmetries can be non-Abelian. However, such a choice is not possible for $1$- or higher-form symmetries and they must therefore always be Abelian.\footnote{If $q$-form symmetries are invertible and thus form a group (like the anyons in this paper), they must form an Abelian group. If we do not demand invertibility, then they can form a commutative fusion category. This is the case for $1$-form symmetries of topologically-ordered phases with non-Abelian anyons.} The $0$-form symmetry operators $\dual \sigma[\Sigma]$ act on the $1$-form symmetry operators $\mscr{W}_d[\gamma]$ by permuting their charges
\begin{equation}
    d=(\ms g, \alpha)\longmapsto \sigma\cdot(\ms g, \alpha)\in\mathcal A, \qquad \sigma\in\mathcal G[\ms G],
\end{equation}
whenever the curve $\gamma$ penetrates the surface $\Sigma$ as in Figure \ref{fig:IntersectionOfSymmetryOperatorInTheBulk}. Together, the $0$-form group $\mathcal G[\ms G]$ and $1$-form group $\mathcal A$ form a mathematical structure called a 2-group\cite{Kapustin:2013uxa, Cordova2019_1,Cordova2019_2, Delcamp:2018wlb, Delcamp:2019fdp}.

\smallskip In the following, we will elaborate more on these two types of topological operators and what kind of structures they holographically induce on a theory living at the $1+1d$ spacetime boundary. We will first focus on the Wilson line operators $\mscr{W}_d[\gamma]$ labeled by charges in $\mathcal A$, then turn to the surfaces $\dual\sigma[\Sigma]$ labeled by the group $\mathcal G[\ms G]$.

\subsection{Boundary Hilbert space from the bulk theory}
\label{Subsec:boundary_Hilbert_space}
Up to this point, we have considered the $2+1d$ bulk  topologically-ordered phase.
Now we consider a setup where the bulk spacetime has a boundary.
We are primarily interested in studying the physics of the $1+1d$ boundary and to use it to learn lessons about generalized quantum spin chains with gobal symmetry $\ms G$. Our formulation is a simple and generalizable example of topological holography.
The $1$-form symmetry operators brought to the boundary become conventional or $0$-form global symmetry for the $1+1d$ spacetime\footnote{Since they are still topological on the boundary, but now the same dimension as a spatial slice, i.e., codimmension-1 in $1+1d$ spacetime.} (see Figure \ref{fig:IntersectionOfSymmetryOperatorInTheBulk}). 
The formulation is based on the algebra of topological operators $\mscr{W}_d[\gamma]$ and $\dual{\sigma}[\Sigma]$ where $\gamma$ and $\Sigma$ are allowed to have boundaries if $\partial\gamma$ and $\partial\Sigma$ are located at the $1+1d$ boundary. 

\smallskip Consider the spacetime manifold\footnote{Here,  $\mbb{R}$ corresponds to the time direction. When computing thermal partition functions, this direction is compactified into a circle $S^1$.} $\mscr{M}_{2+1}=\mbb{R}\times C$, where $C=\mbb{R}_+\times S^1$ denotes a semi-infinite cylinder corresponding to a fixed time-slice at which the bulk Hilbert space is defined. The boundary spacetime manifold is given by $\partial\mscr{M}=\mbb{R}\times S^1$ and we are interested in constructing the quantum system that appears on the boundary. In particular, we would like to construct the generalized spin-chain Hilbert spaces on $S^1$. The choice of the semi-infinite cylinder in the bulk gives rise to all possible twisted sectors on the boundary and reveals subtle global features important when studying effects of dualities.\footnote{We could consider the construction on more general manifolds for example higher-genus surfaces with more boundaries. Such geometries would lead to many copies of the same twisted sectors and therefore do not provide any new information.}

\smallskip Consider Wilson line operators $\mscr{W}_d[\gamma_{a,b}]$ where $\gamma_{a,b}$ is an open curve with end-points $a$ and $b$, both on the boundary.
Such operators correspond to boundary operators which act on the boundary Hilbert space. We will denote such operators as
\begin{equation}\label{eq:the definiton of string operators}
    S_d(a,b)\equiv \mscr{W}_d[\gamma_{a,b}] = \BoundaryOperatorsOnCylinder
\end{equation}
where $a$ and $b$ are the end-points.
When line operators intersect locally, they have a simple commutation relation given pictorially by
\begin{equation} \label{eq:LineOperatorsLocalCrossing_Algebra}
    \begin{aligned}
    \LocalWWCrossingAlgebraOne&= e^{i\theta_{(\ms g,\alpha),(\ms g',\alpha')}}\;\LocalWWCrossingAlgebraTwo.
    \end{aligned}
\end{equation}
Here we have zoomed into a local region as illustrated by the dashed circles, and outside this region the lines are kept unchanged.
Physically this describes the mutual statistics, corresponding to a double exchange of anyons. Equation \eqref{eq:LineOperatorsLocalCrossing_Algebra} is a special case of \eqref{eq:U[X]_intersecting_O(Sigma)} for $1$-form symmetries where the charged line is itself topological. Interestingly, this can be thought of as $1$-form 't Hooft anomalies. Since each line consists of an electric and a magnetic component, labeled by $\alpha$ and $\ms g$, respectively, they are not true lines but rather ribbons. We can represent the twisting of such ribbons by another local rule
\begin{equation}\label{eq:LineOperatorsRibbonTwist_Algebra}
    \LocalRibbonLoop = e^{i\theta_{(\ms g, \alpha)}}\; \LocalRibbonUnloop.
\end{equation}
The phase $e^{i\theta_{(\ms g, \alpha)}}$ is sometimes called a topological spin, as it is gained when an anyon rotates $2\pi$ around itself (as the above graphic encodes). By a version of the spin-statistics theorem, this phase also corresponds to self-statistics which is gained by exchanging two identical anyons. From these two pieces of data, we can form the so-called $S$ and $T$ matrices
\begin{equation}\label{eq:SandT_matrices}
    S_{dd'} = \frac 1{\sqrt{|\mathcal A|}} e^{i\theta_{dd'}}, \qquad T_{dd'}= e^{i\theta_d}\;\delta_{dd'},
\end{equation}
which form a representation of $\tenofo{SL}(2,\mathbb Z)$ and essentially describe all the data about a topological phase we need here. For the finite group gauge theories we consider in this paper, these phases are given by
\begin{equation}
    e^{i\theta_{dd'}} = \ms R_{\alpha}(\ms g')\ms R_{\alpha'}(\ms g), \qquad \ms e^{i\theta_{d}} =  \ms R_{\alpha}(\ms g),
\end{equation}
where $d=(\ms g,\alpha)$ and $d'=(\ms g',\alpha')$ (compare with Section \ref{sec:SymmetryStructureAndBuildingBlocksofLocalOperators}) and $\ms R_{\alpha}(\ms g)$ for a general finite Abelian group of the form \eqref{eq:the generic finite Abelian group} is defined in \eqref{eq:the formula for representation labeled by alpha for a general finite Abelian group}.

\smallskip The operators $S_d(a,b)$, defined in \eqref{eq:the definiton of string operators}, generate a local operator algebra on the boundary that we will call the String Operator Algebra
\begin{equation}
    \mathbb{SOA}[\ms G]\equiv \left\langle S_d(a,b)\,\Big|\,\text{$\forall d\in\mcal{A}$ and $\forall a,b$}\right\rangle.
\end{equation}
From this, we can see that all non-contractible line operators that wrap around the cylinder commute with all boundary operators since they intersect at even number of points with pairwise opposite orientations
\begin{equation}
    \GammaEdgeCommutatorOne \;\;=\; \GammaEdgeCommutatorTwo.
\end{equation}
Line operators corresponding to a non-contractible curve $\gamma$ will be denoted as
\begin{equation}
    \Gamma_d \equiv \mscr{W}_d[\gamma].
\end{equation}
The commutativity of such operators with an string operator can be stated as
\begin{equation}
    \Gamma_d S_{d'}(a,b) = S_{d'}(a,b) \Gamma_d,
\end{equation}
for all $d$, $d'$, $a$ and $b$. In particular, we have pure electric and magnetic non-contractible lines 
\begin{equation}
    \Gamma_{(\ms g, 0)} = \GammaRedCylinder, \qquad 
    \Gamma_{(0, \alpha)} = \GammaBlueCylinder.
\end{equation}
Note that $\Gamma_d\in\mbb{SOA}[\ms G]$ since they are products of boundary operators $S_d(a,b)$, which can be depicted as follows
\begin{equation}\label{eq:GammaOperatorsDecomposedIntoBoundaryOperators}
    \GammaRedCylinder = \GammaRedCylinderDecomposedIntoBoundaryOperators \qquad     \GammaBlueCylinder = \GammaBlueCylinderDecomposedIntoBoundaryOperators.
\end{equation}
The bulk topological order has a topological ground-state degeneracy labeled by the eigenvalues of $\Gamma_{d}$ 
\begin{equation}\label{eq:GammaOperatorEigenstates}
    \Gamma_{d}|d'\rangle=\ms{R}_{\alpha'}(\ms g)\ms{R}_\alpha(\ms g')|d'\rangle,
\end{equation}
where $|d'\rangle=|(\ms{g}',\alpha')\rangle$ denotes the topologically-degenerate ground states. Note that the eigenvalue in \eqref{eq:GammaOperatorEigenstates} is exactly the $S$-matrix \eqref{eq:SandT_matrices} up to normalization, or in other words the mutual braiding statistics between two anyons. Since $\Gamma_d$ commutes with operators in $\mbb{SOA}[\ms G]$, the Hilbert space $\mathcal H$ generated out of boundary operators $S_d(a,b)$ decomposes into super-selection sectors labeled by line operators or anyons $d'$
\begin{equation}\label{eq:SpinChainHilbertSpaceFullDecomposition}
    \mc H = \bigoplus_{d'=(\ms g', \alpha')}\mc H_{d'}.
\end{equation}
The Hilbert space $\mcal{H}$ is a representation space of  $\mbb{SOA}[\ms G]$ (see Appendix \ref{Sec:holographic perspective} for a more detailed construction). These super-selection sectors cannot be connected with any operator in $\mbb{SOA}[\ms G]$. Since boundary operators commute with $\Gamma_d$, we can decompose
\begin{equation}\label{eq:SOAFullDecomposition}
    \mbb{SOA}[\ms G]=\bigoplus_{d'\in\mcal{A}}\mbb{SOA}_{d'}[\ms G],
\end{equation}
into eigenspaces of $\Gamma_d$.
Therefore, the boundary operators $S_d(a,b)$ are block-diagonal and each block lives in $\mbb{SOA}_{d'}[\ms G]$ for a fixed $d'\in\mcal{A}$. 

\smallskip The bulk topological order has additional operators corresponding to semi-infinite lines ending on the boundary
\begin{equation}\label{eq:InfiniteLinesOnBoundaryGraphics}
    \mscr{Y}_{(\ms g,0)} = \WBlueCylinder, \qquad \mscr{Y}_{(0, \alpha)} = \WRedCylinder.
\end{equation}
From \eqref{eq:LineOperatorsLocalCrossing_Algebra}, we see that the following simple algebraic relations hold 
\begin{equation}\label{eq:Graphical commutation relations between gamma and Y}
     \GammaWCommutatorCylinderOne= \ms R_\alpha(\ms g) \GammaWCommutatorCylinderTwo,\qquad\qquad \GammaWCommutatorCylinderThree= \ms R_\alpha(\ms g) \GammaWCommutatorCylinderFour.
\end{equation}
This implies that $\mscr{Y}_d$ operators shift the eigenvalues of the $\Gamma$s
\begin{equation}\label{eq:Gamma Eigenvalues change due to Y}
    \Gamma_{(\ms g,0)}\mscr{Y}_{d''}|d'\rangle=\ms R_{\alpha'+\alpha''}(\ms g)\,\mscr{Y}_{d''}|d'\rangle, \qquad 
    \Gamma_{(0,\alpha)}\mscr{Y}_{d''}|d'\rangle=\ms R_{\alpha}(\ms g'+\ms g'')\,\mscr{Y}_{d''}|d'\rangle,
\end{equation}
and thereby connect super-selection sectors
\begin{gather}
    \begin{aligned}
    \mscr{Y}_{d}:&\; \mc H_{d'}\longmapsto \mc H_{d'+d}.
    \end{aligned}
\end{gather}
The reason $\mscr{Y}_d$ can connect super-selection sectors is that it does not belong to $\mbb{SOA}[\ms G]$. As mentioned earlier, on the boundary $\Gamma_d$ becomes $0$-form symmetries (see Figure \ref{fig:IntersectionOfSymmetryOperatorInTheBulk}) and the String Operator Algebra consists of operators not charged under this symmetry. However, $\mscr{Y}_d$ corresponds to a charged local operator on the boundary. For example
\begin{equation}
    \GammaWPenetration= \ms R_\alpha(\ms g)\;\GammaWNoPenetration. 
\end{equation}
It then follows from these considerations that any state $|d'\rangle\in\mcal{H}_{d'}$ can be constructed from some state $|(0,0)\rangle\in\mcal{H}_{(0,0)}$ in the trivial sector using \eqref{eq:Gamma Eigenvalues change due to Y} 
\begin{equation}\label{eq:any sector from trivial sector by the action of Y}
    |d'\rangle=\mc{Y}_{d'}|(0,0)\rangle.
\end{equation}

\subsection{Mapping to generalized spin-chains}
\begin{figure}[t!]
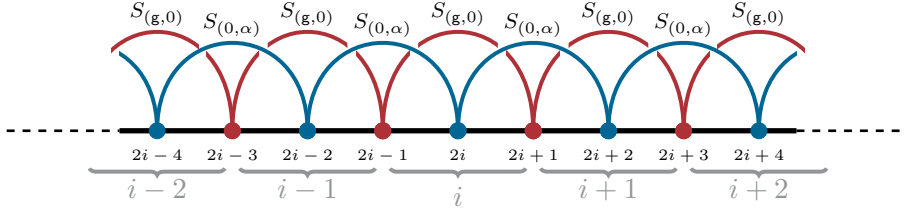

    \centering
    \StringOperatorsBetweenLatticeSites
    \caption{Regularization of the boundary such that magnetic lines $S_{(\ms g,0)}$ can only end on odd sites while electric lines $S_{(0,\alpha)}$ can only end on even sites. Alternatively, this can be thought of as the boundary of a topological gauge theory with alternating topological (gapped) boundary conditions corresponding to the condensation of magnetic and electric lines. With such boundary conditions, electric and magnetic lines are allowed to end on certain segments of the boundary and remain gauge-invariant. These new non-contractible line operators correspond to relative homology cycles and become our boundary operators.}
    \label{fig:string operators between lattice sites}
\end{figure}

In this section, we map the boundary Hilbert space constructed in the previous section to the language of (generalized) spin-chains. In particular, we would like to connect to the standard notation of operators and symmetries discussed in Section \ref{sec:SymmetryStructureAndBuildingBlocksofLocalOperators}. The subset of the non-contractible operators
\begin{equation}\label{eq:UDefinitionInTermsOfGamma}
    \mathcal U_{\ms g} \equiv \Gamma_{(\ms g, 0)} = \GammaRedCylinder,
\end{equation}
form a group isomorphic to $\ms G\subset\mathcal A$ (see equation \eqref{eq:LocalFusionOfLineOperators}). 
This has the following important implications for the $1+1d$ quantum spin chain on the boundary
\begin{itemize}
    \item Since operators $\mathcal U_{\ms g} \equiv \Gamma_{(\ms g, 0)}$ commute with all boundary operators, we can identify it with global symmetries of the boundary.
    \item Conversely, since the operators $\mc U_{\ms g}$ commute with all the operators in $\mathbb{SOA}[\ms G]$, the string operator algebra corresponds to the algebra of $\ms G$-symmetric operators of the boundary theory. 
\end{itemize}
One might wonder about the other non-contractible operators
\begin{equation}\label{eq:TauDefinitionInTermsOfGamma}
    \mathcal T_\alpha \equiv \Gamma_{(0,\alpha)} = \GammaBlueCylinder,
\end{equation}
A natural question is that what do these correspond to in the spin-chain language? A clear way to see what these operators become in the spin-chain language is to decompose the Hilbert space into eigenspaces of $\mathcal T_\alpha$ only, but not of $\mathcal U_{\ms g}$ (unlike \eqref{eq:SpinChainHilbertSpaceFullDecomposition} and \eqref{eq:SOAFullDecomposition})
\begin{equation}\label{eq:SOAandHilbertSpacePartialDecomposition}
    \mathcal H = \bigoplus_{\ms g'\in\ms G}\mathcal H_{\ms g'}, \qquad \mathbb{SOA}[\ms G] = \bigoplus_{\ms g'\in\ms G}\mathbb{SOA}_{\ms g'}[\ms G].
\end{equation}
Within each sector $\mathcal H_{\ms g'}$, these operators are diagonalized and equal to a complex phase $\mathcal T_\alpha\big|_{\ms g'} = \ms R_\alpha(\ms g')\mathbb I$ (see equation \eqref{eq:GammaOperatorEigenstates}).
To make connection with a lattice spin-chain, we introduce a UV regularization such that the boundary is a 1D lattice and $S_{(\ms g,0)}(a,b)$ can only end on odd sites, while $S_{(0, \alpha)}(a,b)$ end on even sites (see Figure \ref{fig:string operators between lattice sites}). 
This regularization removes certain ambiguities related to the ribbon structure of the line operators. For simplicity, we will use this notation for the boundary operators
\begin{equation}\label{eq:Definition of A and B operators}
    \begin{aligned}
    A_i[\ms g] &= \bigoplus_{\ms g'\in\ms G}A^{(\ms g')}_i[\ms g] \equiv S_{(\ms g,0)}(2i-1,2i+1), \\
    B_{i,i+1}[\alpha] &= \bigoplus_{\ms g'\in\ms G}B^{(\ms g')}_{i,i+1}[\alpha] \equiv S_{(0,\alpha)}(2i,2i+2),
    \end{aligned}
\end{equation}
where we have decomposed the boundary operators into the blocks within sectors in \eqref{eq:SOAandHilbertSpacePartialDecomposition}.
We can now make the identifications\footnote{Note that we are using additive notation for $\alpha\in\tenofo{Rep}(\ms G)$ since for the case of groups we are considering, finite Abelian groups, we have $\tenofo{Rep}(\ms G)\simeq\ms G$. Therefore, the operator $\mcal{O}_{-\alpha,i+1}$ is labeled by an element $-\alpha\in\tenofo{Rep}(\ms G)$. For example, consider the group $\mbb{Z}_4=\{0,1,2,3\}$. Then, $-1\sim 3$ since everything is defined modulo $4$.}
\begin{equation}
\label{eq:Wilson_operator_to_lattice_operator}
    A^{(\ms g')}_i[\ms g] = \mathcal O_{\ms g,i} \qquad B^{(\ms g')}_{i,i+1}[\alpha] = \mathcal O_{\alpha,i}\mathcal O_{-\alpha,i+1},
\end{equation}
where $\mathcal O_{\ms g}$ and $\mathcal O_\alpha$ were defined in Section \ref{sec:SymmetryStructureAndBuildingBlocksofLocalOperators}. This is true since the operators satisfy the same algebra as can be explicitly checked (for a more careful approach see Appendix \ref{Sec:holographic perspective}).
Furthermore, these relations translate to (see \eqref{eq:GammaOperatorsDecomposedIntoBoundaryOperators})
\begin{gather}\label{eq:BulkGammaToBoundaryU_T}
    \begin{aligned}
        \Gamma_{(\ms g,0)} &= \prod_i S_{(\ms g,0)}(2i,2i+2)&\longrightarrow\quad \mathcal U_{\ms g}&= \bigotimes_i \mathcal O_{\ms g,i}
        \\
        \Gamma_{(0,\alpha)} &= \prod_i S_{(0, \alpha)}(2i+1,2i+3)&\longrightarrow \quad\mathcal T_{\alpha}&\equiv  \bigotimes_i \mathcal O_{\alpha,i} \mcal{O}_{-\alpha,i+1}
    \end{aligned}
\end{gather}
Since we are on a semi-infinite cylinder and using the fact that $\mcal{O}_{\alpha,i}\mcal{O}_{-\alpha,i}$ is the identity operator, the last operator reduces to $\mc T_\alpha = \mc O_{\alpha, 1}\mc O_{-\alpha, L+1}$ which must be equal to $\ms R_{\alpha}(\ms g')\mathbb I$ in this sector. This can be achieved if we demand
\begin{equation}\label{eq:O operator twisted by rep R}
    \mathcal O_{\alpha,L+1} \equiv \ms R_{\alpha}(\ms g')\mathcal O_{\alpha,1}.
\end{equation}
In other words, if we construct Hamiltonians out of operators $A$ and $B$, then each sector of \eqref{eq:SOAandHilbertSpacePartialDecomposition} is the Hilbert space of a $\ms G$-symmetric spin-chain with symmetry-twisted boundary condition labeled by an element $\ms g'\in\ms G$. Thus the operator \eqref{eq:TauDefinitionInTermsOfGamma} measures the twisted boundary condition of a given spin-chain. From a $1+1d$ spacetime point of view, it corresponds to having inserted a symmetry operator $\mathcal U_{\ms g'}[\gamma]$ along the time direction which causes a twisted boundary condition and can be detected by $\mathcal T_{\alpha}$.

\smallskip Using elements of $\mbb{SOA}[\ms G]$, we can now write any $\ms G$-symmetric spin-chain Hamiltonian in the schematic form
\begin{equation}\label{eq:GeneralGSymmetricHamiltonian}
    {H}[\{t_{\mbs d}\}]=\; \sum_{\mbs d}\sum_{\ell}t_{\mbs d}(\ell)\mc{K}_{\mbs d}(\ell)+ \text{h.c.},
\end{equation}
where $\mcal{K}_{\mbs d}(\ell)$ is some product of string operators $S_{d_1}\cdots S_{d_n}$ on an interval of length $\ell$ on the lattice, ${\mbs d}\equiv (d_1,\cdots,d_n)$ denotes a collection of dyons, and $t_{\mbs d}(\ell)\in\mbb{C}$ are coupling constants of the theory. This can also be written in terms of operators $A$ and $B$ defined in \eqref{eq:Definition of A and B operators}. This Hamiltonian is block-diagonal with respect to the decomposition \eqref{eq:SOAandHilbertSpacePartialDecomposition}, 
\begin{equation} \label{eq:HamiltonianTwistedBC_PartialBlockDiagonalized}
    H=\bigoplus_{\ms g'\in\ms G}H_{\ms g'}=
    \begin{pmatrix}
    H_{\ms g'_1} &  &    
    \\
                   & H_{\ms g'_2} &  
    \\
                               &     &  \ddots   
    \end{pmatrix},
\end{equation}
where $H_{\ms g'}$ corresponds to a spin chain with the symmetry-twisted boundary condition $\ms g'\in\ms G$. For example in the case of $\ms G=\mbb{Z}_2=\{0,1\}$, $H_{\ms g'=0}$ and $H_{\ms g'=1}$ corresponds to periodic and anti-periodic boundary conditions \eqref{eq:O operator twisted by rep R}, respectively. The full decomposition \eqref{eq:SpinChainHilbertSpaceFullDecomposition} is obtained by further decomposing into eigenspaces of the symmetry operator \eqref{eq:UDefinitionInTermsOfGamma}
\begin{equation} \label{eq:HamiltonianTwistedBC_BlockDiagonalized}
    H_{\ms g'} = \bigoplus_{\alpha'\in\tenofo{Rep}(\ms G)}H_{(\ms g',\alpha')}=
    \begin{pmatrix}
    H_{(\ms g', \alpha'_1)} &   
    \\
                   & H_{(\ms g', \alpha'_2)} &
    \\
                               &     &     \ddots
                \end{pmatrix}.
\end{equation}
Now for each anyonic charge $d'\in\mcal{A}$, there is a sector $H_{d'=(\ms g',\alpha')}=P_{\alpha'} H_{\ms g'} P_{\alpha'}=H_{\ms g'}P_{\alpha'}$ corresponding to a symmetry sector labeled by $\alpha'\in\tenofo{Rep}(\ms G)$ and a symmetry-twisted boundary condition labeled by $\ms g'\in\ms G$. We have defined the projection operator into a symmetry sector $\alpha$ by 
\begin{align}
P_{\alpha} \equiv \frac{1}{|\ms{G}|}\sum_{\ms h\in\ms G}\ms R_\alpha^{-1}(\ms h)\,\mathcal U_{\ms h}.     
\end{align}
Note that we have identified $\Gamma_{(\ms g,0)}$ as boundary symmetries while $\Gamma_{(0,\alpha)}$ corresponded to operators related to symmetry-twisted boundary conditions. We could have done it the other way around, this is related to the existence of so-called electric-magnetic dualities that we will discuss shortly. The reader might also wonder why we do not think of both of these as boundary symmetries, this is related to 't Hooft anomaly between these lines. These anomalies also play a related role when considering gapped boundary conditions, as they serve as obstruction to condensing certain anyons simultaneously. We will discuss this point in the following section.

\subsection{Gapped boundaries as gapped phases}
\label{sec:Gapped boundaries in 2+1d as gapped phases in 1+1d}
So far, we have argued that the space of boundary conditions of a bulk topological gauge theory with gauge group $\ms G$ in 2+1 dimensions is equivalent to the space of $\ms G$-symmetric quantum systems in 1+1 dimensions. We can further ask what this correspondence can tell us about $\ms G$-symmetric theories. To answer this question, we remind the reader that the possible gapped boundary conditions of a topological gauge theory with the gauge group $\ms G$ has been classified and correspond to so-called Lagrangian subgroups of $\mcal{A}$. These are maximal subsets\footnote{We have used the same symbol for Lagrangian subgroups and the space of local operators in Section \ref{sec:SymmetryStructureAndBuildingBlocksofLocalOperators}. These should not be confused with each other.} $\mc L\subset \mc A$ of bulk line operators (anyons) that can simultaneously be condensed on the boundary \cite{KapustinSaulina2011,Levin201301,Barkeshlietal_2013}. Physically, the subgroup $\mathcal L$ corresponds to the largest possible subsets of anyons that satisfy: 1) they are all bosons and 2) they all have trivial mutual braiding with each other. Another interesting point of view is that $\mathcal L$ are maximal $1$-form subgroups of $\mathcal A$ with vanishing 't Hooft anomalies \eqref{eq:LineOperatorsLocalCrossing_Algebra} and \eqref{eq:LineOperatorsRibbonTwist_Algebra}. This means that there is no obstruction to condensing those anyons simultaneously \cite{Hsin_Lam_Seiberg, KaidiKomargodskiOhmoriSeifnashriShao102107}. This in turn gives a classification of possible gapped phases of $\ms G$-symmetric quantum theories at the boundary. 
The classification of gapped boundaries is therefore equivalent to the classification of Lagrangian subgroups. This classification for a topological gauge theory with the gauge group $\ms G$ is given by the tuple $(\ms H, \psi)$ where \cite{Ostrik1,Ostrik2}
\begin{equation}
    \ms H\subset \ms G, \qquad [\psi]\in H^2\left(\ms H,\ms U(1)\right).
\end{equation}
Physically, a pair $(\ms H, \psi)$ labelling a gapped boundary condition corresponds to a gapped phase where the $\ms G$ symmetry has been spontaneously broken to a subgroup $\ms H$ but with a possible SPT twist encoded in the 2-cocycle $\psi$. This is consistent with previously known classifications of $1+1$ dimensional $\ms G$-symmetric phases using tensor-network methods \cite{ChenGuWen201008,FidkowskiKitaevSPT2011,PollmannSPT2009A2012,NorbertEtAllSPT2011}.
{
\begin{figure}[t!]
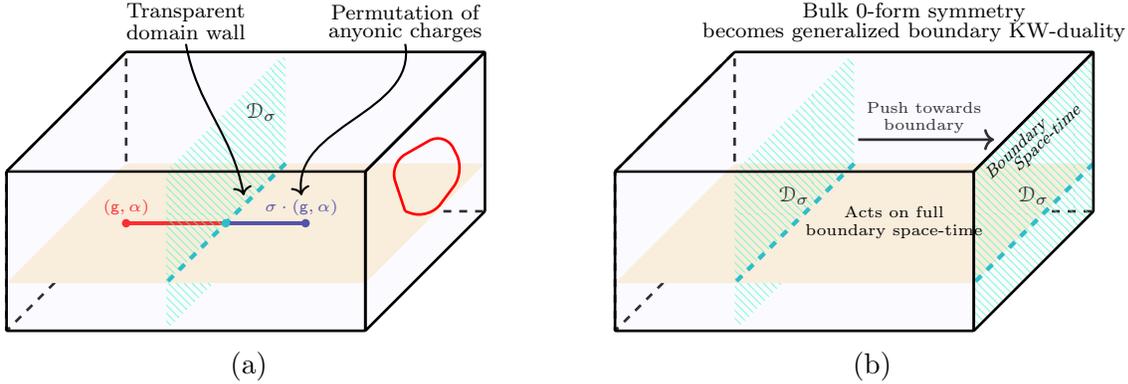
\centering\hspace{-1cm}
    \begin{subfigure}{.4\linewidth}\centering
    \IntersectionOfSymmetryOperatorInTheBulk
    \subcaption{}
    \label{fig:IntersectionOfSymmetryOperatorInTheBulk}
    \end{subfigure}\hspace{1cm}
    ~
    \begin{subfigure}{.4\linewidth}\centering
    \BulkSymmetryOperatorBroughtToTheBoundary
        \subcaption{}
        \label{fig:BulkSymmetryOperatorBroughtToTheBoundary}
    \end{subfigure}
    \caption{Illustration of the $2+1d$ spacetime with a boundary $1+1d$ spacetime. The green hatched surface corresponds to a $0$-form bulk symmetry $\dual\sigma[\Sigma]$, $\sigma\in\mathcal G[\ms G]$, known as anyonic symmetry.
    \textbf{(a)}
    When a $1$-form symmetry is brought to the boundary, it becomes a topological line operator of a $1+1$d theory and thus a $0$-form symmetry.
    Furthermore when the bulk symmetry surface $\Sigma$ is oriented along time, it splits spacetime into two regions and functions as a transparent domain wall. The domain wall is transparent to a line operator $\mscr{W}_{(\ms g, \alpha)}[\gamma]$ as it can penetrate through, although with the cost of transforming its charge to $(\ms g',\alpha')=\sigma\cdot(\ms g, \alpha)$.
    \textbf{(b)}
    When the $0$-form symmetry surface is brought to a boundary, it acts on the full boundary spacetime as a duality transformation.
    }
\end{figure}}
\smallskip Given a Lagrangian subgroup $\mc L$, one can construct a boundary Hamiltonian which only includes the operators $S_d$ where $d\in\mc L$
\begin{equation}\label{eq:the Hamiltonian associated to a Lagrangian subgroup}
    H_{\mc L} = -\sum_{d\in\mc L} \sum_i S_{d}(i,i+1).
\end{equation}
Due to the trivial mutual braiding between any pair of operators in $\mc L$, all terms in this Hamiltonian commute, therefore such a Hamiltonian is a \textit{fixed-point Hamiltonian}. The ground states are in the simultaneous $+1$ eigenspace of the all operators $S_d$ labeled by anyons $d\in\mc L$. 
Additionally $S_{d}$ for any $d\notin \mc L$, does not commute with at least some subset of operators in the Hamiltonian \eqref{eq:the Hamiltonian associated to a Lagrangian subgroup}. 
As a consequence, the  expectation values of string operators $S_d$ are given by
\begin{equation}\label{eq:StringOrderParameterCondensation}
    \langle S_{d}\rangle_{\tenofo{GS}} = 
    \begin{cases}
        1, \qquad \text{if}\qquad d\in\mc L,\\
        0, \qquad \text{if}\qquad d\not\in\mc L.\\
    \end{cases}
\end{equation}
Equivalently, anyons in $\mc L$ are condensed (i.e. they become part of the ground-state of the system at the boundary) while those outside this set are confined and form the excitations at the boundary. Furthermore, this construction gives us the order parameters corresponding to each gapped $\ms G$-symmetric phase \cite{PollmannSPT2009A2012,NorbertEtAllSPT2011, Verresen_2017}. 
These order parameters naturally play an essential role in identifying the tuple $(\ms H, \psi)$ corresponding to each Lagrangian subroup $\mc L$ as described later in Sec.~\ref{subsec:gapped phases}.

\begin{figure}[t!]
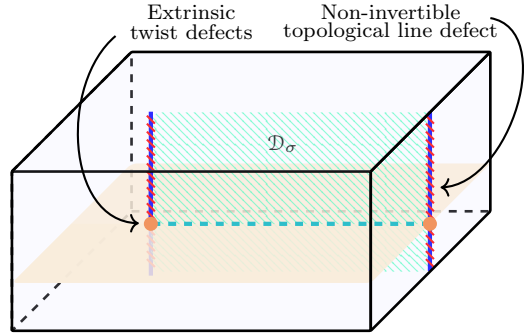

    \centering
    \TwistDefectNoninvertibleDefectLine
    \caption{When the edge of the surface is brought the boundary, its shadow will be a topological line operator on the boundary. However, these line operators will be non-invertible, as boundary operators alone cannot annihilate them.}
    \label{fig:TwistDefectNoninvertibleDefectLine}
\end{figure}

\begin{figure}[H]
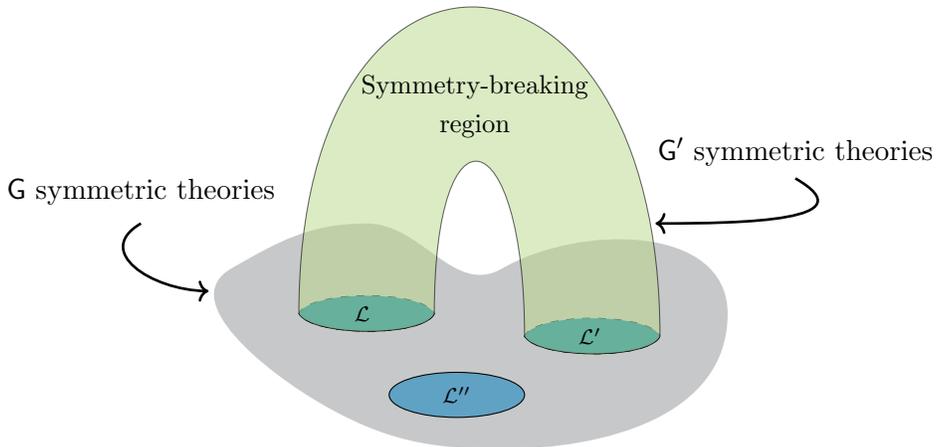

    \centering
    \SymmetryBreakingWormholes
    \caption{The grey surface corresponds to the space of $\ms G$ symmetric theories, while the the regions $\mathcal L$, $\mathcal L'$ and $\mathcal L''$ are three different gapped phases. These phases are considered different, only if $\ms G$-symmetry is preserved. However, if we allow perturbations that break the symmetry down to a subgroup $\ms G'\subset\ms G$, the space of theories will get enlarged. Now it might be possible to connect two previously different phases $\mathcal L$ and $\mathcal L'$ through a symmetry-breaking region (which one might to call a symmetry-breaking ``wormhole''). In such a scenario, $\mathcal L$ and $\mathcal L'$ are the same phase in the space of $\ms G'$-symmetric theories. The notion of gapped phase therefore depends on the space of theories under consideration.
    }
    \label{fig:symmetry-breaking wormhole}
\end{figure}

\smallskip To avoid confusion, let us clarify what we mean by the notion of $\ms{G}$-symmetric gapped phase. We define a gapped phase as follows: any two gapped $\ms{G}$-symmetric Hamiltonians connected by a continuous $\ms{G}$-symmetric gapped path is said to belong to the same phase.\footnote{By a $\ms G$-symmetric path, we mean a deformation of the Hamiltonian in a given gapped phase by a family of Hamiltonians depending on a set of parameters. As we change the parameters, the corresponding Hamiltonian changes producing a path in the space of $\ms G$-symmetric Hamiltonians.} In other words, each topologically-connected component of the space of all gapped $\ms{G}$-symmetric Hamiltonians is a distinct gapped phase. The Lagrangian subgroups discussed above correspond to these phases. Specific $\ms{G}$-symmetric Hamiltonians might have more symmetries and structures beyond $\ms{G}$, for example, it might have global $\ms U(1)$ symmetry. From a physical point of view, there might exist phases related to the spontaneous-breaking of $\ms U(1)$ symmetry without affecting the $\ms{G}$-symmetry. Such phases might correspond to the same Lagrangian subgroups as they might be connected through a $\ms{G}$-symmetric gapped path that explicitly breaks $\ms U(1)$ symmetry (see Figure \ref{fig:symmetry-breaking wormhole}). In order to distinguish those phases, one has to look at the space of $\ms{G}\times\ms U(1)$-symmetric theories and appropriately modify the definition of ``gapped phase" \cite{KaidiKomargodskiOhmoriSeifnashriShao102107}.

\subsection{From bulk symmetries to boundary dualities}\label{subsec:from bulk symmetry to boundary dualities}
We have seen how line operators (or $1$-form symmetries) in the bulk of a $2+1d$ topologically ordered system, corresponds to ($0$-form) global symmetries of $1+1d$ quantum systems. One could wonder about the $0$-form (anyonic) symmetries $\mathcal G[\ms G]$ present in the bulk of a $2+1d$ topological phase, and naturally ask what do they represent in the space of $\ms G$-symmetric $1+1d$ theories? This symmetry group is a subgroup $\mathcal G[\ms G]\subset S_{|\mathcal A|}$ of the group of permutations $S_{|\mcal{A}|}$ of charges in $\mathcal A$ satisfying
\begin{equation}\label{eq:DefinitionOfAnyonicSymmetryGroup}
    \mathcal G[\ms G] = \left\{\sigma\in S_{|\mathcal A|}\;\big|\; S_{\sigma(d),\sigma(d')}=S_{dd'} \text{ and } T_{\sigma(d),\sigma(d')}=T_{dd'}\right\},
\end{equation}
where $S$ and $T$ are defined as in \eqref{eq:SandT_matrices}. In other words, these are all possible ways of permuting charges while preserving all topological properties like fusion rules and braiding statistics. We denote elements in $\mathcal G[\ms G]$ by $\dual\sigma$. 
In particular, for the case of Abelian topological gauge theories, the group of 0-form symmetries $\mc G[\ms G]$ for any generic finite Abelian group $\ms G=\prod_{i=1}^{n}\mbb Z_{N_i}$ are generated by three kinds of symmetry operations \cite{EtingNikshychOstrik200909,NikshychRiepel201309,FuchsPrielSchweigertValentino201404}, 
\begin{enumerate}
    \item {\it Universal kinematical symmetries}: These are induced from the automorphisms of the group $\ms G$. 
    For every automorphism $\varphi: \ms G\to \ms G$, one obtains a 0-form symmetry $\sigma_{\varphi}: \mc A\to \mc A$ in $\mc G[\ms G]$, which acts on a dyon $d=(\ms g, \alpha)\in \mc A$ as
    \begin{equation}
        \sigma_{\varphi}: (\ms g,\alpha)\to (\ms \varphi(\ms g), (\ms {\varphi}^{-1})^{\star}(\alpha)),
    \end{equation}
    where $\varphi^\star:\text{Rep}(\ms G)\rightarrow\text{Rep}(\ms G)$ defined as $[\varphi^\star\ms R_{\alpha}](\ms g)\equiv \ms R_\alpha(\varphi(\ms g))$. This thus maps $\alpha\in\text{Rep}(\ms G)$ to some other $\alpha'\in\text{Rep}(\ms G)$.
    
    \item {\it Universal dynamical symmetries}: These correspond to elements in $H^{2}(\ms G,\ms{U}(1))$, where for $\ms G$ is of the form \eqref{eq:the generic finite Abelian group}. The corresponding cohomology group is
    \begin{equation}
        H^{2}(\ms G,\ms{U}(1))=\prod_{i<j}\mathbb Z_{\text{gcd}(N_i,N_j)}.
    \end{equation}
    For each cohomology class, there is an associated alternating bicharacter\footnote{An alternating bicharacter is a group homomorphism $\mfk{c}:\ms G\times\ms G\to\mbb{C}^*$ in both arguments which satisfies the property $\mfk{c}(\ms g,\ms g)=1$ for all $\ms g\in \ms G$ \cite{FuchsPrielSchweigertValentino201404}.}
    \begin{equation}
        \mfk{c}_{\mbs{\ell}}(\ms g,\ms h)=\exp\left\{2\pi i\sum_{i,j}
        \frac{\ell_{ij}\ms g_i\ms h_j}{\text{gcd}(N_i,N_j)}
        \right\},
        \label{eq:group 2-cocycle}
    \end{equation}
    with $\ell_{ij}=-\ell_{ji}$ and $\ms g=(\ms g_1,\dots, \ms g_n)$, and $\ms h=(\ms h_1,\dots, \ms h_n)$. Given such a bicharacter, one obtains a 0-form symmetry $\sigma_{\mfk{c}}$ which maps the dyon $d=(\ms g,\alpha)$ as 
    \begin{align}
    \sigma_{\mathfrak{c}}: (\ms g,\alpha) \mapsto (\ms g,\alpha^{\mfk{c}}_{\ms g}),
    \end{align}
    where the transformed representation $\alpha^{\mfk{c}}_{\ms g}$ has the form $\ms R_{\alpha^{\mathfrak{c}}_{\ms g}}(\cdot)=\ms R_{\alpha}(\cdot)\mfk{c}_{\mbs\ell}(\ms g,\cdot)$.
    \item {\it Partial electric-magnetic dualities}: These correspond to performing an electric-magnetic duality on a given factor $\mbb Z_{N_j}$ in $\ms G$. 
    Concretely, an electric-magnetic duality acting on the $j^{\text{th}}$ factor transforms the dyons $d\in \mc A$ as
\begin{equation}
    \sigma_{j}:
    \begin{gathered}
        \hspace{.1cm}(\ms g_1,\dots,\ms g_{j-1},\ms g_{j},\ms g_{j+1},\dots \ms g_{n} )\mapsto (\ms g_1,\dots,\ms g_{j-1},\alpha_{j},\ms g_{j+1},\dots \ms g_{n}),
       \\
    (\alpha_1,\dots,\alpha_{j-1},\alpha_{j},\alpha_{j+1},\dots \alpha_{n} ) 
    \mapsto
    (\alpha_1,\dots,\alpha_{j-1},\ms g_{j},\alpha_{j+1},\dots \alpha_{n} ).
    \end{gathered}
\end{equation}
\end{enumerate}
Since the $0$-form symmetry group $\mathcal G[\ms G]$ acts on the $1$-form symmetries $\mathcal A$, the combined symmetry structure forms a 2-group structure. 
Note that even though $\ms G$ and $\mathcal A$ are Abelian, the group $\mathcal G[\ms G]$ can be non-Abelian.

\smallskip Since elements in $\mathcal G[\ms G]$ are standard symmetries in $2+1d$, they are defined on $2d$ surfaces. When $\dual\sigma[\Sigma]$ is put on a time slice as in Figure \ref{fig:ActionOfSymmetryOperatorsOnBulkLines}, they act on the Hilbert space by permuting anyons as shown.  However, we can orient the surface in different ways. For example, when it is oriented as in Figure \ref{fig:IntersectionOfSymmetryOperatorInTheBulk}, it splits the space into two regions and corresponds to a domain wall. Whenever a string operator crosses such a domain wall, it is transformed according to the corresponding anyonic symmetry. For this reason, these domain walls are called transparent domain walls \cite{KitaevKong2012}. These are also called invertible domain walls. The set of all possible gapped domain walls between a topological order and itself can be found using Lagrangian subgroups, with the help of the folding trick \cite{Barkeshlietal_2013}. Only a subset of these are transparent.

\smallskip Now, if we push the surface towards the boundary as in Figure \ref{fig:BulkSymmetryOperatorBroughtToTheBoundary}, we end up with a new kind of object from the boundary point of view. The surface now acts on the entire spacetime, thus acts on the full boundary theory and can in principle change it to a completely different theory. In other words, $0$-form symmetries in $2+1d$ corresponds to dualities between different $1+1d$ theories. It is natural to call these $(-1)$-form symmetry operators (see also \cite{Seiberg_2020}). %
Note that these dualities are a vast generalization of the classic Kramers-Wannier duality \cite{WannierKramers194108a,WannierKramers194108b}.

\smallskip It is instructive to reinterpret this spacetime logic from a spatial Hamiltonian/Hilbert space level. On a fixed-time slice, domain wall in Figure \ref{fig:IntersectionOfSymmetryOperatorInTheBulk} corresponds to a line-like domain wall with the property
\begin{equation}\label{eq:LineGoingThroughDefectTimeSlice}
    \LineGoingThroughDefectTimeSlice.
\end{equation}
This is the standard view of transparent domain walls in the topological order literature \cite{Bombin2010, KitaevKong2012, Barkeshli_2013, Barkeshli_2013b, Teo_2015}. However, if we bring this domain wall to the boundary as in Figure \ref{fig:BulkSymmetryOperatorBroughtToTheBoundary}, then we see that
\begin{equation}
    \ActionOfBulkDefectOnBoundaryOperators,
\end{equation}
i.e. boundary operators maps to other boundary operators

\begin{equation}\label{eq:DualityActing_on_S_operators}
    S_d(a,b) \longmapsto \dualhd\sigma S_d(a,b) \equiv S_{\sigma\cdot d}(a,b).
\end{equation}
This action has the following properties
\begin{gather}\label{eq:the action of duality on sums and products of SOA generators}
    \begin{aligned}
        \dualhd\sigma\mathbb I &= \mathbb I,\\
        \left(\dualhd\sigma S_d(a,b)\right)^\dagger &= \dualhd\sigma S_d(a,b)^\dagger,\\
        \dualhd{\sigma}\sum_{i=1}^n c_{i}\,S_{d_i}(a_i,b_i)&=\sum_{i=1}^nc_i\,\dualhd{\sigma}S_{d_i}(a_i,b_i),
    \\
        \dualhd{\sigma}\prod_{i=1}^nS_{d_i}(a_i,b_i)&=\prod_{i=1}^n\dualhd{\sigma}S_{d_i}(a_i,b_i),
    \end{aligned}
\end{gather}
where $c_i$s are constants. These imply that the map  $\wh{\mscr{D}}_\sigma:\mathbb{SOA}[\ms G]\longmapsto\mathbb{SOA}[\ms G]$ is an algebra homomorphism. Furthermore, since it just permutes the generators of $\mathbb{SOA}[\ms G]$ \eqref{eq:DualityActing_on_S_operators} while preserving \eqref{eq:LineOperatorsLocalCrossing_Algebra} and \eqref{eq:LineOperatorsRibbonTwist_Algebra}, it is a non-trivial automorphism of the algebra of boundary operators $\mathbb{SOA}[\ms G]$. It is useful to also consider how such dualities act on states by assuming the existence of a unitary operator on the Hilbert space with the property
\begin{equation}\label{eq:S operators mapping by unitary that implements the duality action on SOA}
    U_{\sigma}S_d(a,b)U_{\sigma}^\dagger  = \dualhd\sigma S_{d}(a,b) = S_{\sigma\cdot d}(a,b).
\end{equation}
The existence of such unitary operator is guaranteed according to Theorem 3.3 of \cite{BondAlgebraicApproach}. Here $\mathbb{SOA}[\ms G]$ corresponds to our \textit{bond algebra}. Note that all dualities we are discussing here are called self-dualities in the terminology of \cite{BondAlgebraicApproach}, since we are using a bond algebra large enough to contain any $\ms G$-symmetric spin-chain Hamiltonian. We will briefly discuss action of dualities on non-symmetric theories in Section \ref{subsec:action of dualities on non-symmetric theories}.

\smallskip The unitary map preserves the trivial sector in \eqref{eq:any sector from trivial sector by the action of Y} (see Appendix \ref{Sec:holographic perspective}). From \eqref{eq:LineGoingThroughDefectTimeSlice} and \eqref{eq:InfiniteLinesOnBoundaryGraphics}, it then follows that the non-contractible operators transform according to
\begin{equation}\label{eq:TransformationOfGammaAndY}
    U_{\sigma}\Gamma_d U_{\sigma}^\dagger=\Gamma_{\sigma\cdot d}, \qquad U_{\sigma}\mscr{Y}_{d'} U_{\sigma}^\dagger=\mscr{Y}_{\sigma\cdot d'}.
\end{equation}
In particular, when combined with \eqref{eq:any sector from trivial sector by the action of Y}, we see that the duality permutes the super-selection sectors non-trivially
\begin{equation}\label{eq:DualityUnitaryPermutesSectorsOfHilbertSpace}
    U_\sigma: \mathcal H_{d'} \longmapsto \mathcal H_{\sigma\cdot d'},
\end{equation}
the same way as anyons are permuted. On the level of the sectors of the String Operator Algebra we thus have
\begin{equation}
    \dualhd{\sigma}\mbb{SOA}[\ms G]=\bigoplus_{d'\in\mcal{A}} U_{\sigma}\,\mbb{SOA}_{d'}[\ms G]\,U_{\sigma}^\dagger=\bigoplus_{d'\in\mcal{A}} \mbb{SOA}_{\sigma\cdot d'}[\ms G].
\end{equation}
Taking the transformation \eqref{eq:TransformationOfGammaAndY} together with the identifications \eqref{eq:UDefinitionInTermsOfGamma} and \eqref{eq:TauDefinitionInTermsOfGamma}, we see that symmetry sectors and  boundary conditions get non-trivially mixed under these dualities. For example we have that the symmetry operator $\mcal{U}_{\ms g}$ \eqref{eq:UDefinitionInTermsOfGamma} maps to $\mcal{U}_{\ms g^\vee}\mcal{T}_{\alpha^\vee}$ where $(\ms g^\vee,\alpha^\vee)=\sigma\cdot(\ms g,0)$. This is an alternate way of stating \eqref{eq:DualityUnitaryPermutesSectorsOfHilbertSpace}.

\smallskip As a simple example consider the electric-magnetic duality $\sigma\cdot(\ms g,\alpha)=(\alpha,\ms g)$ which exists for any finite Abelian group $\ms G$. Under this duality $\mcal{U}_{\ms g}$ and $\mcal{T}_{\alpha}$ are swapped (see equation \eqref{eq:TransformationOfGammaAndY}). Transforming both states and operators, we find the following relation between matrix elements 
\begin{equation}
    \langle \ms g',\alpha'|\mcal{U}_\ms{g}|\ms{g}'',\alpha''\rangle=\langle \alpha',\ms g'|\mcal{T}_{\alpha=\ms g}|\alpha'',\ms{g}''\rangle.
\end{equation}
In the simplest case of $\ms{G}=\mbb{Z}_2$, this relation can be checked explicitly using
\begin{equation}
    \begin{aligned}
    \mcal{U}&=\begin{blockarray}{ccccc}
    & \matindex{$|0,0\rangle$} & \matindex{$|0,1\rangle$} & \matindex{$|1,0\rangle$} & \matindex{$|1,1\rangle$}
    \\
    \begin{block}{c(cccc)}
    \matindex{$\langle 0,0|$} &  +\mbb{I} & 0 & 0 & 0   \\
    \matindex{$\langle 0,1|$} &  0 & -\mbb{I} & 0 & 0   \\
    \matindex{$\langle 1,0|$} &  0 & 0 & +\mbb{I} & 0   \\
    \matindex{$\langle 1,1|$} &  0 & 0 & 0 & -\mbb{I} \\
    \end{block}
  \end{blockarray}\,\,,
  \\
  \mcal{T}&=\begin{blockarray}{ccccc}
    & \matindex{$|0,0\rangle$} & \matindex{$|0,1 \rangle$} & \matindex{$|1,0 \rangle$} & \matindex{$|1,1\rangle$}
    \\
    \begin{block}{c(cccc)}
    \matindex{$\langle 0,0|$} &  +\mbb{I} & 0 & 0 & 0   \\
    \matindex{$\langle 0,1 |$} &  0 & +\mbb{I} & 0 & 0   \\
    \matindex{$\langle 1,0|$} &  0 & 0 & -\mbb{I} & 0   \\
    \matindex{$\langle 1,1|$} &  0 & 0 & 0 & -\mbb{I} \\
    \end{block}
  \end{blockarray}\,\,.
    \end{aligned}
\end{equation}
The upper two blocks correspond to a periodic $\mbb{Z}_2$-symmetric spin chain in the even and odd symmetry sectors. Similarly, the lower two blocks correspond to an anti-periodic spin chain.

\subsection{Action of dualities on Hamiltonians}
We have seen that how $0$-form symmetries in the bulk act on the boundary Hilbert space, we can now turn to their actions on Hamiltonians. We will see that bulk $0$-form symmetries become dualities in the space of $\ms G$-symmetries theories.
As discussed above, any $\ms G$-symmetric Hamiltonian $H$  in $1+1d$ is constructed out of boundary operators $S_d(a,b)$, defined in \eqref{eq:GeneralGSymmetricHamiltonian}, and hence is an element of $\mathbb{SOA}[\ms G]$. Bringing the domain wall $\mscr{D}_\sigma[\Sigma]$ to the boundary will therefore map $H$ to a dual Hamiltonian
\begin{equation}\label{eq:DualityActing_on_Hamiltonian}
    H \longrightarrow H^\vee \equiv\dualhd\sigma H = U_\sigma H U_\sigma^\dagger,
\end{equation}
where in the second equality, we have used the fact that duality symmetry could be represented by a unitary operator $U_\sigma$ acting on the Hilbert space (see Section \ref{subsec:from bulk symmetry to boundary dualities}). More precisely, the action on a generic Hamiltonian of the form \eqref{eq:GeneralGSymmetricHamiltonian} becomes
\begin{equation}\label{eq:GeneralSymmetricHamiltonianTransformed}
    \begin{aligned}
        H^\vee[\{t_{\mbs d}\}]&=\sum_{\mbs d}\sum_{\ell}t_{\mbs  d}(\ell)\left[\wh{\mscr{D}}_\sigma\cdot \mc{K}_{\mbs d}(\ell)\right]+ \text{h.c.},
        \\
        &=\sum_{\mbs d}\sum_{\ell}t_{\mbs  d}(\ell)\,\mc{K}_{\sigma\cdot\mbs d}(\ell)+ \text{h.c.},
        \\
        &=\sum_{\mbs d}\sum_{\ell}t_{\sigma^{-1}\cdot \mbs d}(\ell)\,\mc{K}_{\mbs d}(\ell)+ \text{h.c.}
        \\
        &=H[\{t_{\sigma^{-1}\cdot \mbs{d}}\}],
    \end{aligned}
\end{equation}
where of course $U_\sigma\mathcal K_{\mbs d}U_\sigma^\dagger = U_\sigma S_{d_1}\dots S_{d_n} U_\sigma^\dagger = S_{\sigma\cdot d_1}\dots S_{\sigma\cdot d_n} = \mathcal K_{\sigma\cdot\mbs d}$, which is the consequence of a nontrivial action of dualities on global sectors, as given in  \eqref{eq:DualityUnitaryPermutesSectorsOfHilbertSpace}. In other words, the duality acts on coupling constants and permutes twisted sectors as
\begin{equation}\label{eq:Dual hamiltonian in terms of original hamiltonian in different sectors}
    H^\vee_{(\ms g,\alpha)}[\{t_{\bs d}\}] = H_{\sigma\cdot(\ms g,\alpha)}[\{t_{\sigma^{-1}\cdot \bs d}\}].
\end{equation}
Since the Hamiltonians are related by a unitary map \eqref{eq:DualityActing_on_Hamiltonian}, their energy spectra are preserved up to a permutation of sectors
\begin{equation}\label{eq:the relation of spectra before and after a duality transformation}
    E_n^{(\ms g,\alpha)}[\{t_{\bs d}\}] = E_n^{\sigma\cdot(\ms g,\alpha)}[\{t_{\sigma^{-1}\cdot \bs d}\}],
\end{equation}
where $n$ labels the eigenvalues.

\smallskip Furthermore, correlation functions of symmetric operators also transform naturally under the action of duality.
Consider a correlation function of the following kind
\begin{equation}
\langle S_{d_1}(\ell_1)S_{d_2}((\ell_2)) \dots \rangle_{\beta, H_d}:= \text{tr}_{\mcal{H}}\left[ S_{d_1}(\ell_1)S_{d_2}(\ell_2) \dots e^{-\beta H_{\ms g}}P_{\alpha}\right],
\end{equation}
where $S_{d}(\ell)$ is the boundary operator obtained from a topological line operator labelled by $d$ of length $\ell$. 
Since the the dyonic labels $d\in \mc A$ transform under duality, one obtains a dual correlation function with an equivalent expectation value 
\begin{align}
    \langle S_{d_1}S_{d_2} \dots \rangle_{\beta, H_d}
    =
    \langle S_{\sigma(d_1)}S_{\sigma(d_2)} \dots \rangle_{\beta, H^{\vee}_{d}}
\end{align}
In this paper, we will show how these dualities are powerful tools to constrain and determine portions of the phase diagrams and compute exact conformal spectra of many non-trivial phase transitions.

\subsection{Action of dualities on non-symmetric theories}\label{subsec:action of dualities on non-symmetric theories}

We have so far considered the action of dualities on the space of $\ms{G}$-symmetric Hamiltonians which are constructed out of string operator algebra $\mbb{SOA}[\ms G]$. This is a subalgebra of the algebra of all linear operators $\mcal{L}(\mcal{H})$ on a Hilbert space $\mcal{H}$, as discussed in Section \ref{sec:SymmetryStructureAndBuildingBlocksofLocalOperators}. On the Hilbert space of the spin chain, string operators with charge $d=(0,\alpha)$ are given by
\begin{equation}\label{eq:the string operator labeled by pure charge}
    S_{(0,\alpha)}(\ell)=\mcal{O}_{\alpha,i}\mcal{O}_{-\alpha,j},
\end{equation}
where $i$ and $j$ are the endpoints of the interval $\ell$. These operators are symmetric and belong to $\mbb{SOA}[\ms G]$. However, $\mcal{O}_{\alpha,i}$ itself is not symmetric and hence does not belong to $\mbb{SOA}[\ms G]$, but it belongs to $\mcal{L}(\mcal{H})$. When a duality acts on $\mbb{SOA}[\ms G]$, it is local. By this, we mean that the support of symmetric operators (for example the interval $\ell$ in \eqref{eq:the string operator labeled by pure charge}) are preserved under duality. Due to this property, $\ms G$-symmetric local Hamiltonians are mapped to dual local Hamiltonians, which are also $\ms G$-symmetric.

\begin{figure}[t!]
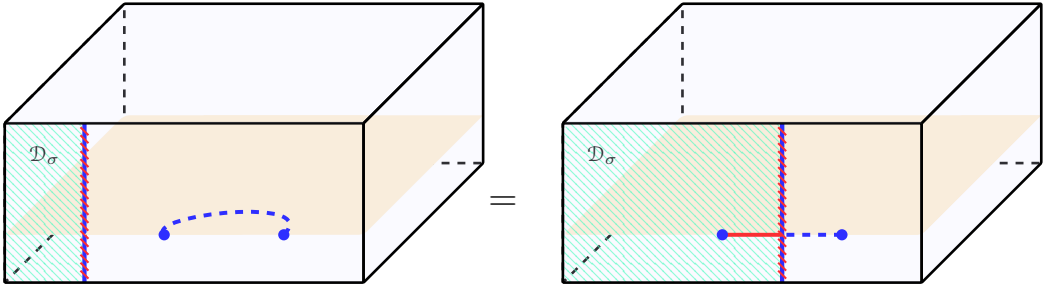
\centering
    \DualityDefectDraggedOverLocalOperatorBulkPicture
    \caption{Duality $\dual\sigma$ acting on part of boundary. When dragged over local non-symmetric operators, $\mathcal O_{\alpha,i}$, the duality becomes non-local. The edge of the duality region can become a (possibly non-invertible) symmetry of the boundary, for boundaries that are self-dual under $\sigma$.}\label{fig:the bulk picture of duality defect dragged over a local operator}
\end{figure}

\smallskip However, nothing prevents us to act with duality on the space of all operators $\mcal{L}(\mcal{H})$. However, it turns out that the mapping of non-symmetric operators becomes non-local. In order to see this, first notice that under a duality $\sigma$, $S_{(0,\alpha)}(\ell)=\mcal{O}_{\alpha,i}\mcal{O}_{-\alpha,j}$ is mapped to $S_{\sigma\cdot(0,\alpha)}(\ell)$. Now consider acting duality on part of spacetime boundary. We see that when the boundary $\partial\Sigma$ of a duality domain wall $\mscr{D}_\sigma[\Sigma]$ is moved across a non-symmetric operator, we find the following relation (see also Figure \ref{fig:the bulk picture of duality defect dragged over a local operator})
\begin{equation}\label{eq:Graphic depicting what happens when non-invertible lines cross local non-symmetric operators}
    \DualityDefectDraggedOverLocalOperatorOneNEW = \DualityDefectDraggedOverLocalOperatorTwoNEW.
\end{equation}
\noindent In other words, as a duality wall is dragged over a local operator, a non-local operator emerges. If the $1d$ quantum spin chain was infinitely long, we could drag $\mcal{O}_{j,-\alpha}$ to $j\to-\infty$ and the non-symmetric local operator would be mapped to an operator with infinite support. For example, under electric-magnetic duality $(\ms g,\alpha)\mapsto(\alpha,\ms g)$, we would have the map 
\begin{equation}
    \mcal{O}_{\alpha=a,i}\quad\longmapsto\quad\mcal{O}^\vee_{\alpha=a,i}=\prod_{j=-\infty}^i\mcal{O}_{\ms g=a,j},
\end{equation}
where local operators $\mcal{O}_{\alpha,i}\mcal{O}_{-\alpha,i+1}$ are mapped to $\mcal{O}^\vee_{\alpha,i}\mcal{O}^\vee_{-\alpha,i+1}=\mcal{O}_{\ms g=\alpha,i+1}$ which is local because the infinite lines cancel each other out. This is the conventional approach to the Kramers-Wannier duality of the Ising spin chain ($\ms G=\mbb{Z}_2$) where the duality mapping is non-local and defined by 
\begin{equation}\label{eq:the conventional, bad, and nonlocal KW-duality mapping}
    \sigma^z_{i}\longmapsto\mu^z_i=\prod_{j=-\infty}^i\sigma^x_j, \qquad \sigma^x_i\longmapsto\mu^x_i=\sigma^z_i\sigma^z_{i+1}. 
\end{equation}
If we restrict ourselves to $\mbb{SOA}[\ms G]\subset\mcal{L}(\mcal{H})$, the dualities are local and unitary-implementable as \eqref{eq:S operators mapping by unitary that implements the duality action on SOA}. Locality is important since local Hamiltonians are mapped to local Hamiltonians, and unitary-implementability is important for the arguments in Section \ref{topological holography: application}, which are crucial for the application of dualities to explore phase diagrams of $\ms G$-symmetric systems. However, if we extend the duality to $\mcal{L}(\mcal{H})$, generically, we lose both of these two properties. For example, non-symmetric Hamiltonians will be generally mapped to non-local and thus unphysical Hamiltonians, as can be seen from \eqref{eq:the conventional, bad, and nonlocal KW-duality mapping}. 

\smallskip Note that non-symmetric operators also exist in the topological holography picture, which are given by line operators with charge $(0,\alpha)$ that are coming from infinity and end on the boundary. The zero-form $\ms G$-symmetry of the boundary $\mcal{U}_{\ms g}[\gamma]$ can detect these operators by
\begin{equation}\label{eq:Infinite Line operator end point links with boundary symmetry operators and transforms accordingly}
    \GammaWPenetration= \ms R_\alpha(\ms g)\;\GammaWNoPenetration. 
\end{equation}
This is exactly how local operators $\mcal{O}_{\alpha,i}$ transform when linking with symmetry operator $\mcal{U}_{\ms g}[\gamma]$ as described in \eqref{eq:U[X]_surface_around_O(x)}. But again, this operator is attached to an infinite line. 

\smallskip In \cite{FrohlichFuchsRunkelSchweigert200404,FrohlichFuchsRunkelSchweigert200607} a very similar construction was presented in the context of conformal field theories. Our discussion here applies to any $G$-symmetric system and in that sense a natural generalization of these results.

\subsection{Emergence of non-invertible symmetries}\label{sec:Emergence of Symmetry}
\smallskip Note that if we apply duality to half of the boundary spacetime as in Figure \ref{fig:the bulk picture of duality defect dragged over a local operator}, we get a domain wall between two different dual theories. 
However, if this is done to a theory $H[\{t^\star_{\mbs{d}}\}]$ self-dual under $\sigma$, the domain wall can become a non-invertible topological line defect.\footnote{Such a construction has recently been used to find non-invertible topological operators in $3+1$ dimensional quantum field theories \cite{Choi:2021kmx, Choi:2022zal}.}
Therefore, there will exist operators commuting with the Hamiltonian $H[\{t^\star_{\mbs{d}}\}]$ corresponding to new emergent symmetries. 
The symmetries of such self-dual Hamiltonians, in general, do not form a group, but rather a fusion category which extends the $\ms G$ global symmetry.
For example, consider the critical point of the transverse-field Ising model which is self-dual under the above-mentioned Kramers-Wannier duality. At this point, besides being invariant under the $\mbb{Z}_2$-symmetry $\mcal{U}=\prod_i\sigma^x_i$, the theory will also be invariant under a new symmetry $\mscr{D}$. Here, $\mscr{D}$ is the edge of the Kramers-Wannier duality interface, or from the bulk point-of-view, a twist defect brought to the boundary. Their product is given by \cite{FrohlichFuchsRunkelSchweigert200404,FrohlichFuchsRunkelSchweigert200909,ChangYingHsuanShaoWangYin201802}
\begin{equation}
    \mscr{D}\times\mscr{D}=\mbb{I}+\mcal{U}.
\end{equation}
The theory now contains three topological operators labelled as $\mbb{I}$, $\mcal{U}$ and $\mscr D$, which form the Ising fusion category, also known as the Tambara-Yamagami category of the group $\mbb{Z}_2$ \cite{TAMBARA1998692}. 
More generally, the $\sigma$ line in \eqref{eq:Graphic depicting what happens when non-invertible lines cross local non-symmetric operators} is topological and can be non-invertible, thus an emergent symmetry in theories that are self-dual under $\sigma$. The action of such lines on local operators as depicted in \eqref{eq:Graphic depicting what happens when non-invertible lines cross local non-symmetric operators} has been studied in the context of conformal field theories \cite{FrohlichFuchsRunkelSchweigert200404,FrohlichFuchsRunkelSchweigert200909,ChangYingHsuanShaoWangYin201802}. However, our description holds for any self-dual theory which is not necessarily conformal or even gapless. 

\smallskip Concretely, the fusion rules of duality defects can be computed as follows. Consider a boundary theory that is self-dual under a duality $\sigma$. Now apply the duality $\dual\sigma[\Sigma]$ on a strip $\Sigma$ on the boundary. The edges of the strip $\partial\Sigma =\gamma_1\cup\gamma_2$ will correspond to two topological line defects $L_\sigma$, $\bar L_\sigma$. The fusion rule $L_\sigma\times \bar L_\sigma$ can be extracted by shrinking the strip to a line or by coarse-graining \cite{Choi:2021kmx, Choi:2022zal}.

\smallskip The above reasoning provides a holographic interpretation of two kinds of topological line defects in $1+1d$ conformal field theories \cite{ChangYingHsuanShaoWangYin201802}, (a) invertible line defects, associated with global symmetry $\ms G$, originating from bulk one-form symmetries (anyonic lines), and (b) non-invertible line defects originating from edges of bulk zero-form symmetries. 

\smallskip A complimentary point of view is to consider a finite bulk domain wall $\dual\sigma[\Sigma]$ with a boundary $\partial\Sigma\not = \emptyset$, which could be seen as an interface between a non-trivial $\dual\sigma$ and a trivial domain wall $\dual 1 = \mathbb I$. Thinking of this situation from a Hamiltonian/spatial point of view (compare to Figure \ref{fig:TwistDefectNoninvertibleDefectLine}) we have 
\begin{equation}
    \TwistDefectImage
\end{equation}
where the end-points of the domain wall correspond to so-called extrinsic twist defects and have been extensively studied in the literature \cite{Bombin2010, Barkeshli_2013, Barkeshli_2013b, Teo_2015}. It is known that they behave like anyons, but their fusion rules can be non-invertible (non-Abelian anyons) even if the underlying topological order is Abelian. From the spacetime point of view, it is clear that twist defects are line-like objects rather than point-like objects. Twist defects in the bulk are thus related to (potentially) non-invertible line defects on the boundary \cite{Burbano:2021loy,KaidiOhmoriZheng202209}, and these are responsible for duality-twisted boundary conditions \cite{Grimm_2002, WeiMoradi2014}.

\section{Topological holography: Applications}
\label{topological holography: application}

\smallskip Thus far we have seen that a topological gauge theory with gauge groups $\ms G$ in $2+1d$, holographically controls symmetry aspects of the space of $\ms G$-symmetric $1+1d$ theories. In particular line-like $1$-form symmetries correspond to $0$-form symmetries in $1+1d$, and surface-like $0$-form symmetries correspond to dualities in the space of $\ms G$-symmetric $1+1d$ theories. 
In this section, we explore a few basic properties of these dualities and see how they can help explore and characterize features of the phase-diagram of $\ms G$-symmetric $1+1d$ theories.
These insights will be used in section \ref{sec:examples} to apply Topological Holography to the study of concrete spin chains with various symmetry groups.

\smallskip The elements of the duality group $\mathcal G[\ms G]$ are by definition automorphisms of $\mathbb{SOA}[\ms G]$ and thus preserve the properties of the local operator algebra. The action of these dualites on Hamiltonians, as described in \eqref{eq:DualityActing_on_S_operators} and \eqref{eq:DualityActing_on_Hamiltonian}, will therefore give rise to dual Hamiltonians with similar properties. 
In particular, the behavior of correlation functions such as exponential and algebraic decay in gapped and critical systems, respectively, are preserved under the duality mapping.
Therefore, if $\ms H$ is gapped or gapless then $\ms H^\vee$ will be also be gapped or gapless, respectively.
However, the duality can act non-trivially on global symmetry properties such as symmetry sectors of the Hilbert space and twisted boundary conditions. 
Therefore, certain aspects of the boundary theory described by $H$ are not preserved.
For example the partition function and hence the spectrum of the theory can change. 
We exploit this to compute the full spectra of conformal field theories describing the critical phenomena of several non-trivial phase transitions.

\subsection{Gapped phases and web of dualities}\label{subsec:gapped phases}
The first question is how are gapped phases mapped to each other? Recall that gapped phases are classified by Lagrangian subgroups $\mathcal L\subset\mathcal A$, which are collections of charges $d=(\ms g,\alpha)\in \mc A$ that simultaneously condense in that phase.
The action of anyonic symmetries $\sigma\in\mathcal G[\ms G]$ on anyonic charges can be lifted to sets of charges by acting elements-wise on each anyonic charge individually. Thus if $\mathcal L=\{d_1, d_2,\dots\}$ then $\sigma\cdot\mathcal L\equiv \{\sigma\cdot d_1, \sigma\cdot d_2, \dots\}$. Note that acting on a Lagrangian subgroup
\begin{equation}
    \mathcal L\longmapsto \mathcal L^\vee = \sigma\cdot\mathcal L,
\end{equation}
gives rise to another set of anyonic charges $\mathcal L^\vee$ which also satisfy the Lagrangian subgroup conditions as the action of $\sigma$ (by definition \eqref{eq:DefinitionOfAnyonicSymmetryGroup}) preserves the braiding and fusion properties of anyons (\eqref{eq:LocalFusionOfLineOperators}, \eqref{eq:LineOperatorsLocalCrossing_Algebra}, and  \eqref{eq:LineOperatorsRibbonTwist_Algebra}).
In other words, maximal non-anomalous subgroups of the 1-form symmetry $\mcal{A}$ (of the bulk topological order) are permuted amongst each other under the action of dualities.
As discussed earlier, for each gapped phase we can write an exactly solvable fixed-point Hamiltonian \eqref{eq:the Hamiltonian associated to a Lagrangian subgroup}.
It then follows from \eqref{eq:DualityActing_on_Hamiltonian} and \eqref{eq:GeneralSymmetricHamiltonianTransformed} that the dualities in $\mathcal G[\ms G]$ map fixed-point Hamiltonians in the gapped phase $\mathcal L$ to a dual Hamiltonian which is the fixed-point Hamiltonian of the gapped phase $\mathcal L^\vee=\sigma\cdot\mathcal L$
\begin{equation}\label{eq:the mapping of fixed-point Hamiltonians under a duality transformation}
     H_{\mc L} \longmapsto  H_{\mc L}^\vee =\dualhd\sigma H_{\mc L} = H_{\mc L^\vee}. 
\end{equation}
 However, how about a more generic non-fixed-point Hamiltonian $H[\{t_{\mbs{d}}\}]$ \eqref{eq:GeneralGSymmetricHamiltonian} that is $\ms G$-symmetric? By definition, each gapped phase is path-connected. This means that we can find a continuous one-parameter family $H_{\mathcal L}(s),\,s\in[0,1]$, of Hamiltonians in the gapped phase $\mcal{L}$ where $H_{\mathcal L}(0)$ is the fixed-point Hamiltonian and $H_{\mathcal L}(1)$ is any arbitrary Hamiltonian in the same phase. After the duality transformation, the path $H_{\mcal{L}}^\vee(s)\equiv \dualhd{\sigma}H_{\mcal{L}}(s)$ is still a continuous function of $s$. This can be seen as follows: $H_{\mathcal L}(s)$ is a linear superposition of products of boundary operators $S_{d}(a,b)$ \eqref{eq:GeneralGSymmetricHamiltonian} where the coefficients $\{t_{\mbs d}(s)\}$ are continuous functions of $s\in[0,1]$. The duality acts term-wise as in \eqref{eq:GeneralSymmetricHamiltonianTransformed} and hence coefficients are merely permuted, i.e.  $t_{\mbs d}(s)$ are replaced by $t_{\sigma^{-1}\cdot \mbs{d}}(s)$, and thus are still continuous functions of $s$. Therefore, the path remains connected after the duality. Furthermore $H^\vee_{\mcal{L}}(s)$ is a one-parameter family of gapped Hamiltonians since duality maps gapped phases onto gapped phases. This means that there are no second-order phase transition along the path after duality. This however does not rule out the possibility of first-order transitions. Note that the duality is unitarily implementable \eqref{eq:DualityActing_on_Hamiltonian} and preserves the spectrum \eqref{eq:the relation of spectra before and after a duality transformation} along the path. Therefore, there cannot be any level-crossing along the path after the duality since there were none before the duality by assumption. 
 This consequently excludes first-order transitions along the path.
 Finally, it follows from \eqref{eq:the mapping of fixed-point Hamiltonians under a duality transformation} that $H_{\mcal{L}}^\vee(0)$ is a fixed-point Hamiltonian in $\mcal{L}^\vee$ because $H_{\mcal{L}}(0)$ is the fixed-point Hamiltonian in $\mcal{L}$. From this argument, we can conclude that the whole path $H_{\mcal{L}}^\vee(s)$ is in $\mcal{L}^\vee$ as there is no phase transition along the path.

\smallskip Note that if a generic Hamiltonian $H[\{t_{\mbs{d}}\}]$ \eqref{eq:GeneralGSymmetricHamiltonian} is gapped, it must either be in a phase labelled by a Lagrangian subgroup $\mcal{L}$ or sits at a first-order transition and has ground-state degeneracies. We will see examples of this later. By gapped phase, we always exclude such first-order transitions.   

\smallskip We will call the group structure of a Lagrangian subgroup $\mc L$ the \textit{fusion structure} of the corresponding gapped phase.%
To preempt any potential confusion, we want to emphasize that the term ``fusion structure" here should not be confused with the fusion category underlying a gapped boundary phase. In the following, we will see that for a given finite Abelian group $G$ and the associated fusion group $\mcal{A}\simeq G\times G$, the Lagrangian subgroups $\mcal{L}\subset \mcal{A}$ may have different group-like structures that are not necessarily isomorphic to $G$ (although as a group, they all have the same order as $G$). %
The name fusion structure refers to the fact that it encapsulates the fusion rules of the condensed anyons associated with the given phase.
This concept proves to be very useful in studying and constraining the action of dualities on various phases, as we will illustrate in the ensuing sections, and thus merits a special designation.
Following the discussion around \eqref{eq:StringOrderParameterCondensation}, we can interpret this fusion structure as the algebra of order parameters for the gapped phase.
From the same line of argument as above, we can see that the two Lagrangian subgroups $\mc L$ and $\mc L^\vee = \sigma\cdot\mc L$ must be isomorphic as groups.
In other words, we can conclude that only gapped phases with the same fusion structure can be dual to each other.
Both of these phases have $\ms G$-symmetric Hamiltonians but the symmetry could be spontaneously broken to subgroups with different SPT twists $(\ms H_{\mcal{L}},\psi_{\mcal{L}})$ and $(\ms{H}_{\mcal{L}^\vee},\psi_{\mcal{L}^\vee})$.
This information can be derived out of $\mcal{L}$ and $\mcal{L}^\vee$ and thus depends on more than their group structure; it depends on their specific embedding inside $\mcal{A}$. 

\smallskip Let us explain how to extract $(\ms H_\mcal{L},\psi_{\mcal{L}})$ from the Lagrangian subgroup $\mcal{L}$. 
Consider the lowest-energy eigen-state  $|\Psi^{(0)}_d\rangle$ of the fixed-point Hamiltonian $H_\mcal{L}$ in the super-selection sector $d=(\ms g,\alpha)\in \mcal{A}$.
Now let us consider the action of a string operator $S_{\bar{d}}(\ell)$ on the state $|\Psi^{(0)}_d\rangle$ where $\bar{d}=(\bar{\ms g},\bar{\alpha})\in\mcal{L}$. 
Since $\bar{d}$ is in the set of condensed dyons in the gapped phase labeled by $\mc L$, the true ground-state must satisfy 
\begin{equation}
    S_{\bar{d}}(\ell)|\Psi^{(0)}_d\rangle=\ms R_{\bar\alpha}(\ms g)\ms R_{\alpha}(\bar{\ms g})|\Psi^{(0)}_d\rangle\overset{!}{=}|\Psi^{(0)}_d\rangle,\qquad \forall \ \bar{d}\in\mcal{L},
\end{equation}
or in other words $\ms R_{\bar\alpha}(\ms g)\ms R_{\alpha}(\bar{\ms g})=1$ (see \eqref{eq:the formula for representation labeled by alpha for a general finite Abelian group} for the definition of $\ms R_{\alpha}(\ms g)$).
This implies that the true ground-states belong to the twisted sectors $d\in\mcal{L}\subset\mcal{A}$. Define the magnetic projector $\Pi:\mcal{A}\simeq\ms G\times\tenofo{Rep}(\ms G)\to \ms G$ given by $\Pi(\ms g,\alpha)=\ms g$. 
In the gapped phase $\mcal{L}$, the symmetry has been spontaneously broken down to $\Pi(\mcal{L})=\ms H_\mcal{L}\subset \ms G$. 
The ground-states are thus given by $|\tenofo{GS}_{\ms h,\alpha}\rangle\equiv |\Psi^{(0)}_{\ms h,\alpha}\rangle$ where $(\ms h,\alpha)\in\mcal{L}$. 
These correspond to the ground-state with symmetry-twisted boundary condition $\ms h\in\ms H_{\mc L}$.
In order to compute the SPT twist $\Psi_\mcal{L}$, consider the expectation value of the remaining-symmetry operators $\mcal{U}_{\bar{\ms h}}$ for $\bar{\ms h}\in\ms H$ in the ground-state
\begin{equation}
    \langle \tenofo{GS}_{\ms h,\alpha}|\mcal{U}_{\bar{\ms h}}|\tenofo{GS}_{\ms h,\alpha}\rangle=\ms R_{\alpha}(\bar{\ms h})=\ms R^{-1}_{\bar{\ms \alpha}}(\ms h).
\end{equation}
The last equality follows from $\ms R_{\bar\alpha}(\ms g)\ms R_{\alpha}(\bar{\ms g})=1$. Let us define the $\ms U(1)$ phase
\begin{equation}\label{eq:definition of phase psiL}
    \begin{aligned}
        \psi_\mcal{L}(\ms h,\bar{\ms h})&\equiv \langle \tenofo{GS}_{\ms h,\alpha}|\mcal{U}_{\bar{\ms h}}|\tenofo{GS}_{\ms h,\alpha}\rangle=\ms R_{\alpha}(\bar{\ms h}).
    \end{aligned}
\end{equation}
One can readily check that this satisfies the group 2-cocycle condition \cite{ChenGuLiuWen201106}. 
In fact, the cohomology class $[\psi_\mcal{L}]$ belongs to the second group-cohomology group $H^2(\ms H,\ms U(1))$, which characterizes the SPT phase $\mcal{L}$ \cite{ChenGuLiuWen201106, Tiwari_2018} and the phase \eqref{eq:definition of phase psiL} is the topological response when the model is coupled to a background $\ms H$-gauge field whose holonomy in the space and time directions are $\bar{\ms h}$ and $\ms h$ respectively.
Since $\ms H$ is finite and Abelian, it takes the general form $\ms H = \prod_i\mathbb Z_{N_i}$ with the corresponding second cohomology group
\begin{equation}
    H^2(\ms H, \ms U(1)) = \prod_{i<j}\mathbb Z_{\tenofo{gcd}(N_i,N_j)}.
\end{equation}
Concretely, any such SPT is labeled by a list of integers $p_{ij}\in\mathbb Z_{\tenofo{gcd}(N_i,N_j)}$. These numbers can be extracted by comparing \eqref{eq:definition of phase psiL} with the following representative 2-cocycle
\begin{equation}
    \psi(\ms h, \bar{\ms h}) = \exp\left(\sum_{i<j}\frac{2\pi i \,p_{ij}}{\tenofo{gcd}(N_i,N_j)}\ms h_i\bar{\ms h}_j\right).
\end{equation}
From this, we see that non-trivial SPT phases appear when the condensed generators of $\mcal{L}$ are dyonic. The integers $p_{ij}$ correspond to which magnetic and electric charges are combined to form the dyons condensed in $\mcal{L}$.

\smallskip Another important observation from our construction is that certain Hamiltonians can have non-trivial emergent symmetries.
First let us define the stabilizer subgroup of a gapped phase
\begin{equation}
    \tenofo{Stab}(\mcal{L})\equiv\left\{\sigma\in\mcal{G}[\ms G]\,\big|\,\sigma\cdot\mcal{L}=\mcal{L}\right\}.
\end{equation}
By construction, any gapped Hamiltonian $H_\mcal{L}[\{t_{\mbs d}\}]$ maps to another Hamiltonian $\dualhd{\sigma}H_\mcal{L}[\{t_{\mbs d}\}]$ in the same phase under dualities $\sigma\in\tenofo{Stab}(\mcal{L})$. However, there can exist points in a gapped phase $\mcal{L}$ that are self-dual
\begin{equation}
    H[\{t^\star_{\mbs{d}}\}]=\dualhd{\sigma}H[\{t^\star_{\mbs{d}}\}],
\end{equation}
for some $\sigma\in\tenofo{Stab}(\mcal{L})$. Such self-dual Hamiltonians will have emergent symmetries beyond the global $\ms G$-symmetry. 
These emergent symmetries can however be non-invertible and form a fusion category.
They are edges of the duality walls coming from the bulk, related to twist defects, from the point of view of topological holography (see Figure \ref{fig:TwistDefectNoninvertibleDefectLine}).
The largest such emergent symmetries happen at points that are self-dual under the full $\tenofo{Stab}(\mcal{L})$.  

\begin{figure}[t]\centering 
    \includegraphics[width=.9\textwidth]{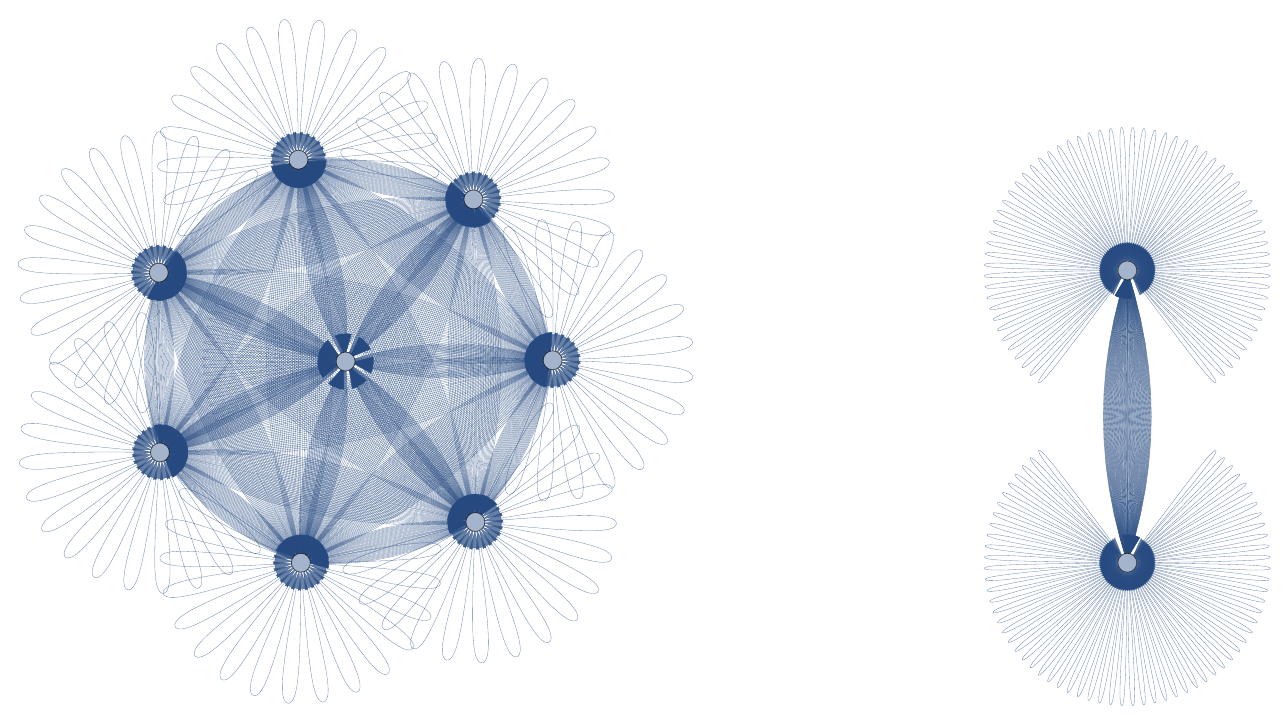}
    \caption{Web of dualities of $\ms G=\mathbb Z_2\times\mathbb Z_4$: each node corresponds to a gapped phase and each line depicts a duality transformation between these phases. The web of dualities has two connected components, corresponding to the two possible fusion structures for theories with $\ms G=\mathbb Z_2\times\mathbb Z_4$ global symmetry.}
    \label{fig:Web_Of_Dualities_Z2xZ4}
\end{figure}

\smallskip As a simple example, consider $\ms G=\mbb{Z}_2\times\mbb{Z}_4$ where we have
\begin{equation}
    \mcal{A}=\left\{(\ms g_1,\ms g_2,\alpha_1,\alpha_2)\,\Big|\,\ms g_1,\alpha_1=0,1\tenofo{ and }\ms g_2,\alpha_2=0,1,2,3\right\}.
\end{equation}
There are ten Lagrangian subgroups labeled as $\mcal{L}_1,\cdots,\mcal{L}_{10}$ (see Section \ref{sec:examples} for details) with the following fusion structures
\begin{equation}
    \mcal{L}_1,\cdots,\mcal{L}_8\simeq \mbb{Z}_2\times\mbb{Z}_4, \qquad \mcal{L}_9,\mcal{L}_{10}\simeq \mbb{Z}_2\times\mbb{Z}_2\times\mbb{Z}_2.
\end{equation}
Therefore, there are two different types of fusion structures. So there only exist dualities between the gapped phases in the set $\{\mcal{L}_1,\cdots,\mcal{L}_8\}$ and similarly between the ones in $\{\mcal{L}_9,\mcal{L}_{10}\}$ but not between these sets. Beyond electric-magnetic dualities, there are other dualities mixing the $\mbb{Z}_2$ and $\mbb{Z}_4$ electric and magnetic charges. The duality group turns out to be $\mcal{G}[\ms G]=(D_4\times D_4)\rtimes \mbb{Z}_2$, which is a non-Abelian group of order $128$. We can draw a \textit{web of dualities} that shows how gapped phases are related by dualities. A web of dualities can be drawn as follows: represent each gapped phase by a node and for each $\sigma\in\mcal{G}[\ms G]$ draw a line connecting gapped phases mapped under this duality. We have illustrated the web of dualities for $\ms G=\mbb{Z}_2\times\mbb{Z}_4$ in Figure \ref{fig:Web_Of_Dualities_Z2xZ4}.
As can be seen, the web of dualities is disconnected and each connected component corresponds to a fusion structure. More generally the set of possible fusion structures is given as follows: any subgroup (up to isomorphism) $\ms F$ of $\mathcal A \simeq\ms G\times\ms G$ satisfying $|\ms F| = |\ms G|$ is a possible fusion structure. Each of these would then correspond to a connected component in a web-of-dualilties graph.\footnote{Our statement here are about dualities obtained from this construction. There might exist dualities beyond the topological holography framework where this might not be true.} 
In Figure \eqref{fig:Web_Of_Dualities_Z2xZ4}, lines that connect a node $\mcal{L}$ to itself correspond to elements of $\tenofo{Stab}(\mcal{L})$. These are the source of emergent non-invertible symmetries as discussed above. 

\smallskip Another example is $\ms G=\mbb{Z}_N$. For any $p$ and $q$ such that $pq=N$, we have a fusion structure $\ms F=\mbb{Z}_p\times\mbb{Z}_q$. Furthermore, there is a Lagrangian subgroup for any $\ms H\subset\ms G$ which means that we have a Lagrangian subgroup given by $\mcal{L}_{p}\simeq \mbb{Z}_p\times\mbb{Z}_{N/p}$. This phase spontaneously breaks $\mbb{Z}_N\to\mbb{Z}_p$. Clearly, any gapped phase $\mcal{L}_p$ comes with a pair $\mcal{L}_{N/p}$ with isomorphic fusion structure: $\mcal{L}_p\simeq\mcal{L}_{N/p}$. Under electric-magnetic duality, these are mapped into each other $\mcal{L}_p\overset{\tenofo{EM}}{\longleftrightarrow}\mcal{L}_{N/p}$. There cannot exist dualities that connect phases in other ways, as they have different fusion structures.

\smallskip We believe that there always exist at least one duality between two gapped phases with the same fusion structure (which implies that number of connected component of a web-of-duality graph is equal to the number of fusion structures). We have implicitly assumed this in the discussion above and will assume this to be true in what follows.

\smallskip Let us summarize the main points
\begin{enumerate}
    \item %
    Properties of $\ms G$-symmetric Hamiltonians $H[\{t_{\mbs{d}}\}]$ such as being gapped, gapless, critical, or at a first-order transition is preserved under duality transformations. 
    
    \item Each gapped phase $\mcal{L}$ is characterized by a pair $(\ms H_{\mcal{L}},\psi_{\mcal{L}})$, corresponding to a spontaneous symmetry-breaking down to a subgroup $\ms H_{\mcal{L}}\subset \ms G$ together with an SPT twist $\psi_{\mcal{L}}\in H^2(\ms H_{\mcal{L}},\ms U(1))$.  
    The data $(\ms H_{\mcal{L}},\psi_{\mcal{L}})$ can be conveniently obtained from the dyons in $\mc L$.

    \item Each gapped phase has a fusion structure corresponding to the algebra of condensed order parameters $\mcal{L}\subset\mcal{A}$. Only gapped phases with isomorphic fusion structures $\mcal{L}\simeq\mcal{L}'$ can be dual to each other. The web of dualities splits into connected components, one for each type of fusion structure.

    \item For any gapped phase $\mcal{L}$, one can associate a stabilizer subgroup $\tenofo{Stab}(\mcal{L})\subset\mcal{G}[\ms G]$. Any Hamiltonian in the gapped phase $\mcal{L}$ will be dual to another Hamiltonian in the same phase under dualities $\sigma\in\tenofo{Stab}(\mcal{L})$. Any such  Hamiltonian at a self-dual point $H[\{t^\star_{\mbs{d}}\}]=\dualhd{\sigma}H[\{t^\star_{\mbs{d}}\}]$ has extra emergent symmetries. These symmetries can be non-invertible which form a fusion category rather than a group. 

\end{enumerate}

\subsection{Gapless phases, phase transitions, and critical points}
\label{subsubsec:phase transitions and critical points}
Having discussed some of the properties of dualities with regards to gapped phases, we will now turn to critical points between gapped phases. All these properties can help constrain and map out phase-diagrams as we will see in examples (see Section \ref{sec:examples}). 

\smallskip It is convenient to define $\tenofo{Crit}[\mathcal L_1,\cdots,\mcal{L}_k]$ as the set of (multi-)critical points between gapped phases $\mcal L_1,\cdots,\mcal{L}_k$. Here is a summary of various properties of dualities when acting on critical points:
\begin{enumerate}
    \item {\it Self-duality}: If $H[\{t^\star_{\mbs{d}}\}]$ is self-dual under a duality $\sigma$ (or subgroup $\mscr{S}\subset\mathcal G[\ms G]$) then it cannot be in any gapped phase $\mcal{L}$ that is not self-dual under the same dualities $\sigma\cdot\mcal{L}\ne\mcal{L}$. In particular, if no gapped phase is self-dual under $\sigma$ (or $\mscr{S}$) then $H[\{t^\star_{\mbs{d}}\}]$ must either be gapless (critical or otherwise) or degenerate/at a first-order transition.
    \begin{enumerate}[label=(\alph*)]
    \item If a critical point is self-dual under a subgroup $\mscr{S}\subset\mcal{G}[\ms G]$ and $\mcal{L}$ is a  gapped phase connected to this critical point, then it is a critical point between at least $|\tenofo{Orb}_\mscr{S}(\mcal{L})|$ number of gapped phases. 
    
    \item If a Hamiltonian $H[\{t^\star_{\mbs{d}}\}]$ (gapped or gapless) is self-dual under a subgroup $\mscr{S}$ of dualities, then this implies that there are topological defect lines commuting with it and therefore has new emergent symmetries. These emergent symmetries can be non-invertible.  
    \end{enumerate}
    
    \item {\it Criticality}: Consider a critical point $\mscr{C}\in\tenofo{Crit}[\mcal{L}_1,\cdots,\mcal{L}_k]$ and a duality $\sigma$ such that $\sigma\cdot\mcal{L}_i=\mcal{L}^\vee_i$. Then, the critical point will be mapped to $\mscr{C}^\vee\equiv \sigma\cdot\mscr{C}\in\tenofo{Crit}[\mcal{L}^\vee_1,\cdots,\mcal{L}^\vee_k]$. There are a few consequences:
    
    \begin{enumerate}[label=(\alph*)]
        \item In fact, there will be a bijection between the set of all critical points $\tenofo{Crit}[\mcal{L}_1,\cdots,\mcal{L}_k]$ and $\tenofo{Crit}[\mcal{L}^\vee_1,\cdots,\mcal{L}^\vee_k]$ induced by any such duality $\sigma$.  
        
        \item The set of critical points $\tenofo{Crit}[\mcal{L}_1,\cdots,\mcal{L}_k]$ depends only on the fusion structures. More precisely, there is a bijection
        \begin{equation}\label{eq:BijectionOfCriticalSpaces}
            \tenofo{Crit}[\mcal{L}_1,\cdots,\mcal{L}_k]\simeq \tenofo{Crit}[\mcal{L}'_1,\cdots,\mcal{L}'_k],
        \end{equation}
        if $\mcal{L}_i\simeq \mcal{L}'_i$ for all $i$ (same fusion structure). By this, we mean that for any critical point between $\mcal{L}_1,\cdots,\mcal{L}_k$, there is one (and only one) critical point between $\mcal{L}'_1,\cdots,\mcal{L}'_k$. If $\tenofo{Crit}[\mcal{L}_1,\cdots,\mcal{L}_k]$ has lines or surfaces of critical points, then so will $\tenofo{Crit}[\mcal{L}'_1,\cdots,\mcal{L}'_k]$. Note that the conformal field theories mapped to each other under this bijection need not be equivalent and can have different scaling operators. But if one has a marginal operator, then so will the other.
        
        \item While a gapped phase is characterized by a fusion structure, a (multi-)critical point is characterized by a collection of fusion structures. The space of $\ms G$-symmetric CFTs can be decomposed based on fusion structures. 
    \end{enumerate}
\end{enumerate}
Let us elaborate on some of these points. The first point follows from the discussion of the previous section. Imagine there exist a Hamiltonian $H[\{t^\star_{\mbs{d}}\}]$ which is self-dual under a subgroup $\mscr{S}$, then which phase does it describe? Since dualities map gapped phases into each other according to \eqref{eq:the mapping of fixed-point Hamiltonians under a duality transformation}, then any gapped phase that is not self-dual under $\mscr{S}$ is forbidden. However, if no gapped phase is self-dual, then $H[\{t^\star_{\mbs{d}}\}]$ cannot be any gapped phase. It must therefore be gapless or at a first-order transition with ground-state degeneracy. 

\smallskip Point (1a) can be seen as follows: consider a path $H_{\mcal{L}}(s)$ connecting a gapped point ($H_\mcal{L}(0)$) in phase $\mcal{L}$  to a critical point $\mscr{C}$ ($H_\mcal{L}(1)$). Under dualities in $\mscr{S}$, $H^\vee_{\mcal{L}}(1)=H_\mcal{L}(1)$, under the assumption of self-duality of $\mscr{C}$. However, the gapped portion of the path can be mapped to other gapped phases. Let us define the orbit of the action of $\mscr{S}$  on $\mcal{L}$ as
\begin{equation}
    \tenofo{Orb}_\mscr{S}(\mcal{L}) \equiv \big\{\sigma\cdot\mcal{L},\,\forall \sigma\in\mscr{S}\big\}. 
\end{equation}
We can therefore connect any pair of gapped phases in $\tenofo{Orb}_\mscr{S}(\mcal{L})$ with a continuous path of Hamiltonians going through $\mscr{C}$. Thus we conclude that $\mscr{C}$ is a multi-critical point between all gapped phases in $\tenofo{Orb}_\mscr{S}(\mcal{L})$. However, there could be other gapped phases attached to $\mscr{C}$ with different fusion structures. 

\smallskip Point (1b) follows from the discussion in previous section, which naturally holds for any self-dual Hamiltonian and not just gapped ones.

\smallskip The second point can be argued as follow. By assumption, we have tuned a Hamiltonian $H_{\mathscr C}[\{t_{\mbs d}\}]$ to a critical point $\mscr{C}\in\tenofo{Crit}[\mcal{L}_1,\cdots,\mcal{L}_k]$. Under a duality transformation $H^\vee_{\mscr C}[\{t_{\mbs d}\}] = \dualhd\sigma H_{\mscr C}[\{t_{\mbs d}\}]$, the  critical point $\mscr{C}$ is mapped to another critical point $\mathscr C^\vee \equiv \sigma\cdot\mscr C$. The question is, which transition does $\mathscr C^\vee$ describe? Consider a path $H(s)$ going from $\mcal{L}_1$, through $\mscr{C}$, to $\mcal{L}_2$. After duality, this path maps into a path going from $\mcal{L}^\vee_1$, through $\mscr{C}^\vee$, to $\mcal{L}^\vee_2$, where $\mcal{L}^\vee_i=\sigma\cdot\mcal{L}$. 
\begin{figure}[H]
    \centering
    \CriticalPointMappingFigure
\end{figure}
Repeating this for all pairs in the set $\{\mcal{L}_1, \cdots,\mcal{L}_k\}$ proves that $\mscr{C}^\vee\in\tenofo{Crit}[\mcal{L}^\vee_1,\cdots,\mcal{L}^\vee_k]$. The above discussion implies that a duality induces a map between spaces $\tenofo{Crit}[\mcal{L}_1,\cdots,\mcal{L}_k]\to\tenofo{Crit}[\mcal{L}^\vee_1,\cdots,\mcal{L}^\vee_k]$. This map was not one-to-one then $\sigma\cdot\mscr{C}_1=\sigma\cdot\mscr{C}_2$. However, this is not possible since the duality just permutes the coefficients \eqref{eq:GeneralGSymmetricHamiltonian} and therefore is trivially invertible. The surjectivity of the map can be argued as follows. Assume that there is phase $\mscr{C}'\in\tenofo{Crit}[\mcal{L}^\vee_1,\cdots,\mcal{L}^\vee_k]$ that is not the image of a point in $\tenofo{Crit}[\mcal{L}_1,\cdots,\mcal{L}_k]$ under a duality $\sigma$. However, applying $\sigma^{-1}$ to $\mscr{C}'$ would necessarily give us a point in the set $\tenofo{Crit}[\mcal{L}_1,\cdots,\mcal{L}_k]$. This can be seen by taking a continuous path from $\mcal{L}^\vee_1$ and $\mcal{L}^\vee_2$ that passes through $\mscr{C}'$. The image of this path under $\sigma^{-1}$ would be a path from $\mcal{L}_1$ to $\mcal{L}_2$ and passes through a critical phase between them. Hence $\sigma^{-1}\cdot\mscr{C}'\in\tenofo{Crit}[\mcal{L}_1,\cdots,\mcal{L}_k]$. This shows point (2a). 

\smallskip To elaborate point (2b) and (2c), let $\tenofo{Fus}[\ms G]=\{\ms F_1,\cdots,\ms F_r\}$ be the set of possible fusion structures of gapped phases of $\ms G$-symmetric systems and define $\tenofo{Crit}[\ms F_{i_1},\cdots,\ms F_{i_k}]$ with $i_1\le\cdots\le i_k$ to be the set of critical phases between gapped phases with fusion structures $\ms F_{i_1},\cdots,\ms F_{i_k}$. This space is bigger than $\tenofo{Crit}[\mcal{L}_1,\cdots,\mcal{L}_k]$ since it also contains all critical points $\tenofo{Crit}[\mcal{L}'_1,\cdots,\mcal{L}'_k]$ satisfying $\mcal{L}_j\simeq\mcal{L}'_j\simeq\ms{F}_{i_j}$. We can decompose the subset of critical points within the space of $\ms G$-symmetric systems as
\begin{equation}
    \tenofo{Crit}[\ms G]=\bigcup_{k\ge 2}\bigcup_{\substack{i_1,\cdots,i_k=1\\ i_1\le\cdots\le i_k }}^r\tenofo{Crit}[\ms F_{i_1},\cdots,\ms F_{i_k}].
\end{equation}
Dualities can only map critical points within each block $\tenofo{Crit}[\ms F_{i_1},\cdots,\ms F_{i_k}]$ but not between different blocks.

\subsection{Conformal spectroscopy from dualities}\label{sec:topological conformal spectroscopy from dualities}

In previous section, we discussed which kinds of global constraints dualities in $\mcal{G}[\ms G]$ impose on the space of $\ms G$-symmetric Hamiltonians. It is instructive to see how these dualities act in a spacetime formulation, or in other words on the level of partition functions. Since partition functions contain the full spectrum of a theory, it can be a powerful tool when combined with dualities. For example, we will see that it can be used to compute the conformal spectrum of many non-trivial phase transitions relatively easily 
if we know the conformal spectrum on one side of the duality.
The conformal spectrum contains information about scaling dimensions of the universality class of the phase transitions .

\begin{figure}[t!]
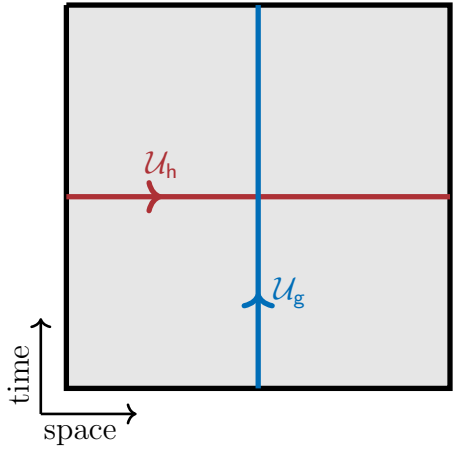

    \centering
    \InsertionOfSymmetryOperators
    \caption{The insertion of symmetry line defects along space and time directions.}
    \label{fig:insertion of symmetry operators}
\end{figure}

\smallskip In order to see the action of dualities on partition functions, it is useful to define a generalized twisted partition function as \cite{PetkovaZuber2001}
\begin{equation}\label{eq:the definition of generalized twisted partition function}
    \mcal{Z}_{\ms g, \ms h} = \text{tr}_{\mathcal H}\left(\mathcal U_{\ms h}\,e^{-\beta H_{\ms g}}\right),
\end{equation}
where $\mcal{U}_{\ms h}$ is the $\ms G$ symmetry operator and $H_{\ms g}$ is a $\ms G$-symmetric Hamiltonian with symmetry-twisted boundary condition labeled by $\ms g$. One can think of this as being a normal partition function with the insertion of a symmetry operator $\mcal{U}_{\ms h}$ in the space direction and a $\mcal{U}_{\ms g}$ in the time direction (see Figure \ref{fig:insertion of symmetry operators}). To see how the presence of a defect in the time direction leads to a twisted boundary condition (see Appendix \ref{sec:symmetry-twsited boundary conditions}). We are considering the generalized partition functions because dualities mix all sectors non-trivially, as we will see.

\smallskip The projection operator to symmetry sector $\alpha\in\tenofo{Rep}(\ms G)$ is given by
\begin{equation}\label{eq:the definition of projection operator to symmetry sector defined in conformal spectrum section}
    P_{\alpha}=\frac{1}{|\ms{G}|}\sum_{\ms h\in\ms G}\ms R_{\alpha}^{-1}(\ms h) \mathcal U_{\ms h}.
\end{equation}
One can readily confirm that $P_\alpha^2=P_\alpha$ and $P^\dagger_\alpha=P_\alpha$. Conversely, the symmetry operator can be written in terms of projectors $P_\alpha$
\begin{equation}\label{eq:the definition of symmetry operator in terms of projectors}
    \mathcal U_{\ms{h}}= \sum_{\alpha\in\tenofo{Rep}(\ms G)}\ms R_{\alpha}(\ms h)P_{\alpha}.
\end{equation}
Using the projectors, $\ms G$-twisted {\it characters} are defined as
\begin{equation}\label{eq:the definition of G-symmetry characters}
    \chi_d = \text{tr}_{\mathcal H}\left(P_{\alpha}\,e^{-\beta H_{\ms g}}\right)
            = \text{tr}_{\mathcal H_{\alpha}}\Big(q^{H_{(\ms g,\alpha)}}\Big),
\end{equation}
where $d=(\ms g,\alpha)$, $q\equiv e^{-\beta}$, and $\mcal{H}_\alpha=P_\alpha\mcal{H}$. Due to the projection operator, this object contains the spectrum of our Hamiltonian with twisted boundary condition $\ms g$ in the symmetry sector $\alpha$. Using \eqref{eq:the definition of projection operator to symmetry sector defined in conformal spectrum section} and \eqref{eq:the definition of symmetry operator in terms of projectors}, we can express the partition functions and characters in terms of one another via
\begin{equation}\label{eq:PartitionFunctionInTermsOfCharacters}
\mathcal{Z}_{\ms g,\ms h} =\; \sum_{\alpha\in\text{Rep}(\ms G)} \ms R_\alpha(\ms h)\chi_{d=(\ms g,\alpha)},
\end{equation}
and 
\begin{equation}\label{eq:the characters in terms of twisted parition functions}
    \chi_d =\; \frac{1}{|\ms G|}\sum_{\ms h\in\ms G}\ms R_{\alpha}^{-1}(\ms h)\mathcal{Z}_{\ms g, \ms h}.
\end{equation}
Under a duality $\sigma$, Hamiltonian $H_{\ms g}$ transforms to $H^\vee_{(\ms g,\alpha)}[\{t_{\mbs d}\}]=H_{\sigma\cdot (\ms g,\alpha)}[\{t_{\sigma^{-1}\cdot\mbs{d}}\}]$ (see equation \eqref{eq:Dual hamiltonian in terms of original hamiltonian in different sectors}). By inserting this in \eqref{eq:the definition of G-symmetry characters}, the characters transform as
\begin{equation}
    \chi_d\quad\longmapsto\quad \chi^\vee_{d}=\tenofo{tr}_{\mcal{H}}\Big(q^{H^\vee_{(\ms g,\alpha)}}\Big)=\tenofo{tr}_{\mcal{H}}\Big(q^{H_{\sigma\cdot(\ms g,\alpha)}}\Big)=\chi_{\sigma\cdot(\ms g,\alpha)}. 
\end{equation}
Therefore, the dual twisted partition functions are given by
\begin{equation}\label{eq:the dual partition function in terms of original parition function}
    \begin{aligned}
        \mathcal{Z}^{\vee}_{\ms g^\vee,\ms h^\vee} & =\sum_{\alpha^{\vee}\in\text{Rep}(\ms G)} \ms R_{\alpha^{\vee}}(\ms h^{\vee})\chi^\vee_{d^{\vee}}
        \\
        &=\sum_{\alpha^{\vee}\in\text{Rep}(\ms G)} \ms R_{\alpha^{\vee}}(\ms h^{\vee})\chi_{(\wh{\ms{g}},\wh{\alpha})},
    \end{aligned}
\end{equation}
where we have defined
\begin{equation}
    (\wh{\ms g},\wh{\alpha})=\sigma\cdot(\ms g^\vee,\alpha^\vee).
\end{equation}
Using \eqref{eq:the characters in terms of twisted parition functions}, this can be written as
\begin{equation}\label{eq:the generic formula for twisted dual partition function}
\begin{aligned}
    \mathcal{Z}^\vee_{\ms g^\vee,\ms h^\vee}&=\frac{1}{|\ms G|}\sum_{\alpha^\vee}\ms R_{\alpha^\vee}(\ms h^\vee)\sum_{\ms h}\ms R^{-1}_{\wh{\alpha}}(\ms h)\,\mathcal{Z}_{\,\wh{\ms g},\ms h}
    \\
    &=\frac{1}{|\ms G|}\sum_{\ms g,\ms h,\alpha^\vee}\ms R_{\alpha^\vee}(\ms h^\vee)\ms R^{-1}_{\wh{\alpha}}(\ms h)\delta_{\ms g,\wh{\ms g}}\,\mathcal{Z}_{\ms g,\ms h}.
\end{aligned}
\end{equation}
In other words, under a duality generalized twisted partition functions transform linearly 
\begin{equation}
\label{eq:Partition funcitions related by eta factor}
    \mcal{Z}^\vee_{\ms g^\vee,\ms h^\vee}\\
    = \frac{1}{|\ms G|} \sum_{\mathsf g,\mathsf h\in \mathsf G} \eta\left(\mathsf g,\mathsf h,\mathsf g^\vee,\mathsf h^\vee\right)\;\mathcal{Z}_{\mathsf g,\mathsf h},
\end{equation}
where the coefficients are 
\begin{equation}\label{eq: eta formula for partition function dualities}
    \eta(\ms g,\ms h,\ms g^\vee, \ms h^\vee)\equiv  %
    \sum_{\alpha^\vee\in\text{Rep}(\ms G)}\ms R_{\alpha^\vee}(\ms h^\vee)\ms R^{-1}_{\wh{\alpha}}(\ms h)\delta_{\ms g,\wh{\ms g}}.
\end{equation}
Starting from a model with periodic boundary condition, the partition function is given by $\mcal{Z}_{0,0}$ and the partition function of the dual theory is given by $\mcal{Z}^\vee_{0,0}$. However, it is clear from \eqref{eq:Partition funcitions related by eta factor} that the dual partition function  $\mcal{Z}^\vee_{0,0}$ depends on all generalized twisted partition functions $\mcal{Z}_{\ms g,\ms h}$ of the original theory. This is another manifestation of how dualities act on twisted sectors. 

\smallskip Physically, the factor $\eta_\sigma$ corresponds to the partition function of a theory living on the duality wall $\dual{\sigma}[\Sigma]$ living in the $2+1d$ bulk TQFT (see Figure \ref{fig:BulkSymmetryOperatorBroughtToTheBoundary}). In terms of $\eta$ factors, the fusion of domain walls \eqref{eq:fusion of surface operators figure} in the bulk is given  
\begin{equation}
    \eta_{\sigma_1\sigma_2}(\ms g,\ms h,\ms{g}'',\ms{h}'')=\frac{1}{|\ms G|}\sum_{\ms g',\ms h'}  \eta_{\sigma_1}(\ms g,\ms h,\ms{g}',\ms{h}')\eta_{\sigma_2}(\ms g',\ms h',\ms{g}'',\ms{h}'').
\end{equation}

\smallskip While \eqref{eq:Partition funcitions related by eta factor} holds for any $\ms G$-symmetric theory, it is particularly powerful at critical points with conformal symmetry. In such cases, the partition functions are expressed in terms of Virasoro characters of the primary fields in the theory. From \eqref{eq:Partition funcitions related by eta factor}, we can therefore directly compute the spectrum of a dual conformal field theory from any other with known partition functions. We will exploit this formula to explicitly predict the conformal spectrum of non-trivial phase transitions and confirm these predictions numerically.

\subsection{Duality: a gauging perspective}

\begin{figure}
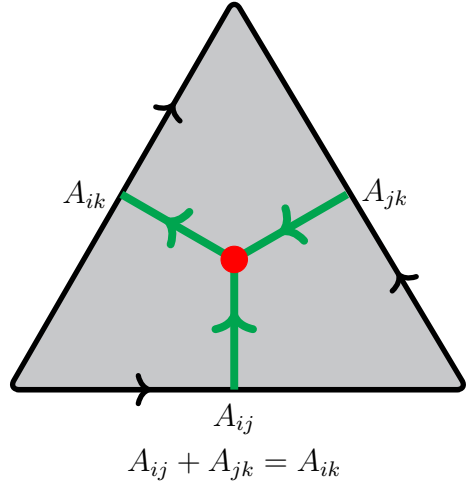

    \centering
    \CocyleDiscreteGaugeField
    \caption{A gauge field corresponding to a finite Abelian group $\ms G$ is given by a $\ms G$ coloring of 1-simplices (in black) or equivalently a network of symmetry domain walls (in green). The cocycle condition for the gauge field $A$ translates to a consistent group composition of the symmetry domain walls.}
    \label{fig:cocycle condition on discrete gauge fields}
\end{figure}

The relation between twisted partition functions $\mc Z_{\ms{g},\ms{h}}$ and dual twisted partition functions $\mc Z^{\vee}_{\ms g^{\vee},\ms h^{\vee}}$ \eqref{eq:Partition funcitions related by eta factor} can often be conveniently expressed as a generalized gauging of the global $\ms G$ symmetry. 
We first note that, $\mc Z_{\ms{g},\ms{h}}$ can be equivalently understood as the partition functions of the model described by the Hamiltonian $H$ coupled to a background $\ms G$ gauge field whose holonomies in the space and time directions of the spacetime torus are $\ms g$ and $\ms h$ respectively.
Since gauge fields for finite groups have a vanishing field strength, the gauge equivalence classes of such gauge fields can be labelled the holonomies around non-contractible cycles of the manifold.
Therefore we may write
\begin{equation}
    \mc Z_{\ms g,\ms h}\equiv \mc Z[A], 
    \qquad
    \mc Z^{\vee}_{\ms g^{\vee},\ms h^{\vee}}\equiv \mc Z[A^{\vee}].
\end{equation}
Then the duality expression, \eqref{eq:the generic formula for twisted dual partition function}, can be expressed as
\begin{align}
    \mc Z[A^{\vee}]= \frac{1}{|\ms G|} \sum_{A}\mc Z[A] \eta(A,A^{\vee}).
\end{align}
We will find that in several duality transformations, the object $\eta(A,A^{\vee})$ reduces to a topological term in terms of the backgrounds $A$ and $A^{\vee}$.
Such topological terms are expressed most conveniently in terms of gauge fields defined on oriented triangulations of the spacetime manifold as illustrated in Fig.~\ref{fig:TriangulationGaugeTransformation}.
A gauge field is defined via a coloring of the 1-simplices by group labels such that the composition of the group labels on the boundary of any 2-simplex or plaquette evaluates to the identity element in $\ms G$.
This is nothing but the cocycle or flatness condition for the gauge field $\ms A$.
The gauge field $A$ can be dualized to obtain a network of $\ms G$ symmetry domain wall. 
Concretely, the gauge field $A_{ij}\in \ms G$ assigned to the 1-simplex $\langle ij\rangle$ dualizes to a symmetry domain wall corresponding to the same group element living on the 1-simplex of the dual triangulation.
Furthermore, the cocycle condition simply translates to the fact that the symmetry domain walls fuse according the group product in $\ms G$ (see Fig.~\ref{fig:cocycle condition on discrete gauge fields}).
Note that, if in $d+1$ dimensions, a $d$-simplex would be dual to a 1-simplex and therefore the symmetry domain walls would be defined on $d$-simplices, which as described earlier, is the correct dimension for 0-form symmetry defects (see Appendix \ref{sec:simplicial calculus} for details). 

\smallskip In two dimensions, the types of topological terms that appear are typically of the form 
\begin{align}
    S_{\text{top}} \propto \int_{\Sigma} A_1 \cup A_{2},
    \label{eq:topological_term_1+1d}
\end{align}
where $A_1$ and $A_{2}$ are some discrete gauge fields.
A generic topological term would contain a sum of terms of the form \eqref{eq:topological_term_1+1d}.
When $\Sigma$ is a two torus, the topological term $S_{\text{top}}$ can be expressed in terms of the holonomies of $A_{1,2}$ as
\begin{align}
    \int_{\Sigma} A_1 \cup A_{2} =
    \left(
    \text{hol}_{x}(A_1)\text{hol}_{t}(A_2)
    -
    \text{hol}_{x}(A_2)\text{hol}_{t}(A_1)
    \right),
\end{align}
where $\text{hol}_{x,t}(A_i)$ is the holonomy of the gauge field $A_{i}$ along the $x,t$ cycle respectively.

\smallskip

\smallskip In Sec.~\ref{subsec:from bulk symmetry to boundary dualities}, we described how the 0-form symmetry group of $2+1d$ topological gauge gauge theories is generated by three kinds of symmetry transformations, namely (i) universal kinematical symmetries, (ii) universal dynamical symmetries and (iii) partial electric-magnetic dualities. 
In turn, there are correspondingly three kinds of generators of the duality group acting on the space of $\ms G$-symmetric quantum systems.
These three families of generators have a natural action on the twisted partition functions $\mc Z[A]$.
\begin{enumerate}
    \item Universal kinematical symmetries: These dualities are labelled by elements of the automorphism group of $\ms G$.
    An automorphism $\varphi$, maps a generator $\ms g_{j}$ of the group $\ms G$ into a new group element $\varphi(\ms g_j)$ of the same order. 
    Correspondingly, the automorphism group acts on the background gauge field by mapping $A_{j}$ to $A^{\vee}_j=\varphi(A_{j})$.
    More precisely, the dual partition function has the form
    \begin{equation}
        \sigma_{\varphi}: \mc Z[A]\longmapsto 
        \mc Z^{\vee}[A^{\vee}]= \sum_{A\in H^1(\Sigma,\ms{G})}\mc Z[A]\delta_{A,\varphi(A^{\vee})},
    \end{equation}
    where $\varphi(A)=(\varphi(A_1),\varphi(A_2),\dots)$.
    \item Universal dynamical symmetries: 
    These dualities are labelled by elements $\mathfrak{c}_{\mbs \ell} \in H^{2}(\ms G,\ms{U}(1))$, which also labels $1+1d$ $\ms G$ SPTs. 
    In particular, the partition function of a $\ms G$-SPT, coupled to a background $\ms G$ gauge field $A$ is (in the topological limit) a pure $\ms{U}(1) $ phase.
    Concretely, for a finite group $\ms G=\prod_{i}\mbb Z_{N_i}$, the group cocyles are given in \eqref{eq:group 2-cocycle}.
    The corresponding topological response theory of the SPT labelled by $\mathfrak{c}_{\ell}$ coupled to a $\ms G$ background $A$ is 
    \begin{align}
    \mc Z_{\text{SPT}-\mathfrak{c}_{\ell}}[A]=\exp\left\{2\pi i\sum_{i<j}\bigintssss_{\Sigma}\frac{\ell_{ij}}{\text{gcd}(N_i,N_j)}A_{i}\cup A_{j}\right\}.
    \end{align}
    Then the duality $\sigma_{\mathfrak c_{\ell}}$ acts on the twisted partition functions as
    \begin{equation}
        \sigma_{\mathfrak{c}_{\ell}}: \mc Z[A] \longmapsto 
         \mc Z[A]\mc Z_{\text{SPT}-\mathfrak{c}_{\ell}}[A].
    \end{equation}
    Therefore the universal dynamical symmetries correspond to pasting $\ms G$-SPTs.
    \item Partial electric-magnetic dualities: These are generalized Kramers-Wannier dualities \cite{FrohlichFuchsRunkelSchweigert200909, BhardwajTachikawa201704, Kapustin_2017} that are obtained by gauging a certain $\mathbb Z_{N_j}$ factor in the group $\ms G$. 
    The resulting theory has a dual (or quantum) $\ms G$ global symmetry. 
    The twisted partition functions transform as follows under such duality operations
    \begin{equation}
        \mcal{Z}^\vee[\cdots,A^\vee_j,\cdots]=\frac{1}{N_j}\sum_{A_j\in H^1(\Sigma,\mbb{Z}_{N_j})}\exp\left(\frac{2\pi i}{N_j}\bigintssss_{\Sigma} A_j\cup A^\vee_j\right)\mcal{Z}[\cdots,A_j,\cdots],
    \end{equation}
\end{enumerate}
More general duality operation can be obtained by taking various combinations of these kinds of duality generators.
These often take the form of generalized gauging in the presence of topological terms. Such interpretations would be discussed in more detail in subsequent sections.
It should be noted that gauging in the presence of an SPT or topological twist has been known in the string theory and conformal field theory literature for more than thirty years as orbifolding in the presence of discrete torsion \cite{Vafa:1986wx, Vafa:1994rv}.

\section{Examples}
\label{sec:examples}
In this section, we study the space of quantum spin chains for specific examples of global symmetry $\ms G$ using the formalism of topological holography introduced in prior sections. In particular, we will describe all gapped phases,  their order parameters, and fundamental excitations. We will write down certain minimal Hamiltonians that capture all the essential (symmetry related) features of such systems. Using dualities, we predict phase diagrams of these models, and numerically verify the theoretical predictions. Finally, we analytically compute and numerically verify the conformal spectrum of various non-trivial topological phase transitions. 

\smallskip The dualities discussed in this paper are particularly powerful for global symmetries $\ms G$ that have large duality groups but very few fusion structures. As explained in Sections \ref{subsec:gapped phases} and \ref{subsubsec:phase transitions and critical points}, gapped phases and critical points with different fusion structures cannot be mapped onto each other under duality. Having a single fusion structure and many dualities can vastly constrain the possible phase diagram. Groups like $\mbb{Z}_N$ belong to the opposite limit where there are few dualities and many fusion structures. Therefore, dualities put modest constraints on possible phase diagrams. Nonetheless, we will start by the examples of $\mbb{Z}_2$ and $\mbb{Z}_N$ as they are the building blocks of more interesting examples. 

\smallskip On the other limit, global symmetries such as $\mbb{Z}_2\times\mbb{Z}_2$ or $\mbb{Z}_3\times\mbb{Z}_3$ have only a single fusion structure but a large duality group. Hence, these examples are particularly well-suited to be studied using topological holography.  

\smallskip For a basic derivation of topological holography using an exactly-solvable model in the bulk, see Appendix \ref{App: Wen-plaquette}. 

\subsection{$\mbb{Z}_2$ symmetric quantum spin chains}

Let us consider the simplest finite Abelian group $\ms G=\mathbb Z_2$. 
We are interested in analyzing the phase diagram of $\mathbb Z_2$ symmetric quantum  systems using the holographic approach outlined in previous sections.
To this end, we consider an auxillary bulk $2+1d$ theory which hosts the $\mathbb Z_{2}$ topological order whose topological line operators are labelled by
\begin{equation}
    \begin{aligned}
    \mc A[\mbb Z_2]=&\; \left\{1,e,m,f\right\}= \mathbb Z_2\times \text{Rep}(\mathbb Z_2) \simeq\mathbb Z_2 \times \mathbb Z_2 ,
\end{aligned}
\end{equation}
where we have associated the group elements $\left\{(0,0),(0,1),(1,0),(1,1)\right\}\in \mc A[\mbb Z_2]$ with the label set $\left\{1,e,m,f\right\}$. 
In other words, the lines corresponding to the group elements $(1,0)$ and $(0,1)$ in $\mc A[\mbb Z_2]$ are the magnetic and electric line operators while the $(1,1)$ element corresponds to the dyonic line obtained via a fusion of the electric and magnetic lines labelled as $f$. 
The $\mathbb Z_{2}$ topological order has a 0-form global symmetry group 
\begin{align}\label{eq:the duality group for Z2}
    \mc G[\mathbb Z_2]=\mathbb Z_2=\left\{id,\sigma\right\},
\end{align}
whose generator acts as an outer automorphism on the 1-form symmetry group $\mathbb Z_{2}\times \mathbb Z_2$, that exchanges $e\leftrightarrow m$ and leaves the product $f=em$ invariant.
The $\mbb{Z}_2$ gauge theory admits two gapped boundaries corresponding to the $\mathbb Z_2$ subgroups of $\mc A[\mbb Z_2]$ generated by $e$ and $m$.
These two gapped boundaries condense the magnetic and electric excitations respectively and will be denoted as $\mc L_e=\{1,e\}$ and $\mc L_m=\{1,m\}$.

\smallskip The phase diagram of $1+1d$ $\mathbb Z_2$ symmetric quantum systems can be captured by an effective spin chain that arises at the lattice on the regularized boundary of the $\mathbb Z_2$ topological order defined on a semi-infinite cylinder \cite{WeiMoradi2014}.
As discussed in Sec.~\ref{Subsec:boundary_Hilbert_space}, the action of the open $e$ and $m$ operators restricted to the boundary can be captured by a quantum spin chain defined on a 1D lattice with a local Hilbert space $\mc H_{i}=\mbb{C}^2$.
From \eqref{eq:Wilson_operator_to_lattice_operator}, it can be seen that the minimal operators subtended from the bulk topological magnetic and electric operators are 
\begin{equation}
    S_m=A_i= \sigma^{x}_i, \quad S_e=B_{i,i+1}=\sigma^z_i\sigma_{i+1}^z,
    \label{eq:local_operators_Z2_spin_chain}
\end{equation}
where $\sigma^{\mu}_i$ are Pauli operators acting on the local Hilbert space $\mc H_i$.
Applying electric-magnetic duality \eqref{eq:the duality group for Z2} swaps electric and magnetic lines $S_m\leftrightarrow S_e$, which in turn becomes the well-known Kramers-Wannier duality on the spin chain, which acts on symmetry operators as \eqref{eq:DualityActing_on_S_operators}
\begin{equation}\label{eq:Z2 Kramers-Wannier duality on spin chain}
    \dualhd\sigma\left(\sigma^z_i\sigma^z_{i+1}\right) = \sigma^x_{i+1} \qquad \dualhd\sigma\left(\sigma^x_i\right) = \sigma^z_i\sigma^z_{i+1}.
\end{equation}
One immediately recognizes these operators as those appearing in the transverse field Ising (TFI) model, which is the paradigmatic quantum spin chain with a global $\mathbb Z_2$ symmetry.
The fixed point Hamiltonians that describe the two gapped phases, i.e., the ferromagnetic and the paramagnetic phase  of the TFI model are obtained as the Hamiltonians corresponding to the electric and magnetic gapped boundaries of the $\mathbb Z_2$ gauge theory
\begin{align}
    H_e=-\sum_{i=1}^L\sigma^z_i\sigma^z_{i+1}, \quad H_{m}=-\sum_{i=1}^L\sum_i\sigma^x_i.
\end{align}
The simplest Hamiltonian capable of capturing the phase diagram of $\mathbb Z_{2}$ symmetric quantum systems is simply a linear combination of these two fixed point Hamiltonians
\begin{align}
\label{eq:Ising_lambda}
    H(\lambda)= \lambda H_m + (1-\lambda)H_e.
\end{align}
Next, we organize the Hilbert space of the TFI model using the bulk topological considerations.
Firstly, since all the non-contractible line operators $\Gamma_d$ with $d\in \mc A[\mbb Z_2]$ commute with all the boundary operators $\mc O_{g,i}$ and $\mc O_{\alpha,i}\mc O_{-\alpha,i+1}$, we may simultaneously diagonalize all the $\Gamma_{d}$ operators, which leads to a decomposition of the boundary Hilbert space into super-selection sectors 
\begin{align}\label{eq:Z2 spin chain decomposition of Hilbert space}
    \mc H= \bigoplus_{d'\in \mc A[\mbb Z_2]}\mc H_{d'},  \quad \text{where} \quad     \Gamma_d\Big|_{\mc H_{d'}}=e^{i\theta_{dd'}}=(-1)^{\ms g\alpha'+\ms g'\alpha}, 
\end{align}
where $\exp\left\{i\theta_{dd'}\right\}$ is the Hopf-linking phase between the $d=(\ms g,\alpha)$ and $d'=(\ms g',\alpha')$ topological line operators (bulk anyons).
Furthermore, we can diagonalize a maximal commuting subalgebra of boundary operators to define a basis for $\mc H_{d'}$.
We choose to diagonalize the operators $\mc O_{\alpha,i}\mc O_{-\alpha,i+1}=\sigma^z_i\sigma_{i+1}^z$.
Doing so, we may construct the basis $\left\{|\bs{a},d'\rangle\right\}$, with $\bs{a} \in C^{1}(\Lambda,\mathbb{Z}_2)$, the space of $\mathbb Z_2$ valued 1-cochains defined on the regularized boundary lattice $\Lambda$.
The various finite local operators in eq.~\eqref{eq:local_operators_Z2_spin_chain} act on the basis states as
\begin{gather}
\begin{aligned}
    \sigma^z_i\sigma_{i+1}^z|\bs{a},d'\rangle=&\;(-1)^{a_{i,i+1}}|\bs{a},d'\rangle, 
    \qquad        
    \prod_i(\sigma^x_i)^{\lambda_i}|\bs{a},d'\rangle=|\bs{a}+\delta \bs{\lambda},d'\rangle,
\end{aligned}
\end{gather}
where $\bs{\lambda} \in C^{0}(\Lambda,\mathbb{Z}_2)$. 
Here, $a_{i,i+1}=0,1$ corresponds to a domain-wall configuration in the spin chain. For example, $a_{i,i+1}=\sigma_{i+1}-\sigma_i$, where $\sigma_i=0,1$ are labels of up and down spins. The action of $\sigma^x_i$ flips a spin $\sigma_i\to\sigma_i+1$ at site $i$, which in turn flips two domain walls attached to it. This is encoded in the notation $\mbs{a}\to\mbs{a}+\delta\mbs{\lambda}$, which explicitly means $a_{i,i+1}\to a_{i,i+1}+\lambda_{i+1}-\lambda_i$. In other word, $\mbs{a}$ can be thought of as a $\mbb{Z}_2$ gauge field and $\mbs{a}\to\mbs{a}+\delta\mbs{\lambda}$ as a gauge transformation.   
Similarly, the non-contractible line operators in the bulk become the symmetry operator $\Gamma_{(\ms g,0)}\equiv\mc U_{\ms g}$ and the boundary condition twist operator $\Gamma_{(0,\alpha)}\equiv\mc T_{\alpha}$ as defined in \eqref{eq:BulkGammaToBoundaryU_T}.
More explicitly, these operators are given by
\begin{equation}\label{eq:U and T for Z2 definition}
    \mcal{U}_{\ms g}=\Bigg(\prod_{i=1}^L \sigma^x_i\Bigg)^{\ms g}, \qquad \mcal{T}_\alpha=\Bigg(\prod_{i=1}^L\sigma^z_i\sigma^z_{i+1}\Bigg)^\alpha.
\end{equation}
Note that the Kramers-Wannier duality $\sigma$ \eqref{eq:Z2 Kramers-Wannier duality on spin chain} swaps these operators. 
These act on the basis states as 
\begin{gather}
    \begin{aligned}
        \mc U_{\ms g}|\bs{a},d'\rangle=&\;(-1)^{\ms g\ms \alpha'}|\bs{a},d'\rangle, \\        
        \mc T_{\alpha}|\bs{a},d'\rangle=&\; (-1)^{\ms g'\ms \alpha}|\bs{a},d'\rangle
        = (-1)^{\text{hol}(\bs{a})\ms \alpha}|\bs{a},d'\rangle,
    \end{aligned}
\end{gather}
where in the second line, we have used $\text{hol}(\bs{a}):=\sum_{i}a_{i,i+1}=\ms g'$.
Here, the holonomy $\tenofo{hol}(\mbs{a})$ can be thought of as the flux $\oint_\gamma \mbs{a}$ of $\mbb{Z}_2$ gauge field $\mbs{a}$, where $\gamma$ is a loop around the spatial direction. The flux $\tenofo{hol}(\mbs{a})=\ms g'$ is created by the insertion of symmetry operator $\mcal{U}_{\ms g'}$ in the time direction, which gives rise to the boundary condition above. 

\smallskip Finally, the endpoint of the semi-infinite operators $\mscr{W}_{d}$ serve to toggle between super-selection sectors. 
This can be readily seen from the commutation relations between the semi-infinite lines $\mscr{Y}_{d}$ and the non-contractible bulk line operators $\Gamma_{d}$ \eqref{eq:Graphical commutation relations between gamma and Y} and \eqref{eq:Gamma Eigenvalues change due to Y}. 
The operator $\mscr{Y}_{e}$, ending on the site $i$ on the effective spin chain, anticommutes with $\sigma_i^{x}$, and therefore effectively acts as the $\sigma_i^{z}$ operator (see \eqref{eq:Infinite Line operator end point links with boundary symmetry operators and transforms accordingly}).
As a consequence, it modifies the charge, i.e., $\mathcal U_{\ms g}$ eigenvalue of the superselection sector.
Similarly, the semi-infinite line $\mscr{Y}_m$ ending on the link $(L,1)$ on the effective spin chain, anticommutes with a single link operator $\sigma^{z}_{L}\sigma^z_{1}$.
As a consequence, this modifies the eigenvalue of the $\mathcal T_{\alpha}$. 
More explicitly, from \eqref{eq:U and T for Z2 definition} we have $\mcal{T}_\alpha=(\sigma^z_{L}\sigma^z_{L+1})^\alpha$. The presence of an infinite $\mscr{Y}_{(\ms g',0)}=(\mscr{Y}_m)^{\ms g'}$ ending on the link $(L,1)$ imposes the twisted boundary condition 
\begin{equation}
    \sigma^z_{L+1}=(-1)^{\ms g'}\sigma^z_{1}.
\end{equation}
To summarize, the semi-infinite line operators act on the Hilbert space sectors as 
\begin{align}
    \mscr{Y}_{d}:\mc H_{d'}\longmapsto \mcal{H}_{d+d'}.
\end{align}
In addition to the boundary Hilbert space, the space of local operators on the spin chain admits a decomposition into superselection sectors labelled by $d\in \mc A$, since the local operators in \eqref{eq:local_operators_Z2_spin_chain} commute with $\Gamma_d$ for all $d$.
In particular, the Hamiltonian admits a decomposition $H(\lambda)=\oplus_d H_d(\lambda)$.

\smallskip The spectrum of Hamiltonian $H_{d}(\lambda)$ can be obtained by coupling the model to a background $\mbb{Z}_2$ gauge field and computing the partition function. 
Different gauge inequivalent configurations of a $\mathbb Z_2$ gauge field $A$ are labelled by the holonomies $(\ms g,\ms h)\in H^{1}(T^2,\mathbb Z_2)$ of $A$ in the space and time direction, respectively. 
The partition function in a symmetry twisted sector or equivalently coupled to a background $\mbb Z_2$ gauge field $A=(\ms g,\ms h)$ is
\begin{align}
    \mc Z[A,\lambda]\equiv\mc Z_{\ms g, \ms h}(\lambda)=\text{Tr}_{\mc H}\left[\mc U_{\ms h}e^{-\beta H_{\ms g}(\lambda)}\right].
\end{align}
Then, the twisted characters, which contain a trace of states in a sector labelled by $d=(\ms g,\alpha)\in \mc A[\mbb Z_2]$ is
\begin{align}
    \chi_{d}(\lambda)=\text{Tr}_{\mc H}\left[P_{\alpha}e^{-\beta H_{g}(\lambda)}\right]=\frac{1}{2}\sum_{\ms h\in \mbb Z_{2}}(-1)^{\alpha\ms h}\mc Z_{\ms g,\ms h},
    \label{eq:twisted_Characters}
\end{align}
where $P_{\alpha}$ is a projector, defined in \eqref{eq:the definition of projection operator to symmetry sector defined in conformal spectrum section}, onto $\alpha \in \text{Rep}(\mathbb Z_{2})$. 
These characters transform under the action of the generator $\sigma$ of the 0-form symmetry group $\mc G[\ms G]$ \eqref{eq:the duality group for Z2} of the bulk as $\chi_d(\lambda)\longmapsto \chi^{\vee}_d(\lambda)$, where
\begin{equation}
\label{eq:dual_characters_Z2}
    \chi^{\vee}_d(\lambda)=\text{Tr}_{\mc H}\left[P_{\alpha^{\vee}}e^{-\beta H^{\vee}_{g^{\vee}}(\lambda)}\right],
\end{equation}
with $\sigma\cdot(\ms g,\alpha)=(\ms g^{\vee},\alpha^{\vee})$. The Hamiltonian $H^{\vee}$ is obtained from the action \eqref{eq:Z2 Kramers-Wannier duality on spin chain} on \eqref{eq:Ising_lambda}. From this we obtain $H^{\vee}(\lambda)=U_P H(1-\lambda) U_P^\dagger$, where $U_P$ is a permutation of super-selection sectors. When decomposed into the sectors \eqref{eq:Z2 spin chain decomposition of Hilbert space}, this explicitly becomes
\begin{equation}
    \begin{pmatrix}
        H^\vee_{P+} & & & \\
        & H^\vee_{P-} & & \\
        & & H^\vee_{A+} & \\
        & & & H^\vee_{A-}
    \end{pmatrix}(\lambda)
    \quad
    = \quad        \begin{pmatrix}
        H_{P+} & & & \\
        & H_{A+} & & \\
        & & H_{P-} & \\
        & & & H_{A-}
    \end{pmatrix}(1-\lambda).
\end{equation}
Here $P$ and $A$ label periodic and anti-periodic boundary conditions, respectively, while $+$ and $-$ label the even and odd $\mathbb Z_2$ symmetry sectors. A physical consequence of the duality is that the spectrum of the Hamiltonians, $H(\lambda)$ and $H^{\vee}(\lambda)$, on either side of the duality are the same but are permuted (see also \eqref{eq:Dual hamiltonian in terms of original hamiltonian in different sectors} and \eqref{eq:the relation of spectra before and after a duality transformation}). From this, we can see that characters transform as
\begin{equation}
    \chi_d(\lambda)\longmapsto \chi^\vee_d(\lambda)=\chi_{\sigma \cdot d}(\sigma\cdot\lambda),\qquad \sigma\cdot\lambda=1-\lambda,
\end{equation}
where $\sigma\cdot (\ms g,\alpha)=(\alpha,\ms g)$. 

\smallskip It is also useful to see how the twisted partition functions transform under duality. 
To do so, we first write the twisted characters $\chi_d$, in terms of the twisted partition functions $\mc Z_{\ms g,\ms h}$ explicitly
\begin{gather}
    \begin{aligned}
    \label{eq:sectors_to_characters_Z2}
        \chi_{1}(\lambda)=&\; \frac{1}{2}\left(\mc Z_{0,0}(\lambda)+\mc Z_{0,1}(\lambda)\right), \quad
        \chi_{e}(\lambda)= \frac{1}{2}\left(\mc Z_{0,0}(\lambda)-\mc Z_{0,1}(\lambda)\right), \\
        \chi_{m}(\lambda)=&\; \frac{1}{2}\left(\mc Z_{1,0}(\lambda)+\mc Z_{1,1}(\lambda)\right),\quad
        \chi_{f}(\lambda)=\frac{1}{2}\left(\mc Z_{1,0}(\lambda)-\mc Z_{1,1}(\lambda)\right),
    \end{aligned}
\end{gather}
which implies the equality of the following combinations of twisted partition functions%
\begin{gather}
\begin{aligned}    
\label{eq:transfo_characters_Z2}
\chi^{\vee}_1(\lambda)=&\; \frac{1}{2}\left(\mc Z^{\vee}_{0,0}(\lambda)+\mc Z^{\vee}_{0,1}(\lambda)\right)=\frac{1}{2}\left(\mc Z_{0,0}(1-\lambda)+\mc Z_{0,1}(1-\lambda)\right)=   \chi_1(\sigma\cdot\lambda) \\
\chi^{\vee}_e(\lambda)=&\; \frac{1}{2}\left(\mc Z^{\vee}_{0,0}(\lambda)-\mc Z^{\vee}_{0,1}(\lambda)\right)=\frac{1}{2}\left(\mc Z_{1,0}(1-\lambda)+\mc Z_{1,1}(1-\lambda)\right)=   \chi_m(\sigma\cdot\lambda) \\
\chi^{\vee}_m(\lambda)=&\; \frac{1}{2}\left(\mc Z^{\vee}_{1,0}(\lambda)+\mc Z^{\vee}_{1,1}(\lambda)\right)=\frac{1}{2}\left(\mc Z_{0,0}(1-\lambda)-\mc Z_{0,1}(1-\lambda)\right)=   \chi_e(\sigma\cdot\lambda) \\
\chi^{\vee}_f(\lambda)=&\; \frac{1}{2}\left(\mc Z^{\vee}_{1,0}(\lambda)-\mc Z^{\vee}_{1,1}(\lambda)\right)=\frac{1}{2}\left(\mc Z_{1,0}(1-\lambda)-\mc Z_{1,1}(1-\lambda)\right)=   \chi_f(\sigma\cdot\lambda).
\end{aligned}
\end{gather}
Finally, the transformation of twisted sectors can be expressed as
\begin{equation}
    \mcal{Z}^\vee_{\ms g^\vee,\alpha^\vee}=\frac{1}{2}\sum_{\ms g,\ms g=0}^1 (-1)^{\ms g^\vee\ms h-\ms h^\vee\ms g}\mcal{Z}_{\ms g,\ms h},
\end{equation}
which is a special case of \eqref{eq:Partition funcitions related by eta factor}. This is nothing but a discrete Fourier transform of twisted partition functions. Written in the language of gauging and cup products, this can be succinctly written as
\begin{align}
    \mc Z^{\vee}[A^{\vee},\lambda]=\frac{1}{\sqrt{|H^{1}(T^2,\mathbb Z_2)|}}\sum_{A}(-1)^{\int_{T^2}A\cup A^{\vee}}\mc Z[A,1-\lambda].
    \label{eq:KW_duality_gauging_formula}
\end{align}
This is a well-known modern perspective on Kramers-Wannier duality \cite{BhardwajTachikawa201704, Kapustin_2017}. 

\smallskip The two Lagrangian subgroups $\mcal{L}_e$ and $\mcal{L}_m$ correspond to ferromagnetic and paramagnetic phases of $\mbb Z_2$-symmetric quantum system. The Kramers-Wannier duality maps $\mcal{L}_e\leftrightarrow\mcal{L}_m$ and thus  ferromagnetic to paramagnetic phase and vice versa. The order parameter $S_e=\sigma^z_i\sigma^z_j$ for condensing $e$ satisfies
\begin{equation}
    \begin{aligned}
        \lim_{|i-j|\to\infty}\langle\tenofo{GS}_{\tenofo{FM}}|\sigma^z_i\sigma^z_{j}|\tenofo{GS}_{\tenofo{FM}}\rangle&\ne 0,
    \\
    \lim_{|i-j|\to\infty}\langle\tenofo{GS}_{\tenofo{PM}}|\sigma^z_i\sigma^z_{j}|\tenofo{GS}_{\tenofo{PM}}\rangle&=0,
    \end{aligned}
\end{equation}
and thus becomes the order parameter for the ferromagnetic phase, detecting the long-range spin order after spontaneous breaking of $\mathbb Z_2$ symmetry. The order parameter $S_m=\prod_{k=i}^j\sigma^x_k$ for condensing $m$ satisfies
\begin{equation}
    \begin{aligned}
    \lim_{|i-j|\to\infty}\langle\tenofo{GS}_{\tenofo{FM}}|\prod_{k=i}^j\sigma^x_k|\tenofo{GS}_{\tenofo{FM}}\rangle&= 0,
       \\
    \lim_{|i-j|\to\infty}\langle\tenofo{GS}_{\tenofo{PM}}|\prod_{k=i}^j\sigma^x_k|\tenofo{GS}_{\tenofo{PM}}\rangle&\ne 0,
        \end{aligned}
\end{equation}
and becomes the disorder parameter for the ferromagnetic phase. As expected, the order and disorder parameters are swapped under Kramer's Wannier duality.

\smallskip Let us now consider the Kramers-Wannier self-dual region between ferromagnetic and paramagnetic phases.
The self-dual point/region of the phase diagram of $\mathbb Z_2$-symmetric quantum systems cannot be in a phase corresponding to either of the two Lagrangian subgroups, i.e, in either of the two gapped phases. 
As a consequence, it must be either gapless, critical or a first-order transition (gapped but with degeneracy).
The minimal self-dual point $\lambda=1/2$ in the Hamiltonian \eqref{eq:Ising_lambda} is well-known to be in universality class of the Ising CFT with central charge $c=\frac{1}{2}$. 
This is the simplest non-trivial minimal model $\mcal{M}(4,3)$.
There are three primary operators in the critical Ising CFT: the identity operator $\mbb{I}$, the spin field $\sigma,$ and the energy operator $\epsilon$ with conformal dimensions $(h,\bar{h})$ being $(0,0), (\frac{1}{16},\frac{1}{16}),$ and $(\frac{1}{2},\frac{1}{2})$, respectively \cite{DiFrancescoMathieuSenechal1997}. Therefore, the partition function in the untwisted sector $(\ms g,\ms h)=(0,0)$ is given by
\begin{equation}
    \mcal{Z}_{0,0}=|\chi_{0}|^2+|\chi_{\frac{1}{2}}|^2+|\chi_{\frac{1}{16}}|^2.
\end{equation}
The Ising CFT has a $\mathbb Z_2$ global symmetry or equivalently a $\mbb Z_{2}$ topological defect line under which the eigenvalues of the three primaries are $+1,-1$ and $+1$, respectively.
Therefore, the partition functions with the insertion   of $\mathbb Z_2$ symmetry line in the spatial direction is given by \cite{PetkovaZuber2001}
\begin{equation}\label{eq:the twisted partition function of the critical Ising model in terms of Virasoro characters pbc}
    \begin{gathered}
         \mcal{Z}_{0,1}=|\chi_{0}|^2+|\chi_{\frac{1}{2}}|^2-|\chi_{\frac{1}{16}}|^2.
    \end{gathered}
\end{equation}
Here $|\chi_i|^{2}=\chi_{i}\bar{\chi}_{i}$ is the (product of) Virasoro characters of the Ising CFT obtained by tracing over the conformal tower obtained from a primary operator of conformal dimension $(i,i)$, not to be confused with the twisted $\mathbb Z_{2}$ characters defined in \eqref{eq:twisted_Characters}.
In terms of the $\mbb Z_2$ twisted characters, these take the form
\begin{equation}
    \chi_1=|\chi_{0}|^2+|\chi_{\frac{1}{2}}|^2, \quad \chi_e=|\chi_{\frac{1}{16}}|^2.       \label{eq:pbc_characters_Ising}
\end{equation}
Additionally, the theory has a $\mathbb Z_2$ twisted Hilbert space, known as the defect Hilbert space in the CFT literature.
In the lattice model, this is realized within the Hamiltonian with anti-periodic boundary conditions.
The defect Hilbert space too has three primary operators denoted as $\psi, \bar{\psi}$ and $\mu$ with conformal dimensions $(\frac{1}{2},0), (0,\frac{1}{2})$ and $(\frac{1}{16},\frac{1}{16})$.
The $\mathbb Z_{2}$ eigenvalues of these three primary operators are $-1,-1$ and $+1$ respectively.
Therefore the $\mathbb Z_2$ twisted sectors with antiperiodic boundary conditions take the form  
\begin{equation}\label{eq:the twisted partition function of the critical Ising model in terms of Virasoro characters apbc}
    \begin{gathered}
         \mcal{Z}_{1,0}=|\chi_{\frac{1}{16}}|^2+\chi_0\bar{\chi}_{\frac{1}{2}}+\chi_{\frac{1}{2}}\bar{\chi}_0,
        \\
        \mcal{Z}_{1,1}=|\chi_{\frac{1}{16}}|^2-\chi_0\bar{\chi}_{\frac{1}{2}}-\chi_{\frac{1}{2}}\bar{\chi}_0.
    \end{gathered}
\end{equation}
The $\mathbb Z_2$ twisted characters, in the anti-periodic sector are
\begin{equation}
       \chi_m=|\chi_{\frac{1}{16}}|^2, \quad \chi_f=\chi_0\bar{\chi}_{\frac{1}{2}}+\chi_{\frac{1}{2}}\bar{\chi}_0.
       \label{eq:apbc_characters_Ising}
\end{equation}
The fact that $\chi_e=\chi_m$ is a direct consequence of the Kramers-Wannier self-duality of the Ising critical point.

\smallskip Finally, we note, that the holographic approach can be used to analyze more general Hamiltonians, i.e., those which are subtended from non-minimal bulk strings.
For instance, consider the Hamiltonian built from, bulk $e$ and $m$ string operators of length 2.
On the boundary Hilbert space, these act as
\begin{align}
    \mathcal O_{\ms g,j}\mathcal O_{\ms g,j+1}=\sigma^x_j\sigma^x_{j+1}, \quad 
   \mathcal O_{\alpha,j}\mathcal O_{-\alpha ,j+2}=\sigma^z_j\sigma^z_{j+2}.
\end{align}
The Hamiltonian obtained upon including these deformations to the transverse field Ising model is known as the anisotropic next-nearest-neighbor Ising (ANNNI) model \cite{SELKE1988213, Affleck_2015, Vidal_2018}
\begin{align}
    H_{\text{ANNNI}}(\lambda, \tau )= \lambda H_m + (1-\lambda)H_e- \tau\sum_{j}\left[\sigma^x_j\sigma^x_{j+1}+\sigma^z_j\sigma^z_{j+2}\right].
\end{align}
Note that the model is self-dual for any $\tau$, when $\lambda=\frac 12$. 
Along the self-dual line, this model has a rich phase diagram which contains an extended $c=1/2$ Ising critical phase as well as the tri-critical Ising fixed point, Lifshitz transition and degenerate gapped regions (first-order transitions). 
This is consistent with the possible phases along the self-dual line as constrained by considerations mentioned in Sec.~\ref{subsubsec:phase transitions and critical points}.
Namely that the gapped phases $\mathcal L_e$ and $\mathcal L_m$ cannot exist along this line.
Instead, this line contains different types of transitions between these two phases. 
In particular, by varying $\lambda$ we can exit the self-dual line and enter one of the gapped phases $\mathcal L_e$ or $\mathcal L_m$.

\smallskip Another class of non-minimal Kramers-Wannier self-dual deformations of the Ising model are generated by the operators 
\begin{align}
    \mathcal O_{\ms g,j}\mathcal O_{\alpha,j+1}\mathcal O_{-\alpha,j+2}+\mathcal O_{\alpha,j}\mathcal O_{-\alpha,j+1}\mathcal O_{\ms g,j+2}=\sigma_{j}^{x}\sigma_{j+1}^{z}\sigma_{j+2}^{z}
    +
    \sigma_{j}^{z}\sigma_{j+1}^{z}\sigma_{j+2}^{x}.
\end{align}
The corresponding model, known as Fendeley-O'Brien model \cite{Fendeley_OBrien_2018, Vidal_2018}, also hosts the tri-critical Ising critical point. 
The tri-critical Ising (TCI) model corresponds to the minimal model $\mathcal M(4,5)$, with central charge $c=7/10$.
These models clearly also have a $\mathbb Z_2$ global symmetry and can therefore be organized a twisted $\mathbb Z_{2}$ characters labelled by $d\in \mathcal A[\mbb Z_2]$.
These twisted characters take the form \cite{Lassig_91,Zou_Vidal_2020}
\begin{gather}
\begin{aligned}
\chi_1^{\tenofo{TCI}}=&\; |\chi_{0}|^{2}+ |\chi_{1/10}|^{2}+|\chi_{\frac{3}{5}}|^{2}+|\chi_{\frac{3}{2}}|^{2}, \\ 
\chi_e^{\tenofo{TCI}}=&\; |\chi_{\frac{3}{80}}|^{2}+|\chi_{\frac{7}{16}}|^{2},\\
\chi_m^{\tenofo{TCI}}=&\; |\chi_{\frac{3}{80}}|^{2}+|\chi_{\frac{7}{16}}|^{2}, \\
\chi_{f}^{\tenofo{TCI}}=&\; \chi_{\frac{6}{10}}\bar{\chi}_{\frac{1}{10}}
+
\chi_{\frac{1}{10}}\bar{\chi}_{\frac{6}{10}}
+
\chi_{\frac{3}{2}}\bar{\chi}_{0}
+
\chi_{0}\bar{\chi}_{\frac{3}{2}}.
\label{eq:twisted_characters_TCI}
\end{aligned}
\end{gather}
The Kramers-Wannier self-duality is manifest from the fact that $\chi_e=\chi_m$.
We note that the TCI model realizes an emergent $\mc N=1$ superconformal algebra.
The algebra is generated by two primaries, i.e., supercharges with scaling dimensions $(3/2,0)$ and $(0,3/2)$.
As can be seen from \eqref{eq:twisted_characters_TCI}, these appear in the $\mathbb Z_{2}$ twisted Hilbert space sector, i.e., in the lattice model with anti-periodic boundary conditions with respect to the $\mathbb Z_{2}$ symmetry and furthermore transform under a non-trivial $\mathbb Z_2$ representation.

\subsection{$\mbb{Z}_N$ symmetric quantum spin chains}
\label{Subsec:ZN example}
As discussed in Sec.~\ref{subsec:gapped phases}, the symmetry group $\ms G=\mbb Z_{N}$ has in general several different fusion structures and few dualities. Therefore, it belongs to the limit of examples where topological holography is not very powerful and constraining. Nonetheless, it is instructive to go through some details as these become useful for more general examples. For a more powerful application of topological holography, the reader can skip to the next example with $\ms G = \mathbb Z_2\times\mathbb Z_2$.

\smallskip Any such finite Abelian group $\ms G$ can be decomposed into a product of finite Abelian groups of prime power order, i.e.,
\begin{align}
\mathbb Z_{N}=\mathbb Z_{p_1^{\ell_1}}\times \mathbb Z_{p_2^{\ell_2}}\times \dots,    
\end{align}
where $p_i$ are all distinct prime numbers and $\ell_i\in \mathbb Z_{\geq 0}$.
In order to organize our understanding of $\ms G$ symmetric quantum systems, we follow the outlined strategy of first studying a $\ms G$-topological order in $2+1d$.
The topological line operators of the $\ms G$ topological order are labelled by elements in the group 
\begin{align}
    \mc A[\mbb Z_N]= \mathbb Z_{N}\times \text{Rep}(\mbb Z_N)\simeq \mathbb Z_{N}\times \mathbb Z_{N} =\langle m,e|m^N=e^N=1\rangle.
\end{align}
The 0-form symmetry group of the $\mbb Z_N$ gauge theory is given by \cite{FuchsPrielSchweigertValentino201404}
\begin{gather}
\begin{aligned}
\mc G[\mathbb Z_{N}]=&\;\text{Aut}(\mathbb Z_{N})\rtimes \mathbb Z_{2} = \mbb{Z}_N^{\times} \rtimes \mbb{Z}_2,
\end{aligned}
\end{gather}
where the $\mathbb Z_2$ subgroup is generated by the electromagnetic duality which acts as
\begin{align}
    \sigma:e \longleftrightarrow m,
\end{align}
while $\mathbb Z_{N}^{\times}$ is the group under multiplication modulo $N$ formed by the set of integers which are coprime with respect to $N$.
Some examples are $\mathbb Z_{4}^{\times}=\mathbb Z_2$, $\mathbb Z_{8}^{\times}=\mathbb Z_2 \times \mathbb Z_2$, $\mathbb Z_{9}^{\times}=\mathbb Z_6$ and $\mathbb Z_{27}^{\times}=\mathbb Z_{18}$.
For any prime number $p$, $\mathbb Z_{p}^{\times}=\mathbb Z_{p-1}$.
A useful relation for later purposes is
\begin{equation}
\text{Aut}(\mbb Z_N)= \left(\text{Aut}(\mathbb Z^{\times}_{p_1^{\ell_1}})\times \text{Aut}(\mathbb Z^{\times}_{p_2^{\ell_2}})\times \dots \right).
\end{equation}
The group $\mbb{Z}_N^{\times}$ acts on $\mbb Z_N$ by mapping its generator to some other element of order $N$.
Concretely, for $\mathbb Z_{N}=\left\{0,1, \dots, N\right\}$, let a generator $s$ of $\mathbb Z_{N}^{\times}$ act as $s(1)=j$, where $\text{gcd}(j,N)=1$.
Then the action of $s$ on the line operators of the $\mathbb Z_N$ gauge theory is given by
\begin{align}
    s:e^{a}m^{b}\longmapsto e^{ja}m^{f(j)b},
\end{align}
where $j\times f(j)=1 \ \text{mod} \ N$.
For the case, where $N$ itself is prime, the generator of $\mathbb Z_{N}^{\times}$ takes the form
\begin{align}
    s:e^{a}m^{b}\longmapsto e^{2a}m^{(\frac{N-1}{2})b},
\end{align}
Note that $(N-1)/2 \in \mathbb Z$ for all primes except $N=2$, for which $\mbb Z_2^{\times}$ is trivial.

\smallskip Next, we move onto the gapped boundaries for $\mbb Z_{N}$ gauge theory. 
As discussed in section \ref{sec:Gapped boundaries in 2+1d as gapped phases in 1+1d}, the possible $\mathbb Z_N$ symmetric gapped phases are in bijective correspondence to the subgroups of $\mathbb Z_N$.
To see this, we write the group in terms of a  product of indecomposable groups  
\begin{align}
    \mbb Z_N=Z_{p_1^{\ell_1}p_2^{\ell_2}\cdots p_M^{\ell_M}}.
\end{align}
Subgroups then take the form
\begin{align}
\label{eq:subgroup_ZN}
 \ms H_{\bs{r}}=Z_{p_1^{r_1}p_2^{r_2}\cdots p_M^{r_M}} \subseteq    Z_{p_1^{\ell_1}p_2^{\ell_2}\cdots p_M^{\ell_M}}=\mbb{Z}_N,
\end{align}
where $0\leq r_{i} \leq \ell_i$.
Consequently, the number of subgroups are $\prod_{i=1}^{M}(\ell_i+1)$ and are labelled by an $M$ component vector $\bs{r}$.
We similarly, can label Lagrangian subgroups by an $\ms r$, which take the form 
\begin{align}
\label{eq:Lagrangian_subgroup_ZN}
    \mc L_{\bs r}=\langle e^{\alpha(\bs{r})}, m^{\msg(\bs{r})}\rangle:=\langle e^{\prod_i p_i^{r_i}}, m^{\prod_i p_i^{\ell_i-r_i}}\rangle.
\end{align}
Having described the line operators (1-form symmetries) labelled by $\mc A[\mbb Z_N]$, the 0-form symmetries $\mc G[\ms G]$ and the set of gapped boundaries of the $\ms G=\mathbb Z_N$ gauge theory, we are now in a position to discuss the phase space of $\mathbb Z_N$ symmetric quantum systems.

\begin{figure}[!t]
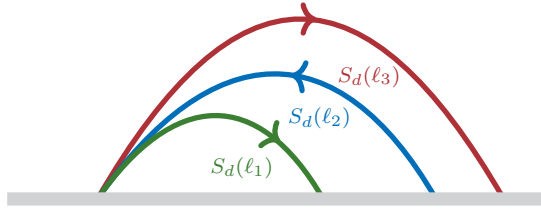

    \centering
    \TMatrixFromThreeStrings
    \caption{Configuration of string operators corresponding to the algebra \eqref{eq:SOA T-matrix} computing self-statistics of bulk anyonic excitations from boundary spin chain.}
    \label{fig:T-matrix from configuration of three strings}
\end{figure}

\begin{figure}[ht]
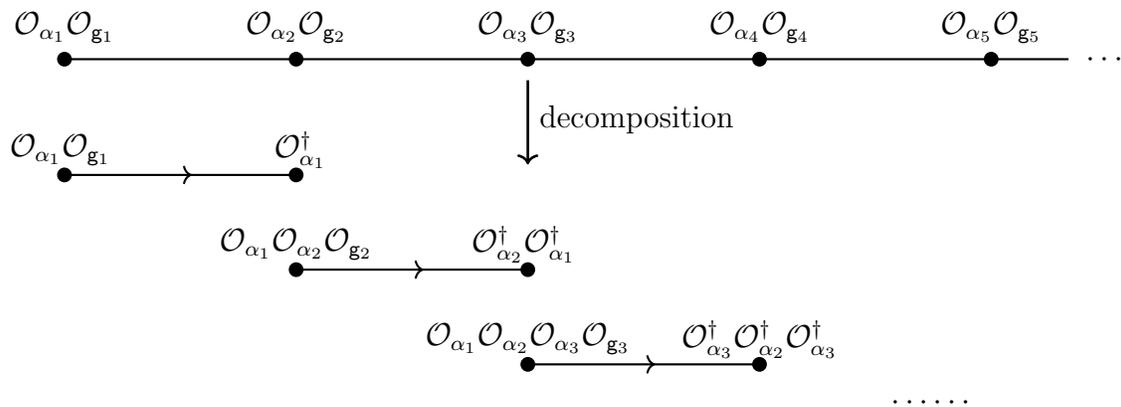
\centering
     \GeneralOperatorConstuction
    \caption{Any $\mathbb Z_N$-symmetric operator acting on a finite region can be decomposed into a product of string operators.\label{fig:G-symmetric operator decomposition}}
\end{figure}

\smallskip Using the approach defined previously, we can obtain an effective spin chain on the $1D$ boundary of the topological gauge theory defined on a spatial semi-infinite cylinder.
The effective spin chain, is defined on a lattice whose sites are endowed with a local Hilbert space $\mc H_{i}\simeq \mathbb C^{N}$.
The space of local operators $\mc{L}(\mc{H}_i)$ are spanned by $X_i$ and $Z_i$, which satisfy the $\mathbb Z_N$ clock algebra (see section \ref{sec:SymmetryStructureAndBuildingBlocksofLocalOperators})
\begin{equation}\label{eq:the ZN clock algebra}
    X^{a}_iZ^{b}_j=\omega_N^{ab\delta_{ij}}Z^{b}_{j}X_{i}^{a},
\end{equation}
where $a,b=0,\cdots,N-1$ and $\omega_N\equiv\exp\left(2\pi i/N\right)$.
The local operator $\mc{O}_{\alpha,i}=Z^{\alpha}_i$ transforms in the representation $\alpha \in \text{Rep}(\mathbb Z_N)$, while the global $\mbb{Z}_N$ symmetry is generated by the operator $\mc U_{\ms g}=\prod_iX^{\ms g}_i$.
The minimal operators subtended from the bulk topological magnetic and electric operators, $m^{\ms g}$ and $e^{\alpha}$ respectively are \eqref{eq:Wilson_operator_to_lattice_operator}
\begin{equation}
    \mc O_{g,i}= X^{\ms g}_i, \quad \mc O_{\alpha,i}\mc O_{-\alpha,i+1}=Z^{\alpha}_i Z_{i+1}^{-\alpha},
    \label{eq:local_operators_Z2Z2_spin_chain}
\end{equation} 
More, generally an open dyonic line operator labelled by $d=(\ms g,\alpha) \in \mc A[\mbb Z_N]$ and with its two ends on the 1D boundary subtends the following string operator on the effective spin-chain 
\begin{align}
    \mc{S}_{(\ms{g},\alpha)}(\ell)=Z^{\alpha}_ {i_1}\left(\bigotimes_{k=1}^{n-1}X^{\ms{g}}_{i_k}\right)Z^{-\alpha}_{i_n}. 
\end{align}
These operators satisfy certain important properties that encode the topological data of the bulk $\mbb Z_N$ gauge theory.
By using the graphical notation for string operators 
\begin{equation}
S_d(\ell) \  = \ \SingleStringLine {8}{d} \ ,    
\end{equation} 
one can readily check that these string operators satisfy the following algebraic relations 
\begin{gather}
\begin{aligned}\label{eq:SOA algebra}
    \BasicStringAlgebra {0}{5}{-5}{5}{d}{d} &=\SingleStringLine{8}{d}
    \\
  \BasicStringAlgebra {0}{6}{0}{6}{d_1}{d_2} &=\,\ms{R}_{\alpha_2^{-1}}(\ms g_1)\ms{R}_{\alpha_1}(\ms g_2)\,\,\,\,\BasicStringAlgebra {0}{6}{0}{6}{d_2}{d_1}
  \\
  \BasicStringAlgebra {0}{6}{-4}{6}{d_1}{d_2} &= e^{i\theta_{d_1d_2}}\;\BasicStringAlgebra {-4}{6}{0}{6}{d_2}{d_1}
  \\
  \BasicStringAlgebra {0}{5}{-5}{5}{d_1}{d_2} &=\ms{R}_{\alpha_2}(\ms g_1)\,\;\BasicStringAlgebra {-5}{5}{0}{5}{d_2}{d_1}
  \\
 \BasicStringAlgebra {0}{5}{0}{7}{d_1}{d_2} &= \ms{R}_{\alpha_2^{-1}}(\ms g_1)\,\;\BasicStringAlgebra {0}{7}{0}{5}{d_2}{d_1}
\end{aligned}
\end{gather}
where $\exp\left\{i\theta_{dd'}\right\}=\ms{R}_{\alpha}(\ms g')\ms{R}_{\alpha'}(\ms g)$. In the above equations we have use the shorthand 
\begin{align}
  \BasicStringAlgebra {0}{8}{0}{8}{d_1}{d_2}  \equiv \mc{S}_{d_2}(\ell)\mc{S}_{d_1}(\ell).
\end{align}
Additionally, if we consider three oriented segments $\ell_{1,2,3}$ as in Figure \ref{fig:T-matrix from configuration of three strings}, which share a common initial point, the corresponding string operators satisfy the algebraic relation \cite{LevinWen2003}
\begin{align} \label{eq:SOA T-matrix}
    \mc{S}_{d}(\ell_1)    \mc{S}_{d}(\ell_2)    \mc{S}_{d}(\ell_3)= e^{i\theta_{d}}
      \mc{S}_{d}(\ell_3)    \mc{S}_{d}(\ell_2)    \mc{S}_{d}(\ell_1),
\end{align}
where $\exp\left\{i\theta_d\right\}=\ms{R}_{\alpha}(\ms{g})$.
The algebra generated by these string operators is the String Operator Algebra (SOA)
\begin{equation}
    \mathbb{SOA}[ \mathbb{Z}_N] = \text{Spac}_{\mbb C} \left\{S_d(\ell)\; \big|\; \forall \  d \text{and } \forall \  \ell \right\},
\end{equation}
which is the algebra of local $\mathbb Z_N$-symmetric operators that are the building blocks of $\mathbb Z_N$-symmetric Hamiltonians. 
To see this, note that the most general $\mathbb Z_N$-symmetric operator on a segment $\ell$ is spanned by operators of the form 
\begin{align}
    \mc K(\ell) = \mc O_{\ms \alpha_1, i_1}\mc O_{\ms g_1, i_1} \dots \mc O_{\ms \alpha_n, i_n}\mc O_{\ms g_n, i_n},
\end{align}
with the constraint that $\sum_{k=1}^n\alpha_{i_k} = 0 \ \text{mod} \ N$.
Such operators are elements of the SOA since we can always decompose these (up to a $\ms{U}(1)$ phase) into a product of string operators as (See also figure \ref{fig:G-symmetric operator decomposition}).

\begin{align}\label{eq:stringDecomposition}
    \mc K(\ell) &= \Bigg(\prod_{k=1}^{n-1}\mc S_{d_k}(\ell_{i_k})\Bigg)S_{(\ms g_n', 1)}(i_n),
\end{align}
where $d_k = (\ms g_k, \sum_{j=1}^k\alpha_j)$ and $\ell_i$ correspond to a single link $\ell_i =\{i,i+1\}$. 
The decomposition of symmetric operators $\mathcal K(\ell)$ into links \eqref{eq:stringDecomposition} may not be the most economical as it assigns a `$d$' label to each link. 
Often $\mathcal K(\ell)$ can be decomposed using longer string operators and thus fewer labels. 
For any symmetric operator $\mathcal K(\ell)$, there is always a minimal decomposition, which contains the minimum number of string operators. Thus any symmetric operator can be labeled by its corresponding set of minimal labels $\boldsymbol d \equiv (d_1, \dots, d_n)$ as
\begin{equation}\label{eq:minimalStringDecomposition}
    \mathcal K_{\boldsymbol d}(\ell) = S_{d_1}(\ell_1)\dots S_{d_n}(\ell_n),
\end{equation}
where $\ell_i$ are non-overlapping sub-segments of $\ell$ of various lengths such that $\ell=\cup_i\ell_i$.  

\smallskip

\smallskip As described in the previous section, the gapped phases/ground states realized in $\ms G$ symmetric quantum systems are in correspondence with gapped boundaries of a $\ms G$ topological gauge theory, classified by Lagrangian subgroups.
Furthermore, we can construct the fixed-point Hamiltonian within each phase directly from the operators corresponding to the line operators allowed to end on a given gapped boundary $\mc L_{\bs r}$ in eq.~\eqref{eq:Lagrangian_subgroup_ZN}.
Such a fixed point Hamiltonian takes the form 
\begin{equation}
    H_{\bs{r}}= -\sum_{j}\left[Z_j^{\alpha(\bs r)}Z_{j+1}^{-\alpha(\bs r)} 
    +
    X_{j}^{\msg(\bs r)} + \text{H.c}
    \right].
\end{equation}
The ground state realized for a Hamiltonian labelled by $\bs{r}$, breaks the global
symmetry down to the subgroup $\ms H_{\bs{r}}$ in eq.~\eqref{eq:subgroup_ZN}.
This can be seen from the fact that the expectation value
\begin{align}
    \lim_{|i-j|\to \infty} \left\langle Z_i^{\alpha(\bs r)}Z_{j}^{-\alpha(\bs r)}\right\rangle  \neq 0,
\end{align}
Furthermore, from the property of cluster decomposition, this implies that 
\begin{align}
    \langle Z_i^{\alpha(\bs r)}\rangle  \neq 0,
\end{align}
which transforms non-trivially under $\mathbb Z_{N}/\ms H_{\bs{r}}$. 
Meanwhile, it can be readily checked that the ground state is invariant under the subgroup $\ms H_{\bs{r}}\subseteq \mathbb Z_{N}$ generated by $\mc U_{\ms g(\bs{r})}$.
A general Hamiltonian capable of realizing all the gapped phases is simply obtained by adding terms corresponding to each Lagrangian subgroup
\begin{align}
\label{eq:generic minimal ZN Hamiltonian}
    H(\bs{\lambda})=\sum_{\bs{r}}\lambda_{\bs{r}}H_{\bs{r}}.
\end{align}
Finally, let us point out that the critical transitions $\mscr{C}_{p,q}$ between two gapped phases $\mc{L}_p$ and $\mcal{L}_q$ labeled by two integers $p$ and $q$ such that $pq=N$ is electric-magnetic self-dual. Using \eqref{eq:the generic formula for twisted dual partition function} and the electric-magnetic duality $\sigma:(\ms g,\ms \alpha)=(\alpha,\ms g)$, we see that ($\omega=\exp\left( \frac{2\pi\mfk{i}}{N}\right)$)
\begin{equation}\label{eq:the EM dual partition function for ZN}
    \mcal{Z}^\vee_{\msg^\vee,\msh^\vee}(\mscr{C}_{p,q})=\frac{1}{N}\sum_{\msg,\msh=0}^{N-1}\omega^{\ms{g}^\vee\ms{h}-\ms{h}^\vee\ms{g}}\mcal{Z}_{\msg,\msh}(\mscr{C}_{p,q}). 
\end{equation}
It is known that the critical transition between ferromagnetic and paramagnetic phases of $N$-state Potts model is described by $\mbb{Z}_N$ parafermion CFT \cite{zamolodchikovFateev1985} and hence $\mcal{Z}_{\msg,\msh}(\mscr{C}_{p,q})$ is the generalized twisted partition function of this CFT. The relation \eqref{eq:the EM dual partition function for ZN} is exactly the sum rule for twisted partition functions of the parafermion CFTs, which reflects the self-duality \cite{GepnerQiu1987,DiFrancescoSaleurZuber1988} (for a brief review, see appendix A of \cite{DiFrancescoSaleurZuber1988}. The relevant sum rule is given in equation (A.8)).

\subsection{$\mathbb Z_2\times\mathbb Z_2$ symmetric quantum spin chains}
\label{Subsec:Z2Z2_example}

The dualities discussed in this paper are most useful when the duality group is large as compared with the number of possible fusion structures. In the case of $\mbb{Z}_N$, we have the opposite limit, there are few dualities but many fusion structures. Hence, dualities are not very powerful in constraining the phase diagram. 

\smallskip In this section, we consider quantum systems with $\ms G=\mathbb Z_2\times \mbb{Z}_2$ global symmetry, which belong to the opposite limit; there is a non-Abelian duality group of order $72$ and six gapped phases with the same fusion structure.
We will see that this can help us to determine the rich phase diagram of models with $\mbb{Z}_2\times\mbb{Z}_2$ global symmetry.
In particular, there will be interesting critical points and lines, as well as regions with large emergent symmetries. 
Besides constraining and predicting a phase diagram using topological holography, our construction provides order parameters that we will use to numerically verify the theoretically obtained phase diagram.

\smallskip Parts of this phase-diagram was analysed long ago using topological holography and non-abelian dualities in \cite{MoradiWei2014,MoradiPhD2018}. The ideas in this paper were inspired by and developed shortly after the work in  \cite{MoradiWei2014,MoradiPhD2018}, though it remained unpublished until now. 

\subsubsection*{Dualities}

In the holographic bulk, we consider a $2+1d$ $\ms G$ topological gauge theory whose line operators are labelled by elements (anyons) in $\mc A$ where 
\begin{align}\label{eq:Z2xZ2 anyon labels}
\mc A[\mbb{Z}_2 \times \mbb{Z}_2]=\mathbb Z_2^2 \times \text{Rep}(\mathbb Z_2) \simeq \mathbb Z_2^2\times \mathbb Z_2^2 =\langle m_L,m_R,e_L,e_R\rangle.
\end{align}
Here the $(\ms g_1, \ms g_2,\alpha_1, \alpha_2)$ charge of each anyon is $m_L=(1,0,0,0)$, $m_R=(0,1,0,0)$, $e_L=(0,0,1,0)$ and $e_R=(0,0,0,1)$.
Let us now explore the 0-form symmetry group $\mc G[\ms G]$. 
Three symmetry operations are immediately obtained as the electromagnetic duality $\sigma_L$ and $\sigma_R$ on the two $\mathbb Z_2$ ``layers" denoted as $L$ and $R$ respectively, along with the layer-swapping $L\leftrightarrow R$ denoted as $f$ such that
\begin{gather}\label{eq:Z2xZ2_firstThreeGenerators}
    \begin{aligned}
    \sigma_L&:\{m_L\mapsto e_L, e_L\mapsto m_L\},
    \\
    \sigma_R&:\{m_R\mapsto e_R, e_R\mapsto m_R\}, 
    \\
    f&:\{m_L\mapsto m_R, m_R\mapsto m_L, e_L\mapsto e_R,e_R\mapsto e_L\},
    \end{aligned}
\end{gather}
where all the other generators of $\mc A[\mbb{Z}_2 \times \mbb{Z}_2]$ that have not been shown are invariant. Together, $\sigma_L,\sigma_R$ and $f$ form the group $D_4=\mathbb Z_4\rtimes\mathbb Z_2$. 
However the duality group turns out to be significantly larger. 
Here, we present the generators that will turn out to be useful in understanding the phase-diagram of the model.
\begin{gather}\label{eq:the hi transformations}
    \begin{aligned}
    h_1&: \{m_R\mapsto m_Lm_R, e_L\mapsto e_Le_R\}, \\
    h_2&: \{m_L\mapsto m_Lm_R, e_R\mapsto e_Le_R\}, \\
    h_3&: \{m_L\mapsto m_R, m_R\mapsto m_L, e_L\mapsto e_R, e_R\mapsto e_L\}, 
    \end{aligned}
\end{gather}
and 
\begin{gather}\label{eq:the ki transformations}
    \begin{aligned}
    k_1&: \{m_L\mapsto m_Le_R, m_R\mapsto m_Re_L\}, \\
    k_2&: \{e_L\mapsto e_Lm_R, e_R\mapsto e_Rm_L\}, \\
    k_3&: \{m_L\mapsto e_R,m_R\mapsto e_L, e_L\mapsto m_R, e_R\mapsto m_L\}. 
    \end{aligned}
\end{gather}
The transformations $h_1$,$h_2h_3$,$k_1$ and $k_2k_3$ generate the group $S_3\times S_3$ as
\begin{gather}\label{eq:The two S3 subgroups of Z2xZ2 duality group}
\begin{aligned}
S_3=&\;\left\langle k_1,k_2k_3 \ \Big| \ k_1^2 = 1, (k_2k_3)^3 = 1, k_1 (k_2k_3)k_1^{-1} = (k_2k_3)^{-1}\right\rangle,\\
S_3=&\;\left\langle h_1,h_2h_3 \ \Big| \ h_1^2 = 1, (h_2h_3)^3 = 1, h_1 (h_2h_3)h_1^{-1} = (h_2h_3)^{-1}\right\rangle,
\end{aligned}
\end{gather}
and $h_ik_j=k_jh_i$ for $i,j=1,2,3$. The previous generators $\sigma_{L,R}$ and $f$ are related to these generators via the relations
\begin{equation}
	h_3 = f, \qquad k_3 = \sigma_L\sigma_Rf,
\end{equation}
implying that $\sigma_R$ and $f$ can be expressed in terms of these generators, and we only need to add $\sigma_L$ to the set of generators. The $\mathbb Z_2$ group generated by $\sigma_L$ act on $S_3\times S_3$, by swapping the two $S_3$ copies 
\begin{equation}
	\sigma_L k_i \sigma_L^{-1} = h_i, \qquad i=1, 2, 3.
\end{equation}
Therefore the full 0-form symmetry group $\mc G[G]$ is 
\begin{gather}
\begin{aligned}
\mathcal G[\mbb{Z}_2\times\mbb{Z}_2]&=  \left\langle \sigma_L, h_1, k_1, h_2h_3, k_2k_3 \right\rangle= (S_3\times S_3)\rtimes\mathbb Z_2.
\end{aligned}
\end{gather}
This is a group of order
\begin{equation}
    |\mathcal G[\mbb{Z}_2\times\mbb{Z}_2]| = (3!)^2\times 2 = 72.
\end{equation}
These bulk $0$-form symmetries in $2+1$ dimensions become dualities in $1+1$ dimensions.
This group can be generated by fewer generators, however we have chosen these as they reveal the structure of the phase-diagram more transparently. 

\subsubsection*{Gapped Phases, Order Parameters, and Excitations}

The bulk topological gauge theory admits a set of six gapped boundaries as enumerated in Table \ref{tab:lagrangian subgroups of Z2*Z2}.
Each gapped boundary can be labelled equivalently by a tuple $(\ms H, \psi)$, where $\ms H \subseteq \mbb{Z}_2\times \mbb{Z}_2 $ and $\psi\in H^{2}(\ms H,\ms{U}(1))$.
The subgroups corresponding to each of the six Lagrangian subgroups are given in Table \ref{tab:lagrangian subgroups of Z2*Z2}.
The cohomology group $H^{2}(\mathbb Z_2,\ms{U}(1))=\mathbb Z_{1}$, while $H^{2}(\mathbb Z_2 \times \mathbb Z_2,\ms{U}(1))=\mathbb Z_{2}$.
Therefore, there are two Lagrangian subgroups $\mc L_5$ and $\mc L_6$ corresponding to $\ms H = \mathbb Z_2 \times \mathbb Z_2$.
In particular, $\mc L_5$ and $\mc L_6$ are labelled by the non-trivial and trivial cohomology class in $H^{2}(\mathbb Z_2 \times \mathbb Z_2,\ms{U}(1))$, respectively.
\begin{table}\centering
    \begin{tabular}{c  c c c c c} \toprule
    \multirow{2}{2cm}{Lagrangian subgroups} & \multirow{2}{2cm}{Generating set} & \multirow{2}{1.8cm}{Image of $\Pi$} & \multirow{2}{1cm}{$\ms H$} & \multirow{2}{1cm}{$\psi(\ms h_1,\ms h_2)$} & \multirow{2}{2cm}{Gapped phase} \\
    \\
          \midrule

      $\mcal{L}_1$ & $e_L,e_R$
      & $1_L,1_R$ & $\mbb{I}$ & 1 & SSB
      \\
      $\mcal{L}_2$ & $m_L,e_R$
      & $m_L$ & $\mbb{Z}^L_2$ & 1 & PSB${}_L$
      \\
      $\mcal{L}_3$ & $e_L,m_R$ 
      & $m_R$ & $\mbb{Z}^R_2$  & 1 & PSB${}_R$
      \\
      $\mcal{L}_4$ & $e_Le_R,m_Lm_R$ 
      & $m_Lm_R$ & $\mbb{Z}^D_2$ & 1 & PSB${}_D$
      \\ 
      $\mcal{L}_5$ & $e_Lm_R,m_Le_R$ 
      & $m_R,m_L$ & $\mbb{Z}^L_2\times\mbb{Z}^R_2$ & $(-1)^{h_{1,L}h_{2,R}}$  & $\text{SPT}_1$
      \\ 
      $\mcal{L}_6$ & $m_L,m_R$  
      & $m_R,m_L$ & $\mbb{Z}^L_2\times\mbb{Z}^R_2$ & 1 & $\text{SPT}_0$
      \\
    \bottomrule
    \end{tabular}
      \caption{Table of Lagrangian subgroups (gapped phases) for $\ms G=\mbb{Z}_2\times\mbb{Z}_2$. Here, $\mbb{Z}_2^L$, $\mbb{Z}_2^R$ and $\mbb{Z}_2^D$ denote the left, the right and the diagonal $\mbb{Z}_2$ subgroups of $\mbb{Z}_2\times\mbb{Z}_2$, respectively. All gapped phases have the same fusion structure $\mcal{L}_i\simeq\mbb{Z}_2\times\mbb{Z}_2$ for all $i$.  $\ms H$ and $\psi$ are the unbroken subgroup of $\ms G$ and the SPT twist of the given phase respectively. These can be extracted from the corresponding Lagrangian subgroup (see Section \ref{subsec:gapped phases}). }
    \label{tab:lagrangian subgroups of Z2*Z2}
\end{table}

Having described the properties of the bulk $\mathbb Z_2\times \mathbb Z_2$ gauge theory that are of interest, we now analyze the phase diagram of 1D quantum systems with $\mbb Z_2\times \mathbb Z_2$ global symmetry.
As before, it is useful to have a quantum spin system in mind,  
both for concreteness as well as to be able to numerically verify various theoretical predictions.
The effective spin chain is obtained at the boundary of a 2D $\ms G$ gauge theory defined on a semi-infinite cylinder.
The Hilbert space of the 1D quantum system decomposes into a tensor product of local Hilbert spaces $\mc H_i\simeq \mbb{C}_L^2\otimes\mbb{C}_R^2$ assigned to the site $i$.
The local operators acting on the spin chain are obtained by bringing minimal-length bulk line operators to the open boundary.
The generators of $\mc A[\mbb{Z}_2 \times \mbb{Z}_2]$ become the following operators when brought to the 1D boundary (see \eqref{eq:Definition of A and B operators} and \eqref{eq:Wilson_operator_to_lattice_operator})
\begin{gather}
\begin{aligned}
e_L \longrightarrow&\; \sigma_{i,L}^z\sigma_{i+1,L}^z, \qquad m_L\longrightarrow \sigma_{i,L}^x \\
e_R \longrightarrow&\; \sigma_{i,R}^z\sigma_{i+1,R}^z, \qquad m_R\longrightarrow \sigma_{i,R}^x.
\end{aligned}
\end{gather}
The symmetry generator is given by
\begin{equation}
    \mcal{U}_{\ms{g}_L,\ms{g}_R}=\bigotimes_i^L\left(\sigma^x_{i,L}\right)^{\ms g_L}\otimes\left(\sigma^x_{i,R}\right)^{\ms g_R}.
\end{equation}
The fixed-point Hamiltonians corresponding to the different gapped phases can be read off from the generators of the corresponding Lagrangian subgroups.
Note that some of the Lagrangian subgroups, in $\mc L_4$ and $\mc L_5$ are generated by dyons which are not in the generator set of $\mc A[\mbb{Z}_2 \times \mbb{Z}_2]$.
We need to additionally specify the operators that are subtended on the boundary by corresponding minimal length bulk dyonic operators.
The generators of $\mc L_4$ can be immediately obtained by taking a product of generators of $\mc A[\mbb{Z}_2 \times \mbb{Z}_2]$, i.e.
\begin{equation}
    e_Le_R\longrightarrow \sigma_{i,L}^z\sigma_{i,R}^z\sigma_{i+1,L}^z\sigma_{i+1,R}^z, \qquad m_Lm_R\longrightarrow
    \sigma_{i,L}^x\sigma_{i,R}^x.
\end{equation}
The generators of $\mc L_5$ require slight care such that the operators corresponding to $e_Lm_R$ commute with the operators corresponding to $e_Rm_L$ regardless of their location. For a more systematic construction of the spin chain and string operators see Appendix \ref{Sec:holographic perspective}. 
The minimal operators take the form
\begin{equation}
e_Lm_R\longrightarrow \sigma_{i,L}^z\sigma_{i+1,L}^z\sigma_{i+1,R}^x, \qquad m_Le_R\longrightarrow
\sigma_{i,L}^x\sigma_{i,R}^z\sigma_{i+1,R}^z.
\end{equation}
Having defined the set of operators, which correspond to the generators of each of the Lagrangian subgroups, we can straightforwardly write down the fixed-point Hamiltonian \eqref{eq:the Hamiltonian associated to a Lagrangian subgroup} for each of the gapped phases
\begin{gather}\label{eq:Z2xZ2 fixed-point hamiltonians}
\begin{aligned} 
H_{1}=& -\sum_{i}\left[\sigma_{i,L}^z\sigma_{i+1,L}^z + \sigma_{i,R}^z\sigma_{i+1,R}^z  \right], \\    
H_{2}=& -\sum_{i}\left[\sigma_{i,L}^x+  \sigma_{i,R}^z\sigma_{i+1,R}^z\right], \\
H_{3}=& -\sum_{i}\left[\sigma_{i,L}^z\sigma_{i+1,L}^z +  \sigma_{i,R}^x \right], \\
H_{4}=& -\sum_{i}\left[\sigma_{i,L}^z\sigma_{i,R}^z\sigma_{i+1,L}^z\sigma_{i+1,R}^z +  \sigma_{i,L}^x\sigma_{i,R}^x \right], \\
H_{5}=& -\sum_{i}\left[\sigma_{i,L}^z\sigma_{i+1,L}^z\sigma_{i+1,R}^x+
\sigma_{i,L}^x\sigma_{i,R}^z\sigma_{i+1,R}^z
\right], \\
H_{6}=& -\sum_{i}\left[\sigma_{i,L}^x+
\sigma_{i,R}^x \right].
\end{aligned}
\end{gather}
From the point of view of the bulk topological order, the different gapped boundaries are characterized by condensation of anyons in their corresponding Lagrangian subgroups \eqref{eq:StringOrderParameterCondensation}. To detect condensation of $e_L$, $e_R$, and $e_Le_R$, we have the order parameters 
\begin{equation}\label{eq:condensationOfes_z2z2}
    \begin{gathered}
         S_{e_L}(i,j)=\sigma^z_{i,L}\,\sigma^z_{j,L}, \qquad
    S_{e_R}(i,j)=\sigma^z_{i,R}\,\sigma^z_{j,R},
    \\
    S_{e_Le_R}(i,j)=\sigma^z_{i,L}\sigma^z_{i,R}\;\sigma^z_{j,L}\sigma^z_{j,R}.
    \end{gathered}
\end{equation}
A non-zero vacuum expectation-value for these operators for $|i-j|\to\infty$ implies long-range spin-spin correlations and thus long-range order. 
Since these operators are bi-local, the property of cluster decomposition of correlation functions induces  genuinely local order parameters.
Thus the local order parameters $\sigma_{i,L}^{z}$, $\sigma_{i,R}^{z}$ and $\sigma_{i,L}^{z}\sigma_{i,R}^{z}$ can be used to detect the spontaneous symmetry-breaking.
Holographically, the condensation of electric excitations, imply that the corresponding symmetry is broken.
Similarly, the condensation of $m_L$, $m_R$, and $m_Lm_R$ are detected by the condensation of the following string operators
\begin{equation}\label{eq:condensationOfms_z2z2}
    \begin{gathered}
    S_{m_L}(i,j)=\prod_{k=i}^j\sigma^x_{k,L}, \qquad
    S_{m_R}(i,j)=\prod_{k=i}^j\sigma^x_{k,R},
    \\
    S_{m_Lm_R}(i,j)=\prod_{k=i}^j\sigma^x_{k,L}\sigma^x_{k,R}.
    \end{gathered}
\end{equation}
These are the disorder parameters, relative to the spin order parameters in \eqref{eq:condensationOfes_z2z2}. Thus the condensation of magnetic excitations imply that the corresponding symmetry is unbroken.
Finally, the condensation of the dyons $e_Lm_R$ and $m_Le_R$ for the gapped phase $\mc L_5$ can be detected by
\begin{equation}
    \begin{aligned}
        S_{e_Lm_R}(i,j)&= \sigma^z_{i,L}\left(\bigotimes_{k=i+1}^{j}\sigma_{k,R}^{x}\right)\sigma^z_{j,L}, 
        \\
        S_{m_Le_R}(i,j)&=\sigma^z_{i,R}\left(\bigotimes_{k=i}^{j-1}\sigma_{k,L}^{x}\right)\sigma^z_{j,R}.
    \end{aligned}
\end{equation}
These are the non-local string order-parameters for SPT phases
\cite{Cirac_2008, Pollmann_2012, PollmannSPT2009A2012}. Thus the condensation of dyonic excitations (for Lagrangian subgroups that are generated by dyonic generators) implies that the corresponding phase is an SPT phase. 
We emphasize that all order parameters for all possible gapped phases naturally emerge out of the holographic construction.

\smallskip The next question is: what are the fundamental excitations of a given gapped phase? For example, consider the gapped phase $\mcal{L}_1$. In this phase, the anyons in $\mathcal L_1=\left\{1,e_L,e_R,e_Le_R\right\}$ are condensed and become the order parameters \eqref{eq:condensationOfes_z2z2}, as explained above. On the other hand, it costs finite energy to create  anyons that are confined on the boundary and thus such confined anyons correspond to excitations.
However, any two anyon that differ by a condensed anyon correspond to identical boundary excitations. The equivalence classes of boundary excitations are thus given by 
\begin{gather}\label{eq:the excitation of gapped boundary L1 in the Z2xZ2 case}
\begin{aligned}
    \phi^{\text{SSB}}_{1}&=\left\{1, e_L,e_R,e_Le_R\right\}, \\
    \phi^{\text{SSB}}_{m_L}&=\left\{m_L, m_Le_L,m_Le_R,m_{L}e_Le_R\right\}, \\
    \phi^{\text{SSB}}_{m_R}&=\left\{m_R, m_Re_L,m_Re_R,m_{R}e_Le_R\right\}, \\
    \phi^{\text{SSB}}_{m_Lm_R}&=\left\{m_Lm_R, m_Lm_Re_L,m_Lm_Re_R,m_{L}m_Re_Le_R\right\},
\end{aligned}
\end{gather}
Together $\phi^{\text{SSB}}_1, \phi^{\text{SSB}}_{m_L}, \phi^{\text{SSB}}_{m_R}$ and $\phi^{\text{SSB}}_{m_Lm_R}$ are the objects of the category of excitations $\text{Vec}_{\mbb Z_2\times \mbb Z_2}$ within the gapped phase corresponding $\mc L_1$. 
Physically, these excitations are nothing but string operators \eqref{eq:condensationOfms_z2z2} corresponding to $m_{L},m_{R}$ and $m_{L}m_{R}$, i.e., the $\mbb Z_2\times \mbb Z_2$ symmetry operators restricted to a finite segment. 
Since the ground-state spontaneously breaks the ${\mbb Z_2\times \mbb Z_2}$ symmetry, these symmetry operators create domain-walls which are the familiar excitations within this phase.
Similarly, the fundamental excitations within other gapped phases $\mcal{L}_i$ can be obtained straightforwardly. 

\subsubsection*{The minimal Hamiltonian and its phase diagram}
A minimal Hamiltonian capable of accessing all the gapped phases arising within $\mathbb{Z}_2^2$ symmetric quantum systems is obtained as a sum of the six fixed point Hamiltonians given in \eqref{eq:Z2xZ2 fixed-point hamiltonians}
\begin{align}\label{eq:the minimal Z2xZ2 Hamiltonian}
    H({\lambda})=\sum_{a=1}^{6}\lambda_{a}H_{a}.
\end{align}
The phase diagram of this model is parameterized by six real parameters $\lambda_1,\cdots,\lambda_6$. 
We can develop an understanding of the topology of the phase diagram as well as the nature of transitions realized by using dualities inherited from the 0-form symmetry group $\mc G[\mbb Z_2\times \mbb Z_2]$ of the corresponding topological order.

\smallskip The action of the duality group $\mc G[\mbb{Z}_2\times \mbb{Z}_2] = (S_3\times S_3)\rtimes\mathbb Z_2$, can be conveniently understood by partitioning the set of gapped phases into 2 disjoint subsets with Lagrangian subgroups $T_1=\left\{\mc L_{2},\mc L_{3}, \mc L_{4}\right\}$ and $T_2=\left\{\mc L_{1},\mc L_{5}, \mc L_{6}\right\}$. 
The first $S_3$ subgroup, generated by $h_i,\,i=1,2,3$ \eqref{eq:The two S3 subgroups of Z2xZ2 duality group}, acts as the permutation of the three Lagrangian subgroups in $T_1$ while leaving the Lagrangian subgroups in $T_2$ invariant. The second $S_3$ subgroup, generated by $k_i,\,i=1,2,3$ \eqref{eq:The two S3 subgroups of Z2xZ2 duality group}, acts as the permutation of the three Lagrangian subgroups in $T_2$ while leaving the Lagrangian subgroups in $T_1$ invariant.

\begin{figure}
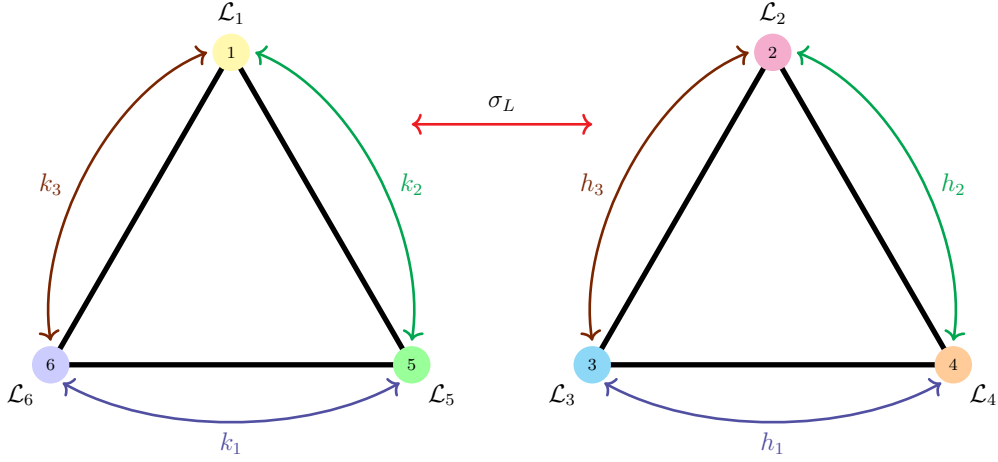
\centering 
    \ZTwoSqAbstractDualityActionOnLs
    \caption{Duality mapping between different gapped phases. 
    The gapped phases labelled by Lagrangian subgroups in the set $T_2 = \left\{\mc L_1,\mc L_5,\mc L_6\right\}$ and $T_1 = \left\{\mc L_2,\mc L_3,\mc L_4\right\}$  transform under duality subgroup $S_{3}=\langle k_1,k_2,k_3\rangle$ (depicted in solid lines) and $S_{3}=\langle h_1,h_2,h_3\rangle$ (depicted in dashed lines) respectively. The order two duality generator $\sigma_L$ maps between the two sets.    
    }
    \label{Fig:duality_on_gapped_phases}
\end{figure}

\smallskip Finally, the transformation $\sigma_{L}$, which generates a $\mathbb Z_{2}$ subgroup of $\mc G[\mbb{Z}_2\times \mbb{Z}_2]$, acts via an outer automorphism on $S_{3}\times S_3$, which exchanges the two copies of $S_3$.
Naturally it also acts by exchanging the two sets of gapped phases, 
\begin{align}
    \mcal{L}_1\overset{\sigma_L}{\longleftrightarrow} \mcal{L}_2, \qquad \mcal{L}_5\overset{\sigma_L}{\longleftrightarrow} \mcal{L}_4, \qquad 
    \mcal{L}_6\overset{\sigma_L}{\longleftrightarrow} \mcal{L}_3.
\end{align}
From this,we see that the action of the duality group on the set of Lagrangian subgroups is geometrically the symmetries of two triangles (See Figure \ref{Fig:duality_on_gapped_phases}). One might wonder whether this abstract symmetry reflects a geometrical structure within the phase diagram. As discussed previously, these symmetries act on the set of anyons which induces an action on operators in the spin chain \eqref{eq:DualityActing_on_S_operators}. In particular, the fixed-point Hamiltonians transform as their corresponding Lagrangian subgroups which in turn induces a permutation of coupling constants $\lambda_a$ in \eqref{eq:the minimal Z2xZ2 Hamiltonian}. 

\smallskip In summary, the dualities of the Hamiltonian \eqref{eq:the minimal Z2xZ2 Hamiltonian} for a fixed set of parameters $\lambda_a$ are as follows: there is a dual Hamiltonian for (1) any permutation of $\{\lambda_2,\lambda_3,\lambda_4\}$, (2) any permutation of $\{\lambda_1,\lambda_5,\lambda_6\}$, and (3) any mapping of parameters between these sets. Since a six-dimensional phase diagram is hard to visualize, we will break it down to interesting subspaces. Let us focus on the three phases in $T_1$ by setting $\lambda_1=\lambda_5=\lambda_6=0$, 
\begin{equation}\label{eq:T1 Hamiltonian}
    H_{T_1}(\lambda_2,\lambda_3,\lambda_4)=\lambda_2H_2+\lambda_3H_3+\lambda_4H_4.
\end{equation}
Here $H_2, H_3$, and $H_4$ are the fixed-point Hamiltonian for the partially spontaneously-broken phases corresponding to $\ms H=\mbb{Z}_2^L$, $\ms H=\mbb{Z}_2^R$, and $\ms H =\mbb{Z}_2^D$ (see Table \ref{tab:lagrangian subgroups of Z2*Z2} and equation \eqref{eq:Z2xZ2 fixed-point hamiltonians}). 

\smallskip This Hamiltonian is dual to any other Hamiltonian related by a permutation of its coefficients. Since the overall scaling of the Hamiltonian has no physical consequences, we can express $\lambda_2,\lambda_3$, and $\lambda_4$ in spherical coordinates \cite{MoradiWei2014} 
\begin{equation}
    \lambda_2=\sin\theta\cos\phi,\qquad 
    \lambda_3=\sin\theta\sin\phi,\qquad 
    \lambda_4=\cos\theta.
\end{equation}
For $(\theta,\phi)=(\pi/2,0)$, $(\theta,\phi)=(\pi/2,\pi/2)$, and $(\theta,\phi)=(0,0)$, we get the fixed-point (exactly-solvable) Hamiltonians for the gapped phases $\mcal{L}_2$, $\mcal{L}_3$, and $\mcal{L}_4$, respectively. 
The quadrant of the sphere spanned by these three points has a particularly nice structure and is easiest visualized through projection of $(\theta,\phi)$ onto the plane $\lambda_2+\lambda_3+\lambda_4=1$.
Any point on this triangle corresponds to a particular set of parameters in the Hamiltonian \eqref{eq:T1 Hamiltonian}, its corners are the fixed-point Hamiltonians and symmetry of the triangle maps parameters $\lambda_a$ to dual parameters $\lambda^\vee_a$.%
We can use the properties of dualities described in Section \ref{topological holography: application} to sketch out a phase diagram.
We assume that only the three gapped phases $\mathcal L_2$, $\mathcal L_3$, and $\mathcal L_4$ can appear in this $2d$-slice of the phase diagram. 

\smallskip Dualities are a powerful tool to make non-perturbative statement which constrain the possible phase diagram.
A possible way of reasoning using dualities is as follows
\footnote{For notational simplicity, we will parameterize Hamiltonians with points lying outside of the $\lambda_2+\lambda_3+\lambda_4=1$ plane. However, through scaling, this parameterization is equivalent to the one for which $\lambda_a$s are lying on the plane.}
\begin{figure}[t!]
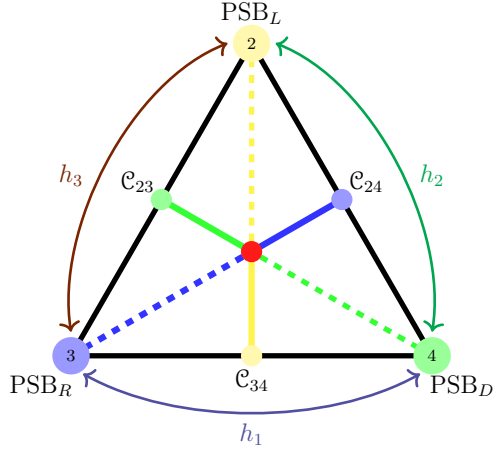
\centering 
    \ZTwoSqPhaseDiagramSelfDualLines
    \caption{Illustration of a triangle on the plane $\lambda_2+\lambda_3+\lambda_4=1$ and $\lambda_1=\lambda_5=\lambda_6=0$. The $S_3$ subgroup of dualities generated $h_1, h_2,$ and $h_3$ acts on the Hamiltonains in this region as the symmetries of the triangle. The colored lines correspond to self-dual points under the various reflections while the red point is self-dual under the full $S_3$ subgroup.}
    \label{Fig:ZTwoSqPhaseDiagramSelfDualLines}
\end{figure}

\begin{figure}[t!]
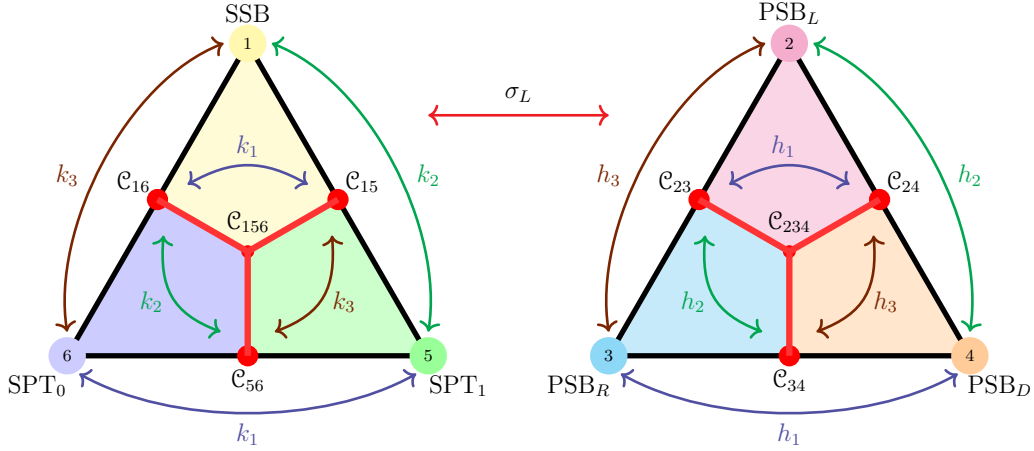
\centering 
    \ZTwoSqPhaseDiagram
    \caption{Phase diagram on two different planes in the six-dimensional parameter space ($T_1$ on the right and $T_2$ on the left). The edges of the triangles are gapped fixed points, the colored regions are gapped phases, while the red points and lines are $c=1$ critical transitions.}
    \label{Fig:ZTwoSqPhaseDiagram}
\end{figure}

\begin{enumerate}
    \item The center of the triangle is self-dual under $S_3\times S_3$, thus it cannot be in any of the six gapped phases and must be a phase transition. 
    
    \item Note that 
    \begin{equation}\label{eq:H T1 at 1 1 0}
      H_{T_1}(1,1,0)=-\sum_i \left[\sigma_{i,L}^x+  \sigma_{i,L}^z\sigma_{i+1,L}^z+\sigma_{i,R}^x+  \sigma_{i,R}^z\sigma_{i+1,R}^z\right],
    \end{equation}
    is the Hamiltonian of the two decoupled copies of the critical Ising chain. Therefore, it is a second-order phase transition described by a conformal field theory with central charge $c=1$ and nine primary fields corresponding to pair-wise combinations of Ising primaries $\mbb{I}$, $\epsilon$, and $\sigma$ with conformal weights $(0,0)$, $(\frac{1}{2},\frac{1}{2})$, and $(\frac{1}{16},\frac{1}{16})$, respectively. This corresponds to the point $\mscr{C}_{23}$ in Figure \ref{Fig:ZTwoSqPhaseDiagramSelfDualLines}. By duality, this point is mapped to the points $\mscr{C}_{24}$ and $\mscr{C}_{34}$. From the properties discussed in Section \ref{topological holography: application}, these two other points must therefore also be critical with central charge $c=1$.
    
    \item The corners of the triangle are fixed-point Hamiltonians and thus are guaranteed to be gapped. The line that goes from the $\mathcal L_2$ corner to $\mscr C_{34}$ is self-dual under the reflection of the triangle that swaps $\mathcal L_3$ and $\mathcal L_4$. Due to this self-duality, only the gapped phase $\mathcal L_2$ or a phase-transition is allowed on this line. 
    Repeating this argument for the other reflection-self-dual lines in Figure \ref{Fig:ZTwoSqPhaseDiagramSelfDualLines}, it is reasonable to expect that the dashed parts of the lines are gapped while the solid parts are phase-transitions, either first-order or second-order. 
    
    \item At the Ising$^2$ transition $\mscr{C}_{23}$, there exist a spinless marginal operator $\epsilon_L\epsilon_R$ with conformal weight $(1,1)$. Deforming the theory with this marginal operator 
    \begin{equation}
        S(\lambda)=S_{\tenofo{Ising}^2}+\lambda \bigintssss \sd^2z\, \epsilon_L(z)\epsilon_R(z),
    \end{equation}
    it remains conformal and all scaling dimensions change continuously as long as the $\epsilon_L\epsilon_R$ perturbation remains marginal. On the spin-chain level, this corresponds to deforming the theory \eqref{eq:H T1 at 1 1 0} from $\mscr{C}_{23}$ towards the center of the triangle $\mscr{C}_{234}$
    \begin{equation}\label{eq:Essentially Ashkin Teller}
        H_{T_1}(1,1,\lambda)=H_{T_1}(1,1,0)-\lambda \sum_{i}\left[\sigma_{i,L}^z\sigma_{i,R}^z\sigma_{i+1,L}^z\sigma_{i+1,R}^z +  \sigma_{i,L}^x\sigma_{i,R}^x \right].
    \end{equation}
    Assuming that this line remains marginal until the center, the solid parts of the lines in Figure \ref{Fig:ZTwoSqPhaseDiagramSelfDualLines} must be critical with $c=1$.
 \end{enumerate}
Note that the same arguments can be made for the triangle $T_2$, though the triangle now lives on the plane $\lambda_2=\lambda_3=\lambda_4=0$ within the six-dimensional phase-diagram
\begin{equation}\label{eq:T2 Hamiltonian}
    H_{T_2}(\lambda_1,\lambda_5,\lambda_6)=\lambda_1H_1+\lambda_5H_5+\lambda_6H_6.
\end{equation}
A possible phase diagram based on these arguments is shown in Figure \ref{Fig:ZTwoSqPhaseDiagram}, where the red lines correspond to a critical line described by $c=1$ conformal field theories and the colored regions are the various gapped phases. The $\mscr C_{56}$ transition corresponds to a critical point that goes beyond the Landau paradigm, since such a transition ought to be ``Landau-forbidden" \cite{TsuiWangLee201511,TsuiHuangJiangLee201701}. However, here the ``Landau-allowed" and ``Landau-forbidden" transitions appear on an equal footing. In particular, there is a duality-mapping between Ising$^2$ transitions $\mscr{C}_{16}$ or $\mscr{C}_{23}$ to the topological criticality $\mscr{C}_{56}$, which we will exploit to compute the exact conformal spectrum of this transition shortly. 

\begin{figure}[t!]
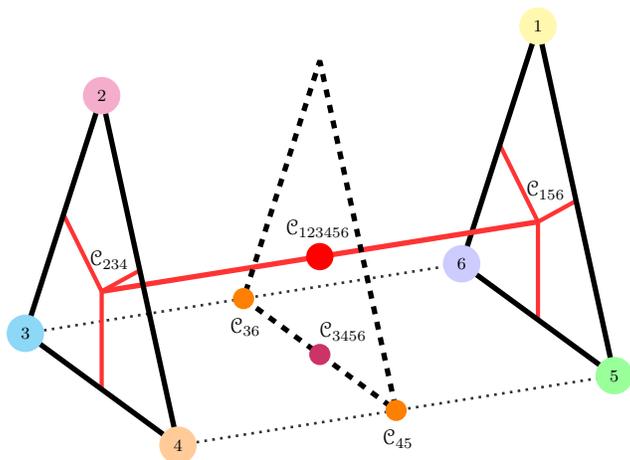
\centering 
    \LineBetweenTriangles{70}{65}
    \caption{Sketch of critical lines in the six-dimensional parameter space for \eqref{eq:the minimal Z2xZ2 Hamiltonian}.  All red lines correspond to $c=1$ conformal field theories. The lines on the left triangle are self-dual under the subgroup of dualities $\mathbb Z_2\times S_3$, generated by $h_1, h_2, h_3$ 
    and one single $k_i$ 
    (depending on line).
    The red lines on the right triangle are self-dual under  $S_3\times\mathbb Z_2$, with $k$ and $h$ swapped.
    The line between the triangles correspond to \eqref{eq:Line between the two Z2xZ2 triangle centers} and is self-dual under $S_3\times S_3$ while the point $\mscr C_{123456}$ is self-dual under the full duality group $\mathcal G[\mathbb Z_2\times\mathbb Z_2] = (S_3\times S_3)\rtimes\mathbb Z_2$. This fully self-dual point is a multicritical point connecting all six gapped phases. The orange points are self-dual under $\mathbb Z_2$ subgroups that map between the triangles and correspond to $c=\frac 12$ transitions. There is such a transition in the six-dimensional parameter space between any two gapped phases belonging to different triangles. These $c=\frac 12$ transitions lie pairwise on triangles (dashed lines) around the fully self-dual point. Finally, the point $\mathscr C_{3456}$ is self-dual under the $D_4$ symmetry of the bottom square. This is a multi-critical point between four gapped phases. There are nine such critical points around $\mathscr C_{123456}$, corresponding to the different squares. Although it is not illustrated in the figure, the line connecting the bottom critical points of the two triangles through $\mscr{C}_{3456}$ is also critical. Crossing the center triangles along different regions also lead to transitions.}
    \label{fig:LineBetweenTriangles}
\end{figure}

\smallskip So far, we have constructed a potential phase diagram on two different planes in six-dimensional parameter space. 
We can in principle go beyond this and explore other regions in the phase diagram.
Dualities prove useful in this regard as well.
Let us look at the known critical point $\mscr{C}_{23}$.
The only $\mbb{Z}_2\times\mbb{Z}_2$ symmetric relevant operators at this critical point are $\epsilon_L$ and $\epsilon_R$ both with conformal weight $(\frac{1}{2},\frac{1}{2})$, which can be thought of as energy-density operators of the form
\begin{equation}
    \begin{aligned}
    \epsilon_L&\sim \sigma^z_{i,L} \sigma^z_{i+1,L}- \sigma^x_{i,L},
    \\
    \epsilon_R&\sim \sigma^z_{i,R} \sigma^z_{i+1,R}- \sigma^x_{i,R}.
    \end{aligned}
\end{equation}
The relative sign comes from the fact that $\epsilon\to -\epsilon$ under Kramers-Wannier duality. Perturbing the $\mscr{C}_{23}$ critical point \eqref{eq:H T1 at 1 1 0} with these two relevant operators 
\begin{equation}
    H=H_{T_1}(1,1,0)+\delta_L \sum_i\left[\sigma^z_{i,L} \sigma^z_{i+1,L}- \sigma^x_{i,L}\right]+\delta_R\sum_i\left[\sigma^z_{i,R} \sigma^z_{i+1,R}- \sigma^x_{i,R}\right],
\end{equation}
we see that this flows to four different gapped phases; (1) $\tenofo{PSB}_L$ $(\delta_L<0$, $\delta_R>0)$, (2) $\tenofo{PSB}_R$ $(\delta_L>0$, $\delta_R<0)$, (3) $\tenofo{SSB}$ $(\delta_L>0$, $\delta_R>0)$, (4) $\tenofo{SPT}_0$ $(\delta_L<0$, $\delta_R<0)$.
Thus, $\mscr{C}_{23}$ is a multi-critical point between four phases and only the first two transitions lie on the plane.
This analysis is done purely using decoupled Ising chains, but all other critical points related to $\mscr{C}_{23}$ by duality, will also be multicritical points with two relevant operators.
Therefore, the gapped phases of the two triangle meet in the region between the triangles.
The structure of the region interpolating between triangles can be further constrained using dualities.
For example, consider the following plane in \eqref{eq:the minimal Z2xZ2 Hamiltonian}
\begin{equation}\label{eq:Line between the two Z2xZ2 triangle centers}
    H(\lambda,\wt{\lambda},\wt{\lambda},\wt{\lambda},\lambda,\lambda)=\lambda(H_1+H_5+H_6)+\wt{\lambda}(H_2+H_3+H_4).
\end{equation}
Since the overall scaling does not matter, without loss of generality, we will restrict ourselves to the line $\lambda+\wt{\lambda}=1$. This line is self-dual under $S_3\times S_3$ and the center point $\lambda=\wt{\lambda}=\frac{1}{2}$ is self-dual under the full duality group $(S_3\times S_3)\rtimes\mbb{Z}_2$. Since no gapped phase is self-dual under these subgroups, the whole line must correspond to phase transitions. In particular, note that this line interpolates between the critical points $\mscr{C}_{156}$ ($\lambda=1$, $\wt\lambda=0$) and $\mscr{C}_{234}$ ($\lambda=0$, $\wt\lambda=1$). Therefore, the centers of the two triangles (see Figure \ref{Fig:ZTwoSqPhaseDiagram}) are connected by a critical line and the center of the line $\mscr{C}_{123456}$ is a multi-critical point connecting all gapped phases (see Figure \ref{fig:LineBetweenTriangles}).

\begin{figure}[t!]\centering
    \begin{subfigure}{.44\linewidth}\centering
    \includegraphics[width=1\textwidth]{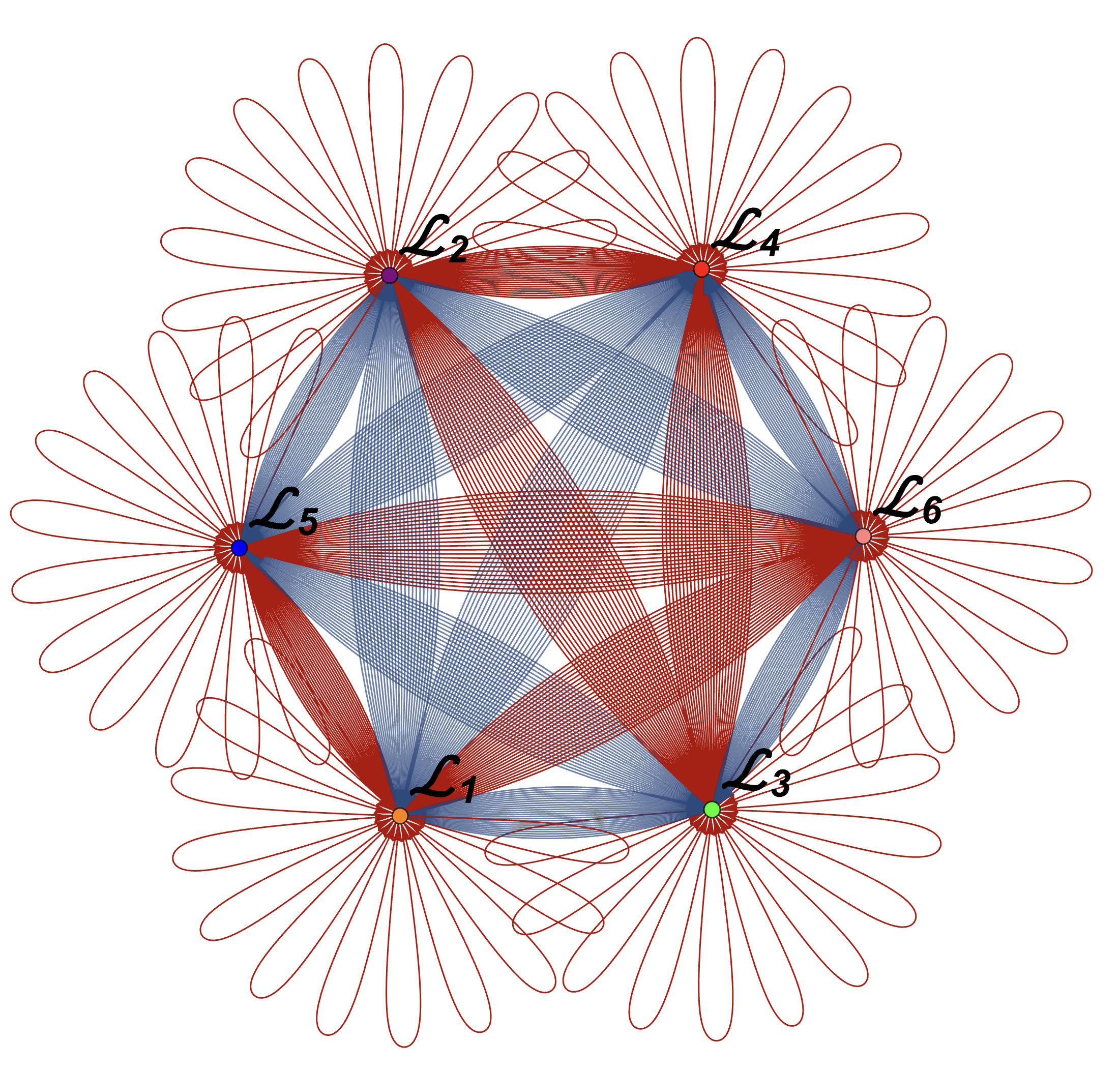}
    \caption{}
    \label{fig:web of dualities for Z2*Z2}
    \end{subfigure}\hspace{1cm}
    ~
    \begin{subfigure}{.4\linewidth}\centering
    \includegraphics[width=1\textwidth]{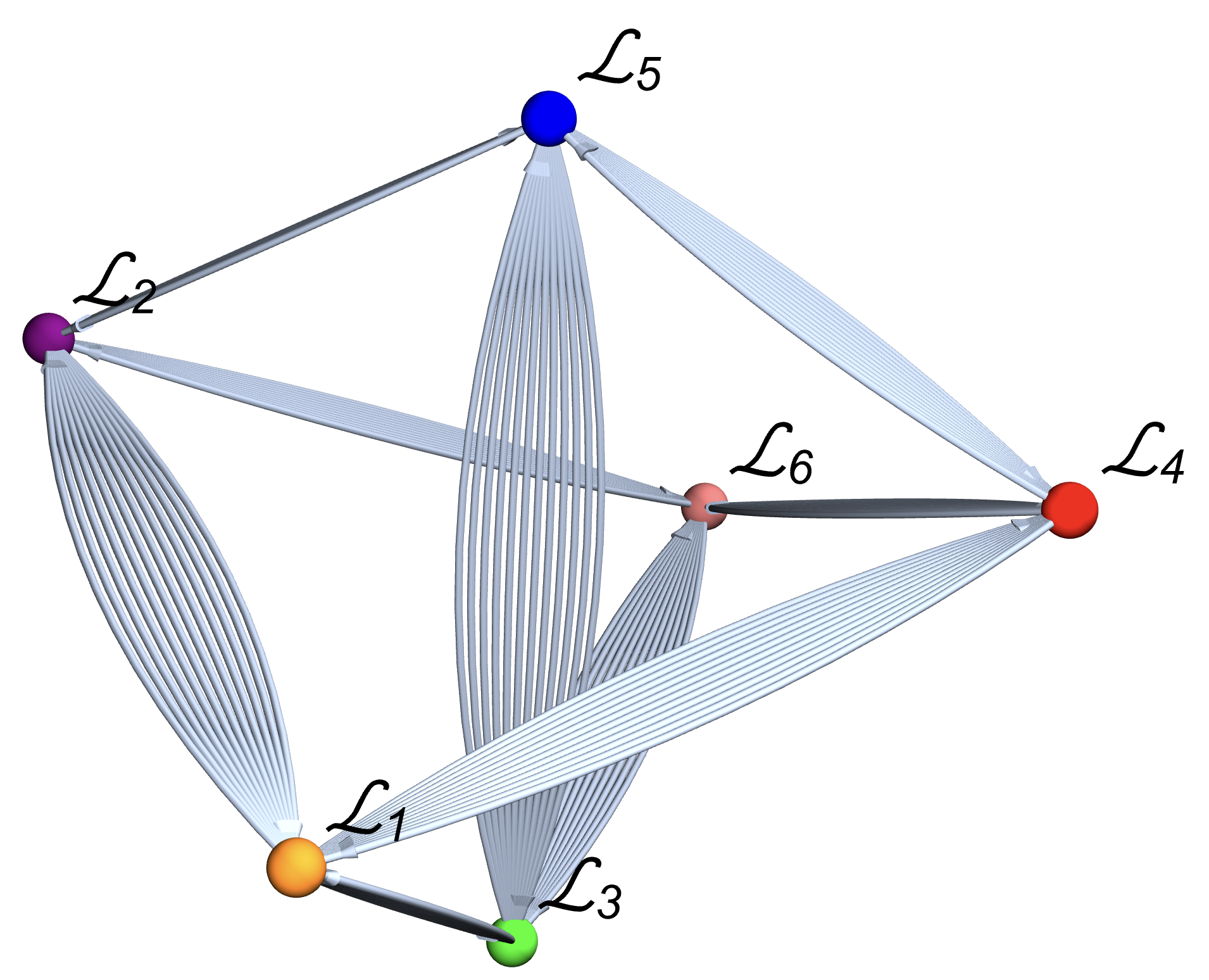}
    \vspace{.8cm}\caption{}
    \label{fig:web of dualities for Z2*Z2 order 4 only}
    \end{subfigure}
    \caption{(a) The web of dualities of $\ms G=\mbb{Z}_2\times\mbb{Z}_2$. Nodes correspond to gapped phases while lines denote the action of dualities on gapped phases. The highlighted red lines correspond to order-$3$ dualities, revealing the asymmetry amongst the six phases. (b) The visualization of order-$4$ dualities.}
\end{figure}

\smallskip One might wonder whether there exist other $S_3$ subgroups of dualities permuting three different gapped phases and thus creating a similar structure in the phase diagram as in Figure \ref{Fig:ZTwoSqPhaseDiagram}.
This possibility can be excluded by examining the web of dualities shown in Figure \ref{fig:web of dualities for Z2*Z2} where the action of dualities on the six gapped phases are illustrated by lines.
The red lines correspond to order-$3$ group elements, from which, it is clear that any other $S_3$ subgroup connect the same sets of gapped phases.

\smallskip In Figure \ref{fig:web of dualities for Z2*Z2 order 4 only}, we have illustrated the action of order-$4$ elements. These order-$4$ elements map between corners of one triangle and the corners of the other triangle, and thus establish connections between the triangles. They are part of $D_4$ subgroups of dualities which are symmetries of squares. One example is as follows
\begin{equation}
    \SquareDuality{75}{55}
\end{equation}
where $\mathcal L_3, \mathcal L_4, \mathcal L_5,$ and $\mathcal L_6$ form a square. There exists a $D_4$ subgroup corresponding to the symmetries of this square ($D_4$ contains two order $4$ elements, pure rotation and reflection-rotation). The two other gapped boundaries just map onto each other under these $D_4$ dualites, thus no gapped phase is self-dual under these dualities. There are 18 order $4$ dualities, corresponding to the 9 different possible squares we can form between the corners of the triangles.

\smallskip We can use self-duality to find critical points between the two triangles. For example, consider these Hamiltonians that are self-dual under a $\mathbb Z_2\subset D_4$ that swap $i\leftrightarrow j$ corners between the two triangles
\begin{equation}\label{eq:H_{ij}}
    H_{ij}=H_i+H_j, \qquad i=2,3,4,\;\; j=1,5,6.
\end{equation}
In particular, consider $H_{21}$ which is nothing but a critical Ising chain with $c=\frac{1}{2}$ coupled to a trivially-gapped chain. Between any pair of gapped phases of the two triangles, there is a $c=\frac{1}{2}$ transition, self-dual under various $\mathbb Z_2$ dualites. One can think of the interpolation of the two triangles in Figure \ref{Fig:ZTwoSqPhaseDiagram} in different ways. Consider $H_{36}$ and $H_{45}$: there exists a plane with a triangle with the $c=\frac 12$ transitions $H_{36}$ and $H_{45}$ as two of the corners and $\mathscr{C}_{123456}$ its center. The last corner is gapped (the transition exists, but on a rotated plane).
This plane can for example be parametrized using Hamiltonian \eqref{eq:the minimal Z2xZ2 Hamiltonian} as 
\begin{equation}\label{eq:Lambda parametrization of plane with center triangle}
    \begin{pmatrix}
        \lambda_1\\ \lambda_2\\ \lambda_3\\ \lambda_4\\ \lambda_5\\ \lambda_6
    \end{pmatrix}
    =
    \begin{pmatrix}
        a+b-1 \\a+b-1 \\ a \\a \\ b \\ b
    \end{pmatrix},
\end{equation}
where $(a,b)=(1,0)$ and $(a,b)=(0,1)$ are the two $c=\frac 12$ points and $(a,b)=(1,1)$ is $\mathscr{C}_{123456}$.
The $c=1$ point $\mathscr{C}_{123456}$ is therefore surrounded by $c=\frac 12$ points, corresponding to various different interpolations between the two triangles in the six-dimensional space. See figure \ref{fig:LineBetweenTriangles}.
Furthermore, there are also points that are self-dual under the full $D_4$ subgroup corresponding to the square formed by $i_1, i_2, j_1, j_2$
\begin{gather}\label{eq:H_{i1i2j1j2}}
    \begin{aligned}
    H_{i_1i_2,j_1j_2}=H_{i_1}+H_{i_2}+H_{j_1}+H_{j_2},
    \end{aligned}\qquad
    \begin{aligned}
     i_1,i_2&=2,3,4,\\
    j_1, j_2&=1,5,6.
    \end{aligned}
\end{gather}
This critical point $\mathscr C_{3456}$ also lies on the plane \eqref{eq:Lambda parametrization of plane with center triangle} with $(a,b) = (\frac 12, \frac 12)$ (see Figure \ref{fig:LineBetweenTriangles}) and its central charge is $c=1$. 

\smallskip As discussed in section \ref{sec:Emergence of Symmetry}, self-dual Hamiltonians will have new emergent symmetries. For example, this implies that the gapped and gapless Hamiltonians along the self-dual lines in Figure \ref{Fig:ZTwoSqPhaseDiagramSelfDualLines} should have emergent symmetries. In particular, let us consider the line from $\mscr{C}_{23}$ to $\mscr{C}_{234}$ (the Hamiltonian in \eqref{eq:Essentially Ashkin Teller}), which is self-dual under $h_3$. This Hamiltonian is clearly symmetric under $\mcal{U}_{L}=\prod_i\sigma^x_{i,L}$, $\mcal{U}_{R}=\prod_i\sigma^x_{i,R}$, and $h_3$, which swaps the two chains. Together, these form the enhanced symmetry group $D_4$. Furthermore, the Hamiltonian is self-dual under $\sigma_L\sigma_R$, which is nothing but a simultaneous Kramers-Wannier duality on both chains. These further enhance the symmetries with non-invertible symmetries, as discussed in section \ref{sec:Emergence of Symmetry}. In fact, the specific Hamiltonian \eqref{eq:Essentially Ashkin Teller} is nothing but the Ashkin-Teller model where it is known to have $D_4$ global symmetry. For $\lambda\in [0,1]$, the model is in the universality class of the $c=1$ $\mbb Z_2$-orbifold compact boson CFT with orbifold radius
\begin{equation}
    R_{O}^{2}=\frac{\pi}{2}\left[\arccos{(-\lambda)}\right]^{-1}.
\end{equation}
For $\lambda=0$, $R_O=1$, which corresponds to the Ising$^2$ point \cite{Ginsparg199108}, while $\lambda=1$ corresponds to $R_O=1/\sqrt{2}$, known as the KT point. 

\smallskip Topological holography implies that there are many more emergent symmetries in these models. We will explore these emergent non-invertible symmetries elsewhere.

\subsubsection*{Numerical verification of phase diagram}

\begin{figure}[t!]\centering
    \begin{subfigure}{.45\linewidth}\centering
    \includegraphics[width=1.0\textwidth]{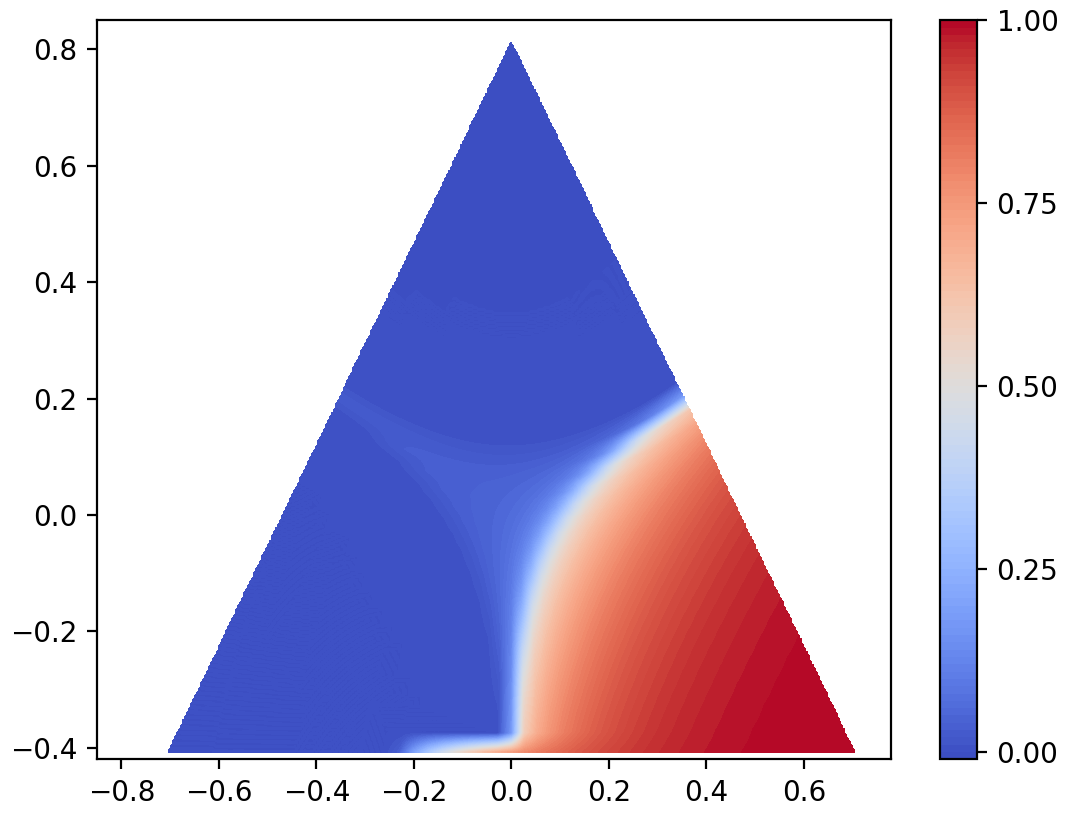}
    \subcaption{$\langle \sigma^z_{i,L}\sigma^z_{i,R}\;\sigma^z_{j,L}\sigma^z_{j,R}\rangle$}
    \end{subfigure}\hspace{1cm}
    ~
    \begin{subfigure}{.45\linewidth}\centering
    \includegraphics[width=1.0\textwidth]{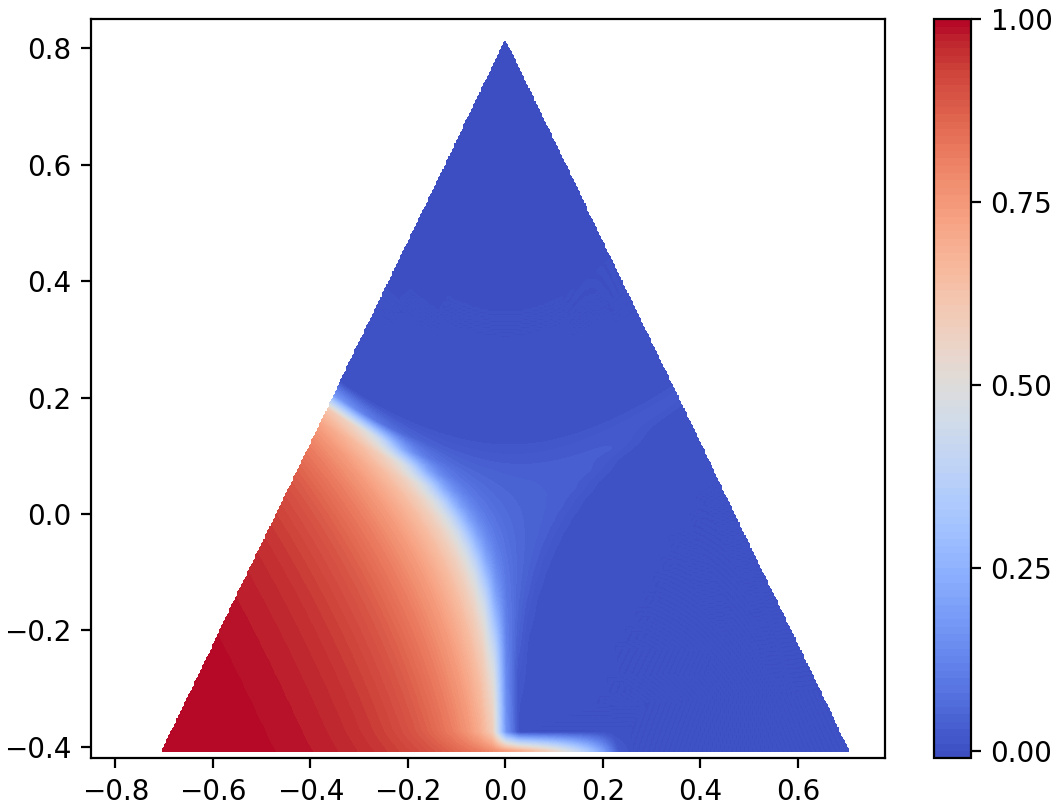}
        \subcaption{$\langle \sigma^z_{i,L}\sigma^z_{j,L}\rangle$}
    \end{subfigure}
    ~
    \begin{subfigure}{.45\linewidth}\centering
    \includegraphics[width=1\textwidth]{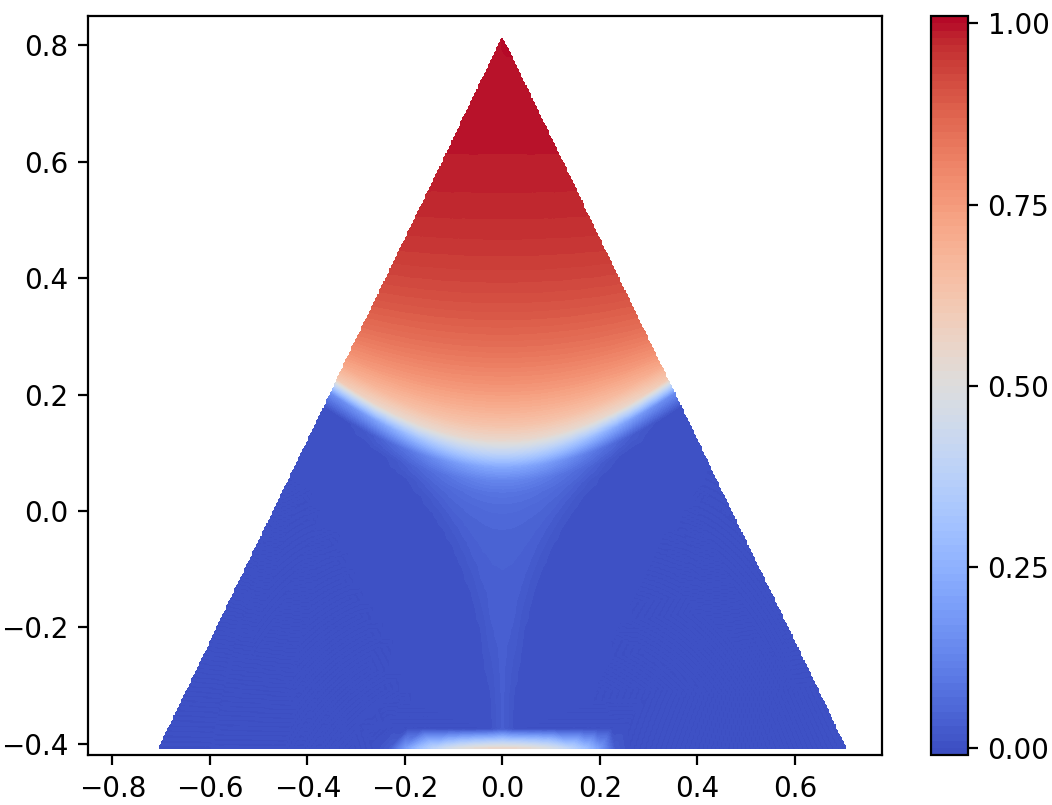}
    \subcaption{$\langle \sigma^z_{i,R}\sigma^z_{j,R}\rangle$}
    \end{subfigure}
    \caption{Numerical calculation of correlation functions of \eqref{eq:T1 Hamiltonian} on the plane $\lambda_2+\lambda_3+\lambda_4=1$ ($0\le\lambda_i\le 1$) using DMRG algorithm. The code was implemented in \texttt{Julia} using \texttt{iTensor} library \cite{itensor}. The simulation was done on a spin chain with $N=2\times 100$ spin-$\frac{1}{2}$ sites with open boundary conditions. In order to avoid cat states in spontaneously-broken phase due to finite-size effects, the symmetry was weakly-broken at the boundary using a small pinning field with strength $\eta<0.05$. In order to capture long-range correlations while avoiding boundary effects, we chose $|i-j|=\frac{1}{2}L$.}
    \label{fig:numerics of the computation of order parameters for Z2xZ2 minimal Hamiltonian}
\end{figure}

So far, we have used dualities to construct a potential phase diagram for the minimal $\mbb{Z}_2\times\mbb{Z}_2$ symmetric lattice Hamiltonian. Since topological holography provides us with various tools such as order parameters for each gapped phase, we can numerically verify the proposed phase diagram. We have performed DMRG calculations on the Hamiltonian \eqref{eq:the minimal Z2xZ2 Hamiltonian} and computed various order parameters on the planes of the triangles \ref{Fig:ZTwoSqPhaseDiagram}. We numerically computed the three order parameters that can distinguish the three gapped phases of the triangle $T_1$ in Figure \ref{Fig:ZTwoSqPhaseDiagram} and the results are shown in Figure \ref{fig:numerics of the computation of order parameters for Z2xZ2 minimal Hamiltonian}. Each gapped phase has two order parameters but we only shown the simplest ones as they are enough to distinguish phases. The points on the triangles denote parameters of the Hamiltonian \eqref{eq:T1 Hamiltonian} on the plane $\lambda_2+\lambda_3+\lambda_4=1$, and the color denotes the value of correlation function. We clearly see that the order parameters associated to each gapped phase is equal to $1$ (or non-zero) in the predicted gapped regions of Figure \ref{Fig:ZTwoSqPhaseDiagram}. From the point of view of topological order, the non-zero expectation values of the order parameters correspond to the condensation of various anyons in the Lagrangian subgroups for the given gapped phase.
\begin{figure}[t!]\centering
    \begin{subfigure}{.43\linewidth}\centering
    \includegraphics[width=1\textwidth]{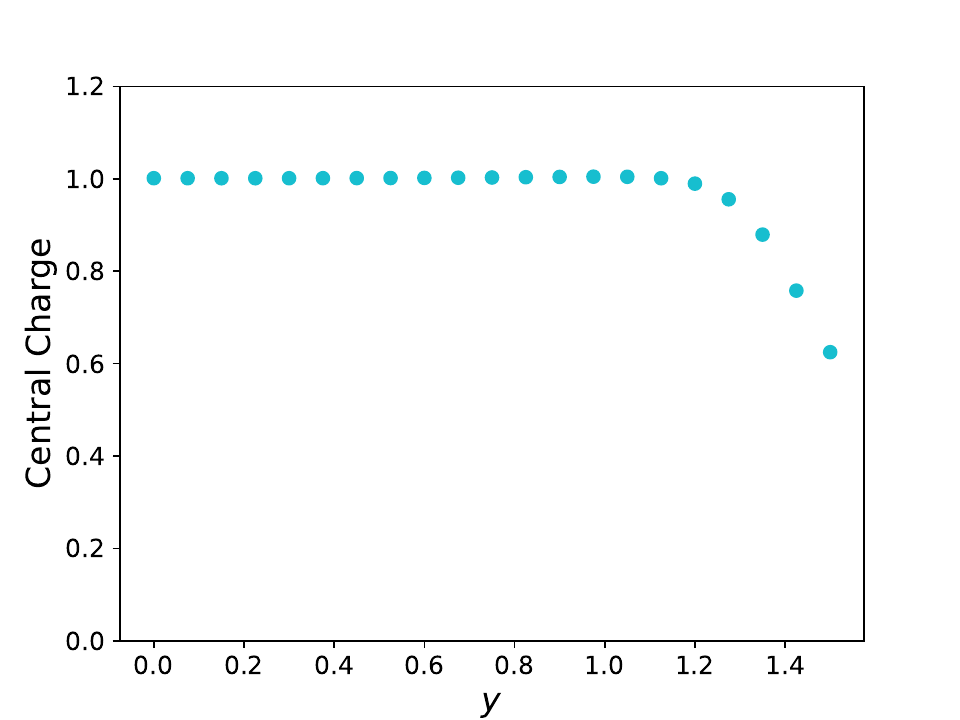}
    \subcaption{$N=2\times 80$ sites}
    \label{fig:Central_charge_line_inside_triangles}
    \end{subfigure}\hspace{1cm}
    ~
    \begin{subfigure}{.43\linewidth}\centering
    \includegraphics[width=1\textwidth]{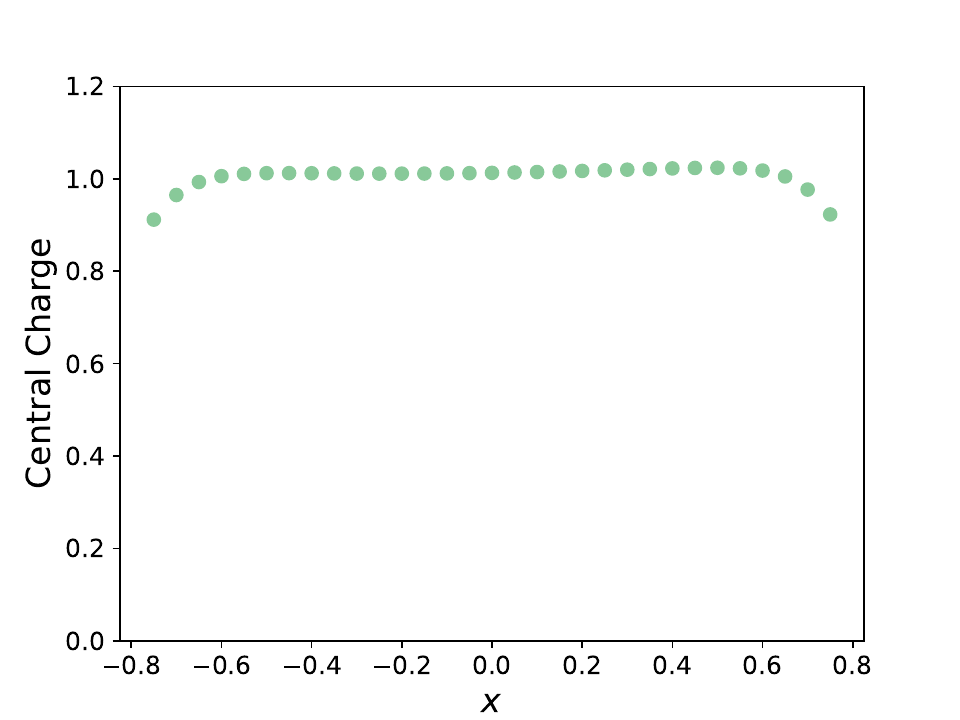}
        \subcaption{$N=2\times 40$ sites}
        \label{fig:Central_charge_line_between_triangles}
    \end{subfigure}
    \caption{Numerical computation of the central charge along two different lines: (a) the line inside a triangle, from $\mscr C_{23}$ ($y=0$) to the center of the triangle $\mscr C_{234}$ ($y=1$) and then a bit further out, and (2) the line between the centers of the two triangles $\mscr C_{234}$ ($x=-0.5$) and $\mscr C_{156}$ ($x=0.5$). The point $x=0$ is the fully self-dual point $\mscr C_{123456}$. In the gapped regions near criticality we have $c=0$, but see a decay on these plots due to finite-size effects and slow convergence of the DMRG algorithm.}
    \label{fig:Numerical calculation of c=1 CFT lines}
\end{figure}

\begin{figure}
    \centering
    \includegraphics[width=0.7\textwidth]{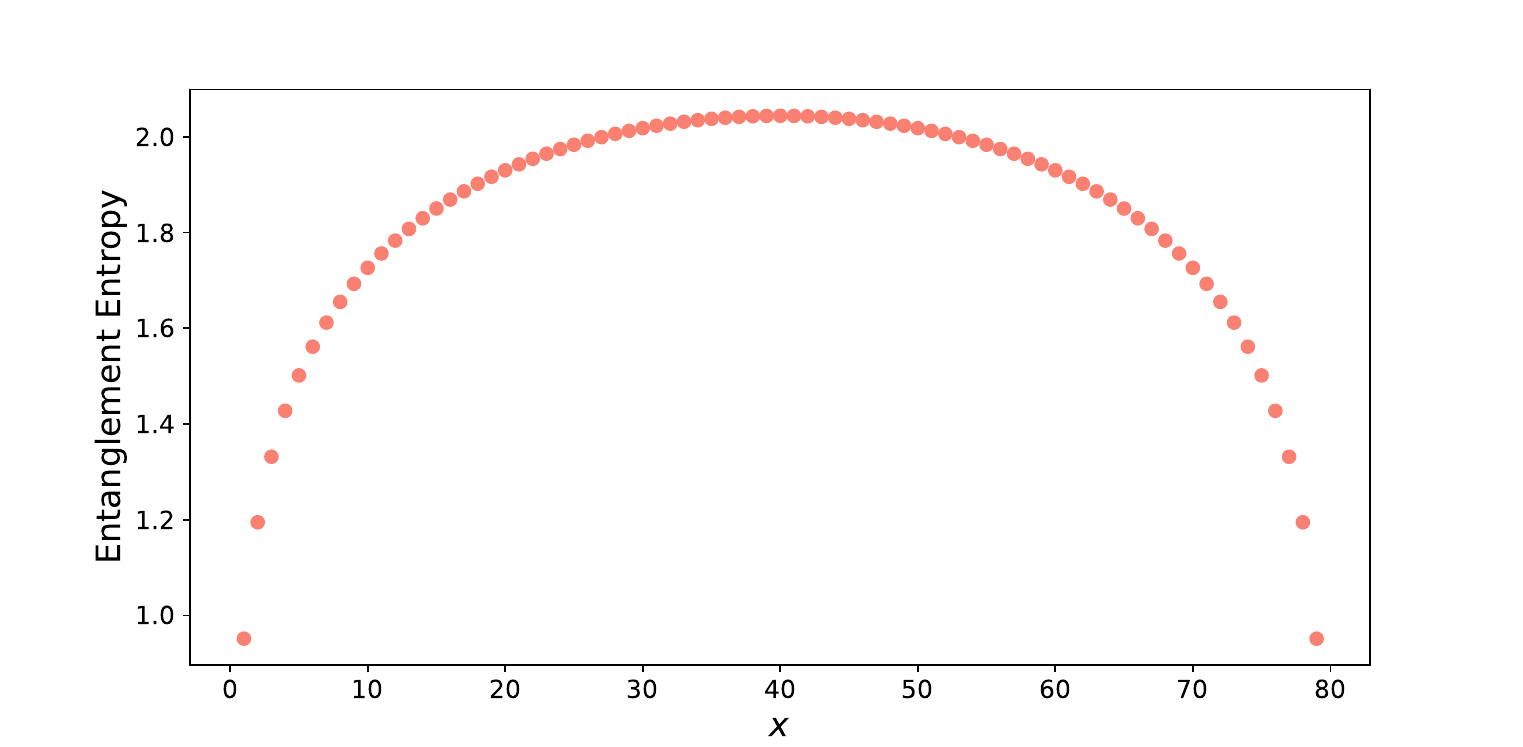}
    \caption{Numerical computation of entanglement entropy for the fully $\mathcal G[\mathbb Z_2\times\mathbb Z_2]=(S_3\times S_3)\rtimes\mathbb Z_2$ self-dual point (equation \eqref{eq:Line between the two Z2xZ2 triangle centers} for $\lambda=\tilde{\lambda}$), using DMRG. The spin-chain has $N=2\times 80$ sites with periodic boundary conditions. When fitted to the function \eqref{eq:Cardy formula PBC}, we find the central charge $c\approx 1.0040$. 
    }
    \label{fig:Z2Z2_fully_self_dual_entanglement_entropy}
\end{figure}
\smallskip Along the predicted gapless lines, there is faint silhouette in the numerical calculation of Figure \ref{fig:numerics of the computation of order parameters for Z2xZ2 minimal Hamiltonian}. These are again consistent with the predictions as the DMRG algorithm converges slower near critical points. We further verify these critical lines by computing their entanglement entropy and extracting their central charge using the Cardy Formula \cite{CalabreseCardy200405}
\begin{equation}\label{eq:Cardy formula PBC}
    S_L(x)=\frac{c}{3}\left[\frac{L}{\pi}\sin\left(\frac{\pi x}{L}\right)\right]+C'.
\end{equation}
Here, $L$ is the length of the chain, $x$ is the entanglement cut, $c$ is the central charge, and $C'$ is a non-universal constant. This formula holds for periodic boundary conditions. DMRG converges slower for periodic boundary conditions but for the computation of central charge, it is known to converge faster.  We have computed the entanglement entropy along two lines: (1) from $\mscr{C}_{23}$ to the center of the triangle $\mscr{C}_{234}$ and then a bit further out, and (2) the line between the centers of the two triangles $\mscr{C}_{234}$ and $\mscr{C}_{156}$, as described below \eqref{eq:Line between the two Z2xZ2 triangle centers} (see figure \ref{fig:LineBetweenTriangles}). In Figure \ref{fig:Numerical calculation of c=1 CFT lines}, the numerically-computed value of the central charge is shown along these two lines. We see that $c$ is equal to $1$ to high level of accuracy, exactly in the regions predicted. As an example of how these values are extracted from entanglement entropy, the entropy of the fully $(S_3\times S_3)\rtimes\mathbb Z_2$ self-dual point $\mscr{C}_{123456}$ ($\lambda =\widetilde\lambda = \frac 12$ in \eqref{eq:Line between the two Z2xZ2 triangle centers}) is shown in figure \ref{fig:Z2Z2_fully_self_dual_entanglement_entropy}.

\subsubsection*{Conformal spectrum of critical points}

In the previous sections, using topological holography, we constructed a potential phase diagram and verified its gapped and critical regions numerically. In this section, we will further use the vast web of dualities to compute the full conformal spectra of various transitions. 

\smallskip Let us remind the reader of the logic of Section \ref{sec:topological conformal spectroscopy from dualities} in a simpler setting. The generalized twisted partition functions of the Hamiltonian \eqref{eq:the minimal Z2xZ2 Hamiltonian} are given by  
\begin{equation}
    \begin{aligned}
        \mcal{Z}_{\ms g,\ms h}(\lambda)&=\tenofo{Tr}_{\mcal{H}}\left(\mcal{U}_{\ms h}e^{-\beta H_{\ms g}(\lambda)}\right)
        \\
        &=\sum_{\alpha_L,\alpha_R= 0}^1 e^{i\pi\left[\alpha_L\ms h_L+\alpha_R\ms h_R\right]}\chi_{d=(\ms g,\alpha)}(\lambda),
    \end{aligned}
\end{equation}
where $\ms g=(\ms g_L,\ms g_R)$ and $\ms h=(\ms h_L,\ms h_R)$ are $\mbb{Z}_2\times\mbb{Z}_2$ elements corresponding to inserting a symmetry line-operator along time and space directions, respectively. $H_{\ms g}(\lambda)$ is the Hamiltonian with twisted boundary condition $\ms g$ and $\mcal{U}_{\ms h}$ is the symmetry operator. There is a sector for each bulk anyon $d=(\ms g,\alpha)\in\mcal{A}$ corresponding to twisted boundary condition $\ms g$ and symmetry sector $\alpha$ and the spectrum in this sector is given by the character $\chi_d(\lambda)$. Under a duality $\sigma\in\mcal{G}[\mbb{Z}_2\times\mbb{Z}_2]=(S_3\times S_3)\rtimes\mbb{Z}_2$, the characters transform as
\begin{align}
    \chi_{d}({\lambda})=\text{Tr}_{\mc H}\left[P_{\alpha}e^{-\beta H_{\msg}({\lambda})}\right]=\text{Tr}_{\mc{H}}\left[P_{\alpha^{\vee}}e^{-\beta H_{\msg^{\vee}}(\sigma \cdot  {\lambda})}\right]=\chi^{\vee}_{\sigma\cdot d}(\sigma \cdot{\lambda}),
    \label{eq:trans_of_characters}
\end{align}
where $\sigma\cdot\lambda$ corresponds to the permuted parameters under duality. Therefore, knowing the spectrum in all twisted sectors at the point $\lambda$, we can compute the spectrum at the point $\sigma\cdot\lambda$ using dualities. Using this, we can also compute the partition function for the dual theory using \eqref{eq:the dual partition function in terms of original parition function} and 
\eqref{eq:the generic formula for twisted dual partition function}.

\smallskip Let us start with the critical point $\mscr{C}_{16}$ between the ferromagnetic (fully-broken) and paramagnetic (unbroken) gapped phases which is described by two decoupled critical Ising chains. The generalized twisted partition functions of a single Ising CFT is given by \cite{DiFrancescoMathieuSenechal1997,PetkovaZuber2001}
\begin{equation}\label{eq:Ising CFT twisted partition funcitions (in Z2xZ2 example)}
    \begin{aligned}
      \mcal{Z}^{\tenofo{Ising}}_{0,0}&=|\chi_0|^2+|\chi_{\frac 12}|^2+|\chi_{\frac 1{16}}|^2,&\qquad \mcal{Z}^{\tenofo{Ising}}_{1,0}&=\chi_0\overline\chi_{\frac 12} + \chi_{\frac 12}\overline{\chi}_0+|\chi_{\frac1{16}}|^2,
      \\
      \mcal{Z}^{\tenofo{Ising}}_{0,1}&=|\chi_0|^2+|\chi_{\frac{1}{2}}|^2-|\chi_{\frac{1}{16}}|^2,&\qquad \mcal{Z}^{\tenofo{Ising}}_{1,1}&=|\chi_{\frac1{16}}|^2 - \chi_0\overline\chi_{\frac 12} - \chi_{\frac 12}\overline\chi_0,
    \end{aligned}
\end{equation}
and the $\mathbb Z_2$ characters in terms of the Virasoro characters are given by
\begin{equation}
\begin{aligned}\label{eq:Ising CFT Z2 characters (in Z2xZ2 example)}
 \chi^{\tenofo{Ising}}_{1}&=|\chi_{0}|^2+|\chi_{\frac{1}{2}}|^2, &\qquad \chi^{\tenofo{Ising}}_{e}&=|\chi_{\frac{1}{16}}|^2,
 \\
 \chi^{\tenofo{Ising}}_{m}&=|\chi_{\frac{1}{16}}|^2,&\qquad \chi^{\tenofo{Ising}}_{em}&=\chi_0\overline\chi_{\frac 12} + \chi_{\frac 12}\overline\chi_{0}.
\end{aligned}
\end{equation}
The generalized twisted partition functions at $\mscr C_{16}$ are thus given by
\begin{equation}
    \mcal{Z}_{\ms g,\ms h}(\mscr{C}_{16})=\mcal{Z}^{\tenofo{Ising}}_{\ms g_L,\ms h_L}\times \mcal{Z}^{\tenofo{Ising}}_{\ms g_R,\ms h_R}.
\end{equation}
Similarly, the $\mbb{Z}_2\times\mbb{Z}_2$ characters can be written in terms of Virasoro characters as 
\begin{equation}
    \chi_{d=(\ms g,\alpha)}(\mscr{C}_{16})=\chi^\tenofo{Ising}_{\ms g_L,\alpha_L}\times\chi^\tenofo{Ising}_{\ms g_R,\alpha_R}. 
\end{equation}
The partition function and spectrum of any critical point related to $\mscr{C}_{16}$ by duality can be readily computed using the above data. 

\smallskip To illustrate our approach, we use this transition and the duality maps $k_2$ and $k_1$ to obtain the universality classes of the $\mscr{C}_{56}$ and $\mscr{C}_{15}$ transitions, respectively (see Figure \ref{Fig:ZTwoSqPhaseDiagram}). 
Since the duality operation $k_2$ leaves $\mc L_{6}$  invariant and maps $\mc L_{1}$ and $\mc L_5$ into each other, it maps between $\mscr{C}_{16}$ and $\mscr{C}_{56}$.
Similarly, $k_1$ leaves $\mc L_{1}$  invariant and maps $\mc L_{6}$ and $\mc L_5$ into each other, therefore it maps $\mscr{C}_{16}$ and $\mscr{C}_{15}$ into each other.
Denoting the $\mbb Z_2\times \mbb Z_2$ characters at the transition $\mscr{C}_{ij}$ as $\chi_d(\mscr{C}_{ij})$ with $d\in \mc A[\mbb Z_2\times \mbb Z_2]$ we have the relations
\begin{equation}
\chi_{d}(\mscr{C}_{56})=\chi_{k_2(d)}(\mscr{C}_{16}), \qquad \chi_{d}(\mscr{C}_{15})=\chi_{k_1(d)}(\mscr{C}_{16}) 
\end{equation}
From \eqref{eq:Ising CFT Z2 characters (in Z2xZ2 example)}, the characters of these two transitions can be be immediately read off.
They take the following explicit form for the $\mscr{C}_{56}$ transition

\begin{gather}\label{eq:Z2xZ2 C_56 characters}
\begin{aligned}
\chi_1(\mscr{C}_{56})&= \chi_{1}(\mscr{C}_{16})=(|\chi_{0}|^2+|\chi_{\frac{1}{2}}|^2)^2, \\ 
\chi_{e_L}(\mscr{C}_{56})&= \chi_{e_Lm_R}(\mscr{C}_{16})=(|\chi_{\frac{1}{16}}|^2)^2, \\
\chi_{e_R}(\mscr{C}_{56})&= \chi_{e_Rm_L}(\mscr{C}_{16})=(|\chi_{\frac{1}{16}}|^2)^2, \\
\chi_{m_L}(\mscr{C}_{56})&= \chi_{m_L}(\mscr{C}_{16})=|\chi_{\frac{1}{16}}|^2(|\chi_{0}|^2+|\chi_{\frac{1}{2}}|^2), \\
\chi_{m_R}(\mscr{C}_{56})&= \chi_{m_R}(\mscr{C}_{16})=|\chi_{\frac{1}{16}}|^2(|\chi_{0}|^2+|\chi_{\frac{1}{2}}|^2), 
\end{aligned}
\end{gather}
etc., and similarly for the $\mscr{C}_{15}$ transition
\begin{gather}
\begin{aligned}
\chi_1(\mscr{C}_{15})&= \chi_{1}(\mscr{C}_{16})=(|\chi_{0}|^2+|\chi_{\frac{1}{2}}|^2)^2, \\ 
\chi_{e_L}(\mscr{C}_{15})&= \chi_{e_L}(\mscr{C}_{16})=|\chi_{\frac{1}{16}}|^2(|\chi_{0}|^2+|\chi_{\frac{1}{2}}|^2), \\
\chi_{e_R}(\mscr{C}_{15})&= \chi_{e_R}(\mscr{C}_{16})=|\chi_{\frac{1}{16}}|^2(|\chi_{0}|^2+|\chi_{\frac{1}{2}}|^2),\\
\chi_{m_L}(\mscr{C}_{15})&= \chi_{m_Le_R}(\mscr{C}_{16})=(|\chi_{\frac{1}{16}}|^2)^2, \\
\chi_{m_R}(\mscr{C}_{15})&= \chi_{m_Re_L}(\mscr{C}_{16})=(|\chi_{\frac{1}{16}}|^2)^2.
\end{aligned}
\end{gather}
Note that the identity sector $\chi_{1}$ is invariant under all dualities since dualities descend from global symmetries of the $\ms G$ topological gauge theory under which the identity (transparent) line operator always remains invariant. We can also explicitly compute the exact partition functions of these critical points using \eqref{eq:the dual partition function in terms of original parition function}. For example, the partition function of the critical point $\mscr{C}_{56}$ for, say, periodic boundary condition, is given by
\begin{equation}\label{eq:the k2-orbifolded partition function}
    \mcal{Z}_{(0,0),(0,0)}(\mscr{C}_{56})=\frac{1}{4}\sum_{\substack{\ms h_L,\ms h_R \\ \alpha'_L,\alpha'_R}}(-1)^{-\alpha'_L\ms h_L-\alpha'_R\ms h_R}\mcal{Z}^{\tenofo{Ising}}_{\alpha'_R,\alpha'_L}\mcal{Z}^{\tenofo{Ising}}_{\ms h_L,\ms h_R}.
\end{equation}
Substituting Ising partition functions in terms of Virasoro characters, this expression gives 
\begin{equation}\label{eq:the k2-orbifolded partition function in terms of Virasoso characters}
    \begin{aligned}
        \mcal{Z}_{(0,0),(0,0)}(\mscr{C}_{56})&=|\chi_0|^4+|\chi_{\frac{1}{2}}|^4+2|\chi_0|^2|\chi_{\frac{1}{2}}|^2
        \\
        &+2|\chi_{\frac{1}{16}}|^4+\chi_0^2\bar{\chi}_{\frac{1}{2}}^2+2|\chi_0|^2|\chi_{\frac{1}{2}}|^2+\bar{\chi_0}^2\chi^2_{\frac{1}{2}}.
    \end{aligned}
\end{equation}
Interestingly, $\mathscr C_{56}$ is a ``Landau-forbidden" transition, corresponding to a topological criticality between trivial and non-trivial SPT phases. The full conformal spectrum, including information about which operator drives this transition was gained from the simple Ising CFT together with a duality.

\smallskip The conformal spectrum of the topological transition $\mscr{C}_{56}$ was recently computed in
\cite{TsuiHuangJiangLee201701} using a lattice map specific to a spin-chain model. They found the following expression
\begin{equation}\label{eq:the partition function of the critical phase between the two Z2*Z2 SPT states}
    \mcal{Z}'_{(0,0),(0,0)}(\mscr{C}_{56})=\frac{1}{2}\sum_{\ms g_L,\ms g_R}\mcal{Z}^{\tenofo{Ising}}_{\ms g_L,\ms g_R}\mcal{Z}^{\tenofo{Ising}}_{\ms g_L,\ms g_R}.
\end{equation}
Despite looking very different from \eqref{eq:the k2-orbifolded partition function}, they are both exactly equal to \eqref{eq:the k2-orbifolded partition function in terms of Virasoso characters}. The reason two very different expression give rise to the same partition function is due to the fact that there are several dualities mapping $\mscr{C}_{16}$ to $\mscr{C}_{56}$. Note that the method used in \cite{TsuiHuangJiangLee201701} is specific to the model while our derivation is not model specific and is therefore applicable in a much wider context.
Similarly, the conformal spectrum of the other transitions $\mscr{C}_{ij}$ can be computed using our method.

\begin{figure}
    \centering
    \includegraphics[width=1.0\textwidth]{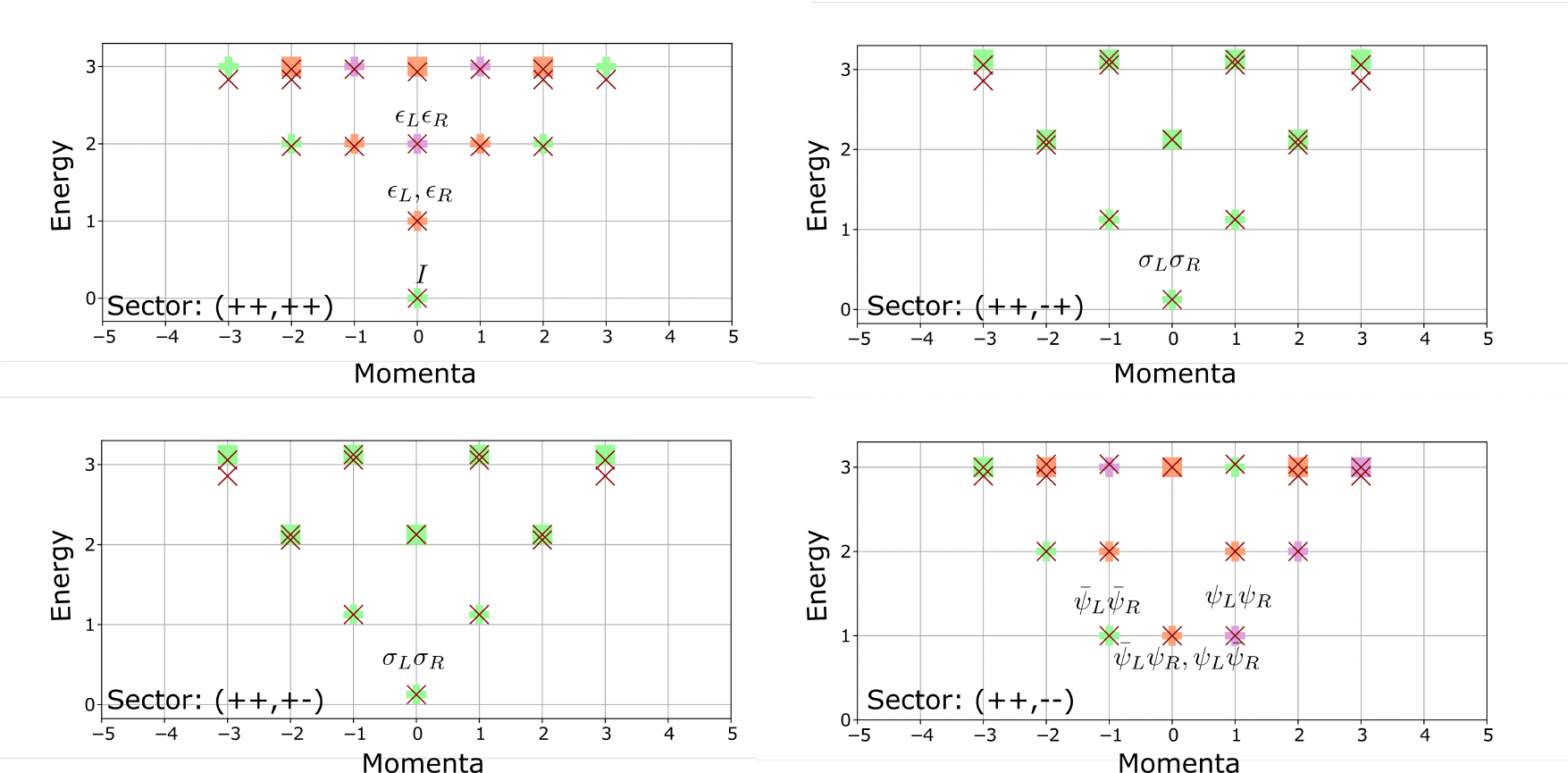}
    \caption{The figure shows a comparison between the analytically-computed conformal spectrum of the transition between the $\mbb Z_2\times \mbb Z_2$ symmetric SPT and PM and the numerically-obtained spectrum for the Hamiltonian $H_{56}$ in \eqref{eq:H_{ij}} using exact diagonalization. 
    All the numerics have been performed for spin chains of length $L=12$, i.e., spin-chains containing $24$ $\mbb Z_2$-spins using the \texttt{Python} package \texttt{QuSpin} \cite{WeinbergBukov201610}.
    The subplots (a)-(d) correspond to  $\mbb{Z}_2 \times \mbb Z_{2}$ eigensectors $(++),(-+),(+-)$ and $(--)$ and with periodic boundary conditions.
    Single and double degenerate states in the conformal towers are depicted using plus signs and squares respectively. For all other boundary conditions, the numerical and analytical results match as well.}
    \label{fig:ED_conformal_spectrum_SPT_transition}
\end{figure}

\smallskip In order to numerically verify the analytical results above, we use exact diagonalization of the Hamiltonian \eqref{eq:the minimal Z2xZ2 Hamiltonian} at various critical points $\lambda=\lambda_c$ and assign a momentum to each eigenstate.  As an example, the spectrum of the topological criticality $\mathscr{C}_{56}$, decomposed into different symmetry eigensectors, is shown in Figure \ref{fig:ED_conformal_spectrum_SPT_transition}.
The crosses are the numerically-computed spectrum while the other points are the analytical results given in \eqref{eq:Z2xZ2 C_56 characters}. As one can see, the two are in good agreement, even for relatively small system sizes. See Appendix.~\ref{sec:conformal spectra at criticality} for a comparison between the numerically obtained and theoretically predicted conformal spectrum at the $\mscr{C}_{15}$, $\mscr{C}_{34}$ and $\mscr{C}_{24}$ transitions.

\smallskip As discussed above, the spectrum along the Ashkin-Teller line is known to belong to the $c=1$ $\mbb Z_2$-orbifold compact boson universality class at various compactification radii.  The conformal spectrum of this CFT is given a $\ms U(1)$ current operator and vertex operators \cite{Ginsparg199108}. From this, we can again use dualities to find the conformal spectrum along any point of the critical lines in Figure \ref{Fig:ZTwoSqPhaseDiagram}.

\subsubsection*{Beyond minimal $\mathbb Z_2\times\mathbb Z_2$ spin chain models}
\smallskip So far we have studied the minimal model, i.e., those containing operators subtended from the shortest possible strings, in quite some detail.
Our construction, however, works for any $\mathbb Z_2\times\mathbb Z_2$ symmetric spin chain, since any such Hamiltonian is an element of the string operator algebra $\mathbb{SOA}[\mathbb Z_2\times\mathbb Z_2]$.
To illustrate this, it is instructive to briefly study the simplest non-minimal $\mbb Z_2\times \mbb Z_2$ symmetric Hamiltonians.

\smallskip For simplicity, let us start with the following minimal Hamiltonian
\begin{equation}\label{eq:H_{16} with string operators}
    H_{16}=-\sum_{\mu=L,R}\sum_{i}\left[S_{e_\mu}(i,i+1)+S_{m_{\mu}}(i)\right],
\end{equation}
which from \eqref{eq:Z2xZ2 fixed-point hamiltonians} and \eqref{eq:H_{ij}} we see are two decoupled critical Ising chains. There are two simple extensions we could consider: (1) by using longer string operators or (2) the products of two minimal operators. Using these operators, let us consider two possible deformations of \eqref{eq:H_{16} with string operators}
\begin{gather}
    \begin{aligned}
    \Delta H_{\text{ANNNI},\;\mu}&=- \sum_{i}\left[S_{e_{\mu}}(i,i+2)+S_{m_\mu}(i,i+1)\right] \\
    &= -\sum_{i}\left[\sigma_{i,\mu}^{z}\sigma_{i+2,\mu}^{z}+\sigma_{i,\mu}^{x}\sigma_{i+1,\mu}^{x}\right], \\
    \Delta H_{\text{F-O'B},\;\mu}&=-
    \sum_{i}\left[S_{e_\mu}(i,i+1)S_{m_\mu}(i+2)+S_{m_\mu}(i)S_{e_\mu}(i,i+1)\right] \\
    &=-\sum_{i}\left[\sigma_{i,\mu}^{z}\sigma_{i+1,\mu}^{z}\sigma_{i+2,\mu}^{x}+\sigma_{i,\mu}^{x}\sigma_{i+1,\mu}^{z}\sigma_{i+2,\mu}^{z}\right],
    \end{aligned}
\end{gather}
where $\mu=L, R$. With these deformations, the model is still self-dual under electric-magnetic dualities and the two chains remain decoupled. These are the ANNNI and the Fendley-O'Brien models, respectively \cite{SELKE1988213, Affleck_2015, Vidal_2018, Fendeley_OBrien_2018}.
On a single chain, these models realize various interesting points such as the $c=7/10$ tri-critical Ising (TCI) universality class, a $c=3/2$ critical phase that realizes decoupled Ising and Luttinger Liquid phases as well as two and four-fold degenerate ground states\cite{Affleck_2015}.
Using these terms, $\mscr{C}_{16}$ has a four-dimensional parameter space of deformations parametrized as
\begin{equation}\label{eq:H_{12} deformed}
    H_{16}^{\text{def}}\left(\{t_\mu, s_\mu\}\right) = H_{16} + \sum_{\mu=L,R} \left(t_\mu\;\Delta H_{\text{ANNNI},\mu} + s_\mu\;\Delta H_{\text{F-O'B},\mu}\right).
\end{equation}
This parameter space therefore contains interesting critical points, Lifshitz transitions, etc between the paramagnetic (SPT$_0$) and ferromagnetic (SSB) phases, realized as decoupled chains. Using dualities, we can find novel transitions between any other gapped phases. As an example, we can use the duality $k_2$ (see Figure \ref{Fig:ZTwoSqPhaseDiagram}), to obtain a four-dimensional space of deformations of the topological critical transition between SPT$_0$ and SPT$_1$ $\mscr{C}_{56}$.
The dual Hamiltonian has the form
\begin{equation}\label{eq:H_{56} deformed}
    H_{56}^{\text{def}}\left(\{t_\mu, s_\mu\}\right) =H_{56} + \sum_{\mu=L,R}\left[t_{\mu}\; \Delta H^{\vee}_{\text{ANNNI},\mu} + s_{\mu}\; \Delta H^{\vee}_{\text{F-O'B},\mu}\right],
\end{equation}
where
\begin{gather}
\begin{aligned}
\Delta H^{\vee}_{\text{ANNNI},L}&=- \sum_{i}\left[S_{e_Lm_R}(i,i+2)+S_{m_L}(i,i+1)\right] \\
&= -\sum_{i}\left[
\sigma^{z}_{i,L}\sigma^{x}_{i+1,R}\sigma^{z}_{i+2,L}\sigma^{x}_{i+2,R}
+
\sigma^{x}_{i,L}\sigma^{x}_{i+1,L}
\right] \\
\Delta H^{\vee}_{\text{ANNNI},R}&=- \sum_{i}\left[S_{m_Le_R}(i,i+2)+S_{m_R}(i,i+1)\right] \\
&= - \sum_{i}\left[
\sigma^{x}_{i,R}\sigma^{x}_{i+1,R}
+
\sigma^{x}_{i,L}\sigma^{z}_{i,R}\sigma^{x}_{i+1,L}\sigma^{z}_{i+2,R}\right] \\
\Delta H^{\vee}_{\text{F-O'B},L}&=- \sum_{i}\left[S_{e_L m_R}(i,i+1)S_{m_L}(i+2)+S_{m_L}(i)S_{e_Lm_R}(i+1,i+2)\right] \\
&= - \sum_{i}\left[
\sigma^{z}_{i,L}\sigma^{z}_{i+1,L}\sigma^{x}_{1+1,R}\sigma^{x}_{i+2,L}
+
\sigma^{x}_{i,L}\sigma^{z}_{i+1,L}\sigma^{z}_{i+2,L}\sigma^{x}_{i+2,R}
\right] \\
\Delta H^{\vee}_{\text{F-O'B},R}&=- \sum_{i}\left[S_{m_L e_R}(i,i+1)S_{m_R}(i+2)+S_{m_R}(i)S_{m_L e_R}(i+1,i+2)\right] \\
&= - \sum_{i}\left[
\sigma^{x}_{i,L}\sigma^{z}_{i,R}\sigma^{z}_{1+1,R}\sigma^{x}_{i+2,R}
+
\sigma^{x}_{i,R}\sigma^{x}_{i+1,L}\sigma^{z}_{i+1,R}\sigma^{z}_{i+2,R}
\right]. 
\end{aligned}
\end{gather}
For any point $(t_L,t_R, s_L,s_R)$ corresponding to an interesting transition in \eqref{eq:H_{12} deformed}, we can find a topological transition in \eqref{eq:H_{56} deformed}. Among the interesting points, consider the doubled tri-critical Ising CFT at four different points
\begin{equation}
    (t_L, t_R,s_L,s_R)\in \left\{(t^*,t^*,0,0),(t^*,0,0,s^*),(0,t^*,s^*,0), (0,0,s^*,s^*)\right\},
    \label{eq:TCI_parameter_set}
\end{equation}
where $t^*\approx 246$ and $s^*=0.428$ \cite{Affleck_2015, Vidal_2018}.
For the above parameters,  the model \eqref{eq:H_{56} deformed} realizes (non-minimal) topological transitions between the trivial and non-trivial SPT phases with central charge $c=7/5$. The conformal spectrum (or partition function) in all the different $\mathbb Z_2\times \mbb{Z}_2$ twisted sectors can readily be obtained using this duality. The conformal spectrum of this topological transition in a few example sectors are given by
\begin{gather}
\begin{aligned}
\chi_1^\vee&=\left(\chi_1^{\tenofo{TCI}}\right)^2=\left(|\chi_{0}|^{2}+ |\chi_{1/10}|^{2}+|\chi_{\frac{3}{5}}|^{2}+|\chi_{\frac{3}{2}}|^{2}\right)^2, \\
\chi_{e_L}^\vee&=\chi_{e}^{\tenofo{TCI}}\chi_{m}^{\tenofo{TCI}}=\left(|\chi_{\frac{3}{80}}|^{2}+|\chi_{\frac{7}{16}}|^{2}\right)^2,
\\
\chi_{e_R}^\vee&=\chi_{e}^{\tenofo{TCI}}\chi_{m}^{\tenofo{TCI}}=\left(|\chi_{\frac{3}{80}}|^{2}+|\chi_{\frac{7}{16}}|^{2}\right)^2, 
\\
\chi_{m_L}^\vee&=\chi_1^{\tenofo{TCI}}\chi_{m}^{\tenofo{TCI}}=\left(|\chi_{\frac{3}{80}}|^{2}+|\chi_{\frac{7}{16}}|^{2}\right)\left(|\chi_{0}|^{2}+ |\chi_{1/10}|^{2}+|\chi_{\frac{3}{5}}|^{2}+|\chi_{\frac{3}{2}}|^{2}\right),
\\
\chi_{m_R}^\vee&=\chi_1^{\tenofo{TCI}}\chi_{m}^{\tenofo{TCI}}=\left(|\chi_{\frac{3}{80}}|^{2}+|\chi_{\frac{7}{16}}|^{2}\right)\left(|\chi_{0}|^{2}+ |\chi_{1/10}|^{2}+|\chi_{\frac{3}{5}}|^{2}+|\chi_{\frac{3}{2}}|^{2}\right),
\\
\chi_{e_Le
_R}^\vee&=\left(\chi_{f}^{\tenofo{TCI}}\right)^2=\left(\chi_{\frac{6}{10}}\bar{\chi}_{\frac{1}{10}}
+
\chi_{\frac{1}{10}}\bar{\chi}_{\frac{6}{10}}
+
\chi_{\frac{3}{2}}\bar{\chi}_{0}
+
\chi_{0}\bar{\chi}_{\frac{3}{2}}\right)^2.
\end{aligned}
\end{gather}
Similarly, we can also compute the exact partition functions of \eqref{eq:H_{56} deformed} at the non-minimal topological critical point. For example, for periodic boundary conditions we have
\begin{equation}
    \begin{aligned}
        \mcal{Z}^\vee_{(0,0),(0,0)}&=\chi^\vee_1+\chi^\vee_{e_L}+\chi^\vee_{e_R}+\chi^\vee_{e_Le_R},
        \\
        &=\left(|\chi_{0}|^{2}+ |\chi_{1/10}|^{2}+|\chi_{\frac{3}{5}}|^{2}+|\chi_{\frac{3}{2}}|^{2}\right)^2+2\left(|\chi_{\frac{3}{80}}|^{2}+|\chi_{\frac{7}{16}}|^{2}\right)^2
        \\
        &+\left(\chi_{\frac{6}{10}}\bar{\chi}_{\frac{1}{10}}+\chi_{\frac{1}{10}}\bar{\chi}_{\frac{6}{10}}+\chi_{\frac{3}{2}}\bar{\chi}_{0}+\chi_{0}\bar{\chi}_{\frac{3}{2}}\right)^2.
    \end{aligned}
\end{equation}
The conformal spectrum can be readily read. The theory has several relevant perturbations, some of these are responsible for driving the topological transition. Using topological holography, we can also find many other interesting transitions. More detailed study will be presented elsewhere.

\subsubsection*{Gauging perspective}
\smallskip Having explored various aspects of the phase diagram of $\mbb Z_2\times \mbb Z_2$ symmetric quantum systems in some detail, we now shift directions towards interpreting the action of dualities in the partition functions in terms of (topological) gauging.
First, we focus on the duality transformations $h_{i}$ which generate an $S_3$ subgroup \eqref{eq:The two S3 subgroups of Z2xZ2 duality group} of the full duality group $\mc G[\mbb Z_{2}\times \mathbb Z_{2}]$.
Notably, the dualities $h_i$ map non-trivially between phases that are labelled by distinct $\mbb Z_2$ subgroup (i.e. $\mbb{Z}_2^L, \mbb{Z}_2^R,$ and $\mbb{Z}_2^D$) of $\mbb{Z}_2\times \mbb{Z}_2$  (see Table \ref{tab:lagrangian subgroups of Z2*Z2}).
These dualities originate from automorphisms  of $\mbb{Z}_2\times \mbb{Z}_2$, which form the group $S_3$.
Twisted partition functions transform as follows under $h_i$ (see \eqref{eq:the dual partition function in terms of original parition function})
\begin{gather}
\begin{aligned}
h_1:&\ \mc Z[A_L,A_R]\mapsto \mc Z[A_L,A_L+A_R], 
\\
h_2:&\ \mc Z[A_L,A_R]\mapsto \mc Z[A_L+A_R,A_R], 
\\
h_3:&\ \mc Z[A_L,A_R]\mapsto \mc Z[A_R,A_L]. 
\end{aligned}
\end{gather}
where $A_{L}$ are $A_{R}$ are background $\mathbb Z_{2}$ gauge fields.
It can be verified that these transforms form the group $S_3$ by choosing the two generators as $h_3h_1$ and $h_2$.
$h_3h_1$ is an order-$3$ element, which acts as
\begin{gather}
\begin{aligned}
(h_3h_1)^{\phantom{2}}:&\ \mc Z[A_L,A_R]\mapsto \mc Z[A_L+A_R,A_L], \\
(h_3h_1)^2:&\ \mc Z[A_L,A_R]\mapsto \mc Z[A_R,A_L+A_R], \\
(h_3h_1)^3:&\ \mc Z[A_L,A_R]\mapsto \mc Z[A_L,A_R].
\end{aligned}
\end{gather}
Similarly, $h_2$ is an order-$2$ element and the group relation $h_2(h_3h_1)h_2^{-1}=(h_3h_1)^2$ can also be explicitly verified.
In order to construct the transformation of the twisted partition functions under the full group $\mc G[\mbb Z_2\times \mbb Z_2]$, we need the action of an additional element $\sigma_L$, which acts simply as electromagnetic or Kramers-Wannier duality on the layer labelled $L$ 
\begin{equation}
 \sigma_L: \mc Z[A_L,A_R] \mapsto \mc Z^{\vee} [A_{L}^{\vee},A_{R}]=
\frac{1}{\sqrt{|H^{1}(\Sigma,\mbb Z_2)|}}\sum_{A_L}\mc Z[A_L,A_R]e^{i\pi \int_{\Sigma}A_{L}\cup A_{L}^{\vee}}.   
\end{equation}
The remaining operations corresponding to $k_i\in \mc G[\mbb{Z}_2\times \mbb{Z}_2]$ can be obtained by using the group relation $k_i=\sigma_L h_i\sigma_L^{-1}$.
They take the explicit form 
\begin{gather}
\begin{aligned}
k_1:& \mc Z[A_L,A_R] \mapsto Z^{\vee} [A_{L}^{\vee},A_{R}^{\vee}] = \mc Z[A_L,A_R]e^{i\pi \int_{\Sigma}A_{L}\cup A_{R}}, \\ 
k_2:& \mc Z[A_L,A_R] \mapsto \mc Z^{\vee} [A_{L}^{\vee},A_{R}^{\vee}]=\frac{1}{\sqrt{|H^{1}(\Sigma,\mbb Z_2^2)|}}\sum_{A_{L},A_{R}}\mc Z[A_L,A_R]e^{i\pi \int_{\Sigma}(A_{L}^{\vee}-A_L)\cup(A_{R}^{\vee} - A_R)}, \\ 
k_3:& \mc Z[A_L,A_R] \mapsto \mc Z^{\vee} [A_{L}^{\vee},A_{R}^{\vee}]=
\frac{1}{\sqrt{|H^{1}(\Sigma,\mbb Z^2_2)|}}\sum_{A_L,A_R}\mc Z[A_L,A_R]e^{i\pi \int_{\Sigma}A_{L}\cup A_{R}^{\vee}+A_{R}\cup A_{L}^{\vee}}.
\label{eq:Z2Z2_duality_group_cup_product_exps}
\end{aligned}
\end{gather}
The duality $k_1$ corresponds to modifying the partition function of the theory by a topological phase.
This topological phase is precisely the response theory corresponding to a $\mbb Z_{2}\times \mbb Z_{2}$ SPT.
Therefore, $k_1$ can be physically interpreted as pasting a non-trivial SPT on the theory.
The remaining dualities are gauging or orbifold like dualities.
It can be shown that the transformations in \eqref{eq:Z2Z2_duality_group_cup_product_exps} satisfy the group relations of $\mc G[\mbb Z_2 \times \mbb Z_{2} ]=(S_3\times S_3)\rtimes \mbb Z_2$ (See Appendix \ref{sec:duality groupoid}).

\subsection{$\mbb{Z}_3\times\mbb{Z}_3$ symmetric quantum spin chains}
To examine a more complicated situation, we could consider $\mbb{Z}_2\times\mbb{Z}_3$. However, this group is isomorphic to $\mbb{Z}_6$, which has been explored in Section \ref{Subsec:ZN example}. 
Therefore, next we consider the example of $\ms G=\mbb{Z}_3\times\mbb{Z}_3$.
As in the previous example, this symmetry group has a single fusion structure and many dualities.
In fact, the order of the duality group is $1152$, therefore this case belongs to the limit where topological holography is particularly powerful.
In this section, we restrict ourselves to study very few details and a deeper analysis will appear elsewhere. 

\smallskip As usual, we consider a $2+1d$ $\mbb Z_3\times \mbb Z_3$ topological gauge theory whose line operators are labelled by elements in the group 
\begin{equation}
    \mcal{A}[\mbb{Z}_3\times\mbb{Z}_3]=\mbb{Z}^L_3\times\mbb{Z}^R_3\times\tenofo{Rep}(\mbb{Z}^L_3\times\mbb{Z}^R_3)\simeq \mbb Z_3^4 \simeq \langle m_L,m_R,e_L,e_R\rangle, 
\end{equation}
where the generators satisfy the following relations $e_L^3=m_L^3=e_R^3=m_R^3=1$. 

\begin{table}[t!]\centering
    \begin{tabular}{c  c c c c c} \toprule
    \multirow{2}{2cm}{Lagrangian subgroups} & \multirow{2}{2cm}{Generating set} & \multirow{2}{1.8cm}{Image of $\Pi$} & \multirow{2}{1cm}{$\ms H$} & \multirow{2}{1cm}{$\psi(\ms h_1, \ms h_2)$} & \multirow{2}{2cm}{Gapped phase} \\\\
    \midrule
    $\mcal{L}_1$ & $e_L,e_R$  &$1_L,1_R$ &  $\mbb{I}$ & $1$ & \tenofo{SSB}
      
      \\
      $\mcal{L}_2$ & $m_L,e_R$  & $m_L$ & $\mbb{Z}^L_3$  & $1$& $\tenofo{PSB}_L$
      
      \\
      $\mcal{L}_3$ & $e_L,m_R$  & $m_R$ & $\mbb{Z}^R_3$ & $1$& $\tenofo{PSB}_R$
      
      \\
      $\mcal{L}_4$ & $e_Le_R^2,m_Lm_R$& $m_Lm_R$ & $\mbb{Z}^D_3$& $1$ & $\tenofo{PSB}_D$
      
      \\
      $\mcal{L}_5$ & $e_Le_R,m_L^2m_R$  & $m_L^2m_R$ & $\mbb{Z}^A_3$ & $1$& $\tenofo{PSB}_A$
      
      \\
      $\mcal{L}_6$ & $m_L,m_R$ &  $m_L,m_R$ & $\mbb{Z}^L_3\times\mbb{Z}^R_3$ & $1$& $\tenofo{SPT}_0$
      
      \\
      $\mcal{L}_7$ & $m_Le_R, e_L^2m_R$ &  $m_L,m_R$ &$\mbb{Z}_3^L\times\mbb{Z}_3^R$ & $\omega^{h_{1,L}h_{2,R}}_3$ & $\tenofo{SPT}_1$
      
      \\
      $\mcal{L}_8$ & $m_Le_R^2,e_Lm_R$  & $m_L,m_R$ & $\mbb{Z}_3^L\times\mbb{Z}_3^R$& $\omega_3^{2h_{1,L}h_{2,R}}$ & $\tenofo{SPT}_2$
      \\
    \bottomrule
    \end{tabular}
    \caption{The Lagrangian subgroups for $\ms G=\mbb{Z}_3\times\mbb{Z}_3$. $\mbb{Z}_3^L, \mbb{Z}_3^R, \mbb{Z}_3^D,$ and $\mbb{Z}_3^A$ are the left, right, diagonal, and anti-diagonal $\mbb{Z}_3$ subgroups of $\mbb{Z}_3\times\mbb{Z}_3$, respectively. $\mcal{L}_i$ for $i=6,7,8$ are the three gapped SPT states.
    The 2-cocycle is evaluated on the group elements $\ms h_i=(h_{i,L},h_{i,R}) \in \ms H$ for $i=1,2$ and 
    $\omega_3=e^{2\pi i/3}$.}
    \label{tab:lagrangian subgroups of Z3*Z3}
\end{table}

\smallskip Following the recipe detailed in Sec.~\ref{Subsec:boundary_Hilbert_space}, we obtain a $1+1d$ quantum spin chain that provides a concrete realization of a $\mbb Z_3\times \mbb Z_3$ symmetric system.
The Hilbert space of the spin chain decomposes into on-site Hilbert spaces 
\begin{equation}
     \mcal H_i= \mc H_{i,L}\otimes \mc H_{i,R}, \qquad \mc H_{i,L} \simeq 
     \mc H_{i,R} \simeq \mbb C^3.
\end{equation}
There is a natural action of generalized Pauli operators $(X_{i,L},Z_{i,L})$ on the on-site Hilbert space $\mc H_{i,L}$ which satisfy the $\mbb Z_3$ clock and shift algebra in \eqref{eq:the ZN clock algebra} with $ N=3$.
Similarly there is a set of operators $(X_{i,R},Z_{i,R})$ that act on $\mc H_{i,R}$ and satisfy an isomorphic algebra. 
The generators of $\mc A[\mbb{Z}_3 \times \mbb{Z}_3]$ become the following operators when brought to the 1D boundary (see \eqref{eq:Definition of A and B operators} and \eqref{eq:Wilson_operator_to_lattice_operator})
\begin{gather}
\begin{aligned}
\label{eq:Z3Z3 bulk to boundary operator map}
e_L \longrightarrow&\; Z^{\pd}_{i,L}Z^{\dagger}_{i+1,L}, \qquad m_L\longrightarrow X_{i,L}, \\
e_R \longrightarrow&\; Z^{\pd}_{i,R}Z^{\dagger}_{i+1,R}, \qquad m_R\longrightarrow X_{i,R}.
\end{aligned}
\end{gather}
The global symmetry operator for $\ms g=(\ms g_L,\ms g_R)\in \mbb Z_3\times \mbb Z_3$ has the form
\begin{equation}
    \mcal{U}_{\ms g}=\bigotimes_i\left(X_{i,L}\right)^{\ms g_L}\otimes \left(X_{i,R}\right)^{\ms g_R}.
\end{equation}
\subsubsection*{Gapped phases, order parameters, and excitations}
There are $8$ Lagrangian subgroups of $\mcal{A}[\mbb{Z}_3\times\mbb{Z}_3]$, which correspond to $8$ distinct gapped phases (see Table~\ref{tab:lagrangian subgroups of Z3*Z3}).
Using the holographic correspondence in \eqref{eq:Z3Z3 bulk to boundary operator map}, we can construct the fixed-point Hamiltonians associated to each of the gapped phases in Table \ref{tab:lagrangian subgroups of Z3*Z3}.
The Hamiltonian associated to the SSB phase is given by\footnote{We label the Hamiltonians with the corresponding gapped phase, described by a Lagrangian subgroup. Therefore, the Hamiltonian associated to the Lagrangian subgroup $\mcal{L}_i$ in Table \ref{tab:lagrangian subgroups of Z3*Z3} is denoted as $H_i$.}
\begin{equation}\label{eq:Z3xZ3 fixed point SSB}
    H_1=-\sum_{i}\left[Z_{i,L}^{\pd}Z_{i+1,L}^{\dagger}+
    Z_{i,R}^{\pd}Z_{i+1,R}^{\dagger}\right]+\tenofo{h.c.}.
\end{equation}
The fixed-point Hamiltonians associated to the four possible partial symmetry-breaking phases are 
\begin{equation}\label{eq:Z3xZ3 fixed points PSB}
    \begin{aligned}
        H_2&=-\sum_i\left[X_{i,L}+Z_{i,R}^{\pd}Z_{i+1,R}^{\dagger}\right]+\tenofo{h.c.},
        \\
        H_3&=-\sum_i\left[Z_{i,L}^{\pd}Z_{i+1,L}^{\dagger}+X_{i,R}\right]+\tenofo{h.c.},
        \\
        H_4&=-\sum_i\left[Z_{i,L}^{\pd}Z_{i,R}^{\dagger}Z_{i+1,L}^{\dagger}Z^{\pd}_{i+1,R}+X_{i,R}X_{i,R}\right]+\tenofo{h.c.},
        \\
        H_5&=-\sum_i\left[Z_{i,L}^{\pd}Z^{\pd}_{i,R}Z_{i+1,L}^{\dagger}Z^{\dagger}_{i+1,R}+X^{\dagger}_{i,L}X^{\pd}_{i,R}\right]+\tenofo{h.c.}.
    \end{aligned}
\end{equation}
Finally, the fixed-point Hamiltonians of SPT phases are given by
\begin{equation}\label{eq:Z3xZ3 fixed points SPT}
    \begin{aligned}
        H_6&=-\sum_i\left[X_{i,L}+X_{i,R}\right]+\tenofo{h.c.}.
        \\
        H_7&=-\sum_i\left[X_{i,L}^{\pd}Z^{\pd}_{i,R}Z^{\dagger}_{i+1,R}+ Z^{\dagger}_{i,L}Z^{\pd}_{i+1,L}X_{i+1,R}^{\pd}\right]+\tenofo{h.c.},
        \\
        H_8&=-\sum_i\left[
        X^{\pd}_{i,L}Z_{i,R}^{\dagger}Z^{\pd}_{i+1,R}
        +
        Z_{i,L}^{\pd}Z_{i+1,L}^{\dagger}X_{i+1,R}^{\pd}\right]+\tenofo{h.c.},
    \end{aligned}
\end{equation}
These are all commuting fixed-point models, a fact that can be easily checked using the clock algebra \eqref{eq:the ZN clock algebra}.

\smallskip As discussed in  \ref{sec:Gapped boundaries in 2+1d as gapped phases in 1+1d}, the order parameter corresponding to the condensation of an anyon $d$ of a Lagrangian subgroup is given by the string operator $S_d(i,j)$ labeled by $d$, where $i$ and $j$ are the endpoints of the operator on the lattice.
A non-zero expectation value for $S_d(i,j)$ in the limit $|i-j|\to\infty$ corresponds to a long-range ordered ground-state.
The holographic perspective suggests that the condensation of charges $e_L$, $e_R$, $e_Le_R$, and $e_Le_R^2$ will be detected by the following string operators 
\begin{equation}\label{eq:The order parameters for condensations of eL, eR, and eLeR in Z3xZ3}
   \begin{aligned}
    S_{e_L}(i,j)&=Z^{\pd}_{i,L}\,Z^{\dagger}_{j,L}, &\qquad
    S_{e_R}(i,j)&=Z^{\pd}_{i,R}\,Z^{\dagger}_{j,R},
    \\
    S_{e_Le_R}(i,j)&=Z^{\pd}_{i,L}Z^{\pd}_{i,R}Z^{\dagger}_{j,L}Z^{\dagger}_{j,R},
    &\qquad 
    S_{e_Le^2_R}(i,j)&=Z^{\pd}_{i,L}Z^{\dagger}_{i,R}Z^{\dagger}_{j,L}Z^{\pd}_{j,R},
    \end{aligned} 
\end{equation}
Alternatively, by cluster decomposition, the local order parameters $Z_{i,L}, Z_{i,R}$, $Z_{i,L}Z_{i,R}$ and $Z_{i,L}Z^{\dagger}_{i,R}$  can be used to detect the spontaneous symmetry-breaking to $\mbb{Z}_3^L$, $\mbb{Z}^R_{3}$, $\mbb{Z}^D_3$, and $\mbb{Z}_3^A$ symmetries in the ground state, respectively (see Table \ref{tab:lagrangian subgroups of Z3*Z3})
Similarly, the condensation of fluxes $m_L$, $m_R$, $m_Lm_R$, and $m_Lm_R^2$ will be detected by the following string operators 
\begin{equation}\label{eq:The order parameters for condensations of mL, mR, and mLmR in Z3xZ3}
   \begin{aligned}
    S_{m_L}(i,j)&=\prod_{k=i}^j X_{k,L},&\qquad
    S_{m_R}(i,j)&=\prod_{k=i}^j X_{k,R},
    \\
    S_{m_Lm_R}(i,j)&=\prod_{k=i}^jX_{k,L}X_{k,R}, &\qquad
    S_{m_Lm^2_R}(i,j)&=\prod_{k=i}^jX_{k,L}X^{\dagger}_{k,R},
    \end{aligned} 
\end{equation}
These are the disorder parameters and their condensation does not break any symmetry. 
Finally, we construct the string order-parameters corresponding to the condensation of dyonic generators of the Lagrangian subgroups. They are given by
\begin{equation}\label{eq:The order parameters for condensations of dyons in Z3xZ3}
    \begin{aligned}
    S_{e_Lm_R^2}(i,j)&=Z^{\pd}_{i,L}\left(\bigotimes_{k=i+1}^j X_{k,R}^{\dagger}\right) Z^{\dagger}_{j,L}, &\ \ S_{m_Le_R}(i,j)&=Z^{\pd}_{i,R}\left(\bigotimes_{k=i}^{j-1} X^{\pd}_{k,L}\right) Z^{\dagger}_{j,R}, 
        \\ 
    S_{e_Lm_R}(i,j)&=Z^{\pd}_{i,L}\left(\bigotimes_{k=i+1}^j X^{\dagger}_{k,R}\right) Z_{j,L}^{\dagger}, &\ \ 
    S_{m^2_L e_R}(i,j)&=Z^{\pd}_{i,R}
     \left(\bigotimes_{k=i}^{j-1} X^{\dagger}_{k,L}\right) Z^{\dagger}_{j,R}, 
    \end{aligned}
\end{equation}
which detect SPT phases.
Denoting the ground-state of the gapped phase $\mcal{L}_i$ by $|\tenofo{GS}_{\mcal{L}_i}\rangle$, we have
\begin{equation}
\lim_{|j_1-j_2|\to\infty}\langle\tenofo{GS}_{\mcal{L}_i}|S_d(j_1,j_2)|\tenofo{GS}_{\mcal{L}_i}\rangle=
    \begin{cases}
    \tenofo{constant}, \qquad &\forall \  d\in\mcal{L}_i,
    \\
    \hphantom{con}0 , \qquad &\forall \  d\ne\mcal{L}_i.
    \end{cases}
\end{equation}
The fundamental excitations within each gapped phase that correspond to the set of confined anyons. 
Any two anyons that only differ by a condensed anyon create the same excitation and are thus identified. 
For example, consider the Lagrangian subgroup $\mcal{L}_6$. 
It is be readily checked that possible equivalence classes of excitations are given by\footnote{Here by $\langle\cdots\rangle$ denotes the set generated by the dyons inside.}
\begin{equation}
    \begin{aligned}
        \phi^{\text{SPT}_0}_{e_L}&=\langle e_Lm_L,e_Lm_R\rangle, &\qquad \phi^{\text{SPT}_0}_{e_L^2}&=\langle e_L^2m_L,e^2_Lm_R\rangle,
        \\
        \phi^{\text{SPT}_0}_{e_R}&=\langle e_Rm_L,e_Rm_R\rangle, &\qquad \phi^{\text{SPT}_0}_{e_R^2}&=\langle e_R^2m_L,e_R^2m_R\rangle,
        \\
        \phi^{\text{SPT}_0}_{e_Le_R}&=\langle e_Le_Rm_L,e_Le_Rm_R\rangle, &\qquad \phi^{\text{SPT}_0}_{e_L^2e_R}&=\langle e_L^2e_Rm_L,e_L^2e_Rm_R\rangle,
        \\
        \phi^{\text{SPT}_0}_{e_Le^2_R}&=\langle e_Le^2_Rm_L,e_Le^2_Rm_R\rangle, &\qquad \phi^{\text{SPT}_0}_{e_R^2e_R^2}&=\langle e_R^2e_R^2m_L,e_R^2e_R^2m_R\rangle,
    \end{aligned}
\end{equation}
which together with the trivial $\phi^{\text{SPT}_0}_{1}=\langle m_L,m_R\rangle$ form the object of the category $\tenofo{Rep}({\mbb{Z}_3\times\mbb{Z}_3})$. The anyon in the subscript of each $\phi$ corresponds to the representative of the confined anyons and the corresponding string operators create fundamental excitations (domain wall) (see \eqref{eq:The order parameters for condensations of eL, eR, and eLeR in Z3xZ3}, \eqref{eq:The order parameters for condensations of mL, mR, and mLmR in Z3xZ3}, and \eqref{eq:The order parameters for condensations of dyons in Z3xZ3}). Similar considerations determine the category of excitations associated to other gapped phases. 

\smallskip We thus see that the holographic perspective could easily determine various gapped phases, their order parameters, and excitations.

\subsubsection*{Dualities, and their action on anyons and gapped phases}
We now study the action of the duality group on the gapped phases.
The duality group $\mcal{G}[\ms G]$ for $\ms{G}=\mbb{Z}^L_3\times\mbb{Z}^R_3$ has order $1152$ and is isomorphic to the Weyl group of the exceptional Lie algebra $F_4$ (see App.~ \ref{appsec:computation of duality group} for details) 
\begin{equation}
    \mcal{G}[\mbb{Z}^L_3\times\mbb{Z}^R_3]=W(F_4).
\end{equation}
The group is generated by $\{\sigma_L,h_1,h_2,k\}$. The generators act on the anyons and gapped phases as follows
\begin{enumerate}[label=(\alph*)]
    
    \item{\bf Universal dynamical dualities}: These are classified by $H^2(\ms G,\ms U(1))=\mbb{Z}_3$ and its actions on anyons is
    \begin{equation}
            k:\{m_L\mapsto m_Le_R^2, m_R\mapsto e_Lm_R\},
    \end{equation}
    and its non-trivial action on gapped phases is
    \begin{equation}
        \mcal{L}_6\overset{k}{\longrightarrow}\mcal{L}_8\overset{k}{\longrightarrow}\mcal{L}_7\overset{k}{\longrightarrow}\mcal{L}_6.
    \end{equation}
    This satisfies $k^3=1$. Physically, $k$ corresponds to a domain wall in the bulk with the SPT$_2$ ($\mathcal L_8$) phase. The duality above is just stacking of $\mathbb Z_3\times\mathbb Z_3$ SPT phases.
    
    \item{\bf Universal kinematical dualities}: These are related to $\tenofo{Aut}(\mbb{Z}_3\times\mbb{Z}_3)=\tenofo{GL}_2(\mbb{F}_3)=Q_8\rtimes S_3$. Their actions on anyons are
    \begin{equation}
        \begin{aligned}
            h_1&:\{m_L\mapsto m_R, m_R\mapsto m_L, e_L\mapsto e_R,e_R\mapsto e_L\},
            \\
            h_2&:\{m_R\mapsto m_Lm_R,e_L\mapsto e_Le_R^2\}.
        \end{aligned}
    \end{equation}
    and hence their nontrivial action on gapped phases are
    \begin{equation}
        \begin{gathered}
        \mcal{L}_2\overset{h_1}{\longleftrightarrow}\mcal{L}_3, \qquad \mcal{L}_7\overset{h_1}{\longleftrightarrow}\mcal{L}_8,
        \\
        \mcal{L}_3\overset{h_2}{\longrightarrow}\mcal{L}_4\overset{h_2}{\longrightarrow}\mcal{L}_5\overset{h_2}{\longrightarrow}\mcal{L}_3.
        \end{gathered}
    \end{equation}
    They act on $k$ as $h_1kh_1^{-1}=k^{-1}$ and $h_2kh_2^{-1}=k$. Together, they generate
    \begin{equation}
        \langle k,h_1,h_2\rangle=\mbb{Z}_3\rtimes\tenofo{GL}_2(\mbb{F}_3). 
    \end{equation}
    
    \item{\bf Partial electric-magnetic duality}: The action on anyons is
    \begin{equation}
        \begin{aligned}
            \sigma_L&:\{m_L\mapsto e_L,e_L\mapsto m_L\},
        \end{aligned}
    \end{equation}
    and the mapping of gapped phases are
    \begin{equation}
        \begin{aligned}
            \mcal{L}_1&\overset{\sigma_L}{\longleftrightarrow}\mcal{L}_2, &\qquad \mcal{L}_3&\overset{\sigma_L}{\longleftrightarrow}\mcal{L}_6,
            &\qquad\mcal{L}_4&\overset{\sigma_L}{\longleftrightarrow}\mcal{L}_8, &\qquad \mcal{L}_5&\overset{\sigma_L}{\longleftrightarrow}\mcal{L}_7.
        \end{aligned}
    \end{equation}
    This satisfies $\sigma_L^2=1$ and generates a $\mbb{Z}_2$ subgroup. Together, $\sigma_L, k, h_1$, and $h_2$ generate the full group $\mcal{G}[\mbb{Z}_3\times\mbb{Z}_3]$. It is convenient to also define $\sigma_R=h_1\sigma_Lh_1$.

\end{enumerate}

\subsubsection*{The minimal Hamiltonian and its phase diagram}
Similar to the $\mbb{Z}_2\times\mbb{Z}_2$ case, we consider the minimal Hamiltonian containing all gapped phases and many interesting transitions of $\mbb{Z}_3\times\mbb{Z}_3$ symmetric theories. This is given by
\begin{equation}\label{eq:Z3xZ3 minimal hamiltonian}
    H=\sum_{a=1}^{8} t_a H_a,\qquad t_a\in\mbb{R},
\end{equation}
where each term is given in \eqref{eq:Z3xZ3 fixed point SSB}, \eqref{eq:Z3xZ3 fixed points PSB}, and \eqref{eq:Z3xZ3 fixed points SPT}.
There are many other interesting models one could write using the minimal string operators, but for this example we will restrict ourselves to this simple model. 
Furthermore, we only make a few observations about the phase-diagram, and a more detailed analysis will appear elsewhere. 

\begin{figure}[t!]\centering
    \begin{subfigure}{.38\linewidth}\centering
    \includegraphics[width=1.0\textwidth]{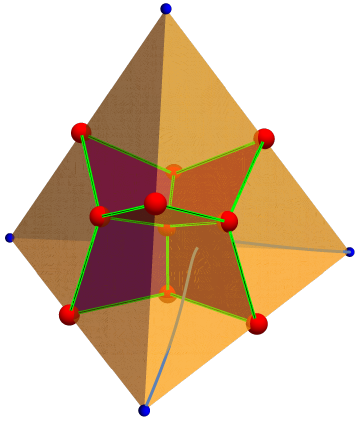}
    \subcaption{}
    \end{subfigure}\hspace{1cm}
    ~
    \begin{subfigure}{.38\linewidth}\centering
    \includegraphics[width=1.0\textwidth]{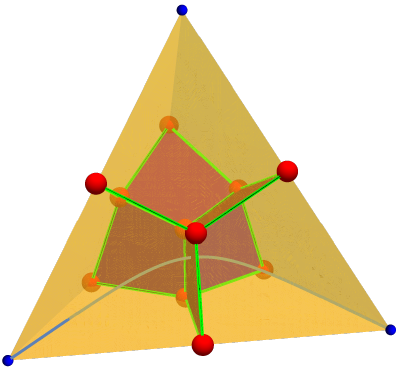}
        \subcaption{}
    \end{subfigure}
    \caption{A proposed phase diagram for the minimal $\mbb{Z}_3\times\mbb{Z}_3$ Hamiltonian at $t_2=t_3=t_4=t_5=0$. The two figures correspond to (different angles of) the subspace $t_1+t_6+t_7+t_8=1$ (or equivalently a quadrant of $S^3\subset\mbb{R}^4$). The blue points are fixed-point Hamiltonians, the yellow regions are gapped phases, the green lines and red surfaces are expected to be first-order transitions while the red points are expected to be second-order transitions. The curve between the fixed-point Hamiltonians $H_1$ and $H_6$ is parameterized as \eqref{eq:parametrization of the curve between H1 and H6 in the Z3xZ3 case}.}\label{fig:Z3xZ3ProposedPhaseDiagram}
\end{figure}

\smallskip The duality group contains many interesting subgroups, one particularly interesting subgroup is generated by the dualities
\smallskip 
\begin{equation}\label{eq:S4 duality subgroup for Z3xZ3 (generators)}
    \begin{aligned}
        a&:\{m_L\mapsto m_R^2, m_R\mapsto m_L^2, e_L\mapsto e_R^2,e_R\mapsto e_L^2\},
           \\
        \sigma_L\sigma_R&:\{m_L\mapsto e_L, m_R\mapsto e_R, e_L\mapsto m_L,e_R\mapsto m_R\},\\
        b&:\{m_L\mapsto m_R, m_R\mapsto m_L^2m_R^2, e_L\mapsto m_L^2e_L^2m_R^2e_R,e_R\mapsto e_L^2m_R^2\}.
    \end{aligned}
\end{equation}
These generate the group of permutations of four objects
\begin{equation}\label{eq:S4 duality subgroup for Z3xZ3}
    \langle a, \sigma_L\sigma_R, b\rangle = S_4 \subset\mathcal G[\mathbb Z_2\times\mathbb Z_2].
\end{equation}
\begin{figure}[t!]
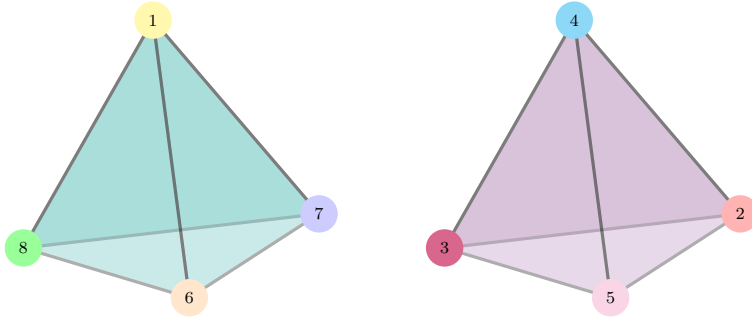

    \centering
    \ZthreeZthreeTetrahedron{-20}{38}
    \caption{The action of the $S_4$ subgroup of dualities \eqref{eq:S4 duality subgroup for Z3xZ3 (generators)} and \eqref{eq:S4 duality subgroup for Z3xZ3} on gapped boundaries can geometrically be thought of as the isometries of two regular tetrahedra as illustrated above. There is a duality/isometry for any permutation of the vertices within each tetrahedron.}
    \label{fig:ZthreeZthreeTetrahedron}
\end{figure}
The action of this group on Lagrangian subgroups give rise to two orbits
\begin{equation}
        \tenofo{Orb}_1 = \{\mathcal L_1,\mathcal L_6,\mathcal L_7,\mathcal L_8\},\qquad
        \tenofo{Orb}_2 = \{\mathcal L_2, \mathcal L_3,\mathcal L_4,\mathcal L_5\},
\end{equation}
and the action of each group element is to permute the gapped phases within these two sets. 
Note that the second orbit consists of all gapped phases in which $\mbb{Z}_3\times\mbb{Z}_3$ symmetry partially breaks down to $\mathbb Z_3$ while the first orbit consists of the SSB phase and all the unbroken (SPT) phases (see Table \ref{tab:lagrangian subgroups of Z3*Z3}).
Geometrically, $S_4$ is the isometry group of the regular tetrahedron and each gapped phases can be thought of as a vertex of a tetrahedron (see Figure \ref{fig:ZthreeZthreeTetrahedron}). 
In order to realize these abstract tetrahedra in the 8-dimensional phase diagram, consider the restrictions of \eqref{eq:Z3xZ3 minimal hamiltonian} to the two different 4-dimensional subspaces
\begin{equation}
    \begin{aligned}
        H_{\tenofo{Orb}_1}&= t_1 H_1 + t_6 H_6 + t_7 H_7 + t_8 H_8,
        \\
        H_{\tenofo{Orb}_2}&= t_2 H_2 + t_3 H_3 + t_4 H_4 + t_5 H_5.
    \end{aligned}
\end{equation}
In \eqref{eq:S4 duality subgroup for Z3xZ3}, there is a duality for any possible permutation of the four parameters in these two Hamiltonians. Since the discussion is the same, we will restrict ourselves to the first Hamiltonian $H_{\tenofo{Orb}_1}$. The overall scale does not matter, and without loss of generality, we can  restrict ourselves to the $3$-dimensional subspace
\begin{equation}
    t_1 + t_6 + t_7 + t_8 = 1.
\end{equation}
This is exactly a tetrahedron with the the four fixed-point Hamiltonians at its vertices $(t_1,t_2,t_3, t_4)=(1,0,0,0)$, $(t_1,t_2,t_3, t_4)=(0,1,0,0)$, etc. Using similar arguments and assumptions that led to Figure \ref{Fig:ZTwoSqPhaseDiagram}, we can construct the simplest possible phase-diagram for $H_{\tenofo{Orb}_1}$, as can be seen in Figure \ref{fig:Z3xZ3ProposedPhaseDiagram}. 

\smallskip One way to argue for such a phase diagram is as follows.
Since $S_4$ dualities act on the tetrahedron parameterizing $H_{\tenofo{Orb}_1}$ as its isometries, they naturally induce maps on any point in the tetrahedron.
The blue vertices are fixed-point Hamiltonians $H_1, H_6, H_7,$ and $H_8$.
The center of the tetrahedron is self-dual under $S_4$ and therefore has to be a transition since no Lagrangian subgroup is self-dual under $S_4$.
Now consider the subgroup $S_3\subset S_4$ corresponding to all permutations of $t_6, t_7,$ and $t_8$ that leave $t_1$ invariant.
The Hamiltonians on the self-dual line must either be transitions\footnote{Recall that we consider anything beyond gapped phases (with non-degenerate ground states) labeled by Lagrangian subgroups as transitions. This in principle could be gapless phases.} or in the gapped phase $\mcal{L}_1$ (assuming that $\mcal{L}_1, \mcal{L}_6, \mcal{L}_7$, and $\mcal{L}_8$ are the only gapped phases in this tetrahedron).
Starting from the fixed-point Hamiltonian $H_1$, we can move along the self-dual line. Assuming that the Hamiltonians remain in the same phase $\mcal{L}_1$, a transition must occur when we reach the $S_4$ self-dual point in the center. 
Also, consider the $\mbb{Z}_2$ subgroup swapping $\mcal{L}_7$ and $\mcal{L}_8$, the plane of reflection is $\mbb{Z}_2$ self-dual and similar arguments can be used. 
Repeating this logic for all the subgroups of $S_4$, the minimal phase diagram is given by Figure \ref{fig:Z3xZ3ProposedPhaseDiagram}. 
The blue points are the four gapped fixed-point Hamiltonians, the yellow regions are gapped phases, red dots, the green lines and red surfaces are transitions.

\smallskip For some of the red points, we can make a slightly stronger statement. The red point at $(t_1,t_6,t_7,t_8)=(1,1,0,0)$ corresponds to two decoupled copies of critical $3$-state Potts model. This critical point is thus described by two copies of the $\mbb{Z}_3$ parafermion CFT with $c=\frac{8}{5}$. Unlike the case of Ising$^2$ CFT, this critical point does not have a marginal operator, therefore, we expect that the green lines and the red surface attached to it to be first-order transitions. All the critical points in the center of the edges (1-simplices) are dual to this point thus also critical with $c=\frac{8}{5}$. Remaining are the four red points at the center of the triangles (2-simplices) that are self-dual under $S_3$ subgroups of $S_4$. Duality does not completely fix the type of transition but it seems reasonable to expect them to be critical.

\smallskip It is important to note that dualities do not completely fix the phase diagram and we have to make a few assumptions to construct the phase diagram in Figure \ref{fig:Z3xZ3ProposedPhaseDiagram}. More complicated scenarios are possible. However, by knowing the phase diagram only in a small corner of the tetrahedron is enough to completely determine the rest of the tetrahedron by dualities.

\begin{figure}[t!]
    \centering
    \includegraphics[width=0.8\textwidth]{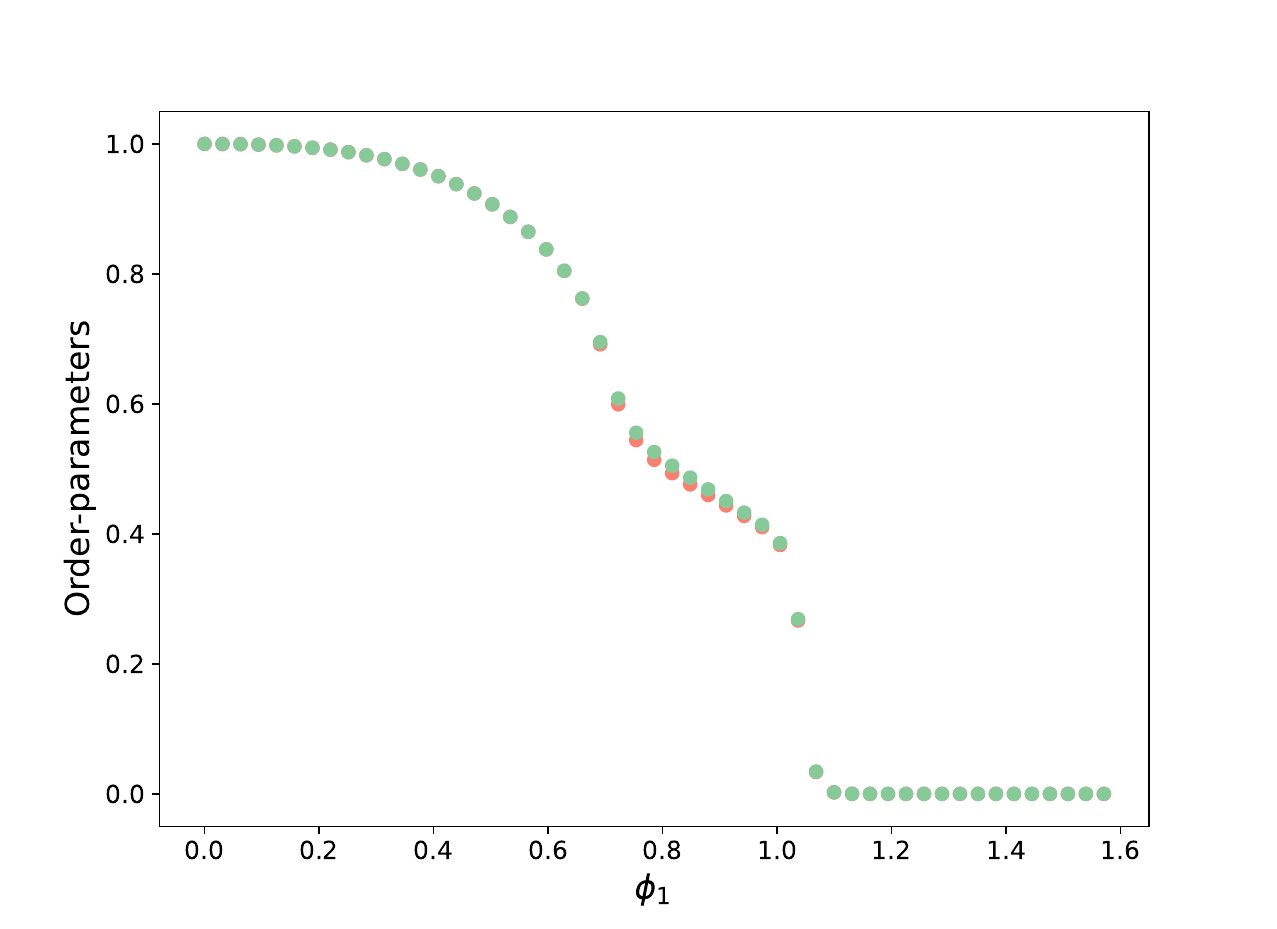}
    \caption{DMRG calculations of order parameters $\langle Z_i^L\rangle$ and $\langle Z_i^R\rangle$ detecting condensation of $e_L$ and $e_R$ along the curve \eqref{eq:parametrization of the curve between H1 and H6 in the Z3xZ3 case}. The calculation was performed for $N=2\times 80$ $\mbb{Z}_3$ spins for open boundary conditions. The transition happened at the expected point $\phi_1=\frac{\pi}{3}\approx 1.0472$.}
    \label{fig:Z3Z3_gamma1_orderparameter}
\end{figure}

\smallskip As a preliminary numerical analysis, we have performed DMRG calculations of some order parameters along a simple curve. Since overall scaling of Hamiltonians does not matter, we can parameterize a three-sphere in four dimensions as 
\begin{equation}
   \begin{pmatrix}
       t_1 
       \\
       t_6
       \\
       t_7
       \\
       t_8
   \end{pmatrix}
   =
   \begin{pmatrix}
       \cos\phi_1
       \\
       \cos\phi_2\sin\phi_1
       \\
       \cos\phi_3\sin\phi_1\sin\phi_2
       \\
       \sin\phi_1\sin\phi_2\sin\phi_3
   \end{pmatrix}.
\end{equation}
Consider the curve 
\begin{equation}\label{eq:parametrization of the curve between H1 and H6 in the Z3xZ3 case}
   \begin{pmatrix}
       \phi_1(s)
       \\
       \phi_2(s)
       \\
       \phi_3(s)
   \end{pmatrix}
   =\frac{\pi}{2}
   \begin{pmatrix}
       s
       \\
       1-s
       \\
       s
   \end{pmatrix}.
\end{equation}
from $\mcal{L}_1$ to $\mcal{L}_6$ crossing our predicted transition surface at $s=\frac{2}{3}$ or $\phi_1=\frac{\pi}{3}\approx 1.0472$ (see Figure \ref{fig:Z3xZ3ProposedPhaseDiagram}). The order parameter for the condensation of $e_L$ and $e_R$ on this curve is shown in Figure \ref{fig:Z3Z3_gamma1_orderparameter}. Exactly at $\phi_1=\frac{\pi}{3}$, we see a sharp transition consistent with a first-order transition, as predicted.  However, further numerical analysis is needed to confirm this more convincingly. We leave a more complete analysis of $\mbb{Z}_3\times\mbb{Z}_3$ symmetric models using topological holography for future work.

\subsubsection*{Predictions of conformal spectra of nontrivial critical points}

We will now briefly turn to the second application of topological holography of computing conformal spectrum of non-trivial transitions. Consider the topological phase transition $\mscr{C}_{67}$ between SPT$_0$ ($\mcal{L}_6$) and SPT$_1$ ($\mcal{L}_7$). The duality $\sigma_Lh_2^2\sigma_Rk^2$
\begin{equation}
    \begin{aligned}
    \mcal{L}_1&\overset{k^2}{\longrightarrow}\mcal{L}_1\overset{\sigma_R}{\longrightarrow}\mcal{L}_3\overset{h^2_2}{\longrightarrow}\mcal{L}_5\overset{\sigma_L}{\longrightarrow}\mcal{L}_7,
    \\
    \mcal{L}_6&\overset{k^2}{\longrightarrow}\mcal{L}_7\overset{\sigma_R}{\longrightarrow}\mcal{L}_4\overset{h^2_2}{\longrightarrow}\mcal{L}_3\overset{\sigma_L}{\longrightarrow}\mcal{L}_6,
    \end{aligned}
\end{equation}
maps $\mscr{C}_{16}$, described by decoupled $\mbb{Z}_3$ parafermion CFTs to the topological transitions $\mscr{C}_{67}$ 
\begin{equation}
    \mscr{C}_{16}\;\xrightarrow{\sigma_Lh_2^2\sigma_Rk^2}\;\mscr{C}_{67}.
\end{equation}
Under this duality, we have
\begin{equation}
(\ms g_L,\alpha_L,\ms g_R,\alpha_R)  \;\xrightarrow{\sigma_Lh_2^2\sigma_Rk^2}\;(2\ms g_R+\alpha_L,2\alpha_R,-2\ms g_L+\alpha_R,\alpha_L).  
\end{equation}
We can now use \eqref{eq: eta formula for partition function dualities} to relate the dual twisted partition functions and the original ones ($\omega=\exp(2\pi i/3)$)
\begin{equation}
    \mcal{Z}^\vee_{\ms g^\vee,\ms h^\vee}=\frac{1}{9}\sum_{\substack{\ms g_L,\ms g_R \\ \ms h_L,\ms h_R}}\omega^{\big[\left(\ms g_L-2\ms g_R^\vee\right)\left(\ms h_L^\vee-2\ms h_L\right)+\left(\ms g_R+2\ms g_L^\vee\right)\left(\ms h_R^\vee-\ms h_R\right)\big]}\mcal{Z}^{\tenofo{3-Para}}_{\ms g_L,\ms h_L}\mcal{Z}^{\tenofo{3-Para}}_{\ms g_R,\ms h_R}.
\end{equation}
where $\mcal{Z}^{\tenofo{3-Para}}_{\ms g,\ms h}$ are the twisted partition functions of $\mbb{Z}_3$ parafermion CFT, $\mcal{Z}^\vee=\mcal{Z}(\mscr{C}_{67})$ and $\ms g^\vee=(\ms g_L^\vee,\ms g_R^\vee)$ and $\ms h^\vee=(\ms h_L^\vee,\ms h_R^\vee)$. As a concrete check, we can see that the partition function in the untwisted sector (periodic boundary conditions) is given by
\begin{equation}\label{eq:partition function in the untwisted sector of C67 in the Z3xZ3 case}
    \begin{aligned}
    \mcal{Z}_{(0,0)(0,0)}(\mscr{C}_{67})&=4|\chi _{\frac{2}{3}}|^4+8|\chi _{\frac{1}{15}}|^2|\chi _{\frac{2}{3}}|^2+4\chi _0\chi_{\frac{1}{15}}
   \chi_{\frac{2}{5}}\bar\chi _{\frac{2}{3}}+4\chi _{\frac{2}{5}} \chi _{\frac{2}{3}}\bar\chi_0
   \chi _{\frac{1}{15}}
   \\
   &+|\chi _0|^4+2| \chi _{\frac{2}{5}}|^2|\chi _0|^2+4 | \chi
   _{\frac{2}{3}}|^2|\chi _0|^2+\left(2 | \chi _{\frac{1}{15}}|^2+| \chi
   _{\frac{2}{5}}|^2\right)^2. 
    \end{aligned}
\end{equation}
\begin{table}[t!]\centering
    \renewcommand{\arraystretch}{1.2}
    \begin{tabular}{c c c c c c c c c c c} 
    \toprule 
    \multicolumn{11}{c}{Topological criticality with no defect}
    \\
    \midrule
    \multirow{1}{2cm}{$h+\bar{h}$} & 0 & $\frac 4{15}$ & $\frac 4{5}$  & $\frac {14}{15}$ & $\frac 85$ & $\frac 43$ & $\frac 83$ & $\frac {22}{15}$ & $\frac {17}{15}$ & $\frac {17}{15}$ 
    \\ 
    \multirow{1}{2cm}{$h-\bar{h}$} & 0 & 0 & 0 & 0 & 0 & 0 & 0 & 0 & $+1$ & $-1$
    \\
    \multirow{1}{2cm}{multiplicity} & 1  & 4 & 2 & 4 & 1 & 4 & 4 & 8 & 4 & 4
    \\
    \multirow{2}{1cm}{\scriptsize $(\alpha_L,\alpha_R)$} & \multirow{2}{0.5cm}{\scriptsize(0,0)} & \multirow{2}{1.2cm}{\scriptsize$(1,0) (0,1)$ $(2,0) (0,2)$} & \multirow{2}{0.5cm}{\scriptsize$(0,0)$} & \multirow{2}{1.2cm}{\scriptsize$(1,1) (1,2)$ $(2,1) (2,2)$} & \multirow{2}{0.5cm}{\scriptsize$(0,0)$} & \multirow{2}{1.2cm}{\scriptsize$(1,1) (1,2)$ $(2,1) (2,2)$} & \multirow{2}{1.2cm}{\scriptsize$(1,0) (0,1)$ $(2,0) (0,2)$} & \multirow{2}{1.2cm}{\scriptsize $(1,0) (0,1)$ $(2,0) (0,2)$} & \multirow{2}{1.2cm}{\scriptsize$(1,1) (1,2)$ $(2,1) (2,2)$}  & \multirow{2}{1.2cm}{\scriptsize$(1,1) (1,2)$ $(2,1) (2,2)$}
    \\\\
    \bottomrule
    \end{tabular}
    \caption{The spectrum of primary fields for topological transitions ($\mscr{C}_{67}, \mscr{C}_{68},$ and $\mscr{C}_{78}$) between $\mbb{Z}_3\times\mbb{Z}_3$ SPT phases for periodic boundary condition. With these boundary conditions, the spectrum and $\mbb{Z}_3\times\mbb{Z}_3$ transformation properties are the same for these three transitions.}
    \label{tab:C67 with no defect, Z3xZ3 case}
\end{table}
In \cite{TsuiHuangJiangLee201701}, the spectrum of the topological transition $\mscr{C}_{67}$ was obtained using a specific lattice mapping giving rise to a diagonal orbifold similar to \eqref{eq:the partition function of the critical phase between the two Z2*Z2 SPT states}. Our result  \eqref{eq:partition function in the untwisted sector of C67 in the Z3xZ3 case} for periodic boundary condition reproduces the result of \cite{TsuiHuangJiangLee201701} (equation E.7). The spectrum of primary fields for this transition with periodic boundary condition is tabulated in Table \ref{tab:C67 with no defect, Z3xZ3 case}. As can be seen in the table, there are three $\mbb{Z}_3\times\mbb{Z}_3$ symmetric relevant operators, two with scaling dimension $\frac{4}{5}$ and one with scaling dimension $\frac{8}{5}$. Hence, this is a multicritical point between up to six different gapped phases. In \cite{TsuiHuangJiangLee201701}, it was shown that one of the operators with scaling dimension $\frac{4}{5}$ is responsible for topological transition $\mscr{C}_{67}$. Similar to the $\mbb{Z}_2\times\mbb{Z}_2$, we expect that the other relevant operators flow to gapped phases in other tetrahedron (see Figure \ref{fig:ZthreeZthreeTetrahedron}). 

\smallskip Similarly, we can obtain the other topological transitions. For example using the duality $\sigma_Rh^2_2\sigma_R:\mscr{C}_{16}\mapsto\mscr{C}_{68}$, we find
\begin{equation}\label{eq:partition function in the untwisted sector of C68 and C78 in the Z3xZ3 case}
    \begin{aligned}
    \mcal{Z}_{\ms g^\vee,\ms h^\vee}(\mscr{C}_{68})&=\frac{1}{9}\sum_{\substack{\ms g_L,\ms g_R \\ \ms h_L,\ms h_R}}\omega^{\big[\left(\ms g_R-\ms g_R^\vee\right)\left(\ms h_L^\vee-\ms h_L\right)+\left(\ms g_L+\ms g_L^\vee\right)\left(\ms h_R^\vee-2\ms h_R\right)\big]}\mcal{Z}^{\tenofo{3-Para}}_{\ms g_L,\ms h_L}\mcal{Z}^{\tenofo{3-Para}}_{\ms g_R,\ms h_R}.
    \end{aligned}
\end{equation}
More transparently, using $k:\mscr{C}_{67}\mapsto\mscr{C}_{68}$ and $k^2:\mscr{C}_{67}\mapsto\mscr{C}_{78}$, we have
\begin{equation}
    \begin{aligned}
        \mcal{Z}_{\ms g^\vee,\ms h^\vee}(\mscr{C}_{68})&=\omega^{\ms g_L^\vee\ms h^\vee_R-\ms g_R^\vee\ms h^\vee_L}\mcal{Z}^{\tenofo{3-Para}}_{\ms g^\vee,\ms h^\vee}(\mscr{C}_{67}),
        \\
        \mcal{Z}_{\ms g^\vee,\ms h^\vee}(\mscr{C}_{78})&=\omega^{2\left(\ms g_L^\vee\ms h^\vee_R-\ms g_R^\vee\ms h^\vee_L\right)}\mcal{Z}^{\tenofo{3-Para}}_{\ms g^\vee,\ms h^\vee}(\mscr{C}_{67}).
    \end{aligned}
\end{equation}
The phase factors are the topological partition functions for the SPT phase on the duality domain-wall labeled by $k$ and $k^2$. Both of these partition functions are equal to $\mscr{C}_{67}$ in the untwisted sector (i.e. $(\ms g^\vee_L,\ms g^\vee_R)=(0,0)$) and one might consider these transition to belong to the same universality class, as has been pointed out in \cite{TsuiHuangJiangLee201701}. However, they differ in the twisted sectors due to the effect of the topological SPT phase. In other words, these CFTs have the same conformal spectra but in the twisted sectors the Virasoro primaries (and their descendants) transform differently under global $\mbb{Z}_3\times\mbb{Z}_3$ symmetry. Therefore, these CFTs belong to the same universality class if we only consider conformal symmetry but they belong to topologically-distinct universality classes if we also consider the global $\mbb{Z}_3\times\mbb{Z}_3$ symmetry. In Table \ref{tab:C67, C68, C78 with U11 defect, Z3xZ3 case}, we have tabulated the conformal spectra of these three transitions with twisted boundary condition $(\ms g_L^\vee,\ms g_R^\vee)=(1,1)$. As can be seen, conformal spectra are identical but the transformation properties of fields under $\mbb{Z}_3\times\mbb{Z}_3$ symmetry are different, illustrating the topological-distinctness of these criticalities.

\smallskip As another application, consider the duality $k:\mscr{C}_{16}\mapsto\mscr{C}_{18}$ and $k^2:\mscr{C}_{16}\mapsto\mscr{C}_{17}$. The corresponding Hamiltonian for transitions are
\begin{equation}
    \begin{aligned}
    H(\mscr{C}_{18})&=-\sum_{i}\left[Z^L_i\left(Z^L_{i+1}\right)^2+Z^R_{i}\left(Z^R_{i+1}\right)^2+Z^L_i(Z^L_{i+1})^2X_{i+1}^R+(X_i^L)^2Z_i^R(Z^R_{i+1})^2\right]+\tenofo{h.c.},
    \\
    H(\mscr{C}_{17})&=-\sum_{i}\left[Z^L_i\left(Z^L_{i+1}\right)^2+Z^R_{i}\left(Z^R_{i+1}\right)^2+Z^L_i(Z^L_{i+1})^2(X_{i+1}^R)^2+X_i^LZ_i^R(Z^R_{i+1})^2\right]+\tenofo{h.c.}.
    \end{aligned}
\end{equation}
These critical spin-chain models couple between the two chains and are in general complicated to analyze. However, using dualities at the level of partition functions, we can find their exact spectra
\begin{equation}
    \begin{aligned}
        \mcal{Z}_{\ms g^\vee,\ms h^\vee}(\mscr{C}_{18})&=\omega^{\ms g_L^\vee\ms h^\vee_R-\ms g_R^\vee\ms h^\vee_L}\mcal{Z}_{\ms g^\vee,\ms h^\vee}(\mscr{C}_{16}),
        \\
        \mcal{Z}_{\ms g^\vee,\ms h^\vee}(\mscr{C}_{17})&=\omega^{2\left(\ms g_L^\vee\ms h^\vee_R-\ms g_R^\vee\ms h^\vee_L\right)}\mcal{Z}_{\ms g^\vee,\ms h^\vee}(\mscr{C}_{16}).
    \end{aligned}
\end{equation}
We again see that $\mscr{C}_{16}, \mscr{C}_{17},$ and $\mscr{C}_{18}$ have identical partition functions for periodic boundary conditions (i.e. $(\ms g^\vee_L,\ms g^\vee_R)=(0,0)$). Naively, one would think that all these transitions are described by two decoupled $\mbb{Z}_3$ parafermion universality class. However, we find that the partition functions do not match in all twisted sectors. In particular, the SPT twist can still be detected in the twisted sectors and therefore, these transitions can be distinguished. We call these criticalities {\it topological spontaneous symmetry-breaking transitions}. 
Using the language of \cite{VerresenThorngrenJonesPollmann201905}, the $\mscr{C}_{68}$, $\mscr{C}_{78}$, $\mscr{C}_{17}$ and $\mscr{C}_{18}$ are nontrivial $\mbb{Z}_3\times\mbb{Z}_3$ symmetry-enriched CFTs. Similar to gapped SPT phases, they can be distinguished by their edge modes.

\begin{table}\centering
    \renewcommand{\arraystretch}{1.2}
    \hspace*{-1cm}
    \begin{tabular}{c c c c c c c c c c c c c c c} 
    \toprule 
    \multicolumn{15}{c}{Topological criticalities $(\mscr{C}_{67},\mscr{C}_{68},\mscr{C}_{78})$ with defect $\mcal{U}_{1,1}$}
    \\
    \midrule
    \multirow{1}{2cm}{$h+\bar{h}$} &  $\frac 4{15}$ & $\frac{14}{15}$ & $\frac{4}{3}$ & $\frac{22}{15}$ & $\frac 83$ & $\frac 23$  & $\frac 45$  & $\frac{22}{15}$ & $\frac 7{15}$ & $\frac 35$ & $\frac 95$ & $\frac{19}{15}$ & $\frac{17}{15}$ & $2$
    \\ 
    \multirow{1}{2cm}{$h-\bar{h}$} & 0 & 0 & 0 & 0 & 0 & $\pm\frac 23$ & $\pm\frac 23$ & $\pm\frac 23$ & $\pm\frac 13$ & $\pm\frac 13$ & $\pm\frac 13$ & $\pm\frac 13$ & $\pm 1$ & $\pm\frac 23$
    \\
    \multirow{1}{2cm}{multiplicity} &2 & 1 & 1 & 4 & 2 & 2  & 4 & 2 & 2 & 4 & 4 & 2 & 2 & 4
    \\
    \multirow{2}{1cm}{\scriptsize $(\alpha_L,\alpha_R)_{\mscr{C}_{67}}$} 
    &\multirow{2}{0.4cm}{\scriptsize$(0,0)$ $(2,1)$}
    &\multirow{2}{0.4cm}{\scriptsize$(1,2)$}
    &\multirow{2}{0.4cm}{\scriptsize$(1,2)$}
    &\multirow{2}{0.4cm}{\scriptsize$(0,0)$ $(2,1)$}
    &\multirow{2}{0.4cm}{\scriptsize$(0,0)$ $(2,1)$}
    &\multirow{2}{0.4cm}{\scriptsize$(0,1)$ $(2,0)$}
    & \multirow{2}{1.45cm}{\scriptsize$(1,1)$ $(0,2)$ $(1,0)$ $(2,2)$}
    &\multirow{2}{0.4cm}{\scriptsize$(0,1)$ $(2,0)$}
    & \multirow{2}{0.4cm}{\scriptsize$(0,1)$ $(2,0)$}
    & \multirow{2}{1.45cm}{\scriptsize$(1,1)$ $(0,2)$ $(1,0)$ $(2,2)$}
    &\multirow{2}{1.45cm}{\scriptsize$(1,1)$ $(0,2)$ $(1,0)$ $(2,2)$}
    &\multirow{2}{0.4cm}{\scriptsize$(0,1)$ $(2,0)$}
    & \multirow{2}{0.4cm}{\scriptsize$(1,2)$}
    & \multirow{2}{1.45cm}{\scriptsize$(1,1)$ $(0,2)$ $(1,0)$ $(2,2)$}
    \\\\
    \multirow{2}{1cm}{\scriptsize $(\alpha_L,\alpha_R)_{\mscr{C}_{68}}$}  &\multirow{2}{0.4cm}{\scriptsize$(0,0)$ $(1,2)$} 
    & \multirow{2}{0.4cm}{\scriptsize$(2,1)$}
    & \multirow{2}{0.4cm}{\scriptsize$(2,1)$}
    & \multirow{2}{0.4cm}{\scriptsize$(0,0)$ $(1,2)$}
    & \multirow{2}{0.4cm}{\scriptsize$(0,0)$ $(1,2)$}
    &  \multirow{2}{0.4cm}{\scriptsize$(1,0)$ $(0,2)$}
    & \multirow{2}{1.45cm}{\scriptsize$(0,1)$ $(1,1)$ $(2,0)$ $(2,2)$}
    & \multirow{2}{0.4cm}{\scriptsize$(1,0)$ $(0,2)$}
    & \multirow{2}{0.4cm}{\scriptsize$(1,0)$ $(0,2)$}
    & \multirow{2}{1.45cm}{\scriptsize$(0,1)$ $(1,1)$ $(2,0)$ $(2,2)$}
    & \multirow{2}{1.45cm}{\scriptsize$(0,1)$ $(1,1)$ $(2,0)$ $(2,2)$}
    & \multirow{2}{0.4cm}{\scriptsize$(1,0)$ $(0,2)$}
    & \multirow{2}{0.4cm}{\scriptsize$(2,1)$}
    & \multirow{2}{1.45cm}{\scriptsize$(0,1)$ $(1,1)$ $(2,0)$ $(2,2)$}
    \\\\
    \multirow{2}{1cm}{\scriptsize $(\alpha_L,\alpha_R)_{\mscr{C}_{78}}$}
    & \multirow{2}{0.4cm}{\scriptsize$(2,1)$ $(1,2)$}
    & \multirow{2}{0.4cm}{\scriptsize$(0,0)$}
    & \multirow{2}{0.4cm}{\scriptsize$(0,0)$}
    & \multirow{2}{0.4cm}{\scriptsize$(2,1)$ $(1,2)$}
    &   \multirow{2}{0.4cm}{\scriptsize$(2,1)$ $(1,2)$}
    & \multirow{2}{0.4cm}{\scriptsize$(1,1)$ $(2,2)$}
    & \multirow{2}{1.45cm}{\scriptsize$(1,0)$ $(0,1)$ $(2,0)$ $(0,2)$}
    & \multirow{2}{0.4cm}{\scriptsize$(1,1)$ $(2,2)$}
    & \multirow{2}{0.4cm}{\scriptsize$(1,1)$ $(2,2)$}
    & \multirow{2}{1.45cm}{\scriptsize$(1,0)$ $(0,1)$ $(2,0)$ $(0,2)$}
    & \multirow{2}{1.45cm}{\scriptsize$(1,0)$ $(0,1)$ $(2,0)$ $(0,2)$}
    & \multirow{2}{0.4cm}{\scriptsize$(1,1)$ $(2,2)$}
    & \multirow{2}{0.4cm}{\scriptsize$(0,0)$}
    &  \multirow{2}{1.45cm}{\scriptsize$(1,0)$ $(0,1)$ $(2,0)$ $(0,2)$}
    \\\\
    \bottomrule
    \end{tabular}
    \caption{The spectrum of primary fields for the topological criticalities $\mscr{C}_{67},\mscr{C}_{68},$ and $\mscr{C}_{78})$ with the twisted boundary condition $(\ms g^\vee_L,\ms g^\vee_R)=(1,1)$. These three critical points have the same conformal spectrum but the action of the global $\mbb{Z}_3\times\mbb{Z}_3$ symmetries differs. As $\mbb{Z}_3\times\mbb{Z}_3$ symmetry-enriched CFTs, these criticalities are topologically distinct and belong to different universality classes.}
    \label{tab:C67, C68, C78 with U11 defect, Z3xZ3 case}
\end{table}

\subsubsection*{Prediction of new nonconformal phase transitions}
There has been convincing recent evidence based on numerical studies that the $\mbb Z_{3}$ symmetric quantum spin systems (see \eqref{eq:generic minimal ZN Hamiltonian} for $N=3$) in general with complex parameters ($\lambda \in \mbb C$) host generic, i.e., non-fine-tuned phase transitions with a dynamical critical exponent $z>1$ \cite{Samajdar_2018}.
The low-energy description of such a transition is not given by a conventional conformal field theory.
In fact, the example of the non-conformal transition realized in the $\mbb Z_3$ spin chain is claimed to be the first such strongly-coupled, generic transition between gapped phases.
Naturally, the $\mbb Z_3\times \mbb Z_3$ model also hosts decoupled doubled non-conformal transitions in the model 
\begin{align}
    H=t_1H^{\tenofo{non-c}}_1+t_2H^{\tenofo{non-c}}_2,
    \label{eq:nonconformal model}
\end{align}
where $H^{\tenofo{non-c}}_i$ is a $\mbb{Z}_3$-symmetric Hamiltonian at a non-conformal transition on the $i$\textsuperscript{th} chain. 
Using the vast web of dualities and the decoupled Hamiltonian \eqref{eq:nonconformal model}, we can in principle find many new non-trivial non-conformal transitions.
These new transitions will generically couple the two chains and correspond to phase transitions between non-trivial, potentially topological, phases. 

\subsection{$\mbb{Z}_2\times\mbb{Z}_4$ symmetric quantum spin chains}

\begin{table}[t!]\centering
    \begin{tabular}{c  c c c c c c} \toprule
    \multirow{2}{2cm}{Lagrangian subgroups} & \multirow{2}{2cm}{Generating set} & \multirow{2}{1.8cm}{Fusion Structure} &\multirow{2}{2cm}{Image of $\Pi$}& \multirow{2}{1cm}{$\ms H$}  & \multirow{2}{1cm}{$\psi(\ms h_1,\ms h_2)$} & \multirow{2}{2cm}{Gapped phase} \\\\
    \midrule
    $\mcal{L}_1$ & $e_L,e_R$ & $\mbb{Z}_2\times\mbb{Z}_4$ & $1_L,1_R$ & $\mbb{I}$ & 1&\tenofo{SSB}
      \\
      $\mcal{L}_2$ & $m_L,e_R$ & $\mbb{Z}_2\times\mbb{Z}_4$ & $m_L$ & $\mbb{Z}^L_2$ & 1&  \tenofo{PSB}$_L$
      
      \\
      $\mcal{L}_3$ & $e_L,m_R$ & $\mbb{Z}_2\times\mbb{Z}_4$ & $m_R$ & $\mbb{Z}^R_4$ & 1&\tenofo{PSB}$_R$
      
      \\ 
      $\mcal{L}_4$ & $e_Le_R,m_Lm_R^2$ & $\mbb{Z}_2\times\mbb{Z}_4$ & $m_Lm_R^2$ & $\mbb{Z}^D_2$ & 1&\tenofo{PSB}$[\mbb{Z}^D_2]$
      
      \\
      $\mcal{L}_5$ & $e_L,e_R^2,m_R^2$ & $\mbb{Z}_2\times\mbb{Z}_2\times\mbb{Z}_2$ & $m_R^2$ & $\mbb{Z}^{(4)}_2$ & 1&\tenofo{PSB}$[\mbb{Z}^{(4)}_2]$
      
      \\
      $\mcal{L}_6$ & $e_Le_R^2,m_Lm_R$ & $\mbb{Z}_2\times\mbb{Z}_4$ & $m_Lm_R$ & $\mbb{Z}^D_4$ & 1&\tenofo{PSB}$[\mbb{Z}^D_4]$
      
      \\
      $\mcal{L}_{7}$ & $m_L,e_R^2,m_R^2$ & $\mbb{Z}_2\times\mbb{Z}_2\times\mbb{Z}_2$ & $m_L,m_R^2$ & $\mbb{Z}_2\times\mbb{Z}^{(4)}_2$ & 1&\tenofo{SPT}$^{(1)}_0$
      
      \\
      $\mcal{L}_{8}$ & $e_Lm_R^2,m_Le_R$ & $\mbb{Z}_2\times\mbb{Z}_4$ & $m_L,m_R^2$ & $\mbb{Z}_2\times\mbb{Z}^{(4)}_2$ &$(-1)^{h_{1,L} h_{2,R}}$& \tenofo{SPT}$^{(1)}_1$
      
      \\ 
      $\mcal{L}_9$ & $m_L,m_R$ & $\mbb{Z}_2\times\mbb{Z}_4$ & $m_L,m_R$ & $\mbb{Z}^L_2\times\mbb{Z}^R_4$ &1& \tenofo{SPT}$^{(2)}_0$
      
      \\
      $\mcal{L}_{10}$ & $e_Lm_R,m_Le_R^2$ & $\mbb{Z}_2\times\mbb{Z}_4$ & $m_R,m_L$& $\mbb{Z}^L_2\times\mbb{Z}^R_4$ &$(-1)^{h_{1,L} h_{2,R}}$& \tenofo{SPT}$^{(2)}_1$
      \\ 
    \bottomrule
    \end{tabular}
    \caption{The Lagrangian subgroups for $\ms G=\mbb{Z}^L_2\times\mbb{Z}^R_4$ and their various properties. $\mbb{Z}_2^D$ and $\mbb{Z}_4^D$ denote the ``diagonal" subgroups, and $\mbb{Z}^{(4)}_2$ denotes the $\mbb{Z}_2$ subgroup of $\mbb{Z}_4$. There are two classes of SPT states denoted as $\tenofo{SPT}^{(1)}$, protected by $\mbb{Z}_2\times\mbb{Z}^{(4)}_2$ symmetry, and $\tenofo{SPT}^{(2)}$, protected by $\mbb{Z}^L_2\times\mbb{Z}^R_4$ symmetry. Since $H^2(\mbb{Z}_2\times \mbb{Z}_4,\ms U(1))\simeq H^2(\mbb{Z}_2\times\mbb{Z}_2,\ms U(1))\simeq\mbb{Z}_2$, each classes of SPT states contains two inequivalent phases denoted as $\tenofo{SPT}^{(1)}_0$ and $\tenofo{SPT}^{(1)}_1$ for the first class and $\tenofo{SPT}^{(2)}_0$ and $\tenofo{SPT}^{(2)}_1$ for the second class. The 2-cocycle is evaluated on the group elements $\ms h_i=(h_{i,L},h_{i,R}) \in \ms H$ for $i=1,2$.}
    \label{tab:lagrangian subgroups of Z2*Z4}
\end{table}

We now turn to the $\mbb{Z}^L_2\times\mbb{Z}^R_4$ global symmetry. Even though this example mixes $\mbb{Z}_2$ and $\mbb{Z}_4$ symmetry, surprisingly there is a large duality group of order 128. This symmetry group has two different fusion structures. Although there is more than one fusion structure, this example still belongs to the limit where topological holography is powerful. A thorough analysis of the phase diagram of such models will appear elsewhere. In this example, we restrict ourselves to computing the conformal spectra of a few interesting phase transitions.

\smallskip The group of excitations of the bulk theory is
\begin{equation}
    \mcal{A}=\mbb{Z}^L_2\times\mbb{Z}^R_4\times\tenofo{Rep}(\mbb{Z}^L_2\times\mbb{Z}^R_4)\simeq \mbb{Z}^L_2\times\mbb{Z}^R_4\times \mbb{Z}^L_2\times\mbb{Z}^R_4\simeq \langle e_L,m_L,e_R,m_R\rangle, 
\end{equation}
where the generators satisfy the following relations
\begin{equation}
    e_L^2=m_L^2=1, \qquad e_R^4=m_R^4=1.
\end{equation}

\smallskip There are $10$ Lagrangian subgroups/gapped phases as tabulated in Table \ref{tab:lagrangian subgroups of Z2*Z4}. The Hilbert space of the corresponding spin chain  is
\begin{equation}
    \mcal{H}=\bigotimes_{i\in\tenofo{sites}}\mbb{C}^2\otimes\mbb{C}^4.
\end{equation}
The symmetry operator
\begin{equation}
    \mcal{U}_{\ms g_L,\ms g_R}=\bigotimes_i\left(\sigma^x_{i}\right)^{\ms g_L}\otimes X_i^{\ms g_R}, \qquad \ms g_L=0,1, \quad \ms g_R=0,1,2,3.
\end{equation}
$X_i$ and $Z_i$ denote the $\mbb{Z}_4$ generalizations of Pauli matrices. The spin-chain operators corresponding to the electric and magnetic bulk excitations are given by (see equation \eqref{eq:Wilson_operator_to_lattice_operator})
\begin{equation}
\begin{aligned}
e_L \longrightarrow&\; \sigma_{i}^z\sigma_{i+1}^z, \qquad m_L\longrightarrow \sigma_{i}^x \\
e_R \longrightarrow&\; Z_{i}Z^3_{i+1}, \qquad m_R\longrightarrow X_{i}.
\end{aligned}
\end{equation}
Spin-chain operators for other dyons can be obtained similarly. The fixed-point Hamiltonians associated to each of the gapped phases can be obtained using \eqref{eq:the Hamiltonian associated to a Lagrangian subgroup}. The Hamiltonian associated to the SSB phase is given by\footnote{We label the Hamiltonians with the corresponding gapped phase, described by a Lagrangian subgroup. Therefore, the Hamiltonian associated to the Lagrangian subgroup $\mcal{L}_i$ in Table \ref{tab:lagrangian subgroups of Z2*Z4} is denoted as $H_i$.}
\begin{equation}
    H_1=-\sum_{i}\left[\sigma_{i}^z\sigma_{i+1}^z+Z_{i}Z^3_{i+1}\right]+\tenofo{h.c.}.
\end{equation}
The fixed-point Hamiltonians associated to the five possible partial symmetry-breaking phases are
\begin{equation}
    \begin{aligned}
        H_2&=-\sum_i\left[\sigma^x_i+Z_iZ^3_{i+1}\right]+\tenofo{h.c.},
        \\
        H_3&=-\sum_i\left[\sigma^z_i\sigma^z_{i+1}+X_i\right]+\tenofo{h.c.},
        \\
        H_4&=-\sum_i\left[\sigma^z_i\sigma^z_{i+1}Z_iZ^3_{i+1}+\sigma^x_iX_i^2\right]+\tenofo{h.c.},
        \\
        H_5&=-\sum_i\left[\sigma^z_i\sigma^z_{i+1}+Z_i^2Z_{i+1}^2+X_i^2X_{i+1}^2\right]+\tenofo{h.c.},
        \\
        H_6&=-\sum_i\left[\sigma^z_i\sigma^z_{i+1}Z_i^2Z_{i+1}^2+\sigma^x_iX_i\right]+\tenofo{h.c.}.
    \end{aligned}
\end{equation}
The fixed-point Hamiltonians corresponding to $\mbb{Z}_2\times\mbb{Z}_2^{(4)}$ SPT phases are
\begin{equation}
    \begin{aligned}
        H_7&=-\sum_i\left[\sigma^x_i+Z_i^2Z_{i+1}^2+X_i^2\right]+\tenofo{h.c.},
        \\
        H_8&=-\sum_i\left[\sigma^z_i\sigma^z_{i+1}X_{i+1}^2+\sigma^x_iZ_iZ^3_{i+1}\right]+\tenofo{h.c.},
    \end{aligned}
\end{equation}
while $\mbb{Z}_2\times\mbb{Z}_4$ SPT phases are
\begin{equation}
    \begin{aligned}
        H_9&=-\sum_i\left[\sigma^x_i+X_i\right]+\tenofo{h.c.},
        \\
        H_{10}&=-\sum_i\left[\sigma^z_i\sigma^z_{i+1}X_{i+1}+\sigma_i^xZ_i^2Z_{i+1}^2\right]+\tenofo{h.c.}.
    \end{aligned}
\end{equation}
Similar to the previous examples, one can study the phase diagram of the simplest minimal model $H=\sum_{a=1}^{10} t_a H_a$ using dualities. However, here we will confine ourselves to the study of few transitions.

\smallskip The duality group is generated by partial electric-magnetic dualities, universal kinematical dualities (elements of $\tenofo{Aut}(\mbb{Z}_2\times\mbb{Z}_4)\simeq D_4)$, and universal dynamical dualities (elements of $H^2(\mbb{Z}_2\times\mbb{Z}_4,\ms U(1))\simeq\mbb{Z}_2$). However, it turns out that the following three dualities generate the full duality group  $\mcal{G}[\mbb{Z}_2\times\mbb{Z}_4]$
\begin{equation}
    \begin{aligned}
    \sigma_L&:\{m_L\mapsto e_L,e_L\mapsto m_L\},
    \\
    \sigma_R&:\{m_R\mapsto e_R,e_R\mapsto m_R\},
    \\
    k&:\{m_L\mapsto m_Le_R^2,m_R\mapsto e_Lm_R\}. 
    \end{aligned}
\end{equation}
Physically $k\in H^2(\mbb{Z}_2\times\mbb{Z}_4,\ms U(1))$ corresponds to a domain wall in the bulk with a $\mbb{Z}_2\times\mbb{Z}_4$ SPT phase. The duality group is 
\begin{equation}
    \mcal{G}[\mbb{Z}_2\times\mbb{Z}_4]=(D_4\times D_4)\rtimes\mbb{Z}_2. 
\end{equation}
The minimal transition $\mscr{C}_{19}$ between $\mcal{L}_1$ and $\mcal{L}_9$ is given by decoupled copies of Ising and $\mbb{Z}_4$ parafermion CFTs (see Appendix \ref{sec:conformal spectra at criticality} for their twisted partition functions). Under the duality $k$, $\mcal{L}_1$ is invariant while $\mcal{L}_9\mapsto\mcal{L}_{10}$ which maps $\mscr{C}_{19}$ to the topological SSB transition $\mscr{C}_{1,10}$. The twisted partition functions of $\mscr{C}_{1,10}$ is given by the following orbifold
\begin{equation}\label{eq:generalized twisted partition functions of C1,10 transition}
    \mcal{Z}_{(\ms g^\vee_L,\ms g^\vee_R),(\ms h^\vee_L,\ms g^\vee_R)}(\mscr{C}_{1,10})=(-1)^{\ms h^\vee_L\ms g^\vee_R-\ms h^\vee_R\ms g^\vee_L}\mcal{Z}^{\tenofo{Ising}}_{\ms g^\vee_L,\ms h^\vee_L}\mcal{Z}^{4-\tenofo{Para}}_{\ms g^\vee_R,\ms h^\vee_R},
\end{equation}
where $\mcal{Z}^{4-\tenofo{Para}}_{\ms g^\vee_R,\ms h^\vee_R}$ denotes the twisted partition function of $\mbb{Z}_4$ parafermion CFT. The phase factor in \eqref{eq:generalized twisted partition functions of C1,10 transition} is the partition function of the SPT phase on the duality domain-wall labeled by $k$. Similar to the previous $\mbb{Z}_3\times\mbb{Z}_3$ example, the partition function with periodic boundary condition is identical to that of the decoupled Ising and $\mbb{Z}_4$ parafermion CFTs ($\mscr{C}_{19}$ transition). However, the twisted sectors of $\mscr{C}_{1,10}$ criticality is twisted by the SPT cocycle and hence differs from $\mscr{C}_{19}$. This is another example of topological SSB or, in the language of \cite{VerresenThorngrenJonesPollmann201905}, a nontrivial $\mbb{Z}_2\times\mbb{Z}_4$ symmetry-enriched CFT.

\smallskip Consider the duality 
\begin{equation}
    \sigma:\{m_L\mapsto e_Lm_R^2,e_L\mapsto m_Le_R^2, m_R\mapsto m_Le_R^3,e_R\mapsto e_Lm_R\},
\end{equation}
which maps $\mcal{L}_1\mapsto \mcal{L}_{10}$ and $\mcal{L}_9\mapsto\mcal{L}_8$. Therefore, this maps $\mscr{C}_{19}$ to $\mscr{C}_{8,10}$ which is the criticality between the two nontrivial SPT phases protected by $\mbb{Z}_2\times\mbb{Z}_2^{(4)}$ and $\mbb{Z}_2\times\mbb{Z}_4$ symmetries, respectively. The  generalized twisted partition functions are
\begin{equation}
    \mcal{Z}_{(\ms g^\vee,\ms h^\vee)}(\mscr{C}_{8,10})=\frac{1}{8}\sum_{\substack{\ms g_L,\ms g_R \\ \ms h_L,\ms h_R}}(-1)^{\left[(\ms g_L-\ms g^\vee_R)(\ms h^\vee_L-2\ms h_R)-\ms h_L\ms g^\vee_L\right]}\omega^{\left[(\ms g_R-2\ms g^\vee_L)(\ms h^\vee_R-2\ms h_L)-3\ms g^\vee_R\ms h_R\right]}\mcal{Z}^{\tenofo{Ising}}_{\ms g_L,\ms h_L}\mcal{Z}^{4-\tenofo{4-Para}}_{\ms g_R,\ms h_R}.
\end{equation}

\subsection{$\mbb Z_{2}\times \mbb Z_{2}\times \mbb Z_{2}$ symmetric quantum spin chains}

Topological holography is extremely powerful for symmetry groups of the form $\mbb{Z}_2\times\cdots\times\mbb{Z}_2$. These groups have a single fusion structure and many dualities. For $\ms G=\mbb{Z}_2\times \mbb{Z}_2 \times \mbb{Z}_2$, the duality group is $\mcal{G}[\ms G]\simeq S_8$ with over forty thousand dualities. For $\ms G=\mbb{Z}_2^{\times 4}$, the duality group is $O_8^+(2)\rtimes\mbb{Z}_2$ with over 350 million dualities. In this example, we will not delve into great details, as the space of $\mbb{Z}_2^{\times 3}$ theories is very rich. Further investigation will appear elsewhere. Here, we will make few simple remarks to highlight the rich geometric structure of the minimal-model phase diagram. 

\smallskip The bulk theory has $64$ anyons
\begin{equation}
    \mc A=\ms G\times \text{Rep}(\ms G)\simeq \ms G^{2}=\langle m_1,m_2,m_3,e_1,e_2,e_3\rangle.
\end{equation}
The duality group is generated by
\begin{enumerate}
    \item \textbf{Partial electric-magnetic dualities}: There are three independent partial em dualities which we denote as $\sigma_{i}$ with $i=1,2,3$, which act on the generators of $\mc A$ as
    \begin{align}
        \sigma_{i}:
        \begin{bmatrix}
            m_{j}\\
            e_{j}
        \end{bmatrix}
        \longmapsto  
                \begin{bmatrix}
            (1-\delta_{ij})m_{j}+\delta_{ij}e_j\\
            (1-\delta_{ij})e_j+\delta_{ij}m_j
        \end{bmatrix}.
    \end{align}
    \item  \textbf{Universal kinematical symmetries}: These dualities are related to authomorphism group of $\mbb{Z}_2^{\times 3}$. They are given by
    \begin{gather}
        \begin{aligned}
        h_1&:\{m_1\mapsto m_2,m_2\mapsto m_1,e_1\mapsto e_2,e_2\mapsto e_1\},
        \\
        h_2&:\{m_1\mapsto m_2,m_2\mapsto m_3,m_3\mapsto m_1,e_1\mapsto e_2,e_2\mapsto e_3,e_3\mapsto e_1\},
        \\
        h_3&:\{m_1\mapsto m_1m_2,e_2\mapsto e_1e_2\}.
        \end{aligned}
    \end{gather}
    Together, $h_1$ and $h_2$ generate the permutation group of three elements $S_3$, while the full group generated by $h_{1}$, $h_{2}$ and $h_{3}$ is the group $\tenofo{PSL}(3,2)$ of order 168. 

    \item  \textbf{Universal dynamical symmetries}: These dualities are classified by $H^{2}(\ms G, \ms{U}(1))=\mbb Z_2^3$ and correspond to bulk duality-walls with SPT phases. They act on anyons as
    \begin{gather}
        \begin{aligned}
        k_1&:\{e_1\mapsto e_1m_2, e_2\mapsto m_1e_2\},
        \\
        k_2&:\{e_2\mapsto e_2m_3,e_3\mapsto m_2e_3\},
        \\
        k_3&:\{e_1\mapsto e_1m_3,e_3\mapsto m_1e_3\}.   
        \end{aligned}
    \end{gather}
\end{enumerate}
It turns out that the full group $\mc G[\ms G]=S_8$ can be generated by $\sigma_1, k_1$ and $k_2$.
\begin{table}[H]\centering
    \begin{tabular}{c  c   c c } \toprule
    \multirow{2}{2cm}{Lagrangian subgroups} & \multirow{2}{2cm}{Generating set} &  \multirow{2}{1cm}{$\ms H$}  & \multirow{2}{1cm}{$\psi(\ms h, \overline{\ms h})$} \\\\
    \midrule
    $\mcal{L}_1$ &    $e_1,e_2,e_3$       &    $\mbb{I}$         & 1 \\     
    $\mcal{L}_2$ &    $m_1,e_2,e_3$       &    $\mbb{Z}_{2}^{(1)}$   & 1 \\
    $\mcal{L}_3$ &    $e_1,m_2,e_3$       &    $\mbb{Z}_{2}^{(2)}$   & 1 \\ 
    $\mcal{L}_4$ &    $e_1,e_2,m_3$       &    $\mbb{Z}_{2}^{(3)}$   & 1 \\
    $\mcal{L}_5$ &    $m_1m_2, e_3, e_1e_2$       &    $\mbb{Z}_{2}^{(1-2)}$   & 1 \\
    $\mcal{L}_6$ &    $m_2m_3, e_1,e_2e_3$        &    $\mbb{Z}_{2}^{(2-3)}$   & 1 \\
    $\mcal{L}_7$ &    $m_1m_3, e_2, e_1e_3$       &    $\mbb{Z}_{2}^{(1-3)}$   & 1 \\
    $\mcal{L}_8$ &    $m_1m_2m_3, e_1e_2, e_2e_3$       &    $\mbb{Z}_{2}^{(1-2-3)}$   & 1 \\
    $\mcal{L}_9$ &    $m_1,m_2,e_3$       &    $\mbb{Z}_{2}^{(1)}\times \mbb{Z}_{2}^{(2)}$   & 1 \\
    $\mcal{L}_{10}$ &    $m_1e_2,m_2e_1,e_3$       &    $\mbb{Z}_{2}^{(1)}\times \mbb{Z}_{2}^{(2)}$   & $(-1)^{h_1\bar{h}_2}$ \\
    $\mcal{L}_{11}$ &    $e_1,m_2,m_3$       &    $\mbb{Z}_{2}^{(2)}\times \mbb{Z}_{2}^{(3)}$   & 1 \\
    $\mcal{L}_{12}$ &    $e_1,m_2e_3,m_3e_2$       &    $\mbb{Z}_{2}^{(2)}\times \mbb{Z}_{2}^{(3)}$   & $(-1)^{h_2\bar{h}_3}$ \\
    $\mcal{L}_{13}$ &    $m_1,e_2,m_3$       &    $\mbb{Z}_{2}^{(1)}\times \mbb{Z}_{2}^{(3)}$   & 1 \\
    $\mcal{L}_{14}$ &    $m_1e_3,e_2,m_3e_1$       &    $\mbb{Z}_{2}^{(1)}\times \mbb{Z}_{2}^{(3)}$   & $(-1)^{h_1\bar{h}_3}$ \\
    $\mcal{L}_{15}$ &    $m_1m_2,e_1e_2,m_3$       &    $\mbb{Z}_{2}^{(1-2)}\times \mbb{Z}_{2}^{(3)}$   & 1 \\ 
    $\mcal{L}_{16}$ &    $m_1m_2e_3,e_1e_2,m_3e_1$       &    $\mbb{Z}_{2}^{(1-2)}\times \mbb{Z}_{2}^{(3)}$   & $(-1)^{h_1\bar{h}_3+h_2\bar{h}_3}$ \\
    $\mcal{L}_{17}$ &    $m_1,m_2m_3,e_2e_3$       &    $\mbb{Z}_{2}^{(1)}\times \mbb{Z}_{2}^{(2-3)}$   & 1 \\ 
    $\mcal{L}_{18}$ &    $m_1e_2, m_2m_3e_1,e_2e_3$       &    $\mbb{Z}_{2}^{(1)}\times \mbb{Z}_{2}^{(2-3)}$   & $(-1)^{h_1\bar{h}_2+h_1\bar{h}_3}$ \\
    $\mcal{L}_{19}$ &    $m_1m_3,m_2,e_1e_3$       &    $\mbb{Z}_{2}^{(1-3)}\times \mbb{Z}_{2}^{(2)}$   & 1 \\ 
    $\mcal{L}_{20}$ &    $m_1m_3e_2, m_2e_1,e_1e_3$       &    $\mbb{Z}_{2}^{(1-3)}\times \mbb{Z}_{2}^{(2)}$   & $(-1)^{h_1\bar{h}_2+h_2\bar{h}_3}$ \\
    $\mcal{L}_{21}$ &    $m_1m_2,m_2m_3,e_1e_2e_3$       &    $\mbb{Z}_2^{(1-2)}\times \mbb{Z}_2^{(2-3)}$   & 1 \\ 
    $\mcal{L}_{22}$ &    $m_1m_2e_3, m_2m_3e_1,e_1e_2e_3$       &    $\mbb{Z}_{2}^{(1-2)}\times \mbb{Z}_{2}^{(2-3)}$   &  $(-1)^{h_1\bar{h}_2+h_2\bar{h}_3+h_3\bar{h}_1}$ \\
    $\mcal{L}_{23}$ &    $m_1,m_2,m_3$       &    $\mbb{Z}_{2}^{(1)}\times \mbb{Z}_{2}^{(2)}\times \mbb{Z}_{2}^{(3)}$   & 1 \\ 
    $\mcal{L}_{24}$ &    $m_1e_2,m_2e_1,m_3$       &  $\mbb{Z}_{2}^{(1)}\times \mbb{Z}_{2}^{(2)}\times \mbb{Z}_{2}^{(3)}$  &  $(-1)^{h_1\bar{h}_2}$ \\
    $\mcal{L}_{25}$ &    $m_1,m_2e_3,m_3e_2$       &  $\mbb{Z}_{2}^{(1)}\times \mbb{Z}_{2}^{(2)}\times \mbb{Z}_{2}^{(3)}$  &  $(-1)^{h_2\bar{h}_3}$ \\    
    $\mcal{L}_{26}$ &    $m_1e_3,m_2,m_3e_1$       &  $\mbb{Z}_{2}^{(1)}\times \mbb{Z}_{2}^{(2)}\times \mbb{Z}_{2}^{(3)}$  &  $(-1)^{h_3\bar{h}_1}$ \\
    $\mcal{L}_{27}$ &    $m_1e_2e_3,m_2e_1,m_3e_1$       &  $\mbb{Z}_{2}^{(1)}\times \mbb{Z}_{2}^{(2)}\times \mbb{Z}_{2}^{(3)}$  &  $(-1)^{h_1\bar{h}_2+h_1\bar{h}_3}$ \\ 
    $\mcal{L}_{28}$ &    $m_1e_2, m_2e_1e_3,m_3e_2$       &  $\mbb{Z}_{2}^{(1)}\times \mbb{Z}_{2}^{(2)}\times \mbb{Z}_{2}^{(3)}$  &  $(-1)^{h_1\bar{h}_2+h_2\bar{h}_3}$ \\
    $\mcal{L}_{29}$ &    $m_1e_3, m_2e_3,m_3e_1e_2$       &  $\mbb{Z}_{2}^{(1)}\times \mbb{Z}_{2}^{(2)}\times \mbb{Z}_{2}^{(3)}$  &  $(-1)^{h_1\bar{h}_3+h_2\bar{h}_3}$ \\
    $\mcal{L}_{30}$ &    $m_1e_2e_3, m_2e_1e_3,m_3e_1e_2$       &  $\mbb{Z}_{2}^{(1)}\times \mbb{Z}_{2}^{(2)}\times \mbb{Z}_{2}^{(3)}$  &  $(-1)^{h_1\bar{h}_2+h_2\bar{h}_3+h_3\bar{h}_1}$ \\
    \bottomrule
    \end{tabular}
    \caption{Lagrangian subgroups for $\ms G=\mbb{Z}_2\times\mbb{Z}_2\times\mbb{Z}_2$. $\mbb{Z}_2^{(i-j)}$ means the diagonal subgroup of $\mbb{Z}_2^i\times\mbb{Z}_2^j$, while $\mbb{Z}_{2}^{(1-2-3)}$ denotes the diagonal subgroup of $\mbb{Z}_2\times\mbb{Z}_2\times\mbb{Z}_2$.
    The 2-cocycles labelling the SPTs are evaluated on group elements $\ms h =(h_1,h_2,h_3)$, $\overline{\ms h} =(\bar{h}_1,\bar{h}_2,\bar{h}_3)$, where $\ms h, \overline{\ms h} \in \ms H$. 
    When $H\subset \mathbb Z_2^3$, for example, $\ms H=\mathbb Z_{2}^{(1)}\times \mathbb Z_2^{(2)}$ for $\mc L_{10}$, we use the same notation but simply set $h_3=\bar{h}_3=0$.}
    \label{tab:lagrangian subgroups of Z2*Z2*Z2}
\end{table}
\noindent All Lagrangian subgroups/gapped phases are tabulated in Table \ref{tab:lagrangian subgroups of Z2*Z2*Z2}. 

\smallskip Similar to the previous examples, we can write a fixed-point Hamiltonian for each of the thirty gapped phases in Table \ref{tab:lagrangian subgroups of Z2*Z2*Z2}. With this, we can consider the minimal Hamiltonian
\begin{equation}
    H=\sum_{a=1}^{30} t_a\,H_a, \qquad t_a\in\mbb{R}.
\end{equation}
The action of the duality group on this Hamiltonian corresponds to certain permutations of the coupling constants $t_a$. The phase is $30$-dimensional and complicated to describe. Here, we will make a few remarks about the phase diagram by attempting to decompose it into smaller known pieces. For example, we find several subgroup of $\mcal{G}[\ms G]\simeq S_8$ isomorphic to $(S_3\times S_3)\rtimes\mbb{Z}_2$. One such subgroups $(S_3\times S_3)\rtimes\mbb{Z}_2=\langle a,b\rangle$ is generated by 
\begin{equation}
    a:\begin{pmatrix}
        m_1 \\ m_2 \\ m_3 \\ e_1 \\ e_2 \\ e_3
    \end{pmatrix}
    \longmapsto
    \begin{pmatrix}
        e_1m_2e_3\\ m_2 \\ m_1m_2m_3e_1e_3 \\ m_1m_2e_3 \\  m_1m_2e_1e_2e_3 \\ e_3
    \end{pmatrix}
    \qquad 
    b:\begin{pmatrix}
        m_1 \\ m_2 \\ m_3 \\ e_1 \\ e_2 \\ e_3
    \end{pmatrix}
    \longmapsto
    \begin{pmatrix}
        m_1m_3\\ m_1m_3e_1e_3 \\ m_1m_3e_1e_2e_3 \\ m_1e_2e_3 \\  m_1m_2e_1e_2e_3 \\ e_1m_2e_3
    \end{pmatrix}
\end{equation}
Upon further study, one finds that the first $S_3$ subgroup permutes the Lagrangian subgroups $T_1=\{\mcal{L}_4,\mcal{L}_{26},\mcal{L}_{27}\}$ while the other $S_3$ subgroup permutes phases in $T_2=\{\mcal{L}_{11},\mcal{L}_{14},\mcal{L}_{16}\}$. Finally, the remaining $\mbb{Z}_2$ subgroup permutes $T_1$ and $T_2$. In other words, these six gapped phases generate a phase diagram identical to Figure \ref{Fig:ZTwoSqPhaseDiagram}. Similarly, there are other $(S_3\times S_3)\rtimes\mbb{Z}_2$ subgroups generating similar triangles often with shared corners generating an interesting geometrical shape. However, there is slight difference to the case of $\ms G=\mbb{Z}_2\times\mbb{Z}_2$: consider the Hamiltonian $H_1+H_{23}$ corresponding to three decoupled critical Ising chains with $c=\frac{3}{2}$. While two decoupled Ising chains had a single marginal operator, this CFT has three marginal operators generating three different $c=\frac{3}{2}$ lines (potentially planes or volumes) of CFTs.

\smallskip This suggests a simple geometric picture of the phase diagram. For $\ms G=\mbb{Z}_2$, the minimal model was essentially corresponded to a 1-simplex with ferromagnetic and paramagnetic fixed-point Hamiltonians as vertices and a $c=\frac{1}{2}$ in the center. For $\ms G=\mbb{Z}_2\times\mbb{Z}_2$, these 1-simplices become boundaries of 2-simplices (triangles in Figure \ref{Fig:ZTwoSqPhaseDiagram}) with interesting new features in the interior of the 2-simplices and a transitions with $c=1$. For $\ms G=\mbb{Z}_2\times\mbb{Z}_2\times\mbb{Z}_2$, it appears that that several 2-simplices are glued together to form higher-dimensional geometric shapes (potentially 3-simplices similar to Figure \ref{fig:Z3xZ3ProposedPhaseDiagram}) with new features in the interior and $c=\frac{3}{2}$ criticalities. It is easy to see such a pattern persists as the symmetry group is extended by more copies of $\mbb{Z}_2$. 

\smallskip This example contains many interesting transitions including Landau-type, topological, topological SSB, and deconfined quantum critical transitions. However, there is a new kind of transition which is both deconfined and topological (for example between $\mcal{L}_{10}$ and $\mcal{L}_{14}$ or between $\mcal{L}_{20}$ and $\mcal{L}_{22}$). We call such transitions {\it Topological Deconfined Criticality}. As far as we aware, such transitions have not been studied in the literature. Using topological holography, we can compute the exact conformal spectra of many of these transitions and study their properties in detail. The in-depth analysis will appear elsewhere.

\section{Summary and future directions}
\label{sec:discussion and future directions}
In this work, we have outlined a framework to systematically study the global symmetry aspects of symmetric quantum systems in a given dimension using a topologically-ordered system in one higher dimension. The main idea of this work can be stated as a fruitful interplay between (1) {\it {\normalfont(}the topological nature of\,{\normalfont)} symmetry}, (2) {\it holography}, and (3) {\it dualities}. This led to a framework which is general in the sense that it can be applied to any dimension and to a broad class of symmetry structures, in particular those with finitely many symmetry operators. As illustrated in Section \ref{sec:examples}, using $2+1d$ topologically ordered systems and $1+1d$ quantum spin chains, topological holography is not merely an abstract framework but can be used to extract physically-interesting information for concrete condensed matter systems. 

\smallskip Let us briefly summarize the main results of this work pertaining to the study of $1+1d$ quantum systems  which are symmetric under a finite Abelian group $\ms G$, using the framework of topological holography.
In Section \ref{Sec:framework}, we built up the basic foundation by combining the two main ingredients in our story, i.e. symmetry and holography.
Using a $2+1d$ topological $\ms G$-gauge theory, we showed that its $1$-form symmetries (generated by topological line operators) can be used to construct the space of $\ms G$-symmetric operators or equivalently the space of $\ms G$-symmetric $1+1d$ theories.
More precisely, the space of $\ms G$-symmetric operators was encapsulated in an algebra dubbed the String Operator Algebra $\mbb{SOA}[\ms G]$, which was used to parameterize the space of $\ms G$-symmetric theories.
To demonstrate the utility of this somewhat abstract holographic framework, we showed how concrete generalized spin models can be efficiently studied using these structures.
Among our results, we could classify all gapped phases and explicitly construct the fixed-point (i.e. exactly-solvable) Hamiltonians and  order parameters for each of them. 

\smallskip We then included the third main ingredient of our story, i.e., dualities.
Within the holographic approach, dualities descend from 0-form symmetries of the $2+1d$ topological gauge theory, i.e., permutations of anyons or topological line operators of the topological gauge theory which preserve all the correlation functions such as braiding (self and mutual) statistics. 
These dualities form a group denoted as $\mathcal{G}[\ms G]$ and each duality operation is associated with a codimension-1 (i.e., a two dimensional surface) topological operator/defect in the holographic bulk.
Bringing the topological surfaces to the boundary leads to the action of dualities on the space of $\ms G$-symmetric theories.
We explicitly worked out the action of dualities on $\mbb{SOA}[\ms G]$.
Since any Hamiltonian can be constructed from elements of $\mbb{SOA}[\ms G]$, we thus deduced the action of dualities on generic symmetric Hamiltonians. 
We argued that the action of dualities restricted to the space of symmetric theories is local while the same action on the space of all operators, including the non-symmetric ones, is in general non-local.
Another important observation was the emergence of new symmetries in self-dual theories.
Such symmetries are associated to certain topological line defects and are generically non-Abelian and sometimes non-invertible.  

\smallskip Remarkably all these non-trivial lessons about $\ms G$ symmetric theories were extracted by simple algebraic manipulations on the set of anyons.

\smallskip In Section \ref{topological holography: application}, we studied some of the implications of the framework we laid out in the previous section. 
It is known that gapped boundaries in $2+1d$ can be labeled by a subgroup $\ms H$ of $\ms G$ and an SPT twist, i.e.,  an element of $H^2(\ms H, \ms U(1))$. We explained how this data can be extracted easily from the condensed set of anyons that label a particular gapped boundary. An important question is---for which symmetry groups is topological holography most powerful? We could answer this question by introducing the notion of fusion structure of a gapped phase. We explained that this concept quantifies the power of applicability of topological holography---less number of fusion structures and larger duality group is the regime in which topological holography is more powerful. Additionally, the fusion structures were shown to impose constraints on possible dualities between gapped phases and critical points/regions. Using these constraints, we uncovered that the space of $\ms G$-symmetric conformal field theories can be decomposed into blocks that are labeled by a subset of possible fusion structures for a given global symmetry $\ms G$. 

\smallskip Possibly, the most powerful aspect of the topological holography is that it provides tools to obtain the spectrum of the dual critical transitions, using the spectra of a known transition as input. We derived a general formula for the generalized symmetry-twisted partition functions of dual transitions, which could be computed using the transformation of anyons under the duality in question. We thus demonstrated that topological holography  can be used to extract stringent constraints on the structure of the phase diagram and compute physical quantities. 

\smallskip Finally, we illustrated this claim in Section \ref{sec:examples}, by applying the above framework to various examples of the symmetry group $\ms G$. We numerically verified our predictions of phase-diagrams and conformal spectra of non-trivial critical transitions based on toplogical holography. We studied various new types of phase-transitions using dualites.

\smallskip There are many open research directions that follow from our work and would enrich the formalism of topological holography. Let us briefly list some of them:
\begin{itemize}
    \item {\bf Non-Abelian and non-invertible symmetries:} In this work, we have focused on elucidating the details of the formalism for the case of finite Abelian group symmetries. 
    An obvious generalization would be to consider symmetries corresponding to a non-Abelian discrete group $\ms G$ or more generally, a non-invertible fusion category $\mathcal C$. 
    In these cases, the bulk topological order would be the Turaev-Viro-Barrett-Westbury statesum model \cite{Turaev:1992hq, Barrett:1993ab} or equivalently the Levin-Wen Hamiltonian models \cite{Levin_Wen_2005} constructed with the fusion category $\text{Vec}_{\ms G}$ and $\mathcal C$ as input respectively (see \cite{Thorngren_anyon}).
    
    \item {\bf Anomalous quantum systems:} Another natural extension of the present work is to holographically study quantum systems with 't Hooft anomalies (or Lieb-Schultz-Mattis constraints) for finite Abelian groups.
    Holographically these symmetry structures correspond to Abelian Dijkgraaf-Witten theory with non-trivial cohomology twists $[\omega]\in H^3(\ms G,\ms U(1))$. 
    
    \item {\bf Fermionic systems}: We have only considered bulk theories which are described by Abelian bosonic TQFTs with gapped boundaries, which are known to be described by Dijkgraaf-Witten theories \cite{KaidiKomargodskiOhmoriSeifnashriShao102107}. Another generalization of the present paper would be to consider $2+1d$ topologically-ordered systems with elementary fermionic excitations whose low-energy description are given by (Abelian or non-Abelian) spin-TQFT \cite{GuWen_2014, Gaiotto:2015zta, Bhardwaj_2017, Putrov:2016qdo} which have a richer structure. 
    One difference is that the boundary of such theories can host topologically non-trivial phases, even in the absence of any global symmetries. 
    Therefore, the classification of beyond-Landau phases is richer in the context of fermionic systems.
    The study of gapped phases and phase transitions from a holographic perspective would be an interesting extension of the present work.
    
    \item {\bf Higher dimensions:} 
    Generalizations to higher dimensions, in particular, to explore the application of topological holography to $2+1$ quantum magnets, which are known to possess rich phase diagrams would be an extremely interesting direction.
    Much less is known about $3+1d$ topological order as compared with $2+1d$ topological order. 
    Moreover $3+1d$ topological orders generically contain topological line and surface operators, i.e., 2-form and 1-form symmetries respectively, which holographically map to 1-form and 0-form symmetries respectively. 
    The boundary theories can therefore have much richer symmetry structures which form some kind of a fusion 2-category. 
    
    \item {\bf Gauging anyonic symmetries:} It is known that gauging anyonic symmetries of $2+1d$ topological orders lead to new  topological phases called twist liquids \cite{Teo_2015, Barkeshli_2019} with generically non-invertible topological operators. 
    On the other hand, gauging anyonic symmetries is related to orbifolding the edge CFT. It would be very interesting to be able to understand the gapped boundaries of these exotic phases in terms of the gapped boundaries of the parent topological order, such as the relation between their phase diagrams, through the lens of topological holography (see \cite{Chen:2017txd,Chen:2017zzx} for related studies pertaining to gapless edges of topological orders in $2+1d$ and $3+1d$ respectively). 
    
    {\bf }
\end{itemize}
We believe that topological holography is a valuable tool in the study of symmetric quantum systems and uncovering its full scope and details would enrich our understanding of the phase diagrams, their possible geometrical interpretation, and in general the nature and the possible unification of the theory of phase transitions between various types of quantum phases of matter. The above-mentioned research directions would pave the way for such an understanding.

\section*{Acknowledgements}
We would like to thank Benjamin Beri, Gunnar M\"oller, Clement Delcamp and Jens H. Bardarson for fruitful discussions. We thank Kevin Costello for critical comments on a draft of this work. HM is supported by Engineering and Physical Sciences Research Council under New Horizon grant award no. EP/V048678/1. The work of SFM is funded by the Natural Sciences and Engineering Research Council of Canada (NSERC) and also in part by the Alfred P. Sloan Foundation, grant FG-2020-13768. AT is supported by the Swedish
Research Council (VR) through grants number 2019-04736 and 2020-00214.

\appendix

\addtocontents{toc}{\protect\setcounter{tocdepth}{1}}

\section{From \texorpdfstring{$\ms G$}{}-topological orders to \texorpdfstring{$\ms G$}{}-symmetric quantum spin chains}
\label{Sec:holographic perspective}

In this appendix, we provide some computational details pertaining to how $2+1$-dimensional $\ms G$ topological gauge theories can be employed as a theoretical tool to study the space of $1+1$-dimensional $\ms G$-symmetric quantum spin chains. 
Our approach takes as an input a discrete Abelian group $\ms G$ and the corresponding $\ms G$ topological gauge theory $\mc T_{\ms G}$ with a trivial action, also referred to as $\ms G$ quantum double. 
We study the theory $\mc T_{\ms G}$ on a spatial semi-infinite cylinder and show that the algebra of bulk line operators restricted to the spatial circle, i.e., the boundary of the cylinder is isomorphic to the algebra of $\ms G$ symmetric operators acting on the Hilbert space of the circle.
We use this insight to flesh out the space of $\ms G$ symmetric quantum spin models. 

\medskip \noindent The theory $\mc T_{\ms G}$ contains a set of anyons or line operators which fall into two classes, namely Wilson lines and 't-Hooft lines. The Wilson lines are labeled by a charge $\alpha \in \text{Rep}(\ms G)$ while the 't-Hooft lines are labeled by a flux $\ms g\in \ms G$. Therefore the most general line operator corresponds to a charge-flux composite also known as a dyon and carries the label $d\equiv(\ms g,\alpha)\in \mc A$. 
\begin{figure}
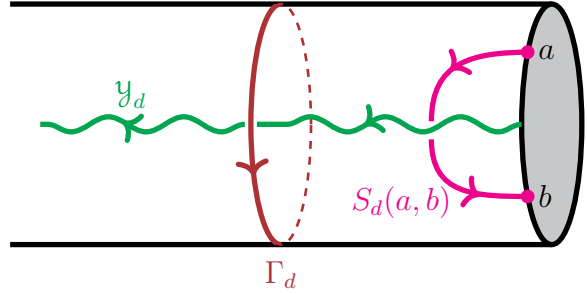
%
  \centering
    \VariousTypesOfBulkLineOperators
  \caption{An illustration of the different kinds of topological line operators for a topological order defined on a semi-infinite cylinder.}
  \label{Fig:open_chain}
\end{figure}
We divide the non-trivial line operators into three families (i) line operators $\Gamma_d $ which wrap the non-contractible cycle $[\ell_x]$ of the cylinder,\footnote{Here $[\cdot]$ refers to a homology class.} (ii) line operators $\mscr{Y}_d$ defined on the line $\ell_{y,a}$ that have one end on the boundary at $a$, the other end at infinity and are oriented away from the boundary and (iii) line operators with both ends on the boundary, denoted as $S_{d}(a,b)$ which have support on some oriented open line $\gamma_{ab}$ from the point $a$ to the point $b$ on the boundary.
In addition, there are bulk line operators defined on contractible loops which act trivially on the boundary theory as they commute with all other operators in the theory.
Note that the line operators of the type $\mscr{Y}_{d}$ and $S_{d}$ are labeled only by the location of the endpoints of the lines they have support on because the precise (bulk) location of these lines can be deformed by the action of the trivial contractible loop operators.
To summarize, the total set of operators that act non-trivially on the boundary Hilbert space is 
\begin{align}
    \Big\{\Gamma_d, \mscr{Y}_d(a), S_{d}(a,b)\Big\}_ {d\in \mc A;a,b \in S^{1}}.
    \label{eq:set_line_operators}
\end{align}
The algebra of these operators takes the form
\begin{align}
    X_{d}(\ell)X_{d'}(\ell')=e^{i\theta_{dd'}
    {\texttt{Intersect}[\ell, \ell']}
    }X_{d'}(\ell')X_{d}(\ell),
\end{align}
where $X_{d}(\ell),X_{d'}(\ell')$ denotes a line operator of the type in eq.~\eqref{eq:set_line_operators} with $d,d' \in \mc A$ and $\ell,\ell'$ one of the three kinds of lines described above. $\texttt{Intersect}[\ell, \ell']$ is the oriented intersection number of the lines $\ell$ and $\ell'$ and $\exp\left\{i\theta_{dd'}\right\}=\ms R_{\alpha}(\ms g')\ms R_{\alpha'}(\ms g)$ which corresponds to the phase accrued when dyons $d$ and $d'$ braid.
Graphically, the commutation relations of the line operators can be extracted from local moves of the form \eqref{eq:LineOperatorsLocalCrossing_Algebra}.
More explicitly, the line operators satisfy the following algebraic relations
\begin{gather} \label{eq:line_operator_algebra}
    \begin{aligned}
    \Gamma_d \Gamma_{d'}=&\;\Gamma_{d'}\Gamma_{d}, \\
    \Gamma_d \mscr{Y}_{d'}(a)=&\;e^{i\theta_{dd'}}\mscr{Y}_{d'}(a)\Gamma_d, \\
     \mscr{Y}_d(a) \mscr{Y}_{d'}(b)=&\;\mscr{Y}_{d'}(b)\mscr{Y}_{d}(a), \\
    S_d(a,b) \mscr{Y}_{d'}(c)=&\;e^{i\theta_{dd'}\texttt{Intersect} [\gamma_{ab},\ell_{y,c}]}\mscr{Y}_{d'}(c)S_d(a,b), \\
    S_d(a,b) S_{d'}(a',b')=&\;e^{i\theta_{dd'}\texttt{Intersect}[\gamma_{ab},\gamma_{a'b'}]}S_{d'}(a',b')S_d(a,b), \\
    S_d(a,b) \Gamma_{d'}=&\;\Gamma_{d'}S_{d}(a,b).
    \end{aligned}
\end{gather}
Notably in the above expressions, the intersection number remains unaltered if we deform the lines $\gamma_{ab}$ and $\ell_{y,a}$ as long as we keep the end-points fixed. Therefore, one is free to choose any representative line in order to evaluate the algebra. 

\begin{figure}
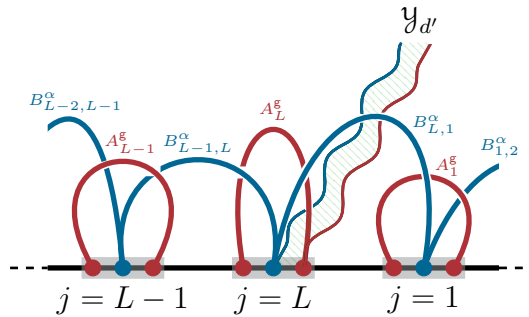

  \centering
  \SectorChangingYOperatorActingOnAPseudoSpinChain
  \caption{Within the holographic construction, one can obtain a sector $\mathcal H_{d'
  }:=:\mathcal H_{(\ms g',\alpha')}$ that transforms in the $\alpha'$ representation of $\ms G$ and has $\ms g'$-symmetry-twisted boundary conditions by acting with a semi-infinite line operator $\mathcal Y_{d'}$.}
  \label{Fig:twisted_bc}
\end{figure}

\medskip \noindent Having described the operator algebra, we now move on to analyzing how one may organize the boundary Hilbert space $\mc H$. Since all the $\Gamma_{d}$ operators mutually commute, we may block-diagonalize $\mc H$ into superselection sectors which are eigenspaces or  of $\Gamma_d $
\begin{align}
\mc H =&\;\bigoplus_{d\in \mc A}\mc H_{d} \nonumber \\
\mc H_{d}=&\; \Big\{ |\psi\rangle \big| \quad  \Gamma_{d'}|\psi\rangle=e^{i\theta_{dd'}}|\psi\rangle \Big\}.
\end{align}
Using the algebraic relations in eq.~\eqref{eq:line_operator_algebra}, it can be seen that the remaining operators $\left\{\mscr{Y}_{d}(a)\right\}$ and $\left\{S_{d}(a,b)\right\}$ act on the sectors $\mc H_{d'}$ as 
\begin{gather}
    \begin{aligned}
    \mscr{Y}_{d}(a):&\; \mc H_{d'}\longmapsto \mc H_{d'\otimes d} \\
    S_{d}(a,b):&\; \mc H_{d'}\longmapsto \mc H_{d'}, \\
    \end{aligned}
\end{gather}
where $d\otimes d'= (\ms g + \ms g', \alpha + \alpha')$. 
Since the subspace $\mc H_{d}$ is left invariant under the action of each $S_{d}(a,b)$, we can span $\mc H_{d}$ using a basis that diagonalizes a maximally commuting subalgebra of the algebra generated by $\left\{S_d(a,b)\right\}$.
For our purposes, it will prove useful to resolve the boundary into $2L$ points where $L$ is some integer. We allow pure charges/fluxes to terminate at odd/even points on the boundary, while more general dyonic operators need to be defined with a choice of framing.
We label the points on the boundary as $x_{i}$ with $i=1,\dots,2L$. 
Any pure charge or flux operator can be decomposed as a product of minimal-length operators
\begin{align}
    A^{\ms g}_{j}\equiv S_{(\ms g,0)}(x_{2j-1},x_{2j+1}), \quad B^{\alpha}_{j,j+1}\equiv S_{(0,\alpha)}(x_{2j},x_{2j+2}). 
\end{align}
Note that the operators $A^{\ms g}$ act on-site on an effective $L$-site pseudo-spin chain while the operator $B^{\alpha}$ acts on two neighboring sites. The algebra satisfied by these operators can be read-off from eq.~\eqref{eq:line_operator_algebra} and takes the form 
\begin{align}
    A^{\ms g}_{k}B^{\alpha}_{j,j+1}=\ms R_{\alpha}(\ms g)^{\delta_{k,j}-\delta_{k,j+1}}B^{\alpha}_{j,j+1}A^{\ms g}_{k}
    \label{eq:AB_algebra}
\end{align}
\begin{figure}
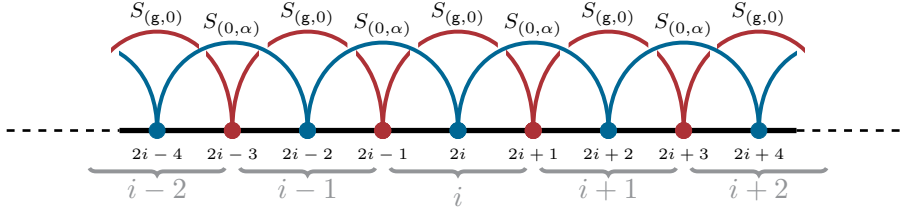
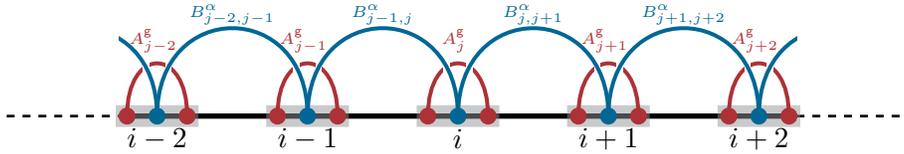
\centering
    \begin{subfigure}{\textwidth}\centering
    \StringOperatorsBetweenLatticeSites
    \caption{The construction of an effective pseudo-spin chain on the boundary of a $2+1d$ topological gauge theory involves regularizing the $1+1d$ boundary such that the electric and magnetic topological lines can end on the even and odd points of the regularized lattice.}
    \end{subfigure}
    \vfill\hfill 
    \begin{subfigure}{\textwidth}\centering
    \PseudoSpinWithAandBOperators
    \caption{The corresponding pseudo-spin chain is obtained by representing the algebra of bulk topological line operators on a tensor product Hilbert space. 
    Every even site of the regularized $1+1d$ boundary of the topological gauge theory realizes a local Hilbert space for the pseudo spin chain, on which the minimal magnetic operators act on site, while the minimal electric operators act on two neighboring sites. 
    }
    \end{subfigure}
    \vfill\hfill 
    \caption{An effective pseudo spin chain on the boundary of a $2+1d$ topological gauge theory.}
    \label{fig:mapping of a spin chain and string operators to pseudo-spin chains and A and B operators}
\end{figure}
with the additional constraint that $\prod_{j}A^g_{\ms j}=\Gamma_{(\ms g,0)}$ and $\prod_{j}B^{\alpha}_{j,j+1}=\Gamma_{(0,\alpha)}$. We identify $A^{\ms g}_{j}$ as the local representative of the global symmetry operator corresponding to the element $\ms g \in  \ms G$. Therefore, $\Gamma_{(\ms g,0)}$ is nothing but the global symmetry operator.  In order to construct the entire Hilbert space, one first starts with a reference state in $\mathcal H_{(0,0)}$ and uses the operator $\mscr{Y}_{d}$ to obtain a locally isomorphic state in the sector $\mathcal H_{d}$ (see Fig.~\ref{Fig:twisted_bc}).
The Hilbert space $\mc H_{d}$ can be spanned by the simultaneous eigenbasis of the operators $B^{\alpha}_{j,j+1}$, such that a given sector is constructed as
\begin{align}
    \mc H_{d}:= \text{Span}_{\mathbb C}\Bigg\{ |\underline{a},d\rangle=W_{d}|\underline{a},0\rangle \Bigg\},
\end{align}
where $\underline{a}=(a_{1,2},a_{2,3},\dots , a_{L-1,L},a_{L,1})$ where $a_{j,j+1}\in \ms G$. More formally $\underline{a}\in C^{1}(M,\ms G)$, i.e. a $\ms G$-valued 1-cochain on the one-dimensional pseudo-spin lattice denoted here as $M$. The eigenstates $|\underline{a},d\rangle$ are simultaneous eigenstates of $ B^{\alpha}_{j,j+1}$ and $\Gamma_{d}$
\begin{align}
 B^{\alpha}_{j,j+1}| \underline{a},d\rangle =&\; \ms R_{\alpha}(a_{j,j+1})|\underline{a},d\rangle,  \quad
\Gamma_{d'}|\underline{a},d\rangle
     = e^{i\theta_{d,d'}}|\underline{a},d\rangle.
\end{align}
Meanwhile, the operators $A^{g}_{j}$ act off-diagonally on the $\left\{|\underline{a},d\rangle\right\}$ basis as
\begin{align}
\prod_{j} A^{\ms g_j}_j| \underline{a},d\rangle 
=&\; |(a_{1,2}+\ms g_{2}-\ms g_1,a_{2,3}+\ms g_{3}-\ms g_2,\dots,a_{L,1}+\ms g_{L}-\ms g_1),d\rangle 
\nonumber \\
=&\;| \underline{a}+\delta \underline{g},d\rangle,    
\end{align}
where $\underline{\ms g}=(\ms g_1,\ms g_2,\dots ,\ms g_L )\in C^{0}(M,\ms G)$ a $\ms G$-values 0-cochain or simply a map from the vertices of the pseudo-spin chain lattice $M$ to the group $\ms G$. 
From this construction one can see that the $B^{\alpha}$ eigenbasis $\left\{|\underline{a},d\rangle \right\}$ is reminiscent of a $\ms G$ gauge theory where the $\underline{a}$ plays the role of the gauge field while the $A^{\ms g}$ operators implement gauge transformations.
This is of course not entirely physically meaningful as in the present case, the ``gauge transformations" are physical unlike in a gauge theory.
In a particular sector, $d=(\ms g, \alpha)$, there is an additional constraint that the holonomy of the ``gauge field" $\prod_{j}a_{j,j+1}=\ms g$.
From a spin-chain perspective, the Hilbert space sector $\mc H_{d}$ contains the states which have $\ms g$ symmetry twisted boundary conditions and transform in the $\alpha$ representation under global $\ms G$ symmetry. 

\begin{figure}
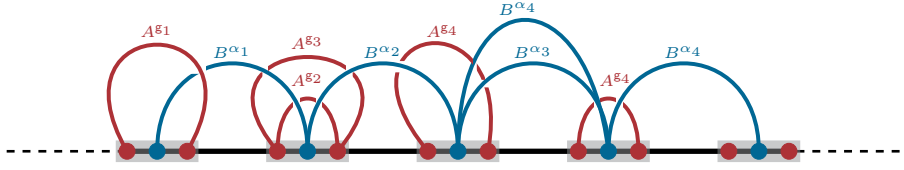
%
  \centering
  \PseudoSpinChainForDyonicsOperators
  \caption{Representation of a dyonic operator (with $\alpha_2=\ms g_5=0$) on the edge quantum spin chain. }
  \label{Fig:general_dyonic_operator}
\end{figure}

\medskip \noindent Next, let us focus on the local operators acting on the boundary pseudo-spin chain. Clearly the algebra of local operators is generated by $\mathbb A:=\left\{A^{\ms g}_j\right\}_{\ms g, j}$ and $\mathbb B:=\left\{B^{\alpha}_{j,j+1}\right\}_{\alpha,j}$ operators. We refer to this algebra as the string operator algebra (SOA), denoted as $\mathbb{ SOA}[\ms G]$. Since every operator in $\mathbb A$ and $\mathbb B$ commutes with $\Gamma_{d}$ for all $d\in \mc A$, we can further block decompose the SOA into eigenspaces of the $\Gamma_d$ operators. Therefore, we get
\begin{align}
 \mathbb {SOA}[\ms G]=\langle \mathbb {A},\mathbb B\rangle = \bigoplus_{d}   \mathbb {SOA}_{d}[\ms G]=\bigoplus_{d}\langle \mathbb {A}_{d}, \mathbb {B}_{d}\rangle, 
\end{align}
where $\mathbb{SOA}\in \mathcal L(\mc H_d)$, the space of linear operators acting on $\mc H_d$.
In order to obtain a more explicit representation of the SOA, one requires a convenient basis of operators that span the algebra. 
In particular, such a basis would play a crucial role in the construction of the space of $\ms G$-symmetric local Hamiltonians and consequently in systematizing the phase diagrams of $\ms G$-symmetric quantum systems. 
We would like to span the algebra $\mathbb{SOA}[\ms G]$ by operators that carry dyonic labels and have properties that encode topological correlations of the bulk dyonic line operators of $\mc T_{\ms G}$.
Let us consider a general Abelian group $\ms G=\prod_{j={\ms a}}^{M}\mathbb Z_{N_{\ms a}}$ with a fixed (arbitrary) ordering on its generators $(\ms{g}_{1},\ms{g}_{2},\dots ,\ms{g}_{M})$ such that $\ms{g}_{{\ms a}}>\ms{g}_{{\ms b}}$ for ${\ms a}>{\ms b}$. %
Then an edge pseudo-spin chain operators $\mathcal O_{d}$ can be constructed that correspond to a general dyonic label $d=(\ms g_1,\alpha_1;\ms g_2,\alpha_2; \dots ; \ms g_M,\alpha_M)\in \mc A$.
We define a minimal length operator $\mc O_{d}$ based at the site $i_0$ as
\begin{align}
    \mathcal O_{d}[i_0]=&\; A^{g_1}_{i_0+\eta_d[1]}B^{\alpha_1}_{i_0+\xi_d[1];i_0+\xi_d[1]+1}%
    \dots
    A^{g_M}_{i_0+\eta_d[M]}B^{\alpha_M}_{i_0+\xi_d[M];i_0+\xi_d[M]+1} \nonumber \\
    =&\; \prod_{\ms a=1}^{M}A^{g_{\ms a}}_{i_0+\eta_d[\ms a]}B^{\alpha_{\ms a}}_{i_0+\xi_d[{\ms a}];i_0+\xi_d[{\ms a}]+1},
    \label{eq:Od_operator}
\end{align}
where the various operators in the product are located to the right of the site $i_0$ where the sites are numbered in increasing order from left to right. A general dyonic operator takes the following form illustrated in Fig.~\ref{Fig:general_dyonic_operator}. The precise location of each operator can be fixed by the following rules: 
\begin{enumerate}
    \item The first non-trivial element in $d$ is located at $i_{0}$. That is: 
    \begin{enumerate}
        \item If $g_{\ms a_0}\neq 0$ and $g_{\ms a}=\alpha_{\ms b}=0$ for all $\ms a,\ms b< \ms a_0$, then the operator string $\mc O_d[i_0]$ starts with the operator $A^{g_{\ms a_0}}_{i_0}$.
        
        \smallskip \hspace{5.5cm} OR
        
        \item If $\alpha_{\ms a_0}\neq 0$ and $g_{\ms a}=\alpha_{\ms b}=0$ for all $\ms a \leq \ms a_0$ and $\ms b< \ms a_0$, then the operator string $\mc O_d[i_0]$ starts with the operator $B^{\alpha_{\ms a_0}}_{i_0,i_0+1}$. 
    \end{enumerate} 
    \item {{An $A$-operator is located at the site where the previous (non-vanishing) $B$-operator ends}.} The operators $A^{\ms g_{\ms a_0}}$ and $A^{\ms g_{\ms a_0+n}}$ are located at the same site iff $\alpha_{\ms b}=0$ for all $\ms a_{0}\leq \ms b < \ms a_0 +n$.%
    \item {{A $B$-operator begins from the site where the previous non-vanishing $A$-operator is located}.} The operators $B^{\alpha_{\ms b_0}}$ and $B^{\alpha_{\ms b_0+n}}$ are located on the same link iff $\ms g_{\ms a}=0$ for all $\ms b_0<\ms a\leq \ms b_0+n$.%
\end{enumerate}
 In order to express $\eta_d$ and $\xi_d$, we need to define a function $\Delta: \ms G\to \mathbb Z_{2}$ such that $\Delta[*]=1-\delta_{*,0}$. $\Delta[\ms g]=0$ if $\ms g$ is identity element in $\ms G$ and $1$ otherwise. The function $\eta(\ms a)$ and $\xi(\ms a )$ can be expressed in terms of the following recursion relations 
\begin{align}
    \eta_d[\ms a]=&\;\sum_{\ms b=1}^{\ms a-1}\Delta[\alpha_{\ms a-{\ms b}}](\xi_d[\ms a-\ms b]+1)\prod_{\ms c=2}^{\ms b}\delta_{\alpha_{\ms a-\ms c+1}} \nonumber \\    
    \xi_d[{\ms a}]=&\;\sum_{\ms b=0}^{\ms a-1}\Delta[\ms g_{\ms a-{\ms b}}]\eta_d[\ms a-\ms b]\prod_{\ms c=1}^{\ms b}\delta_{\ms g_{\ms a-\ms c+1}},
\end{align}
along with the conditions $\eta_d[1]=\xi_d[1]=0$. The length of the operator $\mc O_{d}[i_0]$ is denoted as $\ell_d$ and is defined as the number of pseudo-spin sites on which $\mc O_{d}[i_0]$ acts non-trivially. This is given by the number of sites between $i_0$ and the end-point of the last $B$ operator in the operator string in Eq.~\eqref{eq:Od_operator}. In case, the operator string is only built from $A$ operators, the length of the operator string is 1. We may express $\ell_d$ as
\begin{align}
    \ell_{d}=\sum_{\ms a =0}^{M-1}\Delta[\alpha_{M-a}](\xi_{d}[M-\ms a]+1)\prod_{\ms b=1}^{\ms a}\delta_{\alpha_{M-\ms a+\ms b}}+\prod_{\ms b=1}^{M}\delta_{\alpha_{\ms b}}.
\end{align}

\section{Computation of duality groups}
\label{appsec:computation of duality group}
Given a $2+1d$ topological gauge theory with finite Abelian gauge group $\ms G$ with anyons $\mcal{A}\simeq\ms G\times\ms G$, the anyonic symmetry group (i.e. $0$-form symmetries) is a subgroup $\mathcal G[\ms G]\subset S_{|\mathcal A|}$ of the group of permutations $S_{|\mcal{A}|}$ of charges in $\mathcal A$ satisfying
\begin{equation}
    \mathcal G[\ms G]\equiv \left\{\sigma\in S_{|\mathcal A|}\;\big|\; S_{\sigma(d),\sigma(d')}=S_{dd'} \text{ and } T_{\sigma(d),\sigma(d')}=T_{dd'}\right\},
\end{equation}
where $S$ and $T$ are defined as in \eqref{eq:SandT_matrices}. In other words, these are all possible ways of permuting charges while leaving all topological properties like fusion rules and braiding statistics invariant. In the space of $1+1d$ $\ms G$-symmetric theories, these become the group of dualities. For any generic finite Abelian group $\ms G=\prod_{i=1}^{n}\mbb Z_{N_i}$, the duality group is generated by three kinds of duality operations \cite{EtingNikshychOstrik200909,NikshychRiepel201309,FuchsPrielSchweigertValentino201404}
\begin{enumerate}
    \item {\bf Universal kinematical symmetries}: These are induced from the automorphisms of $\ms G$. 
    For every automorphism $\varphi: \ms G\to \ms G$, one obtains a duality $\sigma_{\varphi}: \mc A\to \mc A$ in $\mc G[\ms G]$, which acts on a dyon $d=(\ms g, \alpha)\in \mc A$ as
    \begin{equation}\label{eq:action of universal kinematical symmetries on anyons}
        \sigma_{\varphi}: (\ms g,\alpha)\to (\ms \varphi(\ms g), (\ms {\varphi}^{-1})^{\star}(\alpha)),
    \end{equation}
    where $\varphi^\star:\text{Rep}(\ms G)\rightarrow\text{Rep}(\ms G)$ defined as $[\varphi^\star\ms R_{\alpha}](\ms g)\equiv \ms R_\alpha(\varphi(\ms g))$. This thus maps $\alpha\in\text{Rep}(\ms G)$ to some other $\alpha'\in\text{Rep}(\ms G)$.
    
    \item {\bf Universal dynamical symmetries}: These correspond to elements in $H^{2}(\ms G,\ms{U}(1))$, where for $\ms G=\prod_{i=1}^{n}\mbb Z_{N_i}$ is
    \begin{equation}
        H^{2}(\ms G,\ms{U}(1))=\prod_{i<j}\mathbb Z_{\text{gcd}(N_i,N_j)}.
    \end{equation}
    Given an element of  $H^{2}(\ms G,\ms{U}(1))$ labeled by $\mbs{\ell}=\left\{\ell_{ij}\right\}$, there is an associated alternating bicharacter $\mfk{c}_{\bs{\ell}}$ given by
    \begin{equation}
        \mfk{c}_{\mbs{\ell}}(\ms g,\ms h)=\exp\left\{2\pi i\sum_{i<j}
        \frac{\ell_{ij}\ms g_i\ms h_j}{\text{gcd}(N_i,N_j)}
        \right\},
        \label{eq:alternating bicharacter (appendix)}
    \end{equation}
    where $\ms g=(\ms g_1,\dots, \ms g_n)$, $\ms h=(\ms h_1,\dots, \ms h_n)$, and $\ell_{ij}=-\ell_{ji}$. An alternating bicharacter is a group homomorphism $\mfk{c}:\ms G\times\ms G\to\mbb{C}^*$ in both arguments which satisfies the property $\mfk{c}(\ms g,\ms g)=1$ for all $\ms g\in \ms G$ \cite{FuchsPrielSchweigertValentino201404}.
    Physically, such an alternating bicharacter is furnished by the torus partition function of an SPT labeled by the corresponding cohomology class. 
    Given such a bicharacter $\mfk{c}_{\bs{\ell}}$, one obtains a duality $\sigma_{\mfk{c}}$ which maps the dyon $d=(\ms g,\alpha)$ as 
    \begin{align}
    \sigma_{\mathfrak{c}}: (\ms g,\alpha) \mapsto (\ms g,\alpha^{\mfk{c}}_{\ms g}),
    \end{align}
    where the transformed representation $\alpha^{\mfk{c}}_{\ms g}$ has the form $\ms R_{\alpha^{\mathfrak{c}}_{\ms g}}(\cdot)=\ms R_{\alpha}(\cdot)\mfk{c}_{\mbs\ell}(\ms g,\cdot)$.
    \item {\bf Partial electric-magnetic dualities}: These correspond to performing an electric-magnetic duality on a given factor $\mbb Z_{N_j}$ in $\ms G$
\begin{equation}\label{eq:partial electric-magnetic duality jth layer}
    \sigma_{j}:
    \begin{gathered}
        \hspace{.1cm}(\ms g_1,\dots,\ms g_{j-1},\ms g_{j},\ms g_{j+1},\dots \ms g_{n} )\mapsto (\ms g_1,\dots,\ms g_{j-1},\alpha_{j},\ms g_{j+1},\dots \ms g_{n}),
       \\
    (\alpha_1,\dots,\alpha_{j-1},\alpha_{j},\alpha_{j+1},\dots \alpha_{n} ) 
    \mapsto
    (\alpha_1,\dots,\alpha_{j-1},\ms g_{j},\alpha_{j+1},\dots \alpha_{n} ).
    \end{gathered}
\end{equation}
\end{enumerate}
The duality group $\mcal{G}[\ms G]$ is in general non-Abelian and often very large for even small Abelian groups $\ms G$. In this appendix, we will compute a few duality groups. Some concrete computations are done using the GAP System \cite{GAPSystem}.

\subsection{$\ms G = \mathbb Z_2$}
We first consider the simplest example which is $\ms{G}= \mathbb Z_2$. The fundamental dyonic excitations are the charge $e=(0,1)$ and flux $m=(1,0)$. The fusion group is then
\begin{equation}
    \mcal{A}[\mbb{Z}]=\{(0,0),(1,0),(0,1),(1,1)\}\equiv\{1,m,e,f\}. 
\end{equation}
The modular $T$- and $S$-matrices are
\begin{equation}
    T=\begin{pmatrix}
                1 & 0 & 0 & 0
                \\
                0 & 1 & 0 & 0
                \\
                0 & 0 & 1 & 0
                \\
                0 & 0 & 0 & -1
    \end{pmatrix}, \qquad 
    S=\frac{1}{2}\begin{pmatrix}
                1 & 1 & 1 & 1
                \\
                1 & 1 & -1 & -1
                \\
                1 & -1 & 1 & 1
                \\
                1 & -1 & -1 & 1
    \end{pmatrix}.
\end{equation}
A duality $\sigma$ is a permutation of anyons that preserves these matrices
\begin{equation}
    \mscr{M}_\sigma T\mscr{M}_\sigma^{-1}=T, \qquad \mscr{M}_\sigma S\mscr{M}_\sigma^{-1}=S,
\end{equation}
here $\mscr{M}_\sigma$ is an invertible integer matrix permuting the set of anyons. Clearly, the only such permutation is the electric-magnetic duality transformation
\begin{equation}
    e\longleftrightarrow m,
\end{equation}
whose matrix is given by
\begin{equation}
    \mscr{M}_{\sigma}=
    \begin{pmatrix}
                1 & 0 & 0 & 0 
                \\
                0 & 0 & 1 & 0 
                \\
                0 & 1 & 0 & 0
                \\
                0 & 0 & 0 & 1
    \end{pmatrix}.
\end{equation}
Therefore, the group of anyonic symmetries is
\begin{equation}
    \mcal{G}[{\mbb{Z}_2}]=\mbb{Z}_2. 
\end{equation}

\subsection{$\ms G = \mathbb Z_2\times\mathbb Z_2$}\label{appsubsec: dyonic symmetry for Z2*Z2}
For $\ms G = \mathbb Z_2\times\mathbb Z_2$, there are 16 anyons and $T$- and $S$-matrices are given 
\begin{equation}
	T=\begin{pmatrix}
			1 & 0 & 0& 0\\
			0 & 1& 0& 0\\
			0 & 0& 1& 0\\
			0& 0&0&-1
      	 \end{pmatrix}\otimes
	 \begin{pmatrix}
			1 & 0 & 0& 0\\
			0 & 1& 0& 0\\
			0 & 0& 1& 0\\
			0& 0&0&-1
      	 \end{pmatrix},
\end{equation}
and
\begin{equation}
S =\frac 14\begin{pmatrix}
			1 & 1 & 1& 1\\
			1 & 1& -1& -1\\
			1 & -1& 1& -1\\
			1& -1&-1&1
      	 \end{pmatrix}
	 \otimes
	 \begin{pmatrix}
			1 & 1 & 1& 1\\
			1 & 1& -1& -1\\
			1 & -1& 1& -1\\
			1& -1&-1&1
      	 \end{pmatrix},
\end{equation}
Dualities $\mcal{G}[\mbb{Z}_2\times\mbb{Z}_2]$ is a subgroup of $S_{16}$ such that
\begin{equation}
	\mscr{M}_\sigma T \mscr{M}_\sigma^{-1}  = T,  \qquad \mscr{M}_\sigma S \mscr{M}_\sigma^{-1}  = S, \qquad \forall \sigma \in \mathcal G[\mbb{Z}_2\times\mbb{Z}_2],
\end{equation}
where $\mscr{M}_\sigma$ is a matrix representing the permutation associated to the duality $\sigma$. The duality group is
\begin{equation}
		\mathcal G[\mbb{Z}_2\times\mbb{Z}_2]= (S_3\times S_3)\rtimes\mathbb Z_2. 
\end{equation}
whose order is $72$. For the details of the generators, see Section \ref{Subsec:Z2Z2_example}.

\subsubsection{Analysis of the outer automorphism group}

We will here briefly describe the construction of the automorphism group of $\mcal{G}_2\equiv \mcal{G}[\mbb{Z}_2\times\mbb{Z}_2]=(S_3\times S_3)\rtimes\mbb{Z}_2$ in terms of its generators. One can readily check that the following two maps are automorphisms
\begin{gather}\nonumber
	\begin{aligned}
		a: \begin{pmatrix}
		\sigma_L\\
		h_1\\
		k_1\\
		h_2h_3\\
		k_2k_3
		\end{pmatrix}
		\longrightarrow
		\begin{pmatrix}
			h_1\\ 
			\sigma_L(h_2h_3)^2(k_2k_3)\\
			\sigma_Lk_1h_1(h_2h_3)(k_2k_3)\\ 
			h_2h_3(k_2k_3)^2\\
			(h_2h_3)^2(k_2k_3)^2
		\end{pmatrix}
		\qquad
		x: \begin{pmatrix}
		\sigma_L\\
		h_1\\
		k_1\\
		h_2h_3\\
		k_2k_3
		\end{pmatrix}
		\longrightarrow
		\begin{pmatrix}
			\sigma_Lk_1h_1(h_2h_3)(k_2k_3)\\ 
			k_1\\
			h_1(h_2h_3)^2\\
			k_2k_3\\
			(h_2h_3)^2
		\end{pmatrix}
	\end{aligned} 
\end{gather}
and they satisfy the following properties
\begin{equation}
	a^8=x^2=1, \qquad xax^{-1} = a^3.
\end{equation}
It is clear that this group is a semidirect product $\mathbb Z_8\rtimes\mathbb Z_2$, but it is distinct from the dihedral group $D_8$ as the action of the $\mathbb Z_2$ is different (the dihedral group has $xax^{-1} = a^4=a^{-1}$). It turns out that this group is called the quasi-dihedral group\footnote{Sometimes called the semi-dihedral group $SD_{16}$. Please note that $D_n$ is a group of $2n$ elements and corresponds to the symmetries of a regular polygon with $n$ corners. This is a geometric notation. In abstract algebra, this group is often notated as $D_{2n}$, to indicate the order of the group. The notation used for $QD_{16}$ follows the abstract algebra notation, rather than the geometry one, as this group has no geometric origin.}
\begin{equation}
	QD_{16} = \left\langle a,x \;\big|\; a^8=x^2=1, \, xax^{-1}=a^3 \right\rangle.
\end{equation}
Furthermore, we have to other automorphisms
\begin{gather}\nonumber
	\begin{aligned}
	b: \begin{pmatrix}
		\sigma_L\\
		h_1\\
		k_1\\
		h_2h_3\\
		k_2k_3
		\end{pmatrix}
		\longrightarrow
	 \begin{pmatrix}
			\sigma_L\\ 
			k_1(k_2k_3)^2\\
			h_1(h_2h_3)^2\\
			h_1\\
			h_2h_3
		\end{pmatrix},
		\qquad
		c: \begin{pmatrix}
		\sigma_L\\
		h_1\\
		k_1\\
		h_2h_3\\
		k_2k_3
		\end{pmatrix}
		\longrightarrow
		 \begin{pmatrix}
			\sigma_L(h_2h_3)(k_2k_3)^2\\ 
			k_1(k_2k_3)^2\\
			h_1(h_2h_3)\\
			k_2k_3\\
			h_2h_3
		\end{pmatrix},
	\end{aligned} 
\end{gather}
generating the product of two decoupled cyclic groups
\begin{equation}
	\mathbb Z_3\times\mathbb Z_3 = \left\langle b, c \;\big|\; b^3=c^3=1, \, bc=cb \right\rangle.
\end{equation}
From the above four automorphisms, one can then check that $QD_{16}$ has the following action on $\mathbb Z_3\times\mathbb Z_3$
\begin{gather}
	\begin{aligned}
		xbx^{-1} &= c,		&	xcx^{-1} &= b,\\
		aba^{-1} &= bc^2,	&	aca^{-1} &=bc.
	\end{aligned}
\end{gather}
In particular, it can be seen that the orbit of the action of $a$ on $\mathbb Z_3\times\mathbb Z_3$, has period eight
\begin{equation}
	b\longrightarrow bc^2\longrightarrow c\longrightarrow bc \longrightarrow b^2\longrightarrow b^2c\longrightarrow c^2\longrightarrow b^2c^2\longrightarrow b
\end{equation}
We, therefore, have the following automorphism group
\begin{equation}
	\text{Aut}\left(\left(S_3\times S_3\right)\rtimes\mathbb Z_2\right) = \left(\mathbb Z_3\times\mathbb Z_3\right)\rtimes QD_{16},
\end{equation}
of order
\begin{equation}
	|\text{Aut}(\mcal{G}_2)| = 144,
\end{equation}
which is exactly twice the order of $\mcal{G}_2$.
Many of these automorphisms are however trivial, in the sense that they correspond to conjugation with group elements
\begin{equation}
	\phi_h:\mcal{G}_2\rightarrow \mcal{G}_2, \qquad \phi(g)=hgh^{-1}.
\end{equation}
The set of such trivial automorphisms form the so-called inner automorphism group and is clearly isomorphic to the group itself
\begin{equation}
	\text{Inn}(\mcal{G}_2) = \left(S_3\times S_3\right)\rtimes\mathbb Z_2.
\end{equation}
This is a normal subgroup of $\text{Aut}(\mcal{G}_2)$ and can be quotient out to get to the so-called outer automorphism group
\begin{gather}
	\begin{aligned}
		\text{Out}(\mcal{G}_2) &= \text{Aut}(\mcal{G}_2)/\text{Inn}(\mcal{G}_2),\\
				    &= \mathbb Z_2,\\
				    &= \left\{[1], [\phi]\right\}.
	\end{aligned}
\end{gather}
Note that these groups generally form a short exact sequence,
\[
	1\longrightarrow \text{Inn}(\mcal{G}_2)\longrightarrow\text{Aut}(\mcal{G}_2)\longrightarrow\text{Out}(\mcal{G}_2)\longrightarrow 1.
\]
This group $\text{Out}(\mcal{G}_2)$ is clearly isomorphic to $\mathbb Z_2$, as the resulting group must be of order two of which there is on only one (up to isomorphism). The automorphisms thus belong to two equivalence classes given by
\begin{equation}
	[1] \approx (S_3\times S_3)\rtimes\mathbb Z_2, \qquad [\phi] = \phi\cdot [1],
\end{equation}
where $\phi$ is a representative of the only non-trivial outer automorphism class
\begin{equation}
	\phi: \begin{pmatrix}
		\sigma_L\\
		h_1\\
		k_1\\
		h_2h_3\\
		k_2k_3
		\end{pmatrix}
		\longrightarrow
		 \begin{pmatrix}
			h_1\\ 
			\sigma_L\\
			\sigma_Lk_1h_1\\
			(k_2k_3)(h_2h_3)^{-1}\\
			(k_2k_3)(h_2h_3)
		\end{pmatrix}.
\end{equation}
The interpretation of this automorphism is clear: it rotates the $S_3\times S_3$ subgroup to a different one consisting of the diagonal and anti-diagonal $S_3\times S_3$.\footnote{If we write elements of $S_3\times S_3$ as $(g_1,g_2)$, then the left and right subgroups are $(g,1)$ and $(1,g)$ corresponding to rotating two triangles independently. The diagonal subgroup is $(g,g)$ where both triangles are rotated together, and the anti-diagonal is $(g,g^{-1})$, where the two triangles are rotated in opposite directions.}

\subsection{$\ms G = \mathbb Z_2\times\dots\times\mathbb Z_2$}
The duality transformations discussed in the last section can easily be generalized to the case of $k$-layers of $\mathbb Z_2$ symmetry. For example, we have the following generators
\begin{gather}
	\begin{aligned}
		\sigma_i&:\left\{e_i\mapsto m_i,\; m_i\mapsto e_i\right\},
		\\
		s_{ij}&:\left\{a_i\mapsto a_j,\; a_j\mapsto a_i \; |\; a = e, m\right\},
		\\
		c&:\left\{a_i\mapsto a_{i+1\mod k} \; |\;a = e, m\right\},
	\end{aligned}
\end{gather}
where $\sigma_i$ is the electromagnetic duality in the $i$\textsuperscript{th} layer, $s_{ij}$ swaps layers $i$ and $j$, and $c$ is a cyclic permutation of the $k$ layers. Note that we only need electromagnetic duality in one layer as the other layers can be gotten by
\begin{equation}
	\sigma_i = s_{1i} \sigma_1 s_{1i}.
\end{equation}
Furthermore, only two generators are needed to generate the full group of permutations of $k$ layers:
$S_k = \langle s_{12}, c\rangle$. We also have a more non-trivial set of dualities
\begin{gather}
		h_{ij}:\left\{e_i\mapsto e_ie_j,\; m_j\mapsto  m_im_j\right\}.
\end{gather}
Note that $h_{ij}$ is not the same as $h_{ji}$ but they are related as $h_{ji}=s_{ij}h_{ij}s_{ij}$. For any pairs of layers $(i,j)$ we can generate the subgroup $S_3 = \left\langle h_{ij}, h_{ji}, s_{ij}\right\rangle$.
These kinds of subgroups are important when analyzing the phase diagram of the system. We can also define
\begin{gather}
	\begin{aligned}
		k_{ij}&:\left\{m_i\mapsto m_ie_j,\; m_j\mapsto  e_im_j\right\}, \\
		\bar k_{ij}&:\left\{e_j\mapsto m_ie_j,\; e_i\mapsto  e_im_j\right\}.
	\end{aligned}
\end{gather}
Note that
\begin{equation}
    k_{ij} = k_{ji}, \qquad \bar k_{ij} = \bar k_{ji}, \qquad \bar{k}_{ij}=\sigma_i\sigma_j k_{ij}\sigma_j\sigma_j.
\end{equation}
Again, for each pairs of layers $(i,j)$ we have another $S_3$ subgroup $S_3 = \left\langle k_{ij}, \bar k_{ij}, \sigma_i\sigma_js_{ij}\right\rangle$,
similar to the two-layer analysis. Another similarity to the two-layer analysis is the following relation
\begin{equation}
    k_{ij}=\sigma_ih_{ij}\sigma_i.
\end{equation}
Therefore, for each pair of indices $(i,j)$, we have the subgroup
\begin{equation}
  \langle h_{ij},s_{ij},k_{ij},\sigma_i,\sigma_j\rangle\simeq (S_3\times S_3)\rtimes \mbb{Z}_2.
\end{equation}
In other words, the phase of diagram of $(\mbb{Z}_2)^{\times k}$-symmetric theories contain many copies of the phase diagram of $\mbb{Z}_2\times\mbb{Z}_2$ symmetric theories, one for each pair of layers $(i,j)$. From the above discussion, we conclude that the full duality group is 
\begin{equation}
	\mathcal G_k = \left\langle \sigma_1, h_{12}, s_{12}, c\right\rangle,
\end{equation}
where we have defined the notation $\mcal{G}_k\equiv \mcal{G} [(\mbb{Z}_2)^{\times k}]$. One can check that these generators leave the $T$- and $S$-matrices invariant. In the language discussed earlier, $\sigma_i$ are partial electric-magnetic dualities, $\{h_{ij},s_{ij},c\}$ are essentially universal kinemaltical symmetries related to automorphism group of $(\mbb{Z}_2)^{\times k}$ while $k_{ij}$ are universal dynamical symmetries related to $H^2((\mbb{Z}_2)^{\times k},\ms U(1))$. For small $k$, this group is isomorphic to the following groups
\begin{equation}
	\mcal{G}_k= \begin{cases}
		\mathbb Z_2 					  &k=1\\
		(S_3\times S_3)\rtimes\mathbb Z_2 	  &k=2\\
		S_8						 	  &k=3\\
		O_8^+(2)\rtimes\mathbb Z_2		  &k=4
	\end{cases}.
\end{equation}
Here $O_8^+(2)  = D_4(2)$ is a so-called Lie type group, more precisely an orthogonal Chevalley group of type D. The order of these groups are 
\begin{equation}
	|D_n(q)| = \frac{q^{n(n-1)}\left(q^n - 1\right)}{gcd(4, q^n-1)}\prod_{i=1}^{n-1}\left(q^{2i}-1\right).
\end{equation}
In particular
\begin{equation}
	|\mcal{G}_4|=|O_8^+(2) \rtimes\mathbb Z_2| = 348\,364\,800.
\end{equation}
In other words, there are almost 350 million dualities for $\ms G = \left(\mathbb Z_2\right)^4$ symmetry. It turns out that this is the same dimension as $W^+(E_8)$, the orientation-preserving subgroup of the Weyl group of the exceptional Lie algebra $E_8$. It is known that $W^+(E_8)$ is also a $\mathbb Z_2$ extension of $O^+_8(2)$. Whether this connection has any significance will be studied elsewhere.

\subsection{$\ms G = \mathbb Z_p\times\mathbb Z_p$}
Another interesting class of symmetries are $\ms G=\mbb{Z}_p\times\mbb{Z}_p$ for prime numbers $p$. Here, the partial electric-magnetic dualities $\sigma_i$ are again given in \eqref{eq:partial electric-magnetic duality jth layer}. The universal dynamical symmetries are given by elements of  $H^2(\mathbb Z_p\times\mathbb Z_p,U(1))=\mathbb Z_p$. The corresponding bicharacters are given by
\begin{equation}
    \beta_{\ell}(\ms g,\ms h)=\exp\left[\frac{2\pi i}{p}\ell(\msg_2\msh_1-\ms g_1\ms h_2)\right],
\end{equation}
where $\ell=0,\cdots,p-1$ lables different cohomology classes. Under these dualities, anyonic charges transform as
\begin{equation}
    (\msg_1,\msg_2,\alpha_1,\alpha_2) \longmapsto (\msg_1,\msg_2,\alpha_1 + \ell \msg_2,\alpha_2-\ell \msg_1). 
\end{equation}
Finally, the universal dynamical symmetries are related to the duality group of $\mbb{Z}_p\times\mbb{Z}_p$, which is
\begin{equation}
    \text{Aut}\left(\mathbb Z_p\times\mathbb Z_p\right) = \tenofo{GL}_2(\mathbb F_p).
\end{equation}
The action of these dualities on anyons is given in \eqref{eq:action of universal kinematical symmetries on anyons}.
Specifically, for $p=2$ we have $\tenofo{GL}_2(\mathbb F_2) = S_3$ and the full duality group becomes $(S_3\times S_3)\rtimes\mbb{Z}_2$, as shown earlier. For $p=3$, we have $\tenofo{GL}_2(\mathbb F_3) = Q_8\rtimes S_3$, and they are generated by
\begin{gather}
    \begin{aligned}
        h_1:& (\ms g_1, \ms g_2;\alpha_1, \alpha_2) \longmapsto (\ms g_2, \ms g_1;\alpha_2, \alpha_1),\\
        h_2:& (\ms g_1, \ms g_2;\alpha_1, \alpha_2) \longmapsto (\ms g_1+\ms g_2, \ms g_2;\alpha_1, 2\alpha_1+\alpha_2).
    \end{aligned}
\end{gather}
The dynamical dualities are generated by
\begin{gather}
    \begin{aligned}
    k:& (\ms g_1, \ms g_2;\alpha_1, \alpha_2) \longmapsto (\ms g_1, \ms g_2;\alpha_1 + \ms g_2, \alpha_2 - \ms g_1),
    \end{aligned}
\end{gather}
with $k^3=1$. We have the relations
\begin{gather}
    \begin{aligned}
    h_1kh_1^{-1} = k^{-1}, \qquad h_2kh_2^{-1} = k.
    \end{aligned}
\end{gather}
Thus $\langle k, h_1, h_2\rangle \simeq \mathbb Z_3\rtimes GL_2(\mathbb F_3)$. When adding electric-magnetic duality, the structure of the group becomes significantly more complicated and it is hard to write the group in a simple way. One possible presentation of the full duality group is
\begin{equation}
    \mathcal G[\mathbb Z_3\times\mathbb Z_3] =\left\{\left[\Big((\mathbb Z_2\times\mathbb Z_2\times\mathbb Z_2)\rtimes(\mathbb Z_2\times\mathbb Z_2)\Big)\rtimes(\mathbb Z_3\times\mathbb Z_3)\right]\rtimes\mathbb Z_2\right\}\rtimes\mathbb Z_2,
\end{equation}
which is not very illuminating. The order of the group is $|G[\mathbb Z_3\times\mathbb Z_3]| =1152$ and one can check that this group is actually
\begin{equation}
    \mathcal G[\mathbb Z_3\times\mathbb Z_3]= W(F_4),
\end{equation}
the Weyl group of the exceptional Lie algebra $F_4$. Whether there is a deeper relation to the Lie algebra $F_4$ will be investigated elsewhere.

\section{From Wen plaquette model to \texorpdfstring{$\mbb{Z}_N$}{}-symmetric spin chains}
\label{App: Wen-plaquette}
In this section we exemplify our approach for the simple example of $\mathbb Z_N$ topological order, using an exactly solvable model. We will show that this $2+1d$ quantum model has a $\mathcal A = \mathbb Z_N\times\mathbb Z_N$ 1-form symmetry and a $\mathcal G = \mathbb Z_2$ $0$-form symmetry related to electric-magnetic duality.
We then show how a spin chain with 
$\mathbb Z_{N}$ global symmetry naturally appears on the $1+1d$ edge of the  $2+1d$ quantum model model and elaborate on various global and topological issues of this bulk to edge correspondence. There are various equivalent formulations of the $\mathbb Z_{N}$ topological gauge theory. Here we find it convenient to work with the so-called Wen plaquette formulation, since (0-form) anyonic symmetries appear as lattice translations.

\paragraph{Bulk Hilbert space and Hamiltonian:}
Consider a square lattice $\Lambda$ with a $N$-state spin degree of freedom on each lattice site and the Hilbert space $\mathcal H_{\Lambda}=\bigotimes_i\mathcal H_i$,
 with the local Hilbert spaces $\mathcal H_i\simeq\mathbb C^{N}$. We define a basis $|\sigma_i\rangle$ with $\sigma_i=0,\dots,N-1$ that spans $\mathcal H_{i}$. The $\mathbb Z_N$ generalization of the Pauli matrices act on $\mc H_{i}$ as
\begin{gather*}
	\begin{split}
		Z_i\ket{\sigma_i} &= \omega^{\sigma_i}\ket{\sigma_i},\\
		X_i\ket{\sigma_i} &=\ket{\sigma_i-1},
	\end{split}
\end{gather*}
where $\omega = e^{\frac{2\pi i}N}$. The operators satisfy the algebraic relations $X_i^N=Z_i^{N}=1$ and $X_iZ_j=\omega^{\delta_{ij}}Z_jX_i$. It will be convenient to use the following graphical notation
\begin{equation}\label{eq:GraphicalNotation}
	\begin{aligned}
		Z_i &= \Z{i}, \qquad&\qquad X_i &= \X{i},
		\\
		X_i^\dagger&= \Xd{i}, \qquad&\qquad Z_i^\dagger &= \Zd{i},
	\end{aligned}
\end{equation}
in terms of which the algebraic relations take the form
\begin{gather*}
	\begin{aligned}
		\XZ &= \omega \ZX, \qquad&\qquad  \XdZ &= \omega^{-1}\ZXd,\\
		\XZd &= \omega^{-1}\ZdX, \qquad&\qquad \XdZd &= \omega \ZdXd.
	\end{aligned}
\end{gather*}
For each plaquette on the lattice, on can define the minimal closed string operator as
\[ \mathcal O_p = \OopNumb{p} = Z_1X_2Z_3^\dagger X_4^\dagger,\]
which again is unitary and satisfy $\mathcal O_p^N=1$ and satisfy $[\mathcal O_p,\mathcal O_{p'}]=0$ for all plaquettes $p$ and $p'$. The $\mathbb Z_N$ generalization of the Wen-plaquette model takes the form
	\begin{equation}\label{eq:ZNWen}
		H_{\mathbb Z_N}^{Wen} = -\frac g2\sum_p\left(\mathcal O_p+\mathcal O_p^\dagger\right) = -\frac g2\sum_p\left(\Oop{p} + \OopDagger{p}\right).
	\end{equation}
Since all plaquettes commute, the eigenstates can be labelled by the $\mathbb Z_{N}$ eigenvalues of the plaquette operators. The ground states subspace is given by
\begin{equation}\label{eq:GroundStateSubspace}
	\mathcal H_{GS}=\big\{\ket\psi\in\mathcal H_\Lambda\;\big|\; \mathcal O_p\ket\psi=\ket\psi, \forall p\big\}\subset \mathcal H_{\Lambda}.
\end{equation}
The ground state degeneracy will depend on the global topology of the spatial manifold the theory lives on. For manifolds without boundary, this degeneracy is protected against any local perturbations that do not close the bulk gap while for manifolds with boundary only a subset of groundstates are protected while the rest correspond to edge states. 

\paragraph{String operators and anyonic excitations}
\begin{figure}\center
\scalebox{.82}{\subfloat[Electric excitations.]{\StringRed}}\hspace{25pt}
\scalebox{.82}{\subfloat[Magnetic Excitations.]{\StringBlue}}\hspace{25pt}
\scalebox{.82}{\subfloat[Dyonic Excitations.]{\StringRedBlue}}\caption{}\label{fig:StringOperators}
\end{figure}
The plaquettes of a square lattice admit a bipartitioning into `odd' and `even' plaquettes (see figure \ref{fig:StringOperators}, where the `odd' plaquettes are shaded gray). Now consider an oriented path on the square lattice that traverses diagonally between plaquettes such that it is fully contained in either the odd or even plaquettes. Any such path is in one-to-one correspondence with a line operator $\mc{W}_{\ell}$ which acts non-trivially on the Hilbert spaces associated to the vertices contained in the path. The operator acting on any given vertex is determined by the orientation of the string using eq.\eqref{eq:GraphicalNotation}. If the path is homologically trivial i.e. contractible, it can be expressed as a product of plaquette operators
\begin{align}
   \mc {W}_{\ell}=\prod_{p\in \partial^{-1}\ell}\mc O_{p}^{\mathfrak {o}(\ell)},
\label{eq:contractible_string}
\end{align}
where $\partial^{-1}\ell$ is the (non-unique) collection of plaquettes whose boundary is $\ell$ and $\mathfrak{o}(\ell)$ is the orientation of $\ell$, which we choose as $\pm 1$ for clockwise and anti-clockwise paths respectively. From eq.~\eqref{eq:contractible_string}, it follows that any closed contractible string operator commutes with the Wen-plaquette Hamiltonian.

\medskip \noindent Consider instead an open path $\ell$ from plaquette $a$ to plaquette $b$ (see figure \ref{fig:StringOperators}). Such an operator commutes with all the plaquette operators except those at $a$ and $b$. More precisely
\begin{equation}\label{eq:OScom}
	\mathcal O_p\mc{W}_{\ell} = \omega^{\delta_{p,a}-\delta_{p,b}} \mc W_{\ell}\mc O_p.
\end{equation}
Therefore, the operator $\mc S_{\ell}$ acting on the ground state creates localized excitations at the plaquettes $a$ and $b$.
 We may deform such a string operator by composing it with an arbitrary collection of plaquette operators which gives rise to modified string $\ell'$ with the endpoints fixed at $a$ and $b$.  In this sense the operators defined on $\ell$ and $\ell'$ are equivalent and we may label them only by the end-points. Since the string operators are contained on either the even (white) or odd (shaded) plaquettes, there are two types of local excitations on plaquettes; magnetic charges $m$ (even) and electric charges $q$ (odd), while more generally we have dyons on both lattices and can they be thought of as living on the link separating them (see fig. \ref{fig:StringOperators}). Thus the set of anyons is given by
\begin{equation}
	\mathcal A = \left\{(q,m) \;|\; q,m=0,\dots N-1\right\},
\end{equation}
with the anyon $(0,0)$ corresponding to the trivial particle.  For compact manifolds excitations can only be created pairwise while for non-compact manifolds we can have single particles since the other end of the string can either be stretched to infinity or onto a boundary. A single dyon with electric and magnetic  charges $q$ and $m$, respectively, cost the energy
\begin{equation}
\Delta E_{(q,m)}= g\left(1-\cos\left(\frac{2\pi}Nq\right)\right)+ g\left(1-\cos\left(\frac{2\pi}Nm\right)\right).	\end{equation}
There are $N^2$ different types of dyonic excitations labelled by $(q,m)$. These different excitations are topologically distinct in the sense that they cannot be turned into another excitation $(\tilde q,\tilde m)\neq (q,m)$ by acting with any local operators.

\paragraph{Global $1$- and $0$-form symmetries}
As discussed in \eqref{eq:contractible_string}, homologically trivial (contractible) loops are given by products of $\mathcal O_p$. Such products can be used to topologically deform these curves, however there will always exist two different types of contractible loops that cannot be deformed into each other
\begin{equation}
    \mathcal W^e_{\ell_{loop}} = \prod_p\OopRed{p}, \qquad \mathcal W^m_{\ell_{loop}} = \prod_p\OopBlue{p}.
\end{equation}
Any such loop commutes with the Hamiltonian and corresponds to $1$-form symmetries isomorphic to $\mathcal A =  \mathbb Z_N\times\mathbb Z_N$. On Manifolds with non-trivial topology like a torus or genus-g surfaces $\Sigma_g$, there will exist non-trivial homology cycles and we can therefore form non-contractible loops. These loops cannot be decomposed into $\mathcal O_p$, but will still commute with the Hamiltonian. Their algebra will lead to non-trivial ground-state degeneracies that only depend on (1) the kind of topological order ($\mathbb Z_N$ in this case) and (2) topology of the manifold (in particular, first homology group). 
This degeneracy is protected against local perturbations, because they are not coming from a $0$-form symmetry but rather from $1$-form symmetries. 

\smallskip Now note that a lattice translation acts on $1$-form symmetries
\begin{equation}
    \mathcal W^e_{\ell} \longleftrightarrow \mathcal W^m_{\ell}.
\end{equation}
It exchanges anything electric and magnetic (lines and excitations), and therefore called electric-magnetic duality. Thus symmetry acts everywhere in space and is therefore a $0$-form symmetry (see figure \eqref{fig:ActionOfSymmetryOperatorsOnBulkLines}). Such symmetry exists in any description of a $\mathbb Z_N$ topological order, the fact that it is generated by lattice translations is a special feature of the Wen plaquette model. If we wanted to put the $0$-form symmetry along time to create a domain wall as in figure \ref{fig:BulkSymmetryOperatorBroughtToTheBoundary}, we would have to create lattice dislocation (see \cite{Bombin2010,YouWen2012}). But note that this object is really a surface in spacetime and not just a line, as it might naively appear in much of the literature on twist defects.

\paragraph{Semi-infinite cylinder:}

\begin{figure}
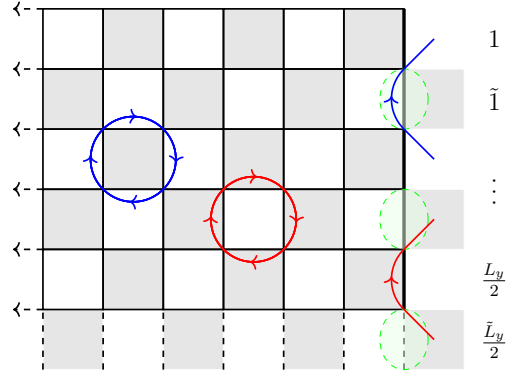
\center
\scalebox{.8}{\LatticeEdge}
\caption{Semi-infinite along $x$-direction, but periodic along $y$. The boundary operators (BO) can be thought of as half of the small loop operators in the bulk. The green ellipses on the edge correspond to the pseudo-spin degree spin of freedom. The blue BO act on a single pseudo-spin, while the red BO act on a pair of nearest-neighbour pseudo-spins.}\label{fig:SemiInfiniteCylinder}
\end{figure}
We will now consider the model \eqref{eq:ZNWen} defined on a semi-infinite cylinder, where the periodic direction is along the $y$-axis with $L_y$ lattice sites (see fig. \ref{fig:SemiInfiniteCylinder}). For now, we will assume that $L_y$ is even and we will label the even plaquettes by $p$ while the odd ones are labelled by $\tilde p$. 

\noindent Besides the bulk plaquette operators $\scalebox{.80}{\OopRed{p}}$ and $\scalebox{.80}{\OopBlue{\tilde p}}$, for this geometry we can consider several other string operators. First we have a set of operators acting near the boundary defined as (see also fig. \ref{fig:SemiInfiniteCylinder})
\[ S_p = \EdgeRed{p},\qquad S_{\tilde p}=\EdgeBlue{\tilde p}. \]
These are nothing but the bulk plaquette operators, cut in half. It is easy to see that both $S_p$ and $S_{\tilde p}$ commute with any $\mathcal O_{p'}$ and $\mathcal O_{\tilde p'}$, but for nearest neightbour $p$ and $\tilde p$ on the boundary we have
\begin{gather}\label{eq:edgeOp_commutationRelations}
	S_pS_{\tilde p} = \begin{cases}
						\omega\, S_{\tilde p}S_p	\quad\quad\text{if} &\quad\TwoBlocks{2},\\								\omega^{-1}\, S_{\tilde p}S_p	\quad\text{if} &\quad\TwoBlocks{1},
					  \end{cases}\quad\text{or graphically}\quad
	\begin{aligned}
		\ComEdge{1}{blue}{red} &= \omega\ComEdge{2}{blue}{red},\\
		\ComEdge{2}{red}{blue} &= \omega^{-1}\ComEdge{1}{red}{blue}.
	\end{aligned}
\end{gather} 
Next we have two types of non-contractable string operators along the periodic direction defined as
\begin{equation}\label{eq:GammaDef}
	\Gamma = \GammaRed,\qquad\tilde\Gamma=\GammaBlue.
\end{equation}
These operators both commute among each other and with all bulk plaquette operators and boundary operators.
Finally we have two operators along the other cycle, strecthing from infinity and ending on the boundary defined as
\begin{equation}
	W = \WRed,\qquad \tilde W = \WBlue.
\end{equation}
Note that if we had a finite cylinder, by gluing it into a torus the above two operators would become non-contractable and contribute to the GSD.
While these commute among themselves and with all bulk plaquette operators, we have the following relations
\begin{gather}
\begin{aligned}
	\tilde\Gamma W &= \omega\, W\tilde\Gamma, \quad & \Gamma\tilde W &= \omega\,\tilde W\Gamma,\\
	\Gamma W &= W\Gamma, \quad & \tilde\Gamma\tilde W &= \tilde W\tilde\Gamma.
\end{aligned}\label{eq:CompleteSet}
\end{gather}
A choice of maximal set of commuting and independent operators is
\begin{equation}\label{eq:maximalBasis_infinite_cylinder}
	\left\{\mathcal O_p,\;\mathcal O_{\tilde p},\;S_{\tilde p},\;\Gamma\right\} =  \left\{\OopRed{p},\;\OopBlue{\tilde p},\;\EdgeBlue{\tilde p},\;\GammaRed\right\}.
\end{equation}
Note that the operators $\tilde\Gamma$ is also included in this set of operators, as it can be generated from the boundary operators $S_{\tilde p}$.
Since the above set commute and is maximal, the corresponding eigenbasis will span the full Hilbert space and is labelled by their eigenvalues. This basis is convenient, since all these states will automatically be eigenstates of the Hamiltonian \eqref{eq:ZNWen}. Given one state in \eqref{eq:GroundStateSubspace}, we can construct all other ground-states by projecting into eigenstates of $S_{\tilde p}$ and $\Gamma$. Since there is one $\Gamma$ and $L_y/2$ $S_{\tilde p}$, we have $N^{\frac{L_y}2+1}$ ground states. However, only $N^2$ of these (from $\Gamma$ and $\tilde \Gamma = \prod_{\tilde p}S_{\tilde p}$) are topologically robust against perturbations, while the rest correspond to edge states and the degeneracy is lifted once we add local perturbations. Since there are roughly $N^{\frac{L_y}2}$ degrees of freedom on the edge, there is $\sqrt N$ degrees of freedom per site. From this we might conclude that the edge contains \textbf{parafermions} (Majorana fermions for $N=2$).

In order to see the emergence of Kramers-Wannier duality and its subtle global effects, it is useful to parametrize the subspace corresponding to edge degrees of freedom carefully and study how boundary operators act on this subspace. Thus we will construct the $N^{\frac{L_y}2+1}$ states explicitly. First note that the eigenstates of $\Gamma$ and $\tilde \Gamma$ cannot be connected to each other using any local operator, since they are only connected through the infinite line operators $\tilde W$ and $W$. We can thus decompose the full Hilbert space into $N^2$ distinct \textbf{superselection sectors} or \textbf{topological sectors}
\begin{equation}
	\mathcal H_{full} = \bigoplus_{\Gamma,\tilde\Gamma = 0}^{N-1} \mathcal H_{\Gamma,\tilde\Gamma},
\end{equation}
corresponding to the eigenspaces of $\Gamma$ and $\tilde\Gamma$.
The projection operators projecting onto the ground state subspace are given by
\begin{equation}
	P_p = \frac 1{|\mathbb Z_N|}\sum_{n=0}^{N-1}\mathcal O_p^n,\qquad P_{\tilde p} = \frac 1{|\mathbb Z_N|}\sum_{n=0}^{N-1}\mathcal O_{\tilde p}^n,
\end{equation}
and a ground state is thus given by
\begin{equation}\label{eq:ProjectedGS_zerozeroSuperselectionSector}
	\ket{\Gamma=0, \tilde\Gamma=0} =\mathcal NP_\Gamma P_{\tilde\Gamma}\prod_p P_p\prod_{\tilde p} P_{\tilde p}\ket{\psi},
\end{equation}
for a choice of reference state $\ket\psi$. We have further projected into the $\Gamma = 0$ and $\tilde\Gamma = 0$ sectors using 
\begin{equation}
	P_\Gamma = \frac 1{|\mathbb Z_N|}\sum_{n=0}^{N-1}\Gamma^n\qquad \text{and} \qquad P_{\tilde\Gamma} = \frac 1{|\mathbb Z_N|}\sum_{n=0}^{N-1}\tilde\Gamma^n.
\end{equation}
All $N^2$ topological sectors are now given by
\begin{equation}
	\ket{\Gamma = a,\tilde\Gamma = b} = \tilde W^aW^b\ket{\Gamma=0,\tilde \Gamma=0},
\end{equation}
with the eigenvalues
\begin{equation}
	\Gamma\ket{\Gamma = a, \tilde\Gamma = b} = \omega^a\ket{\Gamma=a, \tilde\Gamma = b}, \qquad \tilde\Gamma\ket{\Gamma = a, \tilde\Gamma = b} = \omega^b\ket{\Gamma=a, \tilde\Gamma = b}.
\end{equation}
The states are all locally indistinguishable,  differ only globally, thus cannot be mixed under local perturbations. To construct the $N^{\frac{L_y}2-1}$ locally distinguishable edge states within each superselection sector, we can use the following set of projectors
\begin{equation}\label{eq:edgeProjector}
	Q_{\tilde p}(\alpha) = \frac 1{|\mathbb Z_N|}\sum_{n=0}^{N-1}	\omega^{-\alpha n}S^n_{\tilde p}
\end{equation}
satisfying
\[Q_{\tilde p}(\alpha)^\dagger =Q_{\tilde p}(\alpha),\quad Q_{\tilde p}(\alpha)Q_{\tilde p}(\beta) = \delta_{\alpha,\beta}Q_{\tilde p}(\alpha),\quad S_{\tilde p}Q_{\tilde p}(\alpha) = \omega^\alpha Q_{\tilde p}(\alpha).
\]
Here we have $\alpha = 0,\dots, N-1$.
As can be seen, the operator $Q_{\tilde p}(\alpha)$ projects a state down to the $\omega^\alpha$ eigenstate of $S_{\tilde p}$. This implies that we can think of the $\tilde p$ plaquettes on the boundary as a pseudo spin-chain with $L_y/2$ sites (while the $p$ plaquettes is the dual lattice). For later use, we note the following commutation relations (using equation \eqref{eq:edgeOp_commutationRelations})
\begin{gather}\label{eq:edgeOperator_Projector_commutator}
	S_pQ_{\tilde p}(\alpha) =	\begin{cases}
							Q_{\tilde p}(\alpha-1)S_p	\quad\quad\text{if} &\quad\TwoBlocks{2},\\								Q_{\tilde p}(\alpha+1)S_p	\quad\text{if} &\quad\TwoBlocks{1},
					  	\end{cases}
\end{gather}
While they commute for any $p$ and $\tilde p$ that are not nearest neighbours.
A complete basis of $\mathcal H_{ground}$ is thus given by
\[	\ket{\Gamma=a, \tilde \Gamma=b;S_{\tilde 1}=\alpha_{\tilde 1},\dots,S_{\frac{\tilde L_y}2}=\alpha_{\frac{\tilde L_y}2}}_1 = \prod_{\tilde p}Q_{\tilde p}(\alpha_{\tilde p})\tilde W^aW^b\ket{\Gamma = 0, \tilde\Gamma =0}. \]
Each of these states correspond to a boundary state with pseudospin configuration $(\alpha_{\tilde 1}, \alpha_{\tilde 2}, \dots, \alpha_{\tilde L/2})$. The structure of \eqref{eq:GroundStateSubspace} is thus $\mathcal H_{ground} =\bigoplus_{a,b=0}^{N-1}\mathcal H_{a,b}$, with $\text{dim}\,\mathcal H_{ground}=N^{\frac{L_y}2+1}$ and $\text{dim}\,\mathcal H_{a,b}=N^{\frac{L_y}2-1}$. We see therefore that this basis is eigenbasis for the $S_{\tilde p}$ operators, while $S_p$ flip two nearest neighbour spins.

Alternatively, we could use a slightly different basis where $S_p$ are diagonal and $\tilde S_{\tilde p}$ flip spins. For convenience, we will make the following choice of reference state
\begin{equation}
	\ket\psi = \Big|\StateChoice\Big>,
\end{equation}
making the action of $P_p$ and $P_\Gamma$ trivial. The new set of states are now given by
\begin{gather}
\begin{aligned}
	\ket{\Gamma =a,\tilde\Gamma=b;\alpha_{\tilde 1},\dots, \alpha_{\frac{\tilde L_y}2}}_2 \equiv \ket{a,b;\{\alpha_{\tilde p}\}}= \mathcal N\prod_{\tilde p}^{\tilde L_y/2}S_{\tilde p}^{\alpha_{\tilde p}}\ket{\Gamma = a, ,\tilde\Gamma=b}.
\end{aligned}
\end{gather}
In this basis, it is clear that $S_{\tilde p}$ changes only one spin $\alpha_{\tilde p}\rightarrow\alpha_{\tilde p}+1$, while leaving the rest invariant. The action of the other boundary operators are
\[	S_p\ket{a,b;\{\alpha_{\tilde p}\}} = \omega^{\alpha_{\tilde p}-\alpha_{\tilde p+1}}\omega^{\delta_{p,L_y/2}\,a}\ket{a,b;\{\alpha_{\tilde p}\}},\]
where the $\omega^{\delta_{p,L_y/2}\,a}$ factor is due to $\tilde W = \WBlue$. Whether we are using basis $\ket{\dots}_1$ or $\ket{\dots}_2$, it is clear that the boundary degrees of freedom can be interpret as a $\mathbb Z_N$ spin chain with twisted boundary conditions. This pseudospin chain lives on the $\tilde p$ sites in figure \ref{fig:SemiInfiniteCylinder}, of which there are $\tilde L_y/2$ sites (where each psudospin consists of a pair of physical spin). On these boundary pseudo-spins we can make the following identifications of operators
\begin{gather}\label{eq:SpinChainCorrespondence}
	\begin{aligned}
		S_{\tilde p}&=\EdgeBlue{\tilde p}  \rightarrow {\tau^x_{\tilde p}}^\dagger,\\
		S_p&=\EdgeRed{p}  \rightarrow \tau^z_{\tilde p}\tau_{\tilde p+1}^{z\dagger},
	\end{aligned}\qquad
	\begin{aligned}
		\tilde\Gamma &=\GammaBlue  \rightarrow \mathcal S=\prod_{\tilde p}{\tau^x_{\tilde p}}^\dagger,\\
		\Gamma&=\GammaRed  \rightarrow \mathcal T=\omega^a\mathbb I,
	\end{aligned}
\end{gather}
with the following twisted boundary condition
	\begin{equation}\label{eq:SpinChainBC}
		\text{BC:}\qquad \tau^z_{\frac{L_y}2+1} = \omega^{-a}\tau^z_{\tilde 1}.
	\end{equation}
In other words, $\tau^z_{\tilde L_y/2}\tau_{\tilde L_y/2+1}^{z\dagger}=\omega^{a}\tau^z_{\tilde L_y/2}\tau_{\tilde 1}^{z\dagger}$. So each topological sector $\Gamma =a$ can be mapped into a $\mathbb Z_N$ spin chain language with boundary condition \eqref{eq:SpinChainBC}, the $\Gamma=0$ sector corresponds to the usual periodic boundary condition.
In this language, the ground states $\ket{\Gamma=a;\{\alpha_{\tilde p}\}}$ correspond to boundary pseudo-spin configurations.

\paragraph{Boundary Conditions and Emergence of Kramer-Wannier Duality}
Now with a clear understanding of the structure of the Hilbert space and the operators that act on it, let on consider the following boundary conditions of the model on a semi-infinite cylinder
\begin{equation}\label{eq:BC_infinite_cylinder}
	H_{\partial} = -J/2\sum_{p}\EdgeRed{p} -h/2\sum_{\tilde p}\EdgeBlue{\tilde p}_R+ h.c.
\end{equation}
We see that the model (bulk $+$ boundary) is exactly solvable when $J=0$ or when $h=0$, which correspond to the gapped boundaries $L_m$ and $L_e$. However when both $J$ and $h$ are non-zero, the boundary terms do not commute anymore and a some point a phase-transition will happen between these two phases.

Note that the boundary Hamiltonian only changes dynamics of the boundary modes, which are separated from bulk excitations with a large energy. In other words, the boundary Hamiltonian is the low-energy effective Hamiltonian in this case. Decomposing this into the $N^2$ superselection sectors we get
\begin{equation}
	H_{eff} = \bigoplus_{a,b=0}^{N-1}H_{eff}^{\mathcal S=a, \mathcal T=b},
\end{equation}
where in each superselection sector we have the Hamiltonians
\begin{equation}
	H_{eff}^{\mathcal S=a,\mathcal T=b} = P_{\mathcal S}(a)H_{eff}^{\mathcal T=b}P_{\mathcal S}(a),
\end{equation}
where $H_{eff}^{\mathcal T=a}$ is the Hamiltonian in the $\Gamma=\mathcal T=a$ superselection sector (but all $\tilde\Gamma=\mathcal S$ sectors) and is given by
\begin{gather}
	\begin{aligned}
	H_{eff}^{\mathcal T=b} = -J/2\sum_{\tilde p=0}^{\tilde L/2-1}\left(\tau^z_{\tilde p}\tau^{z\dagger}_{\tilde p+1}+\tau^z_{\tilde p+1}\tau^{z\dagger}_{\tilde p} \right) &- h/2\sum_{\tilde p=0}^{\tilde L/2}\left(\tau^{x\dagger}_{\tilde p} + \tau^x_{\tilde p}\right)\\
	 &-\,J/2\left(\omega^b \tau^z_{\tilde L/2}\tau^{z\dagger}_{\tilde 1}+\omega^{-b}\tau^z_{\tilde 1}\tau^{z\dagger}_{\tilde L/2} \right)
	\end{aligned}
\end{gather}
and
 \begin{equation}
 	P_{\mathcal S}(a) = \frac 1{|\mathbb Z_N|}\sum_{n=0}^{N-1}	\omega^{-b n}\mathcal S^n.
\end{equation}
is the projector to the $\Gamma=\mathcal S = a$ symmetry sector (where eigenvalue of $\mathcal S$ is $\omega^a$).
Therefore the low-energy effective Hamiltonian in each superselection sector is nothing but a $\mathbb Z_N$ generalization of the Ising model (Potts model?) with a particular symmetry twisted boundary conditions $\mathcal T=b$ and symmetry sector $\mathcal S = a$. These superselection are sectors naturally inherited from the Bulk topological order and are one-to-one with states created on top of each minimally entangled ground state (MES).

Note that instead of \eqref{eq:maximalBasis_infinite_cylinder}, we could have chosen a different set of maximal commuting set of independent operators. In particular, we could have replaced $S_{\tilde p}$ and $\Gamma$ with $S_p$ and $\tilde\Gamma$. Had we done that, the whole discussion above would have been the same, except the pseudospins would instead live on dual lattice ($p$ instead of $\tilde p$). And this would have changed the map \eqref{eq:SpinChainCorrespondence} to
\begin{gather}
	\begin{aligned}
		S_{\tilde p}&=\EdgeBlue{\tilde p}  \rightarrow  \tau^z_p\tau_{p+1}^{z\dagger},\\
		S_p&=\EdgeRed{p}  \rightarrow{\tau^x_p}^\dagger,
	\end{aligned}\qquad
	\begin{aligned}
		\tilde\Gamma &=\GammaBlue  \rightarrow \mathcal T=\omega^a\mathbb I,\\
		\Gamma&=\GammaRed  \rightarrow \mathcal S=\prod_p{\tau^x_ p}^\dagger.
	\end{aligned}
\end{gather}
This means that when writting the boundary Hamiltonian \eqref{eq:BC_infinite_cylinder} in the pseudospin language we would have to make the following changes

\section{Symmetry-twisted boundary conditions}
\label{sec:symmetry-twsited boundary conditions}
\label{sec:symmetry-twisted boundary conditions}
\begin{figure}[t!]
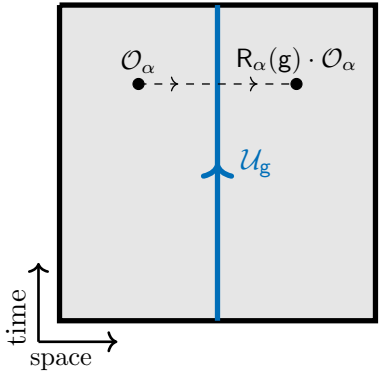

    \centering
    \InsertionOfSymmetryOperatorUg
    \caption{The insertion of $\mcal{U}_{\ms g}$ line defects along time direction. Any charged local operator $\mcal{O}_\alpha$ crossing the symmetry defect will transform under the symmetry.}
    \label{fig:insertion of symmetry operator Ug}
\end{figure}

Symmetry-twisted boundary conditions (STBC) play an important role in dualities that act on the space of $\ms G$-symmetric Hamiltonians.
Let us consider a $1+1$ dimensional $\ms G$-symmetric quantum system to be compact in the spatial direction (quantum model on a circle).
From a space-time point of view, a symmetry-twisted boundary condition corresponds to inserting the topological line defect $\mathcal U_{\ms g}$ along a line in the time-direction (fixed space point) in a partition function (see Figure \ref{fig:insertion of symmetry operator Ug}). 
Any local operator moving along the circle in the spatial direction will eventually cross the $\mathcal U_{\ms g}$ line and transform according to the corresponding symmetry $\ms g\in\ms G$.
This amounts to coupling the quantum system to a background $\ms{G}$ gauge field that has a holonomy $\ms{g}\in \ms{G}$ around the non-contractible spatial cycle.
As is familiar in the context of, say, $\ms{U}(1)$ gauge theories, coupling to a background gauge field requires extending the derivatives to covariant derivatives.
In order to do this on the lattice, consider $\ms G$-symmetric spin chains where $\ms G$ can possibly be non-abelian (or even a Lie group). For simplicity, consider the $\ms G$-symmetric Hamiltonian
\begin{equation}\label{eq:Non-abelian G spin-chain}
    \begin{aligned}
        H&=\sum_{i=1}^L\left[\sum_{\alpha\in\tenofo{Rep}(\ms G)}J_\alpha\,\mathcal{O}_{\alpha,i}^\dagger\cdot\mathcal{O}_{\alpha,i+1}+ \sum_{\ms g'\in\ms G}h_{\ms g'}\,\mathcal O_{\ms g',i}\right]+\tenofo{h.c.},
        \\
        &=\sum_{i=1}^L\left[\sum_{\alpha\in\tenofo{Rep}(\ms G)}J_\alpha\,\mathcal{O}_{\alpha,i}^\dagger\cdot\ms T\mathcal{O}_{\alpha,i}\ms T^{-1}  + \sum_{\ms g'\in\ms G} h_{\ms g'}\,\mathcal O_{\ms g',i}\right] +\tenofo{h.c.},
    \end{aligned}
\end{equation}
where $\mathcal{O}^{I}_{\alpha,i}$ is a local field transforming under some finite-dimensional unitary representation of $\ms G$ (see \eqref{eq:symmetry_action} and \eqref{eq:local symmetry action}) and $\mathcal{O}_{\alpha,i}^\dagger\cdot\mathcal{O}_{\alpha,i+1} = \sum_{I=1}^{\dim\alpha}(\mathcal{O}_{\alpha,i}^{I})^\dagger\mathcal{O}^{I}_{\alpha,i+1}$. In the last equation, we have written the Hamiltonian in terms of the translation operator (see section \ref{sec:SymmetryStructureAndBuildingBlocksofLocalOperators} for notation)
\begin{equation}
    \ms T:
    \begin{cases}
    \mc{O}_{\ms{g},i}\longmapsto \mc{O}_{\ms{g},i+1}, 
    \\
    \mc{O}_{\alpha,i}\longmapsto \mc{O}_{\alpha,i+1},
    \end{cases}.
\end{equation}
In order to twist the boundary condition with a symmetry $\ms g\in\ms G$, we can couple the theory to a background $\ms G$ gauge field $\mbs g$ with holonomy $\ms g$ in the spatial direction corresponding to the insertion of $\mcal U_{\ms g}$ along time. We can think of the background gauge field as distributing group elements on the links of the lattice, $\mbs g = (\ms g_{1,2},\cdots, \ms g_{L,1})$, where $\ms g_{i,i+1}\in \ms G$ is the value of the gauge field on the link $(i,i+1)$. The holonomy is defined as
\begin{equation}\label{eq: holonomy of non-abelian background gauge field}
    \tenofo{hol}(\mbs{g})\equiv \prod_{i}\ms{g}_{i-1,i}=\ms{g}.
\end{equation}
One way to couple the theory to this gauge field is to introduce the modified symmetry-twisted translation operator
\begin{align}
    \ms{T}_{\ms{g}}\equiv \left(\prod_{i=1}^{L}\mc{O}_{\ms{g}_{i-1,i},i}\right)\ms{T}.
\end{align}
Replacing the translation operator in equation \eqref{eq:Non-abelian G spin-chain} we find
\begin{equation}
    \begin{aligned}
        H(\mbs{g}) &=\sum_{i=1}^L\left[\sum_{\alpha\in\tenofo{Rep}(\ms G)}J_\alpha\,\mathcal{O}_{\alpha,i}^\dagger\cdot\ms T_{\ms g}\mathcal{O}_{\alpha,i}\ms T_{\ms g}^{-1}  + \sum_{\ms g'\in\ms G} h_{\ms g'}\,\mathcal O_{\ms g',i}\right] +\tenofo{h.c.},\\
        &=\sum_{i=1}^L\left[\sum_{\alpha\in\tenofo{Rep}(\ms G)}J_\alpha\,\mathcal{O}_{\alpha,i}^\dagger\cdot\ms{R}_{\alpha}(\ms g_{i,i+1})\cdot\mathcal{O}_{\alpha,i+1} + \sum_{\ms g'\in\ms G} h_{\ms g'}\,\mathcal O_{\ms g',i}\right] +\tenofo{h.c.},
    \end{aligned}
\end{equation}
where again $\cdot$ corresponds to matrix product. The appearance of $\ms{R}_{\alpha}(\ms g_{i,i+1})$ in the Hamiltonian above is essentially a Peierls substitution.

\smallskip Note that the above construction has an ambiguity, the choice of $\mbs g$ such that \eqref{eq: holonomy of non-abelian background gauge field} holds is not unique. However, any other choice is gauge equivalent and corresponds to a topological deformation of the symmetry operator $\mcal U_{\ms g}$ along time. In particular, consider the redefinition of charged operators
\begin{equation}
    \mathcal{O}_{\alpha,i}\longmapsto \ms{R}_{\alpha}(\lambda_i)\cdot \mathcal{O}_{\alpha,i}, \qquad \lambda_i\in\ms G.
\end{equation}
This is essentially a local gauge transformation corresponding to $\mbs\lambda = (\lambda_1, \dots, \lambda_L)$. Under this redefinition, the Hamiltonian changes as 
\begin{equation}
    H(\mbs{g})\longmapsto H(\mbs{\lambda}^{-1}\mbs{g}\mbs{\lambda})=\sum_{i,\alpha}\mathcal{O}_{\alpha,i}^\dagger\cdot\ms{R}_{\alpha}(\ms \lambda_i^{-1}g_{i,i+1}\lambda_{i+1})\cdot \mathcal{O}_{\alpha,i+1}+\tenofo{h.c.},
\end{equation}
where $\mbs{\lambda}^{-1}\mbs{g}\mbs{\lambda}=(\lambda_1^{-1}g_{1,2}\lambda_2,\cdots,\lambda_L^{-1}g_{L,1}\lambda_1)$. Clearly, the new transformed background gauge field has the same holonomy
\begin{equation}
    \tenofo{hol}(\mbs{\lambda}^{-1}\mbs{g}\mbs{\lambda})=\tenofo{hol}(\mbs{g})=\ms g.
\end{equation}
Therefore, different choices of $\mbs{g}$ correspond to redefinition of local operators and give rise to unitary-equivalent Hamiltonians
\begin{equation}
    U_{\mbs{\lambda}}^\dagger H(\mbs{g})U_{\mbs{\lambda}}=H(\mbs{\lambda}^{-1}\mbs{g}\mbs{\lambda}),
\end{equation}
where the unitary transformation is given by
\begin{equation}
    U_{\mbs{\lambda}}\equiv \prod_{i}\mcal{O}_{\lambda_i,i}.
\end{equation}
By construction, the Hamiltonian $H(\mbs{g})$ commutes with the symmetry-twisted translation operator, which satisfies
\begin{align}
    \ms{T}_{\ms{g}}^{L}=\mc{U}_{\ms{g}}.
\end{align}
If $\mc{U}_{\ms{g}}$ generates an order-$n$ subgroup of $\ms{G}$, i.e. $\mc{U}_{\ms{g}}^{n}=1$, such a symmetry-twist mathematically corresponds to extending the group $\mathbb Z_{L}$ of translations (generated by $\ms{T}$) by the finite group $\mathbb Z_{n}$ generated by $\mc{U}_\ms{g}$. Practically, this has the implication that the momentum eigenvalues are quantized as $nL^{\text{th}}$ roots of unity. It can be immediately checked that  local operators satisfy symmetry-twisted conditions
\begin{equation}\label{eq:local operators transformation under non-abelian symmetry defect}
    \begin{aligned}
    \mc{O}_{\alpha,i+L}&=\ms{T}_{\ms{g}}^{L}  \mc{O}_{\alpha,i}\ms{T}^{-L}_{\ms{g}} = \ms R_\alpha(\ms{g})\cdot \mc{O}_{\alpha,i},
        \\
    \mc{O}_{\ms{h},i+L}&=\ms{T}_{\ms{g}}^{L}  \mc{O}_{\ms{h},i}\ms{T}^{-L}_{\ms{g}} = \mc{O}_{\ms{h},i},
    \end{aligned}.
\end{equation}
In summary, coupling a spin Hamiltonian with global symmetry $\ms G$ to a background $\ms G$ gauge field $\mbs{g}$ with holonomy $\tenofo{hol}(\mbs{g})=\ms{g}$ in the spatial direction corresponds to the insertion of the symmetry operator $\mcal{U}_{\ms g}$ along the time direction. The ambiguity in the choice of background field $\mbs{g}\mapsto\mbs{\lambda}^{-1}\mbs{g}\mbs{\lambda}$ corresponds to topological deformations of the symmetry operator and has no physical effect. Moving a local operator a full cycle along the spatial direction, it will inevitably cross the symmetry operator and transform as in \eqref{eq:local operators transformation under non-abelian symmetry defect}. Therefore, this procedure will give rise to the symmetry-twisted Hamiltonian $H_{\ms g}$.

\section{Simplicial calculus}
\label{sec:simplicial calculus}
In this work, we often encountered expressions of the form
\begin{equation}
    \bigintssss_M A\cup A^\vee,
\end{equation}
for some gauge fields $A$ and $A^\vee$, some space $M$ and $\cup$ denotes the so-called cup product. We would like to compute these expressions explicitly. We thus need an analog of the theory of integration of differential forms on a differentiable manifold. For triangulated spaces, whose definition we explain below, there is a simplicial analog of the theory of differential forms. In this appendix, we briefly review this simplicial calculus. Standard references are \cite{Munkres2018,Hatcher2002}.

\subsection{Simplicial complexes}
\begin{figure}
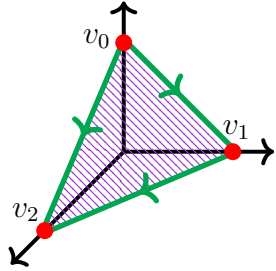

    \centering
    \ExampleOfTwoSimplex
    \caption{An example of a $2$-simplex in $\mbb{R}^3$. The vertices (in red), i.e. $0$-simplices, are denotes as $[v_0]=v_0$, $[v_1]=v_1$, and $[v_2]=v_2$. The $1$-simplices (in green) are $[v_0,v_1]=v_0-v_1$, $[v_1,v_2]=v_1-v_2$, and $[v_0,v_2]=v_0-v_2$. Finally, the only $2$-simplex (in velvet) is $[v_0,v_1,v_2]$.}
    \label{fig:the example simplices in R3}
\end{figure}

Let $\{v_0,\cdots,v_{n}\}$ be a set of geometrically-independent points in $\mbb{R}^N$. This means that if there exists a set of real scalars $\{t_a\}$ such that
\begin{equation}
    \sum_{a=0}^nt_a=0, \qquad \sum_{a=0}^nt_av_a=0,
\end{equation}
imply $t_0=\cdots=t_n=0$. The $n$-simplex $\Delta$ defined by $\{v_0,\cdots,v_{n}\}$ is the set of all points $x\in\mbb{R}^N$ such that
\begin{equation}
    x=\sum_{a=0}^nt_av_a, \qquad \sum_{a=0}^n t_a=1,\qquad \forall t_a\ge 0. 
\end{equation}
The points $v_0,\cdots,v_n$ are called the vertices of $\Delta$ and $n$ is its dimension. For a simplex $\Delta$ defined by vertices $\{v_0,\cdots,v_n\}$, we can define an ordering of its vertices. Two orderings are equivalent if they are different by an even number of permutation of vertices. An oriented simplex is denoted as $[v_0,\cdots,v_n]$. We use the convention that a the orientation of a given $n$-simplex is given by the orientation of its edges in increasing order. An example of simplices in $\mbb{R}^3$ and their orientations are shown in Figure \ref{fig:the example simplices in R3}. We sometimes denote an $n$-simplex by $\Delta^n$, i.e.
\begin{equation}
    \Delta^n\equiv [v_0,\cdots,v_n].
\end{equation}

\smallskip Simplices generated by all proper subsets\footnote{A proper subset of a set is any subset which is not the set itself.} of vertices in $\{v_0,\cdots,v_n\}$ are collectively called subsimplices. Any simplex spanned by a subset of $\{v_0,\cdots,v_n\}$ obtained by removing only one of the vertices is called a face of $\Delta$. The union of all faces is called the boundary of $\Delta$ and it is denoted as $\partial\Delta$. Accordingly, one can define the interior of an $n$-simplex by $\tenofo{Int}(\Delta)=\Delta-\partial\Delta$. 

\begin{figure}
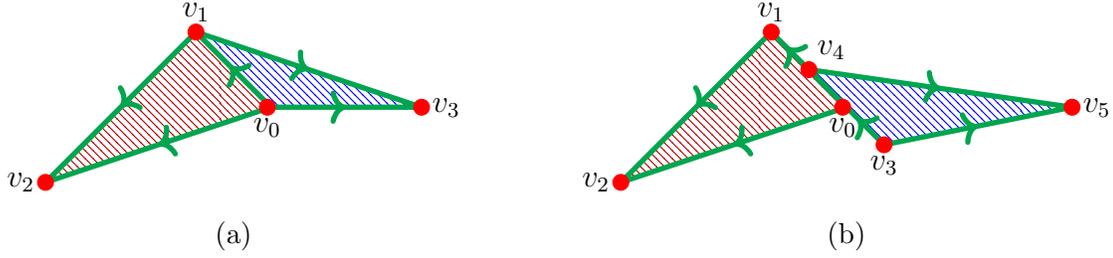

    \centering
    \begin{subfigure}{.45\textwidth}\centering 
    \ProperSimplicialComplex
    \subcaption{}
    \label{subfig:an example of a simplicial complex}
    \end{subfigure}
    ~
    \begin{subfigure}{.45\textwidth}\centering 
    \NonProperSimplicialComplex
    \subcaption{}
    \label{subfig:an example of a nonproper simplicial complex}
    \end{subfigure}
    \caption{(a) This shows a proper simplicial complex consists of two $2$-simplices (filled in red and blue), their edges (in green) and their vertices (in red). The intersection of these $2$-simplices is the edge of either of them and hence belong to the simplicial complex; (b) This shows two $2$-simplices that do not form a simplicial complex. The intersection of them is a subset of the edge of either of them. Hence, the union of the two triangles do not define a simplicial complex.}
    \label{fig:examples of a proper and nonproper simplicial complex}
\end{figure}

\smallskip A simplicial complex $\mcal{S}$ in $\mbb{R}^N$ is a collection of simplices in $\mbb{R}^N$ with two properties: 1) every subsimplex of a simplex $\Delta\in\mcal{S}$ also belongs to $\mcal{S}$; and 2) every intersection of two simplices $\Delta,\Delta'\in\mcal{S}$ is a face of each of them. An example of a simplicial complex (\ref{subfig:an example of a simplicial complex}) and an example which is not a simplicial complex (\ref{subfig:an example of a nonproper simplicial complex}) is shown in Figure \ref{fig:examples of a proper and nonproper simplicial complex} The dimension of a simplicial complex $\mcal{S}$ is simply the maximum dimension of the simplices in $\mcal{S}$.

\begin{figure}
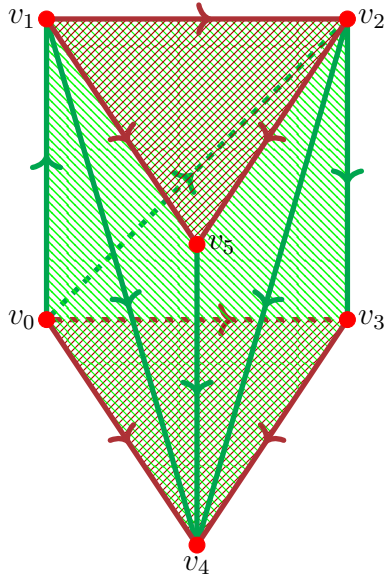

    \centering
    \SimplicialComplexForACylinder
    \caption{A simplicial complex consisting of six $2$-simplices, their faces (edges) and vertices. The underlying space of the top and bottom triangles are circles and the underlying space of the whole simplicial complex is (homeomorphic to) a cylinder. To study simplicial aspects of cylinder, one can instead study similar aspects of this simplicial complex.}
    \label{fig:the simplicial complex whose underlying space is homeomorphic to a cylinder}
\end{figure}

\smallskip An important result\footnote{To make this precise, one has to introduce the notion of cell-complex structure on a space and identify the space as a quotient of disjoint simplices in this cell-complex structure by certain equivalence relation. Here, and to avoid the mathematical details, we will just state the final result.} is that any (reasonably-nice-behaved) space $M$ can be built from $n$-simplices inductively by starting from vertices, and attach edges to them to make a graph, and then attach $2$-simplices, and so on. To make this more precise, let $\mcal{S}$ be a simplicial complex in $\mbb{R}^N$ and let $|\mcal{S}|$ denote the subset of $\mbb{R}^N$ consists of the union of simplices in $\mcal{S}$. This subset admits a natural topology: a subset $S\subset|\mcal{S}|$ is closed if and only if $S\cap\Delta$ is closed for all simplices $\Delta\in\mcal{S}$. Such sets are closed under finite union and arbitrary intersections, and hence define a topology on $|\mcal{S}|$. The subset $|\mcal{S}|$ endowed with this topology is called the {\it underlying space} of $\mcal{S}$ or its {\it polytope}. A map $\phi:|\mcal{S}|\to M$ from $|\mcal{S}|$ to a topological space $M$ is continuous if its restriction $f|_\Delta$ to each $\Delta\in\mcal{S}$ is continuous. Equipped with these definition, we see that the polytope of a simplicial complex can be homeomorphic\footnote{A homeomorphism between two topological spaces is a continuous bijective map with a continuous inverse.} to a topological space. Finally, notice that there might be many simplicial complexes which are homeomorphic to a given topological space $M$. Alternatively, one can choose different homemorphisms from a given polytope to a homeomorphic topological space. A choice of such homeomorphism is called a {\it triangulation} of $M$. After choosing such a homeomorphism, $M$ is called a triangulated space. The study of simplicial aspects of $M$ (simplicial homology and cohomology groups and cup product, etc) can thus be reduced to the study of the simplicial complex underlying its triangulation. We use this correspondence in the following, and often use the simplicial complex $\mcal{S}$ and the space $M$ for which $\mcal{S}$ provides a triangulation interchangeably. For example, the simplicial homology and cohomology groups of a simplicial complex are actually the simplicial homology and cohomology of its underlying space, which is homeomorphic to some topological space. As an example, consider the simplicial complex in Figure \ref{fig:the simplicial complex whose underlying space is homeomorphic to a cylinder} whose underlying space is a cylinder. 

\subsection{Simplicial homology groups}
For a simplicial complex $\mcal{S}$, a $p$-chain is a function $c$ from the set of oriented simplices in $\mcal{S}$ into integers (or more generally any Abelian group $G$) satisfying 1) for two oppositely-oriented simplices $\Delta$ and $\Delta'$, we have $c(\Delta)=-c(\Delta')$; 2) $c(\Delta)=0$ for all but finitely-many oriented $p$-simplices $\Delta$. The set of all $p$-chains naturally forms an Abelian group under addition, which will be denoted as $C_p(\mcal{S},G)$, which is a free Abelian group. For $p<0$ and $p>\dim(K)$, this is a trivial group. One can define the boundary operator $\partial_p$ as the following homomorphism $\partial_p:C_p(\mcal{S},G)\to C_{p-1}(\mcal{S},G)$, which for an oriented $n$-simplex $\Delta=[v_0,\cdots,v_n]$ is defined as
\begin{equation}\label{eq:the definition of boundary operator}
    \partial_p\Delta\equiv \sum_{a=0}^p(-1)^a[v_0,\cdots,\wh{v}_a,\cdots,v_n],
\end{equation}
and $[v_0,\cdots,\wh{v}_a,\cdots,v_n]$ means that the vertex $v_a$ is removed from the set of vertices. The boundary operator has two important properties: 1) For the oppositely-oriented simplex $-\Delta$ of a simplicial complex $\Delta$, we have $\partial_p(-\Delta)=-\partial_p(\Delta)$; 2) it satisfies
\begin{equation}\label{eq:the important property of boundary operator}
    \partial_{p-1}\circ\partial_p=0.
\end{equation}
As an example, consider a $2$-simplex $[v_0,v_1,v_2]$ for which we have $\partial_2[v_0,v_1,v_2]=[v_1,v_2]-[v_0,v_2]+[v_0,v_1]$. The boundary operation defined here is the analog of the boundary operation in the singular homology theory in the context of integration of differential forms on Riemannian manifolds.

\smallskip \eqref{eq:the important property of boundary operator} makes it clear how to define homology group of a simplicial complex. They can be defined by considering the boundary operators $\partial_p:C_p(\mcal{S},G)\to C_{p-1}(\mcal{S},G)$ and $\partial_{p+1}:C_{p+1}(\mcal{S},G)\to C_{p}(\mcal{S},G)$. The kernel of the first is called the group of $p$-cycles $Z_p(\mcal{S},G)$, and the image of the second is called the group of $p$-boundaries $B_p(\mcal{S},G)$. The second property above guarantees that $B_p(\mcal{S},G)\subset Z_p(\mcal{S},G)$. One can then define the $p$\textsuperscript{th} homology group of $\mcal{S}$ by
\begin{equation}
    H_p(\mcal{S},G)\equiv Z_p(\mcal{S},G)/B_p(\mcal{S},G).
\end{equation}
As we have explained above, these are the homology groups of a topological space $M$ which is isomorphic to the underlying space $|\mcal{S}|$ of $\mcal{S}$. In the following, we thus use the notations $C_p(M,G)$ and $C_p(\mcal{S},G)$, $Z_p(M,G)$ and $Z_p(\mcal{S},G)$, $B_p(M,G)$ and $B_p(\mcal{S},G)$, and $H_p(M,G)$ and $H_p(\mcal{S},G)$ interchangeably. 

\smallskip {\bf Example (homology groups of a point with integer coefficients):} Let us consider the simplest possible example, i.e. when $M$ is just a point and $G\simeq\mbb{Z}$. It is clear that $H_p(M,\mbb{Z})=0, \forall p\ge 1$ since there are not $p$-chains with $p\ge 1$ in $M$. On the other hand, $H_0(M,\mbb{Z})$ is the free Abelian group generated by the single point of $M$ and hence $H_0(M,\mbb{Z})\simeq\mbb{Z}$. 

\subsection{Simplicial cohomology groups}
In this section, we define the cohomology groups of simplicial complexes. As it will become clear, these are dual, in some precise sense, to the homology groups. The are  the analog of de Rham cohomology groups in the context of differential forms. 

\smallskip To this end, we consider functions on set of simplicial complexes. For a simplicial complex $\mcal{S}$ the group $C^p(\mcal{S},G)$ of $p$-cochains of $\mcal{S}$ with coefficients in an Abelian group $G$ is defined to be
\begin{equation}\label{eq:the definition of p-cochains}
    C^p(\mcal{S},G)\equiv \tenofo{Hom}(C_p(\mcal{S},G),G),
\end{equation}
where $\tenofo{Hom}(C_p(\mcal{S},G),G)$ denotes the Abelian group  of all homomorphism of $C_p(\mcal{S},G)$ into $G$.\footnote{Note that $C_p(\mcal{S},G)$ are Abelian groups and hence one can define the Abelian group $\tenofo{Hom}(C_p(\mcal{S},G)$ as the group of all homomorphisms between two Abelian groups. The addition in $\tenofo{Hom}(C_p(\mcal{S},G)$ is defined as follows: the addition of two homomorphisms $f_1,f_2\in \tenofo{Hom}(C_p(\mcal{S},G)$ is defined by the addition of their values in $G$.} The value of a $p$-cochain $\varphi^p\in C^p(\mcal{S},G)$ on an oriented $p$-chain $[v_0,\cdots,v_p]\in C_p(\mcal{S},G)$ is denoted by a pairing notation  $\langle\cdot,\cdot\rangle:C^p(\mcal{S},G)\times C_p(\mcal{S},G)\to G$ as
\begin{equation}
 \langle\varphi^p,[v_0,\cdots,v_p]\rangle\in G. 
\end{equation}
To be able to define cohomology groups, one needs to have a notion of a differential on complexes of cochains. In the context of simplicial calculus, this is called the coboundary operator. It is a map $\delta_p:C^p(\mcal{S},G)\to C^{p+1}(\mcal{S},G)$ and it is defined as being the dual of the boundary operator in the following sense
\begin{equation}\label{eq:the definition of coboundary operator}
    \langle \delta_p\varphi^p,[v_0,\cdots,v_{p+1}]\rangle\equiv \langle \varphi^p,\partial_p [v_0,\cdots,v_{p+1}]\rangle. 
\end{equation}
 Using \eqref{eq:the definition of boundary operator} for the boundary operator $\partial_p$, we have
\begin{equation}
    \langle\delta \varphi^p,[v_0,\cdots,v_{p+1}]\rangle=\sum_{a=0}^{p+1}(-1)^a\langle \varphi^p,[v_0,\cdots,\wh{v}_a,\cdots,v_{p+1}]\rangle. 
\end{equation}
Using this formula, one can show the important property 
\begin{equation}
    \delta_{p+1}\circ\delta_p=0.
\end{equation}
This property, which is the dual version of \eqref{eq:the important property of boundary operator}, then makes it clear how to define the cohomology group of simplicial complexes. The kernel of $\delta_p:C^p(\mcal{S},G)\to C^{p+1}(\mcal{S},G)$ is the Abelian group of $p$-cocycles $Z^p(\mcal{S},G)$ and the image of $\delta_p:C^p(\mcal{S},G)\to C^{p+1}(\mcal{S},G)$ is the Abelian group of $p$-coboundaries $B^p(\mcal{S},G)$. The $p$\textsuperscript{th} simplicial cohomology group of a simplicial complex $\mcal{S}$ is defined as
\begin{equation}
    H^p(\mcal{S},G)\equiv Z^p(\mcal{S},G)/B^p(\mcal{S},G).
\end{equation}
The coefficients of all these groups are in $G$, as the notation implies. Since boundary operation $\partial$ and coboundary operation $\delta$ are dual in the sense of \eqref{eq:the definition of coboundary operator}, the simplicial homology and cohomology groups are dual to each other. Finally, as we have explained above, notice that  these cohomology groups are the cohomology groups of a topological space $M$ which is isomorphic to the underlying space $|\mcal{S}|$ of $\mcal{S}$, i.e. we have $H^p(M,G)\simeq H^p(\mcal{S},G)$. In the following, we thus use the notations $C^p(M,G)$ and $C^p(\mcal{S},G)$, $Z^p(M,G)$ and $Z^p(\mcal{S},G)$, $B^p(M,G)$ and $B^p(\mcal{S},G)$, and $H^p(M,G)$ and $H^p(\mcal{S},G)$ interchangeably. 

\smallskip {\bf Example (cohomology groups of a point with integer coefficient):} Let us continue our simple example, i.e. when $M$ is a point and $G\simeq\mbb{Z}$. As we have explained in the previous section, $H_p(M,\mbb{Z})$ vanishes for $p\ge 1$ and $H_0(M,\mbb{Z})\simeq \mbb{Z}$. On the other hand, \eqref{eq:the definition of p-cochains} and the fact that all homology groups are free Abelian tell us that
\begin{equation}
    H^p(M,\mbb{Z})\simeq\tenofo{Hom}(H_p(M,\mbb{Z}),\mbb{Z})
    =
    \begin{cases}
    \tenofo{Hom}(\mbb{Z},\mbb{Z})\simeq\mbb{Z}, &\qquad p=0,
    \\
    \tenofo{Hom}(\mbb{Z},0)\simeq 0, &\qquad p\ne 0.
    \end{cases}
\end{equation}

\subsection{Cup product}
As we have seen in previous section, the theory of simplicial complexes is in direct analogy with the theory of singular complexes, which is used for a rigorous treatment of the theory of differential forms on a manifold culminated in The de Rham Theorem. Therefore, we expect that there should be an analog of wedge product. We will now explain what this operation is. One of the main differences between homology and cohomology is that the latter can be endowed with a natural product. The analog of the wedge product in the simplicial context is called {\it cup product} which turns $H^\bullet\equiv\bigoplus_p H^p(M,R)$ into a ring called the cohomology ring of $M$. 

\smallskip The cup product is defined by considering cohomology with coefficients in a commutative ring $R$. We are mostly interested in the case that $R=\prod_a\mbb{Z}_{N_a}$. For cochains $\varphi^p\in C^p(\mcal{S},R)$ and $\varphi^q\in C^q(\mcal{S},R)$, the cup product $\varphi^p\cup \varphi^q$ defines an element of $C^{p+q}(\mcal{S},R)$ such that whose value on a $(p+q)$-simplex $\Delta=[v_0,\cdots,v_{p+q}]$ is given by
\begin{equation}\label{eq:defining relation of the cup product}
    \langle \varphi^p\cup \varphi^q,\Delta\rangle\equiv \langle \varphi^p,[v_0,\cdots,v_p]\rangle\bullet\langle \varphi^q,[v_p,\cdots,v_{p+q}]\rangle, 
\end{equation}
where $\bullet$ denotes the product operation on the ring $R$. The cup product has some important properties; It is associative, i.e. for $\varphi^{p_i}\in C^{p_i}(\mcal{S},G),\;i=1,2,3$, we have
\begin{equation}
    (\varphi^{p_1}\cup\varphi^{p_1})\cup\varphi^{p_3}=\varphi^{p_1}\cup(\varphi^{p_1}\cup\varphi^{p_3}),
\end{equation}
which can be easily verified using \eqref{eq:defining relation of the cup product}. Furthermore, it is distributive with respect to the addition in the ring $R$. We also have
\begin{equation}
    \delta(\varphi^{p_1}\cup\varphi^{p_2})=\delta\varphi^{p_1}\cup\varphi^{p_2}+(-1)^{p_1}\varphi^{p_1}\cup\delta\varphi^{p_2}.
\end{equation}
Note that the operator $\delta$ reduces to $\delta_{p_i}$ defined in \eqref{eq:the definition of coboundary operator} when acting on $C^{p_i}(\mcal{S},R)$. 

\smallskip Up to now, we have define the cup product for $p$-cochains. However, it can be seen that this cup product induces one at the level of cohomologies. This can be seen by noting that $p$-cocyles are in particular $p$-cochains so one can consider the cup product of cocycles. On the other hand, the cup product of a cocycle and a coboundary is a coboundary, which can be easily seen by noting that $\varphi\cup\delta\varphi'=\delta(\varphi\cup\varphi')$ for $\delta\varphi=0$. We thus see that cup product is a well-defined operation at the level of cohomology, i.e.
\begin{equation}\label{eq:the map that cup product defines in cohomology ring of a space}
    \cup:H^{p_1}(\mcal{S},R)\times H^{p_2}(\mcal{S},R)\longrightarrow H^{p_1+p_2}(\mcal{S},R). 
\end{equation}
Taking into its properties stated above (associativity, distributive with respect to addition in $R$), we see that it indeed endows $\bigoplus_p H^p(M,R)$ with a ring\footnote{Actually cup product endows $H^\bullet$ with the structure of an $R$-algebra and hence an $R$-module.} structure. 

\smallskip {\bf Example (cohomology ring of a point with integer coefficients):} To complete our very basic example, i.e. the case of that $M$ is a point and $G\simeq\mbb{Z}$, let us compute the cohomology ring of this space. We have seen that the cohomology groups of a point vanishes except for $p=0$, which is isomorphic to $\mbb{Z}$. Due to \eqref{eq:the map that cup product defines in cohomology ring of a space}, we thus see that $H^{p_1+p_2}$ vanishes except for $p_1+p_2=0$, which implies $p_1=p_2=0$. Consider $\varphi^0_1, \varphi_2^0\in H^0(M,\mbb{Z})$. They satisfy
\begin{equation}
    \delta\varphi^0_1=\delta\varphi^0_2=0,
\end{equation}
They are thus both constant functions. The value of their cup product $\varphi^0_1\cup\varphi^0_2$ is the product of their values. This is the complete description of the cohomology ring of a point. 

\smallskip In many situation, we are interested in integrating a cup product over a triangulated space $M$. For this purpose, it is enough to know how to compute the integral on the cohomology-ring $H^\bullet(\mcal{S},R)$ of a simplicial complex $\mcal{S}$, as follows. Let $\varphi^{p_i}\in C^{p_i}(\mcal{S},R),\;i=1,\cdots,N$ be a set of $p_i$-cochains. The integral of their cup product over a simplex
\begin{equation}
    \Delta^{p}\equiv \Delta^{p_1+\cdots+p_N}=[v_0,\cdots,v_{p_1+\cdots+p_N}],
\end{equation}
is defined by
\begin{equation}
    \bigintssss_{\Delta^{p}}\varphi^{p_1}\cup\cdots\cup\varphi^{p_N}=\langle\varphi^{p_1}\cup\cdots\cup\varphi^{p_N},\Delta^{p}\rangle. 
\end{equation}
Using \eqref{eq:defining relation of the cup product}, we have
\begin{equation}
    \bigintssss_{\Delta^{p}}\varphi^{p_1}\cup\cdots\cup\varphi^{p_N}=\prod_{a=1}^N\langle  \varphi^{p_a},[v_{p_1+\cdots+p_{a-1}},\cdots,v_{p_1+\cdots+p_{a}}]\rangle,\qquad p_{0}\equiv 0,
\end{equation}
where the product is taken in the ring $R$. Using this result, we can define the cup products of cochains on a given space $M$ which is homeormophic to the underlying space of a simplicial complex $\mcal{S}$. The integral of such cochains are defined by
\begin{equation}\label{eq:the integral of cup products over a triangulated space}
    \bigintssss_M\varphi^{p_1}\cup\cdots\cup\varphi^{p_N}\equiv\sum_{\Delta^{p}\in\mcal{S}}\bigintssss_{\Delta^{p}}\varphi^{p_1}\cup\cdots\cup\varphi^{p_N}
\end{equation}
where the sum is taken over all oriented $p$-simplices. 

\smallskip Let us now get back to the question we posed in the beginning of this appendix, i.e. the explicit computation of expressions of the form
\begin{equation}
    \bigintssss_M A\cup A^\vee,
\end{equation}
for some gauge fields $A$ and $A^\vee$ and some space $M$. This expression can be calculated by what we have explained so far. Since gauge fields are $1$-cochains (and indeed $1$-cocycles) with values $\mbb{Z}_{N}$ (or product of such Abelian groups), $A\cup A^\vee$ is a $2$-cochain. Therefore, it has to be computed on a $2$-chain. Using the $2$-simplex in Figure \ref{fig:the example simplices in R3} and denoting it as $\Delta^2$, we have
\begin{equation}
    \begin{aligned}
    \bigintssss_{\Delta^2}A\cup A^\vee&=\langle A\cup A^\vee,[v_0,v_1,v_2]\rangle
    \\
    &=\langle A,[v_0,v_1]\rangle\langle A^\vee,[v_1,v_2]\rangle,
    \end{aligned}
\end{equation}
where $\langle A,[v_0,v_1]\rangle$ is just the $\mbb{Z}_N$-value of the gauge field on the oriented edge $[v_0,v_1]$ (also denotes as $A_{v_0v_1}$ in the main text) and the same for $\langle A^\vee,[v_1,v_2]\rangle$. Using this basic result, we can now compute the integral of $A\cup A^\vee$ over a triangulated space $M$ using \eqref{eq:the integral of cup products over a triangulated space}.

\section{Group structure of the web of dualities}
\label{sec:duality groupoid}
In this appendix, we collect the computations that verify the group structure at the level of transformation of twisted partition functions for the duality group of $\ms G$-symmetric systems when $\ms G$ is $\mbb Z_{2}\times \mbb Z_{2}$.

\smallskip We first consider the case of $\mbb Z_{2}\times \mbb Z_{2}$.
Recall, that the duality group as described in Sec.~\ref{Subsec:Z2Z2_example} is
\begin{align}
    \mc G[\mbb Z_{2}\times \mbb Z_{2}]=\left\langle \sigma_L, k_1, k_2, k_3 \right\rangle= (S_3\times S_3)\rtimes\mathbb Z_2.
\end{align}
Concretely, $k_2$ and $k_1k_3$ generate one of the $S_{3}$ groups such that $k_2^2=\text{id}$, $(k_1k_3)^{3}=\text{id}$ and $k_2(k_1k_3)k_2^{-1}=(k_1k_3)^{-1}$.
which $\sigma_{L}$ acts on by conjugation to give the other $S_3$ group, which is generated by $h_2$ and $h_1h_3$, where 
\begin{gather}\label{eq:Z2xZ2 anyonic symmetries in partition function appendix}
    \begin{aligned}
    \sigma_L(m_L,m_R; e_L, e_R) &= (e_L, m_R; m_L, e_R), \\
    \sigma_R(m_L,m_R; e_L, e_R) &= (m_L, e_R; e_L, m_R), \\
    h_1(m_L,m_R; e_L, e_R) &= (m_L, m_Lm_R; e_Le_R, e_R), \\
    h_2(m_L,m_R; e_L, e_R) &= (m_Lm_R, m_R; e_L, e_Le_R), \\
    h_3(m_L,m_R; e_L, e_R) &= (m_R, m_L; e_R, e_L), \\
    k_1(m_L,m_R; e_L, e_R) &= (m_Le_R, m_Re_L; e_L, e_R), \\
    k_2(m_L,m_R; e_L, e_R) &= (m_L, m_R; e_Lm_R, e_Rm_L), \\
    k_3(m_L,m_R; e_L, e_R) &= (e_R, e_L; m_R, m_L). 
    \end{aligned}
\end{gather}
At the level of twisted partition functions, these transformations take the form
\begin{gather}
\begin{aligned}
k_1:& \mc Z[A_L,A_R] \mapsto  \mc Z[A_L,A_R]e^{i\pi \int_{\Sigma}A_{L}\cup A_{R}}, \\ 
k_2:& \mc Z[A_L,A_R] \mapsto \mc Z^{\vee} [A_{L}^{\vee},A_{R}^{\vee}]=\frac{1}{\sqrt{|H^{1}(\Sigma,\mbb Z_2^2)|}}\sum_{A_{L},A_{R}}\mc Z[A_L,A_R]e^{i\pi \int_{\Sigma}(A_{L}^{\vee}-A_L)\cup(A_{R}^{\vee} - A_R)}, \\ 
k_3:& \mc Z[A_L,A_R] \mapsto \mc Z^{\vee} [A_{L}^{\vee},A_{R}^{\vee}]=
\frac{1}{\sqrt{|H^{1}(\Sigma,\mbb Z^2_2)|}}\sum_{A_L,A_R}\mc Z[A_L,A_R]e^{i\pi \int_{\Sigma}A_{L}\cup A_{R}^{\vee}+A_{R}\cup A_{L}^{\vee}}, \\
\sigma_L:& \mc Z[A_L,A_R] \mapsto \mc Z^{\vee} [A_{L}^{\vee},A_{R}]=
\frac{1}{\sqrt{|H^{1}(\Sigma,\mbb Z_2)|}}\sum_{A_L}\mc Z[A_L,A_R]e^{i\pi \int_{\Sigma}A_{L}\cup A_{L}^{\vee}}.
\label{eq:Z2Z2_duality_group_cup_product_exps_app}
\end{aligned}
\end{gather}
These can for example be computed using the anyonic symmetries with the formulas \eqref{eq:Partition funcitions related by eta factor} and \eqref{eq: eta formula for partition function dualities}. Here, we demonstrate that these transformations satisfy the group composition rules in $\mc G[\mbb Z_2\times \mbb Z_2]$.
In particular we verify the following relations (1) $k_{1}^2 =\text{id}$, (2) $k_{2}^2 =\text{id}$, (3) $k_{3}^2 =\text{id}$, (4) $\sigma_L^2 =\text{id}$, (5) $(k_1k_3)^3=\text{id}$ and (6) $k_2(k_1k_3)k_2^{-1}=(k_1k_3)^{-1}$.
For simplicity we consider $\Sigma=T^2$ in the formulas below such that $|H^{1}(\Sigma, \ms G)|=|\ms G|^2$.

\begin{enumerate}
    \item \underline{$k_1^2=\text{id}$}: From \eqref{eq:Z2Z2_duality_group_cup_product_exps_app}, twisted partition transform as follows under $k_1^2$
    \begin{equation}
        k_{1}^2:\mc Z[A_{L},A_{R}]\longmapsto \mc Z^{\vee}[A_L,A_{R}]=\mc Z[A_L,A_{R}](-1)^{\int_{\Sigma}2 A_{L}\cup A_{R}} =\mc Z[A_L,A_{R}].
    \end{equation}
    Which verifies $k_{1}^2 =\text{id}$. 
    Physically, $k_{1}$ corresponds to pasting a $\mbb Z_{2}\times \mbb Z_{2}$ SPT, and therefore $k_1^2$ corresponds to pasting two copies of the $\mbb Z_{2}\times \mbb Z_{2}$ SPT.
    Since $\mbb Z_{2}\times \mbb Z_{2}$ SPTs are classified by $\mbb Z_2$, this process is adiabatically connected to not pasting anything at all. 
    \item \underline{$k_2^2=\text{id}$}: Implementing the $k_2$ transformation in \eqref{eq:Z2Z2_duality_group_cup_product_exps_app} twice successively, 
    \begin{equation}
        k_{2}^2: \mc Z[A_L,A_R] \longmapsto \mc Z^{\vee\vee}[A_L^{\vee\vee},A_R^{\vee\vee}],
    \end{equation}
    where
    \begin{gather}
    \begin{aligned}
    \mc Z^{\vee\vee}[A_L^{\vee\vee},A_R^{\vee\vee}]=& \frac{1}{|\ms G|^2}\sum_{A_{L,R},A_{L,R}^{\vee}}\mc Z[A_L,A_R](-1)^{\int_{\Sigma}\left[(A_L^{\vee}-A_L)\cup (A_R^{\vee}-A_R)+(A_L^{\vee\vee}-A^{\vee}_L)\cup (A_R^{\vee\vee}-A^{\vee}_R)\right]} \\
    =&\; \sum_{A_L,A_R}\mc Z[A_L,A_R](-1)^{\int_{\Sigma}[A_{L}\cup A_{R}+A_{L}^{\vee\vee}\cup A_{R}^{\vee\vee}]}\delta_{A_L,A_L^{\vee\vee}}\delta_{A_R,A_R^{\vee\vee}} \\
    =&\; \mc Z[A_{L}^{\vee\vee},A_{R}^{\vee\vee}].
    \end{aligned}
    \end{gather}
    In going to the second line, we have summed over $A_L^{\vee}$ and $A_{R}^{\vee}$\footnote{We use the equation
    \begin{equation}
        \frac{1}{|\ms G|}\sum_{A_{R}^{\vee}}(-1)^{\int_{\Sigma} A_R^{\vee}\cup \left(A+\dots\right)}=\delta(A+\dots).
    \end{equation}}
    \item \underline{$k_3^2=\text{id}$}: Implementing the $k_3$ transformation in \eqref{eq:Z2Z2_duality_group_cup_product_exps_app} twice successively, 
    \begin{equation}
        k_{3}^2: \mc Z[A_L,A_R] \longmapsto \mc Z^{\vee\vee}[A_L^{\vee\vee},A_R^{\vee\vee}],
    \end{equation}
    where
    \begin{gather}
    \begin{aligned}
    \mc Z^{\vee\vee}[A_L^{\vee\vee},A_R^{\vee\vee}]=&
    \frac{1}{|\ms G|^2}\sum_{A_{L,R},A_{L,R}^{\vee}}\mc Z[A_L,A_R]
    (-1)^{\int_{\Sigma}\left[
    A_L\cup A_{R}^{\vee}+A_R\cup A_{L}^{\vee}
    +
    A^{\vee}_L\cup A_{R}^{\vee\vee}+A_R^{\vee}\cup A_{L}^{\vee\vee}
    \right]}. \\
    \end{aligned}
    \end{gather}
    Summing over $A_{L}^{\vee}$ and $A_{R}^{\vee}$, imposes $A_{L}=-A_L^{\vee\vee}$ and $A_{R}=-A_{R}^{\vee}$, which implies 
    \begin{equation}
        \mc Z^{\vee\vee}[A_L^{\vee\vee},A_R^{\vee\vee}]=\mc Z[A_L^{\vee\vee},A_R^{\vee\vee}],
    \end{equation}
    i.e., $k_{3}^2=\text{id}$.
    \item \underline{$\sigma_L^2=\text{id}$}: The duality operation $\sigma_{L}$, corresponds to Kramer's Wannier duality, i.e., gauging the global symmetry corresponding to the background $A_{L}$.
    Performing this operation twice furnishes
    \begin{equation}
        \sigma_{L}^2: \mc Z[A_L,A_R] \longmapsto \mc Z^{\vee\vee}[A_L^{\vee\vee},A_R],
    \end{equation}
    where
    \begin{gather}
    \begin{aligned}
    \mc Z^{\vee\vee}[A_L^{\vee\vee},A_R]=&
    \frac{1}{|\ms G|}\sum_{A_L,A_L^{\vee}}\mc Z[A_L,A_R]
    (-1)^{\int_{\Sigma}\left[
    A_L\cup A_{L}^{\vee}+A_L^{\vee}\cup A_{L}^{\vee\vee}
    \right]}\\
    =&\;\sum_{A_L}\mc Z[A_L^{\vee\vee},A_R]\delta_{A_L,A_L^{\vee\vee}} \\
    =&\; Z[A_L^{\vee\vee},A_R]. 
    \end{aligned}
    \end{gather}
    \item \underline{$(k_1k_3)^3=\text{id}$}: 
    Acting with the $k_1k_3$ duality, transforms the twisted partition function as
    \begin{equation}
        \mc Z[A_{L},A_{R}]\xrightarrow{k_1k_3} 
        \mc Z^{\vee}[A_L^{\vee},A_R^{\vee}]=\frac{1}{|\ms G|}\sum_{A_L,A_R}\mc Z[A_L,A_R](-1)^{\int_{\Sigma}A_L\cup A_R^{\vee}+A_R\cup A_{L}^{\vee}+ A_{L}^{\vee}\cup A_R^{\vee}}.
    \end{equation}
    Acting with $k_1k_3$ again produces the partition function
\begin{gather}
\begin{aligned}
    \mc Z^{\vee\vee}[A_L^{\vee\vee},A_{R}^{\vee\vee}]=&\frac{1}{|\ms G|^2}\sum_{A_{L,R},A_{L,R}^{\vee}}\mc Z[A_L,A_R] \\
    & \times (-1)^{\int_{\Sigma}
    \left[
    A_L\cup A_{R}^{\vee}+A_{R}\cup A_L^{\vee}+A_{L}^{\vee}\cup A_R^{\vee}
    +
    A^{\vee}_L\cup A_{R}^{\vee\vee}+A^\vee_{R}\cup A_L^{\vee\vee}+A_{L}^{\vee\vee}\cup A_R^{\vee\vee}
    \right]}.
\end{aligned}    
\end{gather}
This expression can be simplified by summing over $A_L^{\vee}$ and $A_{R}^{\vee}$.
Doing so, we obtain
\begin{gather}
\begin{aligned}
\label{eq:k1k3square}
\mc Z^{\vee\vee}[A_L^{\vee\vee},A_{R}^{\vee\vee}]=&\frac{1}{|\ms G|}\sum_{A_L,A_R}\mc Z[A_L,A_R](-1)^{\int_{\Sigma}\left[A_{R}\cup A_{L}^{\vee\vee}-A_R \cup A_L -A_L \cup A_R^{\vee\vee}\right]}.
\end{aligned}
\end{gather}
Finally acting with the $k_1k_3$ duality a third time gives 
\begin{gather}
\begin{aligned}
\mc Z^{\vee\vee\vee}[A_L^{\vee\vee\vee},A_R^{\vee\vee\vee}]=&\frac{1}{|\ms G|^2}\sum_{A_{L,R},A_{L,R}^{\vee\vee}}\mc Z[A_L,A_R] \\
& \times (-1)^{\int_{\Sigma}
 \left[
 A_R\cup A_{L}^{\vee\vee}
-A_{R}\cup A_{L}
-A_{L}\cup A_{R}^{\vee\vee}
+A_L^{\vee\vee}\cup A_{R}^{\vee\vee\vee}
+ A_R^{\vee\vee}\cup A_{L}^{\vee\vee\vee}
+A_L^{\vee\vee\vee}\cup A_{R}^{\vee\vee\vee}
 \right]}
\end{aligned}
\end{gather}
summing over $A_{L}^{\vee\vee}$ and $A_{R}^{\vee\vee}$ impose the constraints $A_{R}=A_{R}^{\vee\vee\vee}$ and $A_{L}=A_{L}^{\vee\vee\vee}$ respectively, therefore we get.
\begin{gather}
\begin{aligned}
\mc Z^{\vee\vee\vee}[A_L^{\vee\vee\vee},A_R^{\vee\vee\vee}]=& \sum_{A_L,A_R}\mc Z[A_L,A_R]\delta_{A_{R},A_{R}^{\vee\vee\vee}}
\delta_{A_{L},A_{L}^{\vee\vee\vee}}
(-1)^{\int_{\Sigma}\left[-A_R\cup A_L + A_{L}^{\vee\vee\vee}\cup A_R^{\vee\vee\vee}\right]} \\
=& \mc Z[A_{L}^{\vee\vee\vee},A_R^{\vee\vee\vee}].
\end{aligned}
\end{gather}
Thus establishing that $(k_1k_3)^3=\text{id}$.
\item \underline{$k_2(k_1k_3)^2k_2^{-1}=(k_1k_3)^{-2}$}: This equation translates to 
\begin{equation}
    k_2(k_1k_3)^2k_2(k_1k_3)^{2}=\text{id},
\end{equation}
As a first step, we need to obtain how the twisted partition functions transform under $k_2(k_1k_3)^2$.
Using \eqref{eq:k1k3square} and \eqref{eq:Z2Z2_duality_group_cup_product_exps_app}, following the same steps as above, it can be shown that
\begin{equation}
    k_2(k_1k_3)^2:\mc Z[A_L,A_R]\longmapsto \mc Z^{\vee}[A_L^\vee, A_R^{\vee}]=\frac{1}{|\ms G|}\sum_{A_L,A_R}\mc Z[A_L,A_R](-1)^{\int_{\Sigma}\left[A_{L}\cup A_{R}^{\vee}+ A_{L}^{\vee}\cup A_R \right]}.
\end{equation}
Acting with the $k_2(k_1k_3)^2$ duality again gives
\begin{gather}
\begin{aligned}
\mc Z^{\vee\vee}[A_{L}^{\vee\vee},A_R^{\vee\vee}]=& \frac{1}{|\ms G|^2}\sum_{A_{L,R},A^{\vee}_{L,R}}\mc Z[A_L,A_R](-1)^{\int_{\Sigma}\left[
A_L \cup A_R^{\vee} + A_{L}^{\vee} \cup A_R
+
A^{\vee}_L \cup A_R^{\vee\vee} + A{L}^{\vee\vee} \cup A^{\vee}_R
\right]} \\
=& \mc Z[A_L^{\vee\vee},A_R^{\vee\vee}].
\end{aligned}
\end{gather}
\end{enumerate}
We have thus established that the transformation \eqref{eq:Z2Z2_duality_group_cup_product_exps_app}, form a representation of $\mc G[\mbb Z_2\times \mbb Z_{2}]$.

\section{Conformal spectra of various phase transitions}
\label{sec:conformal spectra at criticality}
In this appendix, we collect conformal spectra of various critical points for some of the quantum spin chains studied in Section \ref{sec:examples}. 

\subsection{Brief recap of $\mbb{Z}_N$ parafermion CFT}
Let us briefly recap some of the main features of the CFT describing the critical point between ferromagnetic (fully-broken) and paramagnetic (unbroken) phases of $N$-state Potts model, i.e. the $\mbb{Z}_N$ parafermion CFT \cite{GepnerQiu1987}.  This is a diagonal RCFT described by the coset model $\mfk{su}(2)_N/\mfk{u}(1)_{2N}$ with central charge 
\begin{equation}
    c=\frac{2(N-1)}{N+2}.
\end{equation}
The primary states of the theory are labeled by a pair of integers $(l,m)$ whose range is as follows: $l=0,\cdots,N$ labels $\mfk{su}(2)_N$ primary fields and $-l+2\le m\le l$ labels the primaries of $\mfk{u}(1)_{2N}$. Furthermore, these integers must satisfy the constraint $l=m\mod 2$. From this, one can conclude that there are $N(N+1)/2$ primary fields, which we denote as $\varphi\indices{_{l,m}}$. The conformal dimensions of these fields are given by
\begin{equation}
    h\indices{_{lm}}=\bar{h}\indices{_{lm}}=\frac{l(l+2)}{4(N+2)}-\frac{m^2}{4N}.
\end{equation}
We also have the following identification of primary fields
\begin{equation}
    \varphi_{l,m}=\varphi_{N-l,m-N}=\varphi_{N-l,m+N}. 
\end{equation}
The components of modular $T$-matrix of the theory are given
\begin{equation}\label{eq:components of the modular T-matrix of ZN parafermion CFT}
    T\indices{_{lm}^{l'm'}}=e^{2\pi\mfk{i}h_{lm}}\delta\indices{_l^{l'}}\delta\indices{_m^{m'}},
\end{equation}
while those of the modular $S$-matrix are
\begin{equation}\label{eq:components of the modular S-matrix of ZN parafermion CFT}
    S\indices{_{lm}^{l'm'}}=\frac{2}{\sqrt{N(N+2)}}\sin\left[\frac{\pi(l+1)(l'+1)}{N+2}\right]e^{\pi\mfk{i}mm'/N}.
\end{equation}
We are often interested in knowing twisted partition functions for $\mbb{Z}_N$ parafermion CFT in various symmetry sectors and with twisted boundary conditions (see \eqref{eq:the definition of generalized twisted partition function}) in terms of Virasoro characters. The twisted partition functions in the sector $(\ms g,\ms h)$ in terms of Virasoro characters $\chi_{l,m}(\tau)$ is given by \cite{GepnerQiu1987}
\begin{equation}\label{eq:generalized twisted partition functions of parafermion CFT}
    \mathcal{Z}_{\msg,\msh}=\sum_{l,m}e^{2\pi \mfk{i}\frac{\ms h(m-\ms g)}{N}}
    \chi_{l,m}\bar{\chi}_{l,m}. 
\end{equation}
The modular group $\Gamma\simeq\tenofo{PSL}(2,\mbb{Z})$ is generated by $T:\tau\mapsto \tau+1$ and $S:\tau\mapsto-\frac{1}{\tau}$. Under a modular transformation
\begin{equation}
    \tau\longmapsto \frac{a\tau+b}{c\tau+d}, \qquad
    \begin{pmatrix}
    a & b
    \\
    c & d
    \end{pmatrix}\in\Gamma,
\end{equation}
partition functions \eqref{eq:generalized twisted partition functions of parafermion CFT} transform as
\begin{equation}
    \mathcal{Z}(\msg,\msh)\longmapsto \mathcal{Z}(a\msg+b\msh,c\ms g+d\ms h). 
\end{equation}

\subsection{Some Results for $\ms G=\mbb Z_2\times \mbb Z_{2}$ transitions}

In this section, we collect numerical data showing the agreement between theoretically predicted conformal spectrum and numerical data for several quantum phase transitions realized in $\mbb Z_2\times \mbb Z_2$ symmetric quantum spin chains.

\subsubsection{Some basic facts about Ising CFT}

To compute the spectra analytically, let us first write down the modular $T$- and $S$-matrices for Ising CFT. Using \eqref{eq:components of the modular T-matrix of ZN parafermion CFT} and \eqref{eq:components of the modular S-matrix of ZN parafermion CFT} and ordering the three primary fields of the model as $\{\varphi_{0,0}\equiv\mbb{I},\varphi_{1,1}\equiv\sigma,\varphi_{2,0}\equiv\psi\}$, whose conformal dimension are $(h_{lm},\bar{h}_{lm})=\{(0,0),(\frac{1}{16},\frac{1}{16}),(\frac{1}{2},\frac{1}{2})\}$, we see that these matrices are given by
\begin{equation}
    T^{\tenofo{Ising}}=\begin{pmatrix}
    1 & 0 & 0
    \\
    0 & e^{\frac{\pi\mfk{i}}{8}} & 0
    \\
    0 & 0 & 1
    \end{pmatrix}, \qquad 
    S^{\tenofo{Ising}}=\frac{1}{2}\begin{pmatrix}
    1 & \sqrt{2} & 1
    \\
    \sqrt{2} & 0 & -\sqrt{2}
    \\
    1 & -\sqrt{2} & 1
    \end{pmatrix}, \qquad 
\end{equation}
The spectrum of CFT is tabulated in Table \ref{tab:primary fields of Ising cft}.
\begin{table}[H]\centering
    \begin{tabular}{c c c c} \toprule
    \multirow{2}{2cm}{Primary fields} & \multirow{2}{2cm}{$\mbb{I}$} &
    \multirow{2}{2cm}{$\epsilon$} &
    \multirow{2}{2cm}{$\sigma$} 
    \\\\
    \midrule
    $(h,\bar{h})$ & $(0,0)$ & $(\frac{1}{2},\frac{1}{2})$ & $(\frac{1}{16},\frac{1}{16})$
    \\
    \bottomrule
    \end{tabular}
    \caption{Primary fields of the Ising CFT describing the phase transition between ferromagnet and paramagnetic phases of Ising chain.}
    \label{tab:primary fields of Ising cft}
\end{table}

\subsubsection{Analytical and numerical results}

The theoretical predictions are made using a known transition and various duality mappings.
The known transition is in the Ising ${}^2$ universality class and is realized between the minimal realizations of either the fully symmetric paramagnet ($\mc L_6$) and the symmetry broken phase ($\mc L_1$) or between two partial symmetry broken phases ($\mc L_2$ and $\mc L_3$).
More precisely in the Hamiltonians $H_{16}$ and $H_{23}$ as described in \eqref{eq:H_{ij}} which realize the transitions $\mscr{C}_{16}$ and $\mscr{C}_{23}$.
\begin{enumerate}
    \item The transition $\mscr{C}_{15}$ between the symmetry broken phase and the the SPT described by the Lagrangian subgroups $\mc L_{1}$ and $\mc L_5$ respectively can be obtained from the the transition $\mscr{C}_{16}$ using the duality transformation $k_1$. The conformal spectrum in the different symmetry eigensectors $\alpha\in \text{Rep}(\mbb Z_2 \times \mbb Z_2) $ and with different symmetry twisted boundary conditions $\ms g\in \mbb Z_2\times \mbb Z_2$ are given in Table.~\ref{Tab:conformal spectra C_15}. The comparison between the numerically obtained spectrum and the analytically predicted conformal spectrum can be found in Figure \ref{fig:ED_C15}.
\begin{table}[t!]
\centering
\begin{tabular}{c c } \toprule
{Symmetry sector $(\ms g,\alpha)$} & {Conformal spectrum} 
\\
\midrule
$1=(0,0;0,0)$ & $(|\chi_0|^{2}+|\chi_{1/2}|^{2})^2$ \\ 
$m_L=(1,0;0,0)$ & $(|\chi_{1/16}|^{2})^2$ \\ 
$m_R=(0,1;0,0)$ & $(|\chi_{1/16}|^{2})^2$ \\ 
$m_Lm_R=(1,1;0,0)$ & $(\chi_0\bar{\chi}_{1/2}+\bar{\chi}_0\chi_{1/2})^2$ \\ 
$e_L=(0,0;1,0)$ & $(|\chi_0|^{2}+|\chi_{1/2}|^{2})|\chi_{1/16}|^2$ \\ 
$m_Le_L=(1,0;1,0)$ & $(\chi_0\bar{\chi}_{1/2}+\bar{\chi}_0\chi_{1/2})|\chi_{1/16}|^2$ \\ 
$m_Re_L=(0,1;1,0)$ & $(|\chi_0|^{2}+|\chi_{1/2}|^{2})|\chi_{1/16}|^2$ \\ 
$m_Rm_Le_L=(0,1;1,0)$ & $(\chi_0\bar{\chi_{1/2}}+\bar{\chi}_0\chi_{1/2})|\chi_{1/16}|^2$ \\
$e_R=(0,0;0,1)$ & $(|\chi_0|^{2}+|\chi_{1/2}|^{2})|\chi_{1/16}|^2$ \\ 
$m_Le_R=(1,0;0,1)$ & $(|\chi_0|^{2}+|\chi_{1/2}|^{2})|\chi_{1/16}|^2$ \\ 
$m_Re_R=(1,0;0,1)$ & $(\chi_0\bar{\chi}_{1/2}+\bar{\chi}_0\chi_{1/2})|\chi_{1/16}|^2$ \\ 
$m_Lm_Re_R=(1,1;0,1)$ & $(\chi_0\bar{\chi}_{1/2}+\bar{\chi}_0\chi_{1/2})|\chi_{1/16}|^2$ \\ $e_Le_R=(0,0;1,1)$ & $(|\chi_{1/16}|^2)^2$ \\ 
$m_Le_Le_R=(1,0;1,1)$ & $(\chi_0\bar{\chi}_{1/2}+\bar{\chi}_0\chi_{1/2})(|\chi_0|^{2}+|\chi_{1/2}|^{2})$ \\ 
$m_Re_Le_R=(0,1;1,1)$ & $(\chi_0\bar{\chi}_{1/2}+\bar{\chi}_0\chi_{1/2})(|\chi_0|^{2}+|\chi_{1/2}|^{2})$ \\ 
$m_Lm_Re_Le_R=(1,1;1,1)$ & $(|\chi_{1/16}|^{2})^2$ \\ 
\bottomrule
\end{tabular}
\caption{Conformal spectrum for the critical point $\mscr{C}_{15}$ realized as the minimal transition between the spontaneous symmetry breaking phase (labelled by $\mc L_1$) and the symmetry protected topological phase (labelled by $\mc L_5$).}
\label{Tab:conformal spectra C_15}
\end{table}
\begin{figure}[H]
    \centering
    \includegraphics[width=1.0\textwidth]{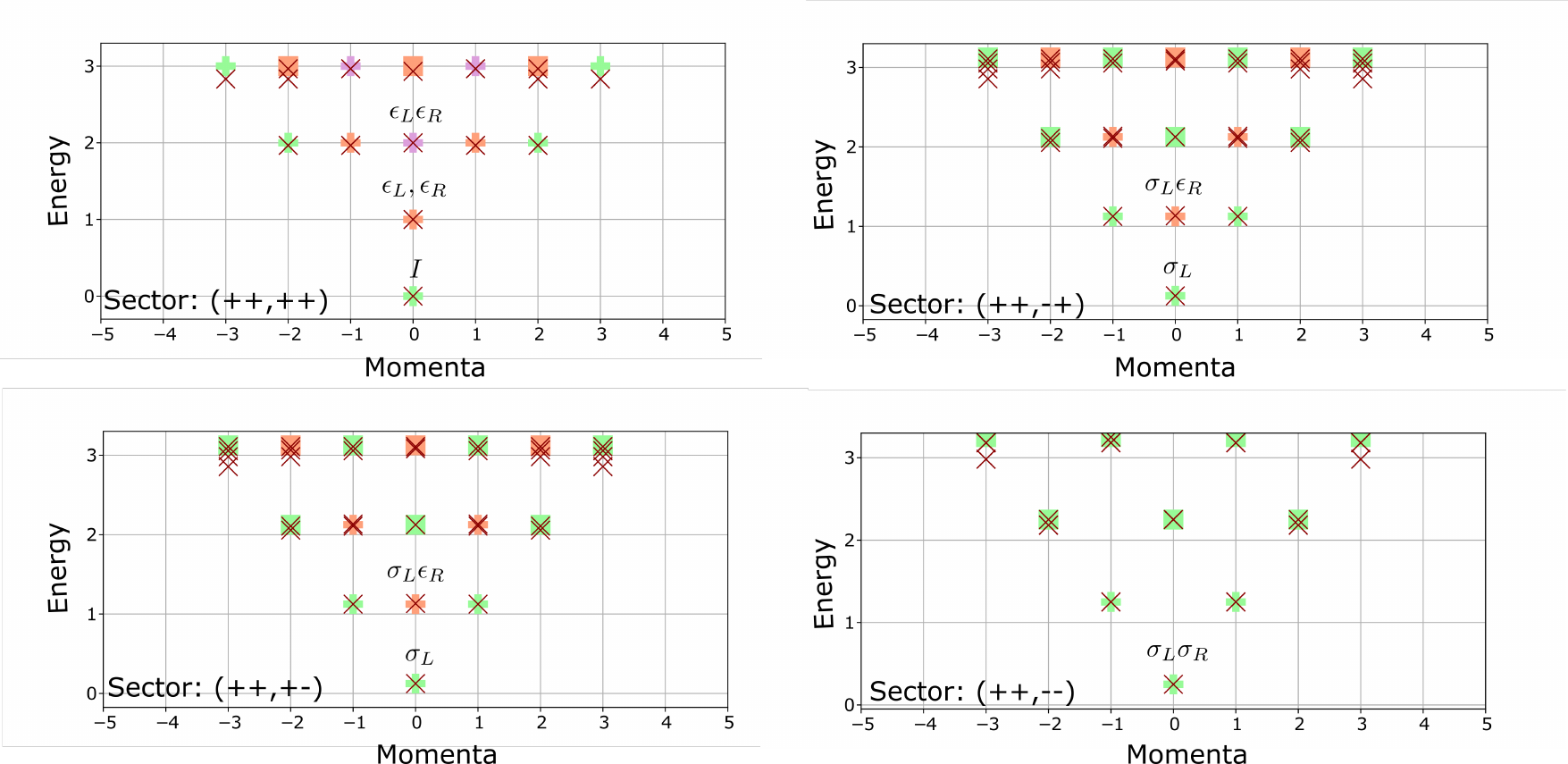}
    \caption{The figure shows shows a comparison between the analytically computed conformal spectrum of the transition $\mscr{C}_{15}$ and the numerically obtained spectrum for the Hamiltonian $H_{15}$ in \eqref{eq:H_{ij}} using exact diagonalization (depicted with crosses). 
    All the numerics have been performed for spin chains of length $L=12$, i.e., spin-chains containing $24$ $\mbb Z_2$-spins.
    The subplots (a)-(d) correspond to  $\mbb \times \mbb Z_{2}$ eigensectors $(++),(-+),(+-)$ and $(--)$ and with periodic boundary conditions.
    }
    \label{fig:ED_C15}
\end{figure}

\item The transition $\mscr{C}_{34}$ between the partial symmetry symmetry broken phases labelled by the Lagrangian subgroups $\mc L_{3}$ and $\mc L_4$ respectively can be obtained from the the transition $\mscr{C}_{23}$ using the duality transformation $h_2$. The conformal spectrum in the different symmetry eigensectors $\alpha\in \text{Rep}(\mbb Z_2 \times \mbb Z_2) $ and with different symmetry twisted boundary conditions $\ms g\in \mbb Z_2\times \mbb Z_2$ are given in Table.~\ref{Tab:conformal spectra C_34}. 
The comparison between the numerically obtained spectrum and the analytically predicted conformal spectrum can be found in Figure \ref{fig:exact diagonalization C34}.

\begin{table}[H]
\centering
\begin{tabular}{c c } \toprule
{Symmetry sector $(\ms g,\alpha)$} & {Conformal spectrum} 
\\
\midrule
$1=(0,0;0,0)$ &  $(|\chi_0|^{2}+|\chi_{1/2}|^{2})^2$  \\ 
$m_L=(1,0;0,0)$ &  $(|\chi_{1/16}|^2)^2$ \\ 
$m_R=(0,1;0,0)$ & $(|\chi_0|^{2}+|\chi_{1/2}|^{2})|\chi_{1/16}|^2$ \\ 
$m_Lm_R=(1,1;0,0)$ &  $(|\chi_0|^{2}+|\chi_{1/2}|^{2})|\chi_{1/16}|^2$ \\ 
$e_L=(0,0;1,0)$ &  $(|\chi_0|^{2}+|\chi_{1/2}|^{2})|\chi_{1/16}|^2$ \\ 
$m_Le_L=(1,0;1,0)$ & $(\chi_0\bar{\chi}_{1/2}+\bar{\chi}_0\chi_{1/2})|\chi_{1/16}|^2$ \\ 
$m_Re_L=(0,1;1,0)$ & $(|\chi_{1/16}|^2)^2$  \\ 
$m_Rm_Le_L=(0,1;1,0)$ & $(\chi_0\bar{\chi}_{1/2}+\bar{\chi}_0\chi_{1/2})(|\chi_0|^{2}+|\chi_{1/2}|^{2})$ \\ 
$e_R=(0,0;0,1)$ & $(|\chi_{1/16}|^2)^2$  \\ 
$m_Le_R=(1,0;0,1)$ &  $(\chi_0\bar{\chi}_{1/2}+\bar{\chi}_0\chi_{1/2})^2$ \\  
$m_Re_R=(1,0;0,1)$ &  $(\chi_0\bar{\chi}_{1/2}+\bar{\chi}_0\chi_{1/2}) |\chi_{1/16}|^2$ \\  
$m_Lm_Re_R=(1,1;0,1)$ & $(\chi_0\bar{\chi}_{1/2}+\bar{\chi}_0\chi_{1/2}) |\chi_{1/16}|^2$ \\  
$e_Le_R=(0,0;1,1)$ & $(|\chi_0|^{2}+|\chi_{1/2}|^{2})|\chi_{1/16}|^2$ \\  
$m_Le_Le_R=(1,0;1,1)$ &  $(\chi_0\bar{\chi}_{1/2}+\bar{\chi}_0\chi_{1/2}) |\chi_{1/16}|^2$  \\  
$m_Re_Le_R=(0,1;1,1)$ & $(\chi_0\bar{\chi}_{1/2}+\bar{\chi}_0\chi_{1/2})(|\chi_0|^{2}+|\chi_{1/2}|^{2})$  \\  
$m_Lm_Re_Le_R=(1,1;1,1)$ & $(|\chi_{1/16}|^2)^2$ \\  
\bottomrule
\end{tabular}
\caption{Conformal spectrum for the critical point $\mscr{C}_{34}$ realized as the minimal transition between the two partial symmetry breaking gapped phases labelled by $\mc L_{3}$ and $\mc L_4$.}
\label{Tab:conformal spectra C_34}
\end{table}
\begin{figure}[H]
    \centering
    \includegraphics[width=1.0\textwidth,scale=.9]{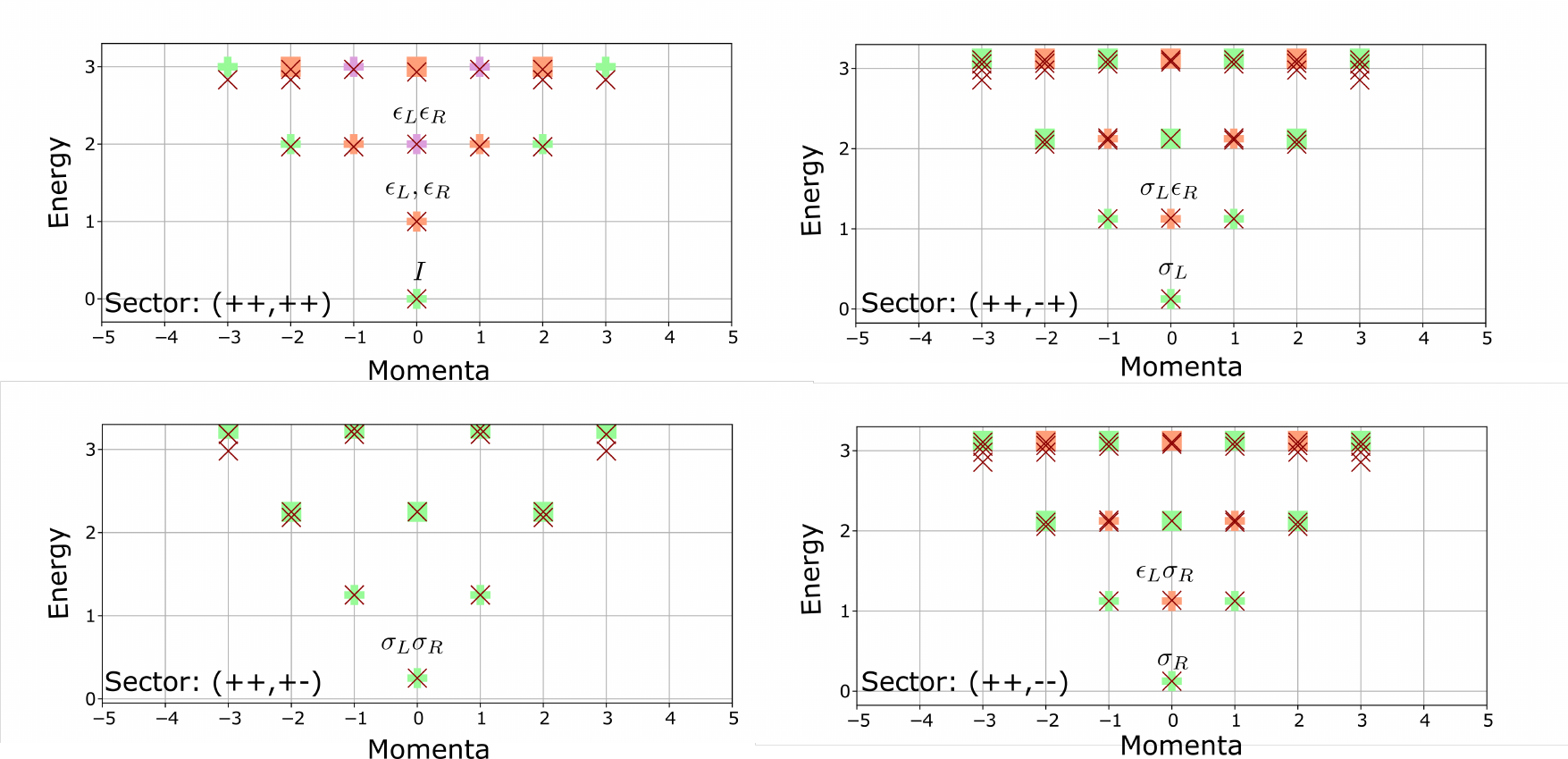}
    \caption{The figure shows shows a comparison between the analytically computed conformal spectrum of the transition $\mscr{C}_{34}$ and the numerically obtained spectrum for the Hamiltonian $H_{34}$ in \eqref{eq:H_{ij}} using exact diagonalization (depicted with crosses). 
    All the numerics have been performed for spin chains of length $L=12$.
    The subplots (a)-(d) correspond to  $\mbb \times \mbb Z_{2}$ eigensectors $(++),(-+),(+-)$ and $(--)$ and with periodic boundary conditions.
    }
    \label{fig:exact diagonalization C34}
\end{figure}

\item The transition $\mscr{C}_{24}$ between the partial symmetry symmetry broken phases labelled by the Lagrangian subgroups $\mc L_{2}$ and $\mc L_4$ respectively can be obtained from the the transition $\mscr{C}_{23}$ using the duality transformation $h_1$. The conformal spectrum in the different symmetry eigensectors $\alpha\in \text{Rep}(\mbb Z_2 \times \mbb Z_2) $ and with different symmetry twisted boundary conditions $\ms g\in \mbb Z_2\times \mbb Z_2$ are given in Table.~\ref{Tab:conformal spectra C_24}.
The comparison between the numerically obtained spectrum and the analytically predicted conformal spectrum can be found in Figure \ref{fig:ED_C24}.

\begin{table}[H]
\centering
\begin{tabular}{c c } \toprule
{Symmetry sector $(\ms g,\alpha)$} & {Conformal spectrum} 
\\
\midrule
$1=(0,0;0,0)$ &   $(|\chi_0|^{2}+|\chi_{1/2}|^{2})^2$  \\ 
$m_L=(1,0;0,0)$ & $(|\chi_0|^{2}+|\chi_{1/2}|^{2})|\chi_{1/16}|^2$  \\  
$m_R=(0,1;0,0)$ & $(|\chi_{1/16}|^2)^2$   \\  
$m_Lm_R=(1,1;0,0)$ & $(|\chi_0|^{2}+|\chi_{1/2}|^{2})|\chi_{1/16}|^2$ \\ 
$e_L=(0,0;1,0)$ & $(|\chi_{1/16}|^2)^2$ \\ 
$m_Le_L=(1,0;1,0)$ &  $(\chi_0\bar{\chi_{1/2}}+\bar{\chi}_0\chi_{1/2})|\chi_{1/16}|^2$ \\ 
$m_Re_L=(0,1;1,0)$ &  $(\chi_0\bar{\chi}_{1/2}+\bar{\chi}_0\chi_{1/2})^2$ \\ 
$m_Rm_Le_L=(0,1;1,0)$ & $(\chi_0\bar{\chi_{1/2}}+\bar{\chi}_0\chi_{1/2})|\chi_{1/16}|^2$  \\ 
$e_R=(0,0;0,1)$ &  $(|\chi_0|^{2}+|\chi_{1/2}|^{2})|\chi_{1/16}|^2$ \\ 
$m_Le_R=(1,0;0,1)$ & $(|\chi_{1/16}|^2)^2$  \\ 
$m_Re_R=(1,0;0,1)$ & $(\chi_0\bar{\chi_{1/2}}+\bar{\chi}_0\chi_{1/2})|\chi_{1/16}|^2$ \\ 
$m_Lm_Re_R=(1,1;0,1)$ & $(\chi_0\bar{\chi}_{1/2}+\bar{\chi}_0\chi_{1/2})(|\chi_0|^{2}+|\chi_{1/2}|^{2})$ \\ 
$e_Le_R=(0,0;1,1)$ &  $(|\chi_0|^{2}+|\chi_{1/2}|^{2})|\chi_{1/16}|^2$ \\ 
$m_Le_Le_R=(1,0;1,1)$ & $(\chi_0\bar{\chi}_{1/2}+\bar{\chi}_0\chi_{1/2})(|\chi_0|^{2}+|\chi_{1/2}|^{2})$ \\ 
$m_Re_Le_R=(0,1;1,1)$ &  $(\chi_0\bar{\chi_{1/2}}+\bar{\chi}_0\chi_{1/2})|\chi_{1/16}|^2$ \\ 
$m_Lm_Re_Le_R=(1,1;1,1)$ & $(|\chi_{1/16}|^2)^2$  \\ 
\bottomrule
\end{tabular}
\caption{Conformal spectrum for the critical point $\mscr{C}_{24}$ realized as the minimal transition between the two partial symmetry breaking gapped phases labelled by $\mc L_{2}$ and $\mc L_4$.}
\label{Tab:conformal spectra C_24}
\end{table}
\end{enumerate}

\begin{figure}[H]
    \centering
    \includegraphics[width=1.0\textwidth]{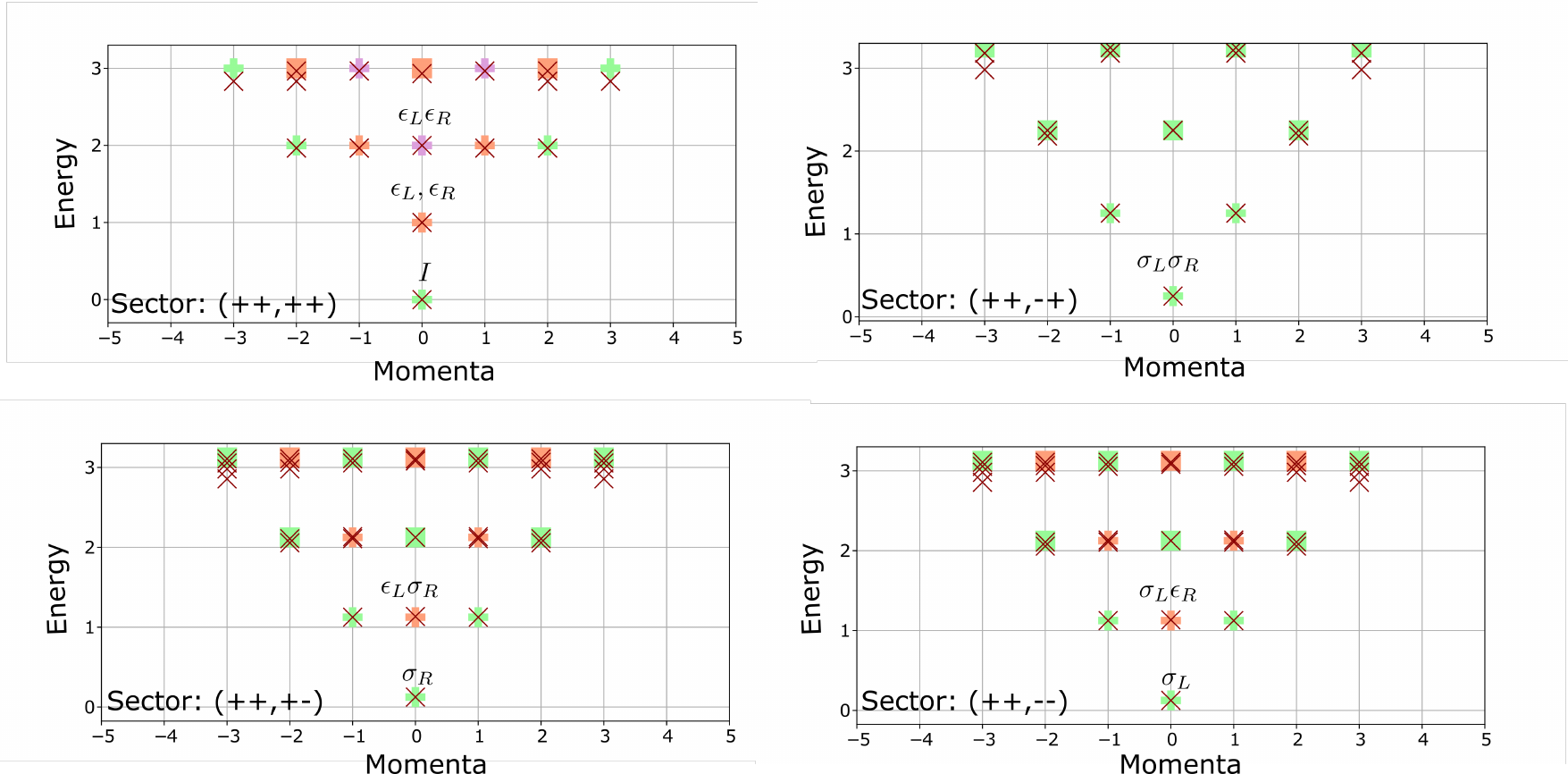}
    \caption{The figure shows shows a comparison between the analytically computed conformal spectrum of the transition $\mscr{C}_{24}$ and the numerically obtained spectrum for the Hamiltonian $H_{24}$ in \eqref{eq:H_{ij}} using exact diagonalization (depicted with crosses). 
    All the numerics have been performed for spin chains of length $L=12$, i.e., spin-chains containing $24$ $\mbb Z_2$-spins.
    The subplots (a)-(d) correspond to  $\mbb \times \mbb Z_{2}$ eigensectors $(++),(-+),(+-)$ and $(--)$ and with periodic boundary conditions.
    }
    \label{fig:ED_C24}
\end{figure}

Using \eqref{eq:PartitionFunctionInTermsOfCharacters}, we can compute the CFT partition functions in various twisted sectors. First consider the criticality $\mscr{C}_{15}$ (the phase transition between fully symmetry-broken phase $\mathcal{L}_1$ and SPT$_1$ phase $\mathcal{L}_5$). We have
\begin{equation}
    \begin{aligned}
    \mcal{Z}_{(1,0),(0,1)}(\mscr{C}_{15})&=|\chi_{\frac{1}{16}}|^4-|\chi_0|^2|\chi_{\frac{1}{16}}|^2-|\chi_{\frac{1}{2}}|^2|\chi_{\frac{1}{16}}|^2
    \\
    &+\left(|\chi_0|^2|+|\chi_\frac{1}{16}|^2-|\chi_{\frac{1}{2}}|^2\right)\chi_0\bar{\chi}_{\frac{1}{2}}+\left(|\chi_0|^2|+|\chi_\frac{1}{16}|^2-|\chi_{\frac{1}{2}}|^2\right)\bar\chi_0{\chi}_{\frac{1}{2}},
    \\
    \mcal{Z}_{(0,1),(0,1)}(\mscr{C}_{15})&=|\chi_{\frac{1}{16}}|^4+|\chi_0|^2|\chi_{\frac{1}{16}}|^2+|\chi_{\frac{1}{2}}|^2|\chi_{\frac{1}{16}}|^2
    \\
    &+\left(|\chi_0|^2|-|\chi_\frac{1}{16}|^2-|\chi_{\frac{1}{2}}|^2\right)\chi_0\bar{\chi}_{\frac{1}{2}}+\left(|\chi_0|^2|-|\chi_\frac{1}{16}|^2-|\chi_{\frac{1}{2}}|^2\right)\bar\chi_0{\chi}_{\frac{1}{2}},
    \\
    \mcal{Z}_{(1,1),(1,0)}(\mscr{C}_{15})&=2|\chi_0|^2|\chi_{\frac{1}{2}}|^2+\bar{\chi_0}^2\chi_{\frac{1}{2}}^2+\chi_0^2\bar{\chi}_{\frac{1}{2}}^2-|\chi_{\frac{1}{16}}|^4.
    \\
    \mcal{Z}_{(1,1),(1,0)}(\mscr{C}_{15})&=|\chi_{\frac{1}{16}}|^4-2|\chi_{\frac{1}{16}}|^2\left(\bar{\chi_0}\chi_{\frac{1}{2}}+\chi_0\bar{\chi}_{\frac{1}{2}}\right)+\chi_0^2\bar{\chi}_{\frac{1}{2}}^2+2|\chi_0|^2|\chi_{\frac{1}{2}}|^2+\bar{\chi}_0^2\chi_{\frac{1}{2}}^2
    \end{aligned}
\end{equation}
Similarly, for the $\mscr{C}_{56}$ criticality between SPT$_0$ and SPT$_1$,  we have
\begin{equation*}
    \begin{aligned}
    \mathcal{Z}_{(1,0),(0,0)}(\mscr{C}_{56})&=2\left(|\chi_0|^2+|\chi_{\frac{1}{2}}|^2+\chi_0\bar{\chi}_{\frac{1}{2}}+\bar{\chi}_0\chi_{\frac{1}{2}}\right)|\chi_{\frac{1}{16}}|^2,
    \\
    \mathcal{Z}_{(1,0),(1,0)}(\mscr{C}_{56})&=2|\chi_0|^2|\chi_{\frac{1}{16}}|^2-2\chi_0\bar{\chi}_{\frac{1}{2}}|\chi_{\frac{1}{16}}|^2-2\bar{\chi}_0\chi_{\frac{1}{2}}|\chi_{\frac{1}{16}}|^2+2|\chi_{\frac{1}{2}}|^2|\chi_{\frac{1}{16}}|^2,
    \\
    \mathcal{Z}_{(0,0),(1,0)}(\mscr{C}_{56})&=|\chi_0|^4-\chi_0^2\bar{\chi}_{\frac{1}{2}}^2-\bar\chi_0^2\chi_{\frac{1}{2}}^2+|\chi_{\frac{1}{2}}|^2,
    \\
    \mathcal{Z}_{(1,1),(1,1)}(\mscr{C}_{56})&=2|\chi_{\frac{1}{16}}|^4-2|\chi_0|^2\left(\chi_0\bar{\chi}_\frac{1}{2}+\bar{\chi}_0\chi_{\frac{1}{2}}\right)-2|\chi_{\frac{1}{2}}|^2\left(\chi_0\bar{\chi}_\frac{1}{2}+\bar{\chi}_0\chi_{\frac{1}{2}}\right)
    \end{aligned}
\end{equation*}

\subsection{Some results for $\ms G=\mbb Z_3\times \mbb Z_{3}$ transitions}
The primary states of the $\mbb{Z}_3$ parafermion CFT are tabulated in Table \ref{tab:primary fields of 3-state Potts model}.
\begin{table}[H]\centering
    \begin{tabular}{c c c c c c c} \toprule
    \multirow{2}{2cm}{Primary fields} & \multirow{2}{2cm}{$\mbb{I}$} &
    \multirow{2}{2cm}{$\psi\bar{\psi}$} &
    \multirow{2}{2cm}{$\psi^\dagger\bar{\psi}^\dagger$} &
    \multirow{2}{2cm}{$E=\epsilon\bar\epsilon$} &
    \multirow{2}{2cm}{$S=\sigma\bar{\sigma}$} &
    \multirow{2}{2cm}{$S^\dagger=\sigma^\dagger\bar{\sigma}^\dagger$} 
    \\\\
    \midrule
    $(h,\bar{h})$ & $(0,0)$ & $(\frac{2}{3},\frac{2}{3})$ & $(\frac{2}{3},\frac{2}{3})$ &
    $(\frac{2}{5},\frac{2}{5})$ &
    $(\frac{1}{15},\frac{1}{15})$ & $(\frac{1}{15},\frac{1}{15})$ 
    \\ 
    \bottomrule
    \end{tabular}
    \caption{Primary fields of the $\mbb{Z}_3$ parafermion CFT describing the phase transition between ferromagnet and paramagnetic phases of $3$-state Potts model}
    \label{tab:primary fields of 3-state Potts model}
\end{table}
The modular $T$- and $S$-matrices are given by
\begin{equation}
    T^{\mbb{Z}_3}=
    \begin{pmatrix}
    1 & 0 & 0 & 0 & 0 & 0
    \\
    0 & e^{\frac{-2\pi\mfk{i}}{3}} & 0 & 0 & 0 & 0
    \\
    0 & 0 & e^{\frac{-2\pi\mfk{i}}{3}} & 0 & 0 & 0
    \\
    0 & 0 & 0 & e^{\frac{4\pi\mfk{i}}{5}} & 0 & 0
    \\
    0 & 0 & 0 & 0 & e^{\frac{2\pi\mfk{i}}{15}} & 0
    \\
    0 & 0 & 0 & 0 & 0 & e^{\frac{2\pi\mfk{i}}{15}}
    \end{pmatrix}, 
\end{equation}
and ($\phi=\frac{1+\sqrt{5}}{2}$ and $\omega=\exp\left(\frac{2\pi\mfk{i}}{3}\right)$)
\begin{equation}
    S^{\mbb{Z}_3}=\frac{1}{\sqrt{6+2\phi}}
    \begin{pmatrix}
    1 & 1 & 1 & \phi & \phi & \phi
    \\
    1 & \omega^2 & \omega & \phi & \omega^2\phi & \omega\phi
    \\
    1 & \omega & \omega^2 & \phi & \omega\phi & \omega^2\phi
    \\
    \phi & \phi & \phi & -1 & -1 & -1
    \\
    \phi & \omega^2\phi & \omega\phi & -1 & -\omega^2 & -\omega 
    \\
    \phi & \omega\phi & \omega^2\phi & -1 & -\omega & -\omega^2 
    \end{pmatrix}
\end{equation}
A typical twisted partition functions for the SPT$_0$ and SPT$_1$ phases is
\begin{equation}
    \begin{aligned}
    \mcal{Z}_{(0,1),(0,0)}(\mscr{C}_{67})&=+|\chi_{\frac{1}{15}}|^4+|\chi_{\frac{2}{3}}|^4+\chi_0^2\bar{\chi}_{\frac{2}{3}}^2+\bar{\chi}_0^2\chi_{\frac{2}{3}}^2+\chi_{\frac{2}{5}}^2\bar{\chi}_{\frac{1}{15}}^2+\bar{\chi}_{\frac{2}{5}}^2\chi_{\frac{1}{15}}^2
    \\
    &\hphantom{=\,\,}+2\left(\chi_0\bar{\chi}_{\frac{2}{3}}+\bar{\chi}_0\chi_{\frac{2}{3}}+|\chi_{\frac{2}{3}}|^2+|\chi_{\frac{2}{5}}|^2+\chi_{\frac{2}{5}}\bar{\chi}_{\frac{1}{15}}+\bar{\chi}_{\frac{2}{5}}\chi_{\frac{1}{15}}\right)|\chi_{\frac{1}{15}}|^2
    \\
    &\hphantom{=\,\,}+2\left(\chi_0\bar{\chi}_{\frac{2}{3}}+\bar{\chi}_0\chi_{\frac{2}{3}}+\chi_{\frac{1}{15}}\bar{\chi}_{\frac{2}{5}}+\bar{\chi}_{\frac{1}{15}}\chi_{\frac{2}{5}}+|\chi_{\frac{2}{5}}|^2\right)|\chi_{\frac{2}{3}}|^2
    \\
    &\hphantom{=\,\,}+2\left(|\chi_{\frac{1}{15}}|^2+|\chi_{\frac{2}{3}}|^2\right)|\chi_0|^2+2\chi_0\bar{\chi}_{\frac{2}{3}}\chi_{\frac{2}{5}}\bar{\chi}_{\frac{1}{15}}+2\bar{\chi}_0\chi_{\frac{2}{3}}\bar{\chi}_{\frac{2}{5}}\chi_{\frac{1}{15}}.
    \end{aligned}
\end{equation}
Here, we have demonstrated the method for few simple example. Similarly, one can obtain the twisted partition functions in any sector for any finite Abelian group of the form \eqref{eq:the generic finite Abelian group} using \eqref{eq:the generic formula for twisted dual partition function}.

\clearpage

\bibliography{References.bib}
\bibliographystyle{JHEP}

\end{document}